\documentclass[a4paper,12pt]{article}
\usepackage{jheppub,esint,shuffle,psfrag}
\usepackage[utf8]{inputenc}
 \usepackage{epsfig,amssymb,amsmath,psfrag,subfigure,rotate,color,wasysym,xcolor}
\usepackage[bbgreekl]{mathbbol}
\usepackage{empheq}
\usepackage{array}
\usepackage{hyperref}

\usepackage{tikzit}

\tikzstyle{blue dot}=[scale=0.3, draw=blue, shape=circle, fill=blue]
\tikzstyle{textdot}=[shape=circle, scale=0.7]
\tikzstyle{textdot2}=[shape=circle, scale=0.4]
\tikzstyle{point}=[fill=black, draw=black, shape=circle, inner sep=0pt, minimum size=3pt]
\tikzstyle{lr_point}=[fill=red, draw=red, shape=circle, inner sep=0pt, minimum size=2pt]
\tikzstyle{sh_point}=[fill={rgb,255: red,128; green,128; blue,128}, draw={rgb,255: red,128; green,128; blue,128}, shape=circle, inner sep=0pt, minimum size=3pt]
\tikzstyle{sh_lr_point}=[fill={red!40}, draw={red!40}, shape=circle, inner sep=0pt, minimum size=2pt, tikzit fill={rgb,255: red,255; green,128; blue,0}, tikzit draw={rgb,255: red,255; green,128; blue,0}]
\tikzstyle{blue_point}=[fill=blue, draw=blue, shape=circle, inner sep=0pt, minimum size=3pt]
\tikzstyle{circle dot}=[fill=white, draw=black, shape=circle, dashed, inner sep=0pt, minimum size=120pt]

\tikzstyle{new edge style 0}=[-, dashed]
\tikzstyle{arrow end}=[->]
\tikzstyle{dashed_st}=[-, dashed]
\tikzstyle{arrow}=[->]
\tikzstyle{dashed_grey}=[-, draw=black, dotted]
\tikzstyle{blue_line}=[draw=blue, ->]
\tikzstyle{energy}=[draw={rgb,255: red,255; green,128; blue,0}, decoration={{snake,amplitude=1pt,segment length=6pt,post length=1pt}}, decorate, ->]
\tikzstyle{blue_line_0}=[-, draw=blue]
\tikzstyle{redline}=[-, draw=red]

\allowdisplaybreaks

\DeclareSymbolFontAlphabet{\mathbbm}{bbold}
\DeclareSymbolFontAlphabet{\mathbb}{AMSb}%
  
\def \Jone {j_2}
\def \Jtwo {j_1}
\def \be  {\begin{equation}}
\def \ee  {\end{equation}}
\def \ba  {\begin{eqnarray}}
\def \ea  {\end{eqnarray}}
\def \baa {\begin{eqnarray*}}
\def \eaa {\end{eqnarray*}}
\def \lab #1 {\label{#1}}

\newcommand\re[1]{(\ref{#1})}
\def\d{\hbox{{d}\kern-.20em\hbox{l}}}
\def \qqquad {\qquad\quad}
\def \qqqquad {\qquad\qquad}
\def \matrix #1 {\left(\begin{array}{cc} #1 \end{array}\right)}
\def \Tr {\mathop{\rm Tr}\nolimits}
\def \tr {\mathop{\rm tr}\nolimits}

\def \e  {\mathop{\rm e}\nolimits}
\newcommand\lr[1]{{\left({#1}\right)}}
\newcommand \widebar [1] {\overline{#1}}

\newcommand \vev [1] {\langle{#1}\rangle}

\newcommand \ket [1] {|{#1}\rangle}
\newcommand \bra [1] {\langle {#1}|}

\def\1{\hbox{{1}\kern-.25em\hbox{l}}}

\newcommand{\ft}[2]{{\textstyle\frac{#1}{#2}}}

\newcommand{\p}[1]{(\ref{#1})}

\def \norm  {}

\newcommand{\cD}{{\cal D}}
\newcommand{\cM}{{\cal M}}

\newcommand{\cN}{{\cal N}}

\newcommand{\cG}{{\cal G}}
\newcommand{\cO}{{\cal O}}
\newcommand{\cQ}{{\cal Q}}
\newcommand{\cE}{{\cal E}}

\newcommand{\cX}{{\cal X}}

\newcommand{\m}[1]{\mathbb{#1}}

\newcommand{\nt}{\notag\\} 
\newcommand{\pa}{\partial}
\newcommand{\ep}{\epsilon}
\newcommand{\om}{\omega}

\renewcommand{\a}{\alpha}
\renewcommand{\b}{\beta}
\newcommand{\g}{\gamma}
\newcommand{\s}{\sigma}
\newcommand{\la}{\lambda}
\newcommand{\tl}{\tilde\lambda}

\newcommand{\da}{{\dot\alpha}}
\newcommand{\db}{{\dot\beta}}

\newcommand{\q}{\theta}

\newcommand{\tx}{\tilde{x}}

\newcommand{\qq}{\,, \qquad }

\newcommand{\nn}{\nonumber}
\newcommand{\ho}{\hat\omega}
\newcommand{\Om}{\Omega}
\newcommand{\zb}{\bar{z}}
\newcommand{\ty}{\tilde{y}}

\newcommand{\yb}{\bar{y}}

\def\be#1\ee{\begin{align}#1\end{align}}

\begin{document}

\begin{flushleft}
 \hfill \parbox[c]{40mm}{CERN-TH-2021-090\\
  {IPhT--T21/031}\\
  LAPTH-022/21}
\end{flushleft}

\title{Generalizing event shapes:\\
In search of lost collider time}

\author{Gregory P. Korchemsky$^{a}$, Emery Sokatchev$^{b,c}$ and Alexander Zhiboedov$^b$ }

\affiliation{
$\null$
$^a${Institut de Physique Th\'eorique\footnote{Unit\'e Mixte de Recherche 3681 du CNRS}, Universit\'e Paris Saclay, CNRS, CEA, F-91191 Gif-sur-Yvette, France}  \\
$\null$
$^b$CERN, Theoretical Physics Department, CH-1211 Geneva 23, Switzerland\\
$\null$
$^c$Universit\'e Grenoble Alpes, USMB, CNRS, LAPTh,  F-74000 Annecy, France
}

\begin{abstract}{
We introduce a new class of collider-type observables in conformal field theories which we call generalized event shapes. They are defined as matrix elements of light-ray operators that are sensitive to the longitudinal, or time-dependent, structure of the state produced in the collision. Generalized event shapes can be studied using both correlation functions and scattering amplitudes. They are infrared finite and smoothly transit over to the familiar event shapes. We compute them in planar ${\cal N}=4$ super-Yang-Mills theory at weak and strong coupling, and study their physical properties. 
 We show that at strong coupling both the stringy and quantum-gravitational corrections to the energy-energy correlation exhibit longitudinal broadening that manifests itself through the presence of long-time tails in the energy flux measured by  the detectors. 
 }
\end{abstract}

\maketitle

\setcounter{tocdepth}{1}

\section{Introduction}
\label{sec:introduction}
  
In this paper we study a new class of observables in the familiar conformal collider physics setting \cite{Hofman:2008ar}.  They characterize the Lorentzian evolution of  a state created by a local operator acting on the vacuum. As time passes, the state evolves in a nontrivial way. Its constituents interact with each other and propagate outwards until they reach the detectors situated at a macroscopic distance  from the  collision point. 

One of the simplest detectors that we can consider is the energy calorimeter. It measures the energy deposited in a given direction  at null infinity on the celestial sphere. The  observables obtained by considering multiple  calorimeters are called energy correlations~\cite{Sveshnikov:1995vi,Korchemsky:1997sy,Korchemsky:1999kt,Hofman:2008ar}.  In field theory terms,  the operation of measuring the energy deposited in a certain  direction  on the celestial sphere corresponds to the insertion of an  energy flow operator.~\footnote{Up to a conformal   transformation it is equivalent to the so-called {\it averaged null energy condition} (ANEC) operator \cite{Hartman:2016lgu}.\label{conf}} In $d=4$ dimensions it is defined as 
\cite{Sveshnikov:1995vi,Korchemsky:1997sy,Korchemsky:1999kt,Hofman:2008ar}  
 \be
\label{eq:energydet}
\cE(n) =  \int_{-\infty}^\infty d u  \lim_{r \to \infty} r^{2} T_{0i} n^i (t=u+r, r \vec n)\,,
\ee
where the stress-energy tensor is sent to infinity in the direction of the null vector $n^\mu = (1, \vec n)$, with the unit vector  $\vec n^2 = 1$ specifying the location of the energy calorimeter on the celestial sphere. The integral in \eqref{eq:energydet} has the meaning of an average over the infinite working time $u$ of the detector.  The positivity of the ANEC operator in a unitary QFT, proved in \cite{Faulkner:2016mzt,Hartman:2016lgu}, corresponds to the  intuitive statement that over an infinite time the energy calorimeter measures  a non-negative energy flux in any state.

A hallmark of the detector defined  \eqref{eq:energydet} is that it probes the {\it transverse} structure of the state. In other words, if the time evolution produces energy fluxes collimated around particular angular directions, known as jets, these will manifest themselves as peaks in the energy correlations. At the same time, in strong coupling physics the energy calorimeters  \p{eq:energydet} produce,  to leading order,  a featureless, spherically symmetric energy  distribution \cite{Hofman:2008ar}.  

One may wonder, on the other hand, what is the {\it longitudinal} or time-dependent structure of the state? Do energy correlations arise as radiation arrives at future null infinity as a short pulse, or does the dynamical evolution lead to the broadening of the signal in time? The detectors \eqref{eq:energydet} are not suited for addressing this question. Indeed, by averaging over time in \eqref{eq:energydet} any time-dependent features of the energy flux get lost. 

In order to probe the longitudinal aspects of the state, the detector operators should have a finite time resolution. This question has been addressed by the authors of Ref.~\cite{Hatta:2012kn}. They considered  deep inelastic   {scattering}, where the state was probed with a stress-energy tensor $T_{\mu \nu}$ localized in a certain region of spacetime and carrying  some characteristic four-momentum. The issue of longitudinal broadening was discussed in \cite{Hatta:2012kn} as well, both at weak and at strong coupling. While the setup of \cite{Hatta:2012kn} is very physical, the observable in question is quite complicated and it is not directly connected to the more familiar event shapes such as energy correlations. 

Inspired by \cite{Hatta:2012kn}, in this paper we would like, on the one hand, to probe the longitudinal structure of the state, and on the other hand, to stay as close as possible to the well-studied event shapes, such as energy correlations. To this end we consider the following generalization of the energy calorimeter 
\be
\label{eq:energydetom}
\cE ({\hat \omega},  n) = \int_{-\infty}^\infty d u \, e^{- i \hat \omega u}  \lim_{r \to \infty} r^{2} T_{0i} n^i (t=u+r, r \vec n)\, ,
\ee
which probes the state at the characteristic timescale ${1 / \hat\omega}$.  {The deformation parameter $\ho$ is interpreted as the energy (or  `frequency') transferred from the detector to the particles that it probes (see \p{eq:momentumdef} below). } We call the matrix elements of  products of operators \eqref{eq:energydetom} {\it generalized event shapes}.
 Their study is the main subject of the present paper.\footnote{This operator appeared recently in \cite{Belin:2020lsr}, where $\hat \om$ played the role of a regulator in the computation of various commutators of  light-ray operators. Our motivation here is different. In particular, in this paper we only study the matrix elements of light-ray operators of the type \eqref{eq:energydetom} at non-coincident points, for which all the commutators vanish, see e.g. \eqref{eq:commutativityE}.}

The generalized event shapes are not  standard differential cross sections. To elucidate this point, let us denote the quantum state by $| \Psi \rangle$, and write the corresponding density matrix $\rho_\Psi = | \Psi \rangle  \langle \Psi |$. Physical observables are then  associated to hermitian operators $X = X^\dagger$ and are computed as their expectation values,
\be
\label{eq:observableIntro}
\langle X \rangle_\Psi \equiv \Tr \left[ \rho_\Psi X \right] \,. 
\ee
In fact, the observables normally studied in  collider experiments, the so-called event shapes, are more restrictive than \eqref{eq:observableIntro}. To understand this, it is useful to 
write the density matrix in a multi-particle basis $| n \rangle$,
\be
\label{eq:rhodens}
\rho_\Psi =\sum_{n, n'} | n \rangle \langle n' | \times \langle n  | \Psi \rangle  \langle \Psi | n' \rangle \,.
\ee
A standard event shape is then defined by restricting to a measurement protocol, or detector, $X$ that is diagonal in the multi-particle basis $X_{\text{standard}} = \sum_{n} x_n  | n \rangle \langle n |$. Such detectors are natural since they can be easily realized in an experimental setting. These measurements are classical in the sense that they only probe the probability $| \langle \Psi | n \rangle |^2$, also known as differential cross section, and are not sensitive to the phases $\phi_{\Psi,n}$ of the scattering amplitudes $\langle \Psi | n \rangle = | \langle \Psi | n \rangle | e^{i \phi_{\Psi,n}}$.~\footnote{The question to what extent the phase of an amplitude is uniquely fixed by the differential cross section has a long history in the $S$-matrix bootstrap, see \cite{Martin:2020jlu} for a recent review. In this context it was rigorously shown that in some special situations, namely elastic scattering of massive particles at low energies where no particle production is possible and with some extra technical assumptions, knowing the differential cross section does uniquely specify the phase. For such cases the observables discussed here can be computed in the standard collider setting by first reconstructing the phase of the amplitude from the measured differential cross section and then evaluating \eqref{eq:observableIntro}.}  In the language of conformal field theory, the standard event shape  distributions, or weighted cross sections, correspond to the matrix elements of the so-called light-ray operators \cite{Kravchuk:2018htv}, of which the ANEC operator \eqref{eq:energydet} is an important example. Thinking about event shapes in terms of the matrix elements of light-ray operators has proved very fruitful and has led to new insights into the structure of event shapes even in theories without conformal symmetry such as  QCD, see e.g. \cite{Chen:2020vvp,Dixon:2019uzg,Chen:2021gdk}.

Conceptually nothing prevents us from considering more general detectors $X$
with non-vanishing non-diagonal elements $\vev{n|X|n'}$ in the multi-particle basis. The corresponding observables \re{eq:observableIntro}  are sensitive to the phases of the scattering amplitudes $\langle n  | \Psi \rangle $.~\footnote{More precisely, they are sensitive to the relative phases of the scattering amplitudes since the overall phase of the S-matrix is not observable.} In fact, in a  measurement of finite duration this is essentially unavoidable.

The consequences of this fact are probably most dramatic in a gapless theory.  
 Indeed, due to the uncertainty principle, any finite-time $T$ measurement necessarily perturbs the system at the energy scale $\delta E \sim {1 / T}$. This causes transitions inside the physical detector $\langle n| X | n' \rangle$ with $E_n - E_{n'} \sim {1 / T}$. The observables of this type with a finite time resolution are the main subject of the present paper. To realize such observables in practice, one would need to devise an experiment that measures the phases of the scattering amplitudes $\langle \Psi | n \rangle$ and not just the cross sections $| \langle \Psi | n \rangle |^2$, since from \eqref{eq:rhodens} we have $\rho_\Psi =| n \rangle \langle n' |  \cdot | \langle \Psi | n \rangle | |  \langle \Psi | n' \rangle | e^{i (\phi_{\Psi,n'} - \phi_{\Psi,n})}$.~\footnote{This is different from the interference effect between different intermediate states that contribute to the  same final state, recently discussed in \cite{Chen:2020adz}.} This is precisely the effect of introducing a non-zero $\hat \om$ in \eqref{eq:energydetom}.

Still, the generalized detector operators \eqref{eq:energydetom} share many properties with the standard energy calorimeter \eqref{eq:energydet}. For example, they commute when placed at different points on the celestial sphere,  
\be
\label{eq:commutativityE}
[\cE ({\hat \omega}_1,  n_1), \cE ({\hat \omega}_2,  n_2)] =0\,,\qqqquad n_1 \neq n_2 \,.
\ee
This follows directly from the arguments of \cite{Kologlu:2019bco}. 
For non-positive (non-negative) $\hat \om$ the operators \eqref{eq:energydetom}  annihilate the right (left) vacuum  
\be
\label{eq:annihilationE}\nn
\cE ({\hat \omega},  n)  | \Omega \rangle = 0,  \qqqquad \hat \om \leq 0 \,,
\\
 {\langle\Omega|\cE ({\hat \omega},  n)   = 0,  \qqqquad \hat \om \geq 0} \,.
\ee
The relations \eqref{eq:commutativityE} and \eqref{eq:annihilationE} are familiar from the study of ordinary event shapes.

{Finally,} a new feature of \eqref{eq:energydetom} compared to \eqref{eq:energydet} is that $\cE ({\hat \omega},  n)$ carries a nonzero momentum $\hat \om n^\mu$, 
\be
\label{eq:momentumdef}
[P^{\mu}, \cE ({\hat \omega},  n)] = \hat \om n^\mu \cE ({\hat \omega},  n) \,.
\ee
In other words, when an excitation goes through the detector \eqref{eq:energydetom}, it acquires an energy `kick'.   In particular, such detectors can  {use part of their energy to} create or annihilate particles. This leads to various new and interesting effects that we describe below. 
 
The generalized event shapes have a much richer structure than the standard ones. This is not surprising since no information is lost from the stress-energy tensor after the Fourier transform in \eqref{eq:energydetom}, which is to be contrasted with the averaging over time in \eqref{eq:energydet}. From this point of view the generalized event shapes are nothing but the familiar Wightman functions understood through the prism of a collider-type experiment. We believe that this point of view is useful since it reveals the dynamical information contained in the Wightman function in a more intuitive form. 

Importantly,  the generalized event shapes are IR finite if the underlying  undeformed event shapes are IR finite. In fact, we expect them to be IR finite even when the undeformed event shapes are not. We present explicit examples of this type in the paper. This might be interesting in the context of QCD, where by considering generalized event shapes one can study a broader class of observables.~\footnote{A natural example  is the generalized charge-charge correlation which is not IR safe in the ordinary setup.}  In weakly coupled theories, generalized event shapes can be  computed using  either the standard scattering amplitudes techniques, or equivalently from  correlation functions.  In a strongly coupled CFT only the correlation function picture is available. 

After considering the general properties of the generalized event shapes, we analyze them in the planar ${\cal N}=4$ SYM theory both at weak and at strong coupling. For $\ho_i=0$ these event shapes were studied in a series of papers \cite{Belitsky:2013xxa,Belitsky:2013bja,Belitsky:2013ofa,Henn:2019gkr}, of which the present paper is a natural continuation. More precisely, we consider two-point generalized event shape distributions
which are given by an expectation value of the product of two detector operators over the state created out of the vacuum by a local operator.
We work in Mellin space and obtain a concise representation for these distributions in the form of a convolution of the Mellin amplitude for four-point correlation function and certain kinematical kernel depending on the choice of the detector.
We use ${\cal N}=4$ supersymmetry Ward identities to relate  to each other the results for different detectors, generalizing the $\ho_i =0$ analysis of \cite{Belitsky:2013xxa,Belitsky:2013bja,Belitsky:2014zha,Korchemsky:2015ssa}.  

Below we quote our results for the energy-energy correlation (EEC) at weak and at strong coupling. 
It is convenient to introduce dimensionless frequencies $\om_i$ (related to $\hat \om_i$ above by a simple rescaling {$\omega_i ={2 \hat{\omega}_i (q n_i)  / q^2}$}) and the conformal version of the angle between the detectors $z$,\footnote{In the rest frame of the source  $z = (1 - (\vec n_1 \vec n_2))/2=\sin^2(\theta/2)$, where  $\theta$ is the angle between $\vec n_1$ and $\vec n_2$.} see Section~\ref{sec:SSCcorrelator} for the detailed definitions. 
\begin{itemize}
\item
At weak coupling, to leading order in the 't Hooft coupling $a=g^2N_c/(4\pi^2)$  we get for $0< z<1$  
\begin{align}\notag
& \text{EEC}_{+-}^{\text{weak}} =  {a\over 4}\bigg[
-\frac{\omega_1^2 \omega_2^2}{36(\omega_2 +1)z}
\log\lr{z(\omega_2+1)\over \omega_2 z+1}
\\\notag
&+\frac{ \left( \omega_1^2 \omega_2^2 z^2+6 \omega_1  \omega_2 z \left(\omega_1+\omega_2+3\right)
   + 6 \left(\omega_1^2+\omega_2^2+3 \om_1 \om_2 +6 (\omega
   _1+\om_2)+6\right) \right) \log\lr{ \omega_2 z+1\over 1-z}}{36(1-z)
   z^2} 
\\ &   
   + \frac{(\omega_2+1) (\omega_2^2+6\omega_2+6) 
   \left( {\omega_1^2(  \omega_2 z+1)^2\over (\omega_2+1)^2}+{6 \omega_1  (\omega_2 z+1)\over (\omega_2+1)}
   +6 
  \right)\log \left(\frac{\omega_1 \omega_2 z+\omega_1+\omega_2+1}{(\omega_2+1)^2}\right)}{72(1-z) z (\omega_2 z+1)} \bigg] \,. 
\end{align} 
\item
At strong coupling,  we get from  tree-level supergravity for $0< z<1$ 
\be
\text{EEC}_{+-}^{\text{sugra}} &= \frac{ (\omega_2+1)^2 \left(\omega_2^2+6 \omega_2+6\right) \left(\frac{\omega_1^2
   (\omega_2 z+1)^2}{(\omega_2+1)^2}+\frac{6 \omega_1 (\omega_2
   z+1)}{\omega_2+1}+6\right)}{72(\omega_2 z+1)^3}\,. 
\ee
\end{itemize}
The subscript `$+-$' in the formulas above refers to the particular choice  {$\om_1 \geq 0$ and $-1 < \om_2 \leq 0$}. Setting $\om_i =0$ we reproduce the previously known results for the (undeformed) energy-energy correlation \cite{Hofman:2008ar,Belitsky:2013xxa,Belitsky:2013bja}. We will show that the expressions for EEC$(\om_1,\om_2)$ are sensitive to the signs of $\om_i$ and  are not analytic around $\om_i = 0$.~\footnote{This property is not obvious from the expressions above but can be seen from the results collected in Section~\ref{sec:summaryofresults}.}  
This is a common feature  of the generalized event shapes. The complete set of our results is summarized in Section~\ref{sec:summaryofresults}. The expressions above should be completed by contact terms at $z=0$, i.e. when the detectors are on top of each other. Such terms require a more careful treatment and will be the subject of a separate paper \cite{contact}. At weak coupling there are also terms localized in the back-to-back region $z=1$ which we discuss below.

As we mentioned above, the generalized event shapes can be computed even in the cases in  {when} the ordinary event shapes are not well defined. One notable example is the computation of stringy and quantum-gravitational (QG) corrections to the energy-energy correlation at strong coupling.  More precisely,  computing the $1/\lambda$ (stringy) or $1/c_T$ (quantum-gravitational) corrections to the correlation function of local operators, one finds that they produce an infinite contribution to the energy-energy correlation.~\footnote{In a gauge theory $c_T$ is related to the number of colors $N_c$. In what follows, in the framework of ${\cal N}=4$ SYM we set $c_T={(N_c^2 -1)/4}$.} On the other hand, based on general grounds, the energy-energy correlation should be finite at finite $c_T$ and finite $\lambda$, see \cite{Kologlu:2019bco}. Introducing non-zero $\om_i$ solves the problem and makes the corrections to the generalized event shapes finite and systematically computable. 

A distinctive feature of the stringy and QG corrections to the generalized event shapes is the presence of  divergences in the $\om_i \to 0$ limit. By means of a Fourier transform these can be interpreted as long-time effects in the detectors, which continue to detect the non-zero energy flux at arbitrarily late times. This is an artifact of perturbation theory because in the complete theory the energy flux eventually goes to zero. From this point of view, the correct physical interpretation of the $\om_i \to 0$ divergences is the presence of a long-time tail  of radiation measured by the detectors.  As we will see below, the presence of such long-time tails is typical for theories with a gravity dual. This is very different from what is measured by the detectors at weak coupling where the radiation is localized on timescales determined by the characteristic size of the wave packet that creates the state. Moreover, by focusing on the different types of time (or, equivalently, frequency) dependence of the energy flux one can distinguish different underlying physics: classical gravity, stringy corrections, and quantum-gravitational effects.

Choosing only one non-vanishing $\om_i$, say $\om_2 \neq 0$, and making use of the recent progress in the understanding of  gravitational loops in AdS \cite{Alday:2017xua,Aprile:2017bgs,Alday:2017vkk},  in Section~\ref{sec:QGsection} we analyze the leading correction to the energy-energy correlation in ${\cal N}=4$ SYM from the one-loop supergravity Mellin amplitude \cite{Alday:2018kkw}. As expected, the correction is divergent in the $\om_2 \to 0$ limit. We compute all the divergent terms explicitly,   the result given by \eqref{eq:resoneloopsugra}.  To better understand the origin of the divergences (and how they disappear at finite $c_T$ and $\lambda$) we use the Mellin space dispersion relations \cite{Penedones:2019tng} to derive a dispersive representation for the energy-energy correlation, see \eqref{eq:dispersiveB}.

Equipped with the perturbative results at $\om_i \neq 0$, in Section~\ref{sec:finiteN} we discuss corrections to the undeformed event shapes ($\om_i =0$) at finite $\lambda$ and finite $c_T$. Closely related to the discussion of long-time effects above, a characteristic feature of the perturbative corrections to the undeformed event shapes at strong coupling is that they are enhanced compared to the perturbative expansion of the correlation function. We review the stringy contributions to the energy-energy correlation, emphasizing on the enhancement of the leading stringy correction to the energy-energy correlation compared to the analogous correction to the correlation function. We point out that these corrections have a long-time origin and are drastically different from the pattern of radiation produced in classical gravity. Finally, we discuss the finite $c_T$ corrections to the ordinary event shapes. Here again we observe that the explicit one-loop computation from Section~\ref{sec:QGsection} predicts the existence of both $\lambda$- and $c_T$-enhanced terms in the expansion of the energy-energy correlation. We speculate on the possible form of the leading $1/c_T$ correction to the energy-energy correlation, see \eqref{eq:leadingQG}.

The structure of the paper is as follows. In Section~\ref{sec:definition} we introduce the generalized detector operators as the $\omega-$deformed light-ray transform of local operators. We then define the generalized event shapes and discuss their general properties in a conformal field theory. In Section~\ref{sect:amp}, we compute one- and two-point correlations at weak coupling using the amplitude approach. In Section~\ref{sec:SSCcorrelator}, we employ the correlation function approach to study the properties of the scalar-scalar correlation both at weak and strong coupling in $\mathcal N=4$ SYM. In Section~\ref{sN4WI} we extend our analysis to more complicated correlation functions involving the conserved $R-$current and the stress-energy tensor. 
We also establish an interesting relation between the event shape correlations involving energy and charge detectors. Section~\ref{sec:summaryofresults} contains a summary of our results for various event shapes in $\mathcal N=4$ SYM at weak and strong coupling. In Section~\ref{sec:QGsection} we discuss the corrections to the generalized energy-energy correlation at strong coupling coming from gravitational loops. In Section~\ref{sec:finiteN} we discuss stringy and gravitational corrections to the undeformed energy-energy correlation at finite $\lambda$ and finite $c_T$. Section~\ref{sect:conc} contains concluding remarks and a list of interesting open questions to pursue. 
The technical details are presented in several appendices. 

\section{Definition of the observables}
\label{sec:definition}
 
In this section we define the generalized event shapes that will be the main subject of the paper.  We start by introducing the $\omega$-deformed light-ray transform (which is a one-parameter generalization of the usual light transform of \cite{Kravchuk:2018htv}) and review its basic properties in the context of $d$-dimensional CFT.\footnote{In the following sections we will focus on the generalized event shapes in  $d=4$ ${\cal N}=4$ SYM.} The new objects have slightly unusual conformal transformation properties but preserve some of the salient features of the familiar, $\hat \omega = 0$, light-ray operators. Physically, the main new feature of the $\omega$-deformed detectors is that they can create and annihilate particles depending on the sign of $\hat \om$. This leads to various new interesting effects.

\subsection{$\omega$-deformed light transform}\label{se2.1}

Eventually, we will be interested in placing symmetric traceless primary operators at future null  infinity  and integrating them over time with a plane wave profile. It is appropriate to start with a general discussion of the conformal properties of such objects and specify their insertion at null infinity later.

To describe the conformal properties of the $\omega$-deformed calorimeters it is convenient to use the embedding space formalism in which  Minkowski space is realized as a subset of the projective null cone in $\mathbb{R}^{2,d}$.  We denote $(d+2)-$dimensional vectors in the embedding space as  $X=(X^+,X^-,X^\mu)$ where  $X^\pm = X^{d+2} \pm X^{d+1}$ and $X^\mu=X^0, \ldots , X^{d-1}$.  The scalar product in these coordinates looks as
\be
X \cdot X = X^+ X^- + (X^0)^2 - (X^1)^2 - \dots - (X^{d-1})^2 \,.
\ee
The projective null cone is defined as $X \cdot X = 0$ modulo rescaling $X \sim \lambda X$, where $\lambda \in \mathbb{R}_+$. Minkowski space arises as the locus $(X^+,X^-,X^\mu) = (1, -x^2, x^\mu)$, where $x^2 = (x^0)^2 - (\vec x)^2$ and $\vec x=(x^1, \ldots, x^{d-1})$. 
 
To describe local operators with spin we introduce the familiar index-free notation
 \be\label{Oz}
O(x,z) \equiv O_{\mu_1 ... \mu_J} (x) z^{\mu_1} ... z^{\mu_J} \,,
\ee
where the Lorentz indices of the operator are contracted with a future-pointing null polarization vector $z^\mu$.~\footnote{The notation  $z^{\mu}$ for the polarization vector is quite common. Later in this section $z$ will also denote an angular variable, see Eq.~\re{z}, but this should not lead to  confusion. }
In the embedding space $O(x,z)$ gets lifted to a {\it homogeneous} (see \p{eq:primaryA} below) function $O(X,Z)$, where $X^2 = Z^2 = X \cdot Z = 0$. It is related to the  operator $O(x,z)$  as follows
\be
\label{eq:embpoinc}
O(X,Z) = (X^+)^{-\Delta} O\left(x = {X \over X^+} , z=Z-{Z^+ \over X^+} X\right)\,.
\ee
Here  on  the left-hand side   $O(X,Z)$ is written in terms of  the $(d+2)-$dimensional vectors $X$ and $Z$, while on the right-hand side   $O(x, z)$ depends on the $d$-dimensional {coordinate} $x^\mu = X^\mu/X^+$ and the auxiliary $d$-dimensional null polarization vector $z^\mu=Z^\mu-X^\mu {Z^+ / X^+} $.
Up to gauge fixing, the two {representations} encode the same information. We will switch between the embedding and physical coordinates interchangeably, hoping that this will not cause confusion.

The conformal transformations act linearly on the vectors  $X$ and $Z$  by multiplying them by $SO(2,d)$ matrices.\footnote{See appendix~\ref{AppXZ} for futher details.} The primary operators $O(X, Z)$ are invariant under such transformations. They are homogeneous functions,  
\be
\label{eq:primaryA}
O(\lambda X, \rho Z) = \lambda^{-\Delta} \rho^J O(X, Z) \qq \la,\ \rho >0\,,
\ee
and the degrees of homogeneity $\Delta$ and $J$  {define} the quantum numbers {of the operator} (or representation labels) -- dimension and spin, respectively. In addition, 
$O(X, Z)$ is invariant under the following shifts
\be
\label{eq:primaryB}
O(X, Z + \beta X) = O(X, Z)\,.
\ee

{The light transform of a primary operator $O(X, Z)$ is defined as \cite{Kravchuk:2018htv}}
\be
\label{eq:lighttransform}
 {\bf L} [O](X,Z)  := \int_{- \infty}^{\infty} d \alpha \ O(Z-\alpha X,-X)\,.
\ee
{The light-ray operator ${\bf L} [O](X,Z)$ defined in such a way is a primary operator with the scaling dimension $1-J$ and spin $1-\Delta$}.   The conventional event shapes can be understood as matrix elements of the light-ray operators inserted at spatial infinity $X_\infty$ \cite{Kologlu:2019bco}, see Section~\ref{sec:insertionatspatialinfinity} for the precise definition. 

As discussed in the introduction, in this paper we are interested in generalized event shapes {which are} sensitive to the longitudinal structure of the state. We would therefore like to consider a generalization of \eqref{eq:lighttransform}, where instead of simply integrating over $\alpha$ with weight $1$ we insert some nontrivial weight function. As we will see below, a natural choice is to consider the Fourier transform with respect to $\alpha$.\footnote{We trank Petr Kravchuk for discussions related to the material in this section.}
We thus define the $\omega$-deformed light transform as follows 
\be
\label{eq:detector}
{\bf L}_{\hat{\omega}} [O](X,Z) \equiv \int_{- \infty}^{\infty} d \alpha \,e^{-i \hat{\omega} \alpha} O(Z-\alpha X,-X)\,. 
\ee
In the context of a collider experiment the insertion of  the phase $e^{-i \hat{\omega} \alpha}$ in \eqref{eq:detector} leads to two important new properties of the detectors. Firstly, the detector carries a non-vanishing  null momentum, which is quite natural if the source and the sink are momentum eigenstates. Secondly, the phase $e^{-i \hat{\omega} \alpha}$ improves the convergence of the integral at large $\alpha$, which is related to the IR safety of the observables in question \cite{Kologlu:2019bco}. 
 
Let us find out how the $\hat{\omega}$-deformed light-ray operators ${\bf L}_{\hat{\omega}} [O](X,Z)$ transform under 
{\eqref{eq:primaryA} and \eqref{eq:primaryB}}.     Using the definition \eqref{eq:detector} it is easy to check that  
\be
\label{eq:transformationA}
& {\bf L}_{\hat{\omega}}[O](\lambda X, \lambda Z)  = \lambda^{J - \Delta} {\bf L}_{\hat{\omega}}[O](X, Z) , \\[2mm]
& {\bf L}_{\hat{\omega}}[O](X, Z+ \beta X)  = e^{-i \hat{\omega} \beta} {\bf L}_{\hat{\omega}}[O]( X,  Z) \,.
\label{eq:transformationB}
\ee
In addition, we use \p{eq:primaryA} to find  
\be
\label{eq:zresccaling}
{\bf L}_{\hat{\omega}}[O](X, \gamma Z) &= \int_{- \infty}^{\infty} d \alpha e^{-i \hat{\omega} \alpha} O(\gamma Z-\alpha X,-X)  
= \gamma^{1-\Delta} {\bf L}_{\gamma \hat{\omega}} [O](X, Z) \,. 
\ee
Comparing \eqref{eq:transformationA}, \eqref{eq:transformationB} and \eqref{eq:zresccaling} to the transformation properties of the primary operators \eqref{eq:primaryA} and \eqref{eq:primaryB}, one deduces that for $\hat\omega \neq 0$ the operator ${\bf L}_{\hat{\omega}}[O](X, Z)$ does not transform like an ordinary primary operator.  {At the same time, for} $\hat{\omega} = 0$   {it is} a primary {operator} with the quantum numbers $(1- J, 1- \Delta)$. 

To better understand {the properties of ${\bf L}_{\hat{\omega}} [O](X,Z)$ for $\hat{\omega} \neq 0$} let us rewrite \eqref{eq:detector} in Poincar\'e coordinates. Using the correspondence formula \eqref{eq:embpoinc} and the scaling property \p{eq:primaryA},  and fixing the gauge as  $X=(1, -x^2, x^\mu)$ and $Z=(0,-2 (z \cdot x) , z)$,  we get
\be
\label{eq:definitionpoincare}
{\bf L}_{\hat{\omega}} [O](x,z) = \int_{- \infty}^\infty d \alpha  (- \alpha)^{- \Delta - J} e^{-i \hat{\omega} \alpha} O\left(x - {z \over \alpha}, z\right) \,.
\ee
Let us now act on this operator with the generators {of conformal transformations} in $x-$space. We focus on their action  at the origin $x=0$, since ${\bf L}_{\hat{\omega}} [O](x,z) = e^{i P \cdot x} {\bf L}_{\hat{\omega}} [O](0,z) e^{-i P \cdot x}$.  
For the generator of the special conformal transformation $K_\mu$ we get  
\be
\label{eq:specialconformal}
  [K_{\mu}, {\bf L}_{\hat{\omega}} [O](0,z)] =2 \hat{\omega} z_{\mu} {\bf L}_{\hat{\omega}} [O](0,z) \,. 
\ee
To derive this relation, we applied the special conformal transformation to the local operator under the integral in  \eqref{eq:definitionpoincare}, used the identity  
\be\label{eee1.13}
\left[K_{\mu}, O\left(- {z \over \alpha}, z\right) \right] &=-i \left[2 z_{\mu} \pa_\alpha O\left(- {z \over \alpha}, z\right) - 2 (\Delta+J) {z_\mu \over \alpha}  O\left(- {z \over \alpha}, z\right)  \right] ,
\ee
and finally integrated by parts.\footnote{An alternative derivation of \p{eq:specialconformal} and  \p{eq:dilatationFIN} within the embedding formalism is presented in appendix~\ref{AppXZ}.} 

It follows  {from \re{eq:specialconformal} that the light-ray operators ${\bf L}_{\hat{\omega}} [O](0,z)$} are eigenstates of the special conformal generator with a non-zero eigenvalue $2 \hat{\omega} z_{\mu}$.  
The significance of this fact is that, when placed at spatial infinity as discussed in detail in the next section, ${\bf L}_{\hat{\omega}} [O](\infty,z)$ becomes an eigenstate of the momentum generator $P_{\mu}$, or in other words it carries a definite momentum. This   is what motivates our choice of the factor $e^{- i \hat \omega \alpha}$ in \eqref{eq:detector} versus some more general weight function $f(\a)$. In particular, when studying event shapes or matrix elements of the $\omega$-deformed light transform operators between  momentum eigenstates, we will find a simple selection rule or, equivalently, an overall momentum-preserving  delta function. 

For the dilatations we get   
\be
\label{eq:dilatationFIN}
\lambda^{i D} {\bf L}_{\hat{\omega}} [O](0,z) \lambda^{-i D}  &=\int_{- \infty}^\infty d \alpha  (- \alpha)^{- \Delta - J} e^{-i \hat{\omega} \alpha}  \lambda^{i D} O\left(- {z \over \alpha}, z\right) \lambda^{-i D} \nn \\
&=\int_{- \infty}^\infty d \alpha  (- \alpha)^{- \Delta - J} e^{-i \hat{\omega} \alpha}  \lambda^{\Delta} O\left(-  \lambda {z \over \alpha}, z\right)  \nn \\
&= \lambda^{1-J} \int_{- \infty}^\infty d \alpha  (- \alpha)^{- \Delta - J} e^{-i \lambda \hat{\omega} \alpha} O\left(- {z \over \alpha}, z\right) = \lambda^{1-J} {\bf L}_{\lambda \hat{\omega}} [O](0,z) \,.  
\ee
Finally, under the Lorentz transformation by an identical argument we get
\be
\label{eq:Lorentz}
U_\Lambda  {\bf L}_{\hat{\omega}} [O](0, z) U_\Lambda^{-1} =   {\bf L}_{\hat{\omega}} [O](0, \Lambda z) \,.
\ee

It is easy to check that the {relations} \eqref{eq:specialconformal}, \eqref{eq:dilatationFIN} and \eqref{eq:Lorentz} are consistent with the conformal algebra,  $[D, K_{\mu}] = - i K_{\mu}$ and $[D, L_{\mu \nu}] = 0$, where $L_{\mu \nu}$ are the generators of the Lorentz transformations. 
In the standard discussion of the represesentations of the conformal group, see for example \cite{Qualls:2015qjb}, it is assumed that operators placed at the origin transform in a finite-dimensional representation of the Lorentz group.
 {In this case,} applying Schur's lemma {one would} conclude from $[D, L_{\mu \nu}]=0$ that $D$ has to be proportional to the identity matrix. The light-ray operators discussed here transform in an infinite-dimensional representation of the Lorentz group \eqref{eq:Lorentz} and therefore the {above} argument does not apply. 
 {Indeed, it follows from \eqref{eq:dilatationFIN} that $D$ acts nontrivially on ${\bf L}_{\hat{\omega}} [O](0,z)$. Moreover, if $D$ were proportional to the identity matrix, one would deduce from $[D, K_{\mu}] = - i K_{\mu}$ that $K_{\mu}$ has to annihilate the operator ${\bf L}_{\hat{\omega}} [O](0,z)$, in contradiction to \re{eq:specialconformal}.}

\subsection{Familiar frames: the null plane}
\label{sec:nullplane}
 
To get a better intuition regarding (\ref{eq:detector}) it is useful to write it in various conformal frames.  Let us first consider the insertion at past null infinity,  
\be\label{past}
X_0 &= \left(0,0,-{1\over 2}, \frac12, \vec 0\right) ,  \qquad 
Z_0  = (1,\vec y^2, 0,0 , \vec y) \,,
\ee
where $ \vec y=(y^2,y^3, ... \ , y^{d-1} )$.  In this frame we get  (defining $u = t- y^1$, $v = t+y^1$)  
\be
\label{eq:nullplanerepr}
{\bf L}_{\hat{\omega}} [O](X_0,Z_0) &= \int_{- \infty}^{\infty} d \alpha \, e^{-i \hat{\omega} \alpha} O(Z_0-\alpha X_0,-X_0) \nn \\
&=  \int_{- \infty}^{\infty} d u \, e^{-i \hat{\omega} u} O_{u ... u} (u,v=0, \vec y) ,
\ee
where we applied the relations \eqref{eq:embpoinc} and \re{Oz}
and used that $O_u = {1 \over 2} O^v = {1 \over 2} (O^t + O^{y^1}) $. In this way we have recovered a definition of the light-ray operator on the null hyperplane $v=0$.

{It follows from \eqref{eq:nullplanerepr} that the operator  $ {\bf L}_{\hat{\omega}} [O](X_0, Z_0)$ annihilates the right (left) vacuum for $\hat{\omega} \leq 0$ ($\hat{\omega} \geq 0$), e.g.
\be
\label{eq:annihil}
 {\bf L}_{\hat{\omega}} [O](X_0, Z_0) | \Omega \rangle = 0 , \qquad \hat{\omega} \leq 0  \,.
\ee
Indeed the state on the left-hand side of this relation has the momentum $P_u =\hat{\omega}$. In a unitarity QFT it has to satisfy the condition $P_{u} > 0$ or vanish otherwise.}

\subsection{Familiar frames: insertion at  spatial infinity}
\label{sec:insertionatspatialinfinity}

In this paper   we will be interested in a setup {when} the detectors {are placed} at spatial infinity. In terms of ${\bf L}_{\hat{\omega}} [O](X,Z)$ this corresponds to (see \re{eq:embpoinc})
\be
\label{eq:infinity}
X_{\infty}  = (0, 1 , 0^\mu)  , \qqqquad
Z_\infty &= (0, 0 , n^\mu)   ,
\ee
where $n^2=0$. {It is easy to verify} that  $Z_\infty \cdot X_\infty = 0$, as well as $Z_\infty^2 = X_{\infty}^2 = 0$.

It is convenient to place the detector at finite distance in the Poincar\'e coordinates and then approach the spatial infinity point \eqref{eq:infinity} by taking a  `detector limit'.  
To achieve this, we consider the following $r$-dependent points
\be
\label{eq:infinityLim} 
 X_{r}  \equiv \left(0, 1 , -{\bar n \over 2 (n  \bar n) r}\right)\,,  \qqqquad
Z_{r} &\equiv \left({1 \over r}, 0 , n\right)  \,,
\ee
satisfying $Z_r \cdot X_r =Z_r^2 = X_r^2 = 0$.
Here we introduced an arbitrary null vector $\bar n^2 =0$ such that $(n  \bar n) \neq 0$.  Eq.~\eqref{eq:infinityLim} becomes \eqref{eq:infinity} in the detector  limit $r \to \infty$  and the dependence on $\bar n$ drops out.  
With this choice of  {the conformal frame}  we get from \re{eq:detector} 
\be\label{2.22}
{\bf L}_{\hat{\omega}} [O](X_\infty,Z_\infty) &= \lim_{r \to \infty} {\bf L}_{\hat{\omega}} [O](X_r,Z_r) \nn \\
&= \int_{- \infty}^{\infty} d \alpha\, e^{- 2i (n \bar n) \hat{\omega} \alpha}  \lim_{r \to \infty} {r^{\Delta-J}  \over [2 (n  \bar n)]^{J-1}}  O_{\mu_1 ... \mu_J} (r n + \alpha \bar n) \bar n^{\mu_1} \dots \bar n^{\mu_J} ,
\ee
where we substituted  \eqref{eq:infinityLim} into 
 $ O(Z_{r}-\alpha X_{r},-X_{r})$ and applied \eqref{eq:embpoinc} and \re{Oz} for $x^\mu = r n^\mu + {\alpha \bar n^\mu \over 2 (n \cdot \bar n)}$  and $z^\mu ={\bar n^\mu \over 2 (n  \bar n) r}$.  
 
 The careful reader might have noticed that in the last line of \eqref{2.22} we exchanged the order of the integration and the limit $r \to \infty$. The justification is as follows. Physically, we  imagine placing the detectors at some finite distance $r$ from the collision point, integrating over working time of the detectors and, then, taking the limit $r \to \infty$. In practice, however, it is much easier to send the detectors to infinity first and only then integrate over time. This is what we do in the present paper. We expect that the order of operations does not matter as long as the detectors do not communicate with each other, namely for $n_i \neq n_j$.
 This is the regime that we study in the present paper. In other words, exchanging the limit and the integration in \re{2.22} can lead to results that differ by contact terms localized at $n_i =n_j$, see \cite{contact}.

In the sections below we consider a slightly different definition of the $\omega$-detectors. We change $\hat{\omega} \to { \hat{\omega} / 2}$ in the formula above and consider $O_J (\hat \omega, n)  \equiv 2^{J-1} {\bf L}_{{\hat{\omega} \over 2}} [O](X_\infty,Z_\infty)$, so that 
\be
\label{eq:definitiondet}
 \cO_J (\hat \omega, n) \equiv  \int_{- \infty}^{\infty} d \alpha \, e^{-i (n  \bar n) \hat{\omega} \alpha}  \lim_{r \to \infty} {r^{\Delta-J}  \over (n   \bar n) ^{J-1}}  O_{\mu_1  ... \mu_J} (r n + \alpha \bar n) \bar n^{\mu_1} \dots \bar n^{\mu_J}\,,
\ee
where $n^2 = \bar n^2 = 0$.
Let us emphasize again that the precise choice of $\bar n$ is immaterial since the result does not depend on it. We will always assume that $n^\mu$ is a {\it future-pointing} null vector, namely $n^0 > 0$.
The advantage of rescaling $\hat{\omega}$ by a factor of ${1 / 2}$ comes from noticing that 
\be
\label{eq:momentumeigenstate}
[P^\mu , \cO_J (\hat \omega, n)  ] = \hat{\omega} n^\mu \cO_J (\hat \omega, n)  , 
\ee
which is an immediate consequence of \eqref{eq:specialconformal} {because the} special conformal transformations at infinity correspond to translations around the origin.  In other words, the detectors \eqref{eq:definitiondet} carry a definite null momentum $\hat{\omega} n^\mu$.~\footnote{Our convention for the symmetry generators is such that $[P^{\mu}, O(q)] = q^{\mu} O(q)$, where $O(q) \equiv \int d^d x e^{-i q \cdot x} O(x)$ and here we work in the mostly minus signature.}

The momentum operator acts on the local operator as $[P^\mu ,O_{\mu_1  ... \mu_J} (x)] = i \partial^\mu O_{\mu_1  ... \mu_J} (x)$. For $x=r n + \alpha \bar n$ and $r\to\infty$ the leading contribution only comes from the light-cone component $n^\mu (P\bar n)/(n\bar n)=i n^\mu \partial_\alpha/(n\bar n) $. The contribution of  the remaining components of $P^\mu$ is suppressed by a power of $1/r$. Integrating by parts in \p{eq:definitiondet} we obtain \p{eq:momentumeigenstate}.

From \eqref{eq:zresccaling} we derive the scaling property
\be
\label{eq:transformationproperty}
\cO_J \Big( \lambda^{-1} \hat \omega , \lambda n \Big)  =  \lambda^{1-\Delta} \cO_J (\hat \omega, n)  \,.  
\ee
Setting $\hat{\omega} =0$ and $n^\mu = (1, \vec n)$ in \eqref{eq:definitiondet} {and replacing $O_{\mu_1  ... \mu_J}$ with} a conserved current   or to the stress-energy tensor we recover the familiar charge ${\cal Q}(n)$ and energy ${\cal E}(n)$ detector operators from \cite{Hofman:2008ar,Belitsky:2013xxa,Belitsky:2013bja}.
From \eqref{eq:momentumeigenstate} it follows immediately that
\be
\label{eq:generalizedannihil}
\cO_J (\hat \omega, n)  | \Omega \rangle = 0 , ~~~ \hat{\omega} \leq 0 , \nn \\[2mm]
\langle \Omega | \cO_J (\hat \omega, n)  = 0, ~~~ \hat{\omega} \geq 0 ,
\ee
where $| \Omega \rangle$ is the CFT vacuum. Eq. \eqref{eq:generalizedannihil} is an obvious generalization of the corresponding property of light-ray operators at $\hat{\omega} = 0$, see \cite{Kravchuk:2018htv} for the detailed argument.

\subsection{Signature}
\label{sec:signature}

Every relativistic quantum field theory has an anti-unitary symmetry which we denote by $\mathsf{CRT}$, see e.g. \cite{Witten:2018zxz}. This transformation reverses time $\mathsf{T}$, one spatial direction $\mathsf{R}$, as well as all charges $\mathsf{C}$. Combining $\mathsf{CRT}$ with Hermitian conjugation we can classify operators into eigenspaces of $\big( (\mathsf{CRT}) ... (\mathsf{CRT})^{-1} \big)^\dagger $ \cite{Kravchuk:2018htv}. The corresponding eigenvalue $\pm 1$ is called the signature of the operator.  

Let us examine the action of $\mathsf{CRT}$ on the $\omega$-deformed light-ray transforms \re{eq:nullplanerepr}. $\mathsf{CRT}$  is an anti-unitary symmetry that acts as
\be
(u,v,\vec y) &\ \to\  (-u,-v,\vec y).
\ee
{Combining} $\mathsf{CRT}$ with Hermitian conjugation we get 
\be
\label{eq:CRT}
\Big( (\mathsf{CRT}) \int_{-\infty}^\infty du \, e^{-i \hat \omega u} O_{u \cdots u}(u,0,\vec y) (\mathsf{CRT})^{-1} \Big)^\dagger &=\eta_O (-1)^{J} \int_{-\infty}^\infty du \, e^{i \hat \omega u} O_{u \cdots u}(-u,0,\vec y) \nn \\
&=\eta_O (-1)^{J}\int_{-\infty}^\infty du\, e^{-i \hat \omega u} O_{u \cdots u}(u,0,\vec y) ,
\ee
where $\eta_O$ is an arbitrary phase factor.  Depending on {the} definition of the operator, the latter  can be set to $1$.\footnote{However this is not always a natural choice. For example,  {for} the conserved current in a free scalar theory, $J_\mu = (J_\mu)^\dagger = i (\bar \phi \partial_\mu \phi - \partial_{\mu} \bar \phi \phi)$ we get $\eta_{J_\mu} = -1$. In fact this is the choice that we make for the charge detector operator ${\cal Q}(\hat \om, n)$ in this paper.} For $\hat \omega =0$ the eigenvalue $\eta_{O} (-1)^J$ is called the `signature' of the light-ray operator.  {As follows from \re{eq:CRT},} the signature of the operator does not depend on $\hat \om$.

We review the implications of $\mathsf{CRT}$ for the one- and two-point generalized event shapes in appendix \ref{app:CRT}. 

\subsection{Generalized event shapes}

Having {defined} the $\omega$-deformed light-ray operators \eqref{eq:definitiondet}, we can now consider their matrix elements over
physical states
\be
\label{eq:quantumeventshapes}
\langle \phi_2(\tilde q)  \cO_{J_1}( \hat{\omega}_1, n_1)  ... \cO_{J_k}( \hat{\omega}_k, n_k)   \phi_1(q) \rangle &= (2 \pi)^d \delta^d (\tilde q - \sum_{i=1}^k \hat{\omega}_i n - q) \theta(q) \theta( q + \sum_{i=1}^k \hat{\omega}_i n) \nn \\
&\times \langle \langle \phi_2(\tilde q)  \cO_{J_1}( \hat{\omega}_1, n_1)  ... \cO_{J_k}( \hat{\omega}_k, n_k)   \phi_1(q) \rangle \rangle \,.
\ee
Here $\cO_{J_i}(\hat{\omega}_i, n_i) $ are identified as the `detectors', $\phi_1(q)$ as the `source' and $ \phi_2(\tilde q)$ as the `sink'. To describe a nontrivial physical state, $q$ and $\tilde q$ have to be future-pointing timelike momenta.
The overall momentum conservation delta function in \re{eq:quantumeventshapes} takes into account the fact that  each detector carries momentum $\hat{\omega}_i n_i^\mu$. The step function 
\be
\label{eq:thetaconstraint}
\theta(q) \equiv \theta(q^0) \theta(q^2) 
\ee 
ensures  that the relevant momenta are future-pointing time-like. This puts a constraint on the values of $\hat{\omega}_i$ for which the correlation function \re{eq:quantumeventshapes} is non-zero.

Below we choose the source and sink  to be described by a scalar primary operator $\phi_1 = \phi_2^\dagger = \phi$ with the scaling dimension $\Delta_\phi$. We  define the properly normalized correlation {function} 
\be
\label{eq:eventshapesQ}
\langle  \cO_{J_1}( \hat{\omega}_1, n_1)  ... \cO_{J_k}( \hat{\omega}_k, n_k) \rangle_q &\equiv {(q^2)^{\Delta_\phi -{d \over 2}} \over \sigma_{tot}(q)}\langle \langle \phi^\dagger (\tilde q)  \cO_{J_1}( \hat{\omega}_1, n_1)  ... \cO_{J_k}( \hat{\omega}_k, n_k)  \phi(q) \rangle \rangle \,,
\ee
which we shall call a {\it generalized event shape}. 
Here we made the left-hand side of \eqref{eq:eventshapesQ} independent on the normalization of the operator $\phi(x)$ by dividing the right-hand side by the Fourier transform of the Wightman two-point function 
\begin{align}\label{2.32}
 \sigma_{tot}(q) &\equiv \int d^d x\, e^{-i q \cdot x} \langle  \phi^\dagger (x) \phi(0)   \rangle \,. 
\end{align}
 It has the meaning of the total cross section of the process $\text{source}\to \text{anything}$. Choosing $\langle  \phi^\dagger (x) \phi(0)   \rangle=1/(-x^2+i0 x^0)^{\Delta_\phi}$ we get 
\begin{align} 
\sigma_{tot}(q) =  
\theta(q)  {2\pi^{d/2+1} (q^2/4)^{\Delta_\phi-d/2}\over \Gamma(\Delta_\phi)\Gamma(\Delta_\phi+1-d/2)}\,.
\end{align}

\subsubsection*{One-point function}\label{seOne}

Let us consider some simple examples of \re{eq:eventshapesQ}. We start with the one-point event shape. {Its form is fixed by symmetry up to an arbitrary function of dimensionless argument}
\be
\label{eq:onepoint}
\langle  \cO_J (\hat \om, n) \rangle_q =  {{1\over S_{d-2}}} {(q^2)^{\Delta_\phi - {d \over 2}  + {J + \Delta_{O} \over 2} - 1} \over (q n)^{\Delta_{O} - 1}}   f_{\cO_J} \Big(  \hat{\omega}{2  (q n)  \over q^2} \Big),
\ee
where the power of $(qn)$ and $q^2$ is dictated by the relation  \eqref{eq:transformationproperty} and by the scaling dimension of the observable, respectively. In this relation $(\Delta_O,J)$ are the quantum numbers of the detector operator. The normalization factor involving   the area of the unit sphere $S_{d-2} =  {2 \pi ^{\frac{d-1}{2}}}/{\Gamma \left(\frac{d-1}{2}\right)}$  was introduced  for convenience (see e.g. \re{f+} and \re{f-} below). 

It is convenient to replace the argument of the function $f_{\cO_J}$ in \p{eq:onepoint} by  the dimensionless frequency
\be
\label{eq:rescaledfreq}
  \omega ={2  (q n)  \over q^2}  \hat{\omega} \,.
\ee
The  constraint \eqref{eq:thetaconstraint} that the sink carries a time-like momentum   becomes $\theta(1 +  \omega)$, or $ \omega \geq - 1$. 
The rescaled frequency  makes the transformation property \eqref{eq:transformationproperty} look simpler,
\be
\cO_J (\om, \lambda n) =  \lambda^{1-\Delta} \cO_J( \omega , n)  \,.
\ee 
In what follows we  will interchange freely between $\hat \om$ and $\om$, keeping in mind the relation \eqref{eq:rescaledfreq}  between them. We hope this will not cause any confusion. 

In the present paper we consider the detector operators \eqref{eq:definitiondet} made out of Hermitian local operators. We then have the following relation
\be\label{dagg}
\Big[ \cO_J (\hat \om, n) \Big]^\dagger =  \cO_J (- \hat \om, n)  \,,
\ee
which relates the event shapes for positive and negative $\om$. In the case of the one-point function we have
\be
\label{eq:hermitian1}
\Big[ \langle  \cO_J (\hat \om, n)  \rangle_q  \Big]^* = \langle \Big[ \cO_J (\hat \om, n)  \Big]^\dagger \rangle_{q + \hat \omega n} = \langle \cO_J (-\hat \om, n) \rangle_{q + \hat \omega n} \,. 
\ee
We can also derive a similar relation using the $\mathsf{CRT}$ transformation and the Hermitian conjugation \eqref{eq:CRT}, see appendix \ref{app:CRT} for details,
\be
\label{eq:CRT1}
\mathsf{CRT}: ~~~\langle  \cO_J (\hat \om, n)  \rangle_q =  \pm  \langle \cO_J (-\hat \om, n) \rangle_{q + \hat \omega n} \,,
\ee
where the sign $\pm$ on the right-hand side corresponds to the signature of the detector operator $\cO_J (\hat \om, n)$ discussed in Section~\ref{sec:signature}. 
{Combining together \eqref{eq:hermitian1} and \eqref{eq:CRT1}  we find that that $\langle \cO_J (\hat \om, n)  \rangle_q$, or equivalently the function $f_{\cO_J}( \omega)$ in \re{eq:onepoint},  is real for the  detector operators of positive signature.}

We can now use \eqref{eq:onepoint} to rewrite the relation \eqref{eq:CRT1} as follows
\be
 {(q^2)^{\Delta_\phi - {d \over 2}  + {J + \Delta_{O} \over 2} - 1} \over (q  n)^{\Delta_{O} - 1}}   f_{\cO_J} \Big(\hat\omega {2  (q  n)  \over q^2} \Big) =  {(\hat q^2)^{\Delta_\phi - {d \over 2}  + {J + \Delta_{O} \over 2} - 1} \over (\hat q  n)^{\Delta_{O} - 1}}   f_{\cO_J} \Big(- \hat \omega {2  (\hat q n)  \over \hat q^2} \Big),  
\ee
where $\hat q = q + \hat \omega n$.  
It simplifies to
\begin{empheq}[ ]{equation}
\label{eq:onepoincrt}
    f_{\cO_J}( \omega) = (1 +  \omega)^{\Delta_\phi - {d \over 2}  + {J + \Delta_{O} \over 2} - 1} f_{\cO_J}\left(-{ \omega \over  \omega + 1}\right) .
\end{empheq}
Some explicit examples of $f_{\cO_J}( \omega)$ can be found in Section~\ref{sec:specialdetectors}.

\subsubsection*{Two-point functions}

The two-point correlation \re{eq:eventshapesQ} takes the following general form 
\be
\label{eq:twopointgeneral}
\langle   \cO_{J_1} (\om_1, n_1)   \cO_{J_2} (\om_2, n_2)   \rangle_q = {1\over S_{d-2} S_{d-3}} {(q^2)^{\Delta_\phi - {d \over 2}  + \sum_{i=1}^2 {J_i + \Delta_{O_i} - 2 \over 2} } \over (q   n_1)^{\Delta_{O_1} - 1}  (q   n_2)^{\Delta_{O_2} - 1}}  f_{\cO_{J_1} \cO_{J_2} }(\omega_1, \omega_2 , z), 
\ee
where the scaling function $f_{\cO_{J_1} \cO_{J_2} }$ depends on the rescaled dimensionless frequencies \eqref{eq:rescaledfreq} of the detectors and on the cross-ratio
\be\label{z}
z = {q^2 (n_1 n_2) \over 2 (q n_1) (q n_2)} \ , \qqqquad 0 \leq z \leq 1 \,.
\ee
It captures the dependence on the relative angle between the detectors.
To derive the condition $0 \leq z \leq 1$ recall that $q^\mu$ is a future-pointing time-like vector and $n_i^\mu$ are future-pointing null vectors. 
Like in  \re{eq:onepoint},
the normalization factor  in \re{eq:twopointgeneral} helps simplify the expressions for the scaling functions $f_{\cO_{J_1} \cO_{J_2} }$. 

Like for the one-point function, the condition that both the sink and the source carry time-like momenta, 
$q$ and $\tilde q =  q + \hat \om_1 n_2 + \hat \om_2 n_2$, respectively, restricts the values of $ \omega_i$ as follows 
\be
\label{eq:thetafunctiontilde}
\theta\Big( q + \sum_{i=1}^2 \hat \omega_i n_i \Big)  : \qquad 1 +  z \omega_i\ge 0\,, \qquad 1 +  \omega_1 +  \omega_2 + z  \omega_1  \omega_2 \geq 0 \,. 
\ee
These  relations are obtained from \p{eq:thetaconstraint}  by switching to $ \om_i = 2 (q n_i)/q^2\,  \ho_i   $ and using  \eqref{z}.

Repeating the argument in Section~\ref{seOne},  namely considering the Hermitian conjugation and $\mathsf{CRT}$ transformation of the positive signature detector operators,  we get (see appendix \ref{app:CRT} for details)
\be\nn
\label{eq:hermitiantwop}
\Big[ \langle   \cO_{J_1} (\om_1, n_1)   \cO_{J_2} (\om_2, n_2)   \rangle_q  \Big]^* 
& =\langle   \cO_{J_1} (\om_1, n_1)   \cO_{J_2} (\om_2, n_2)  \rangle_q 
\\& 
= \langle   \cO_{J_1} (-\om_1, n_1)   \cO_{J_2} (-\om_2, n_2)    \rangle_{q + \hat \omega_1 n_1 + \hat \omega_2 n_2} \,.
\ee
Combining this relation with  \eqref{eq:twopointgeneral}  we find
\begin{align}\label{eq:CRT2}
 f_{\cO_{J_1} \cO_{J_2}}(\omega_1, \omega_2 , z) = {(1+\omega_1 +\omega_2 + \omega_1 \omega_2 z )^{\Delta_\phi - {d \over 2}  + \sum_{i=1}^2 {J_i + \Delta_{O_i} - 2 \over 2} } \over (1 + \omega_2 z)^{\Delta_{O_1}-1} (1 + \omega_1 z)^{\Delta_{O_2} - 1}}  f_{\cO_{J_1} \cO_{J_2} } ( \omega_1', \omega_2', z' ) \,,
\end{align}
where $f_{\cO_{J_1} \cO_{J_2}}(\omega_1, \omega_2 , z)$ is real for the $\omega$-deformed light transform of the hermitian operators $O_i$ of positive signature
and
\begin{align}\notag\label{w'}
\omega_1' &=   -\omega_1 {1 + \omega_2 z \over 1+\omega_1 +\omega_2 + \omega_1 \omega_2 z}\,,
\\\notag
\omega_2' &=  -\omega_2  {1 + \omega_1 z \over 1+\omega_1 +\omega_2 + \omega_1 \omega_2 z}\,, 
\\[1.5mm]
z' &= z {1+\omega_1 +\omega_2 + \omega_1 \omega_2 z \over (1 + \omega_1 z) (1 + \omega_2 z)}\,.
\end{align}
{The quantity $1+\omega_1 +\omega_2 + \omega_1 \omega_2 z$ is the dimensionless version of $\tilde q^2 = (q + \hat \om_1 n_2 + \hat \om_2 n_2)^2 $, which already appeared above in \eqref{eq:thetafunctiontilde}.}

The crucial feature of \eqref{eq:CRT2} is that it relates the two-point functions {defined for} $\omega_i$ and $\omega_i'$ of  opposite signs. In the context of the $S$-matrix this {relation} corresponds to the crossing transformation which exchanges particles and  anti-particles. We can therefore call \eqref{eq:CRT2} a {\it crossing relation}. It will serve as a nontrivial consistency check for our computations in the following sections. Note that for $\omega_i = 0$ the relation \eqref{eq:CRT2} becomes trivial.

Finally, {if the detectors are identical, the following relation holds}
\be
 f_{\cO_J \cO_J}(\omega_1, \omega_2 , z) =  f_{\cO_J \cO_J}(\omega_2, \omega_1 , z) \qq z \neq 0 \,. 
\ee
It expresses the fact that the $\omega$-deformed event shape \re{eq:eventshapesQ} is invariant under the permutation of the detectors. For $z=0$ the detectors do not necessarily commute, see \cite{contact}.
 
\subsection{Special detectors: S, Q, and E}
\label{sec:specialdetectors}
 
In the framework of $d=4$, ${\cal N}=4$ SYM we consider three types of detector operators: 
\begin{itemize}
\item  {\it Scalar detector}  
 $\cO (\hat \omega, n)$, denoted by S,  obtained from the local operator with $J=0$ and $\Delta = 2$; 
\item   {\it Charge detector} 
$\cQ(\hat \omega, n)$, denoted by Q, obtained from the  $R-$current with $J=1$ and $\Delta = 3$;
\item {\it Energy detector}  
$\cE(\hat \omega, n)$, denoted by E, obtained from the stress-energy tensor with $J=2$ and $\Delta =4$.
\end{itemize}
They all  have positive signature and the corresponding generalized event shapes will be real, in agreement with the general discussion in the previous subsection. When elaborating the detailed form of the correlators in ${\cal N}=4$ SYM we will need to carefully specify the $R$-symmetry structure. This will be done in Section~3 below but for the purpose of this section it is not  important.

The one-point functions of such detectors, evaluated in a state created by a scalar primary with  scaling dimension $\Delta_\phi$, are completely fixed by conformal invariance (see Appendix~\ref{secOnept}).  Applying \re{eq:onepoint} for $d=4$ and $\Delta_O=J+2$ we get 
\be\label{1pt-exp}
\langle  \cO_J (\hat \om, n) \rangle_q = {1\over 4\pi} {(q^2)^{\Delta_\phi -2  + J} \over (q n)^{J+1}}   f_{\cO_J} (\omega)\,,
\ee
 where $\omega$ is defined in \re{eq:rescaledfreq}.
 
The calculation  in Appendix~\ref {app:1pt} shows that the scaling function  $f_{\cO_J}$ takes  different form for positive and negative $\omega$: 
\begin{align}\notag
\om>0:~~~& f_\text{S}(\omega) = 1\,,\qquad
\\[2mm]\notag
& f_\text{Q}(\omega) = 1+\frac{\omega}{2} (\Delta_\phi -1) \,,\qquad
\\
& f_\text{E}(\omega) = 1 + \frac{\omega}{2}\Delta_\phi +\frac{\omega ^2}{12} (\Delta_\phi -1) \Delta_\phi \,; \label{f+}\\
&\nt
-1< \om<0:~~~~& f_\text{S}(\omega) = (1+\omega )^{\Delta_\phi -2}\,,\qquad  \notag
\\[2mm]\notag
& f_\text{Q}(\omega) =  (1+\omega )^{\Delta_\phi -2} \left[1-\frac{\omega}{2}(\Delta_\phi -3)\right]\,,\qquad  \notag
\\
& f_\text{E}(\omega) =  (1+\omega )^{\Delta_\phi -2} \left[1-\frac{\omega}2 (\Delta_\phi -4)+ \frac{\omega^2}{12}(\Delta_\phi -4)
   (\Delta_\phi -3) \right].  \label{f-}
\end{align}
 For $\omega=0$ we have $ f_{\cO_J} (0)=1$. In this case, in the rest frame of the source, for $q^\mu=(q,\vec 0)$, the integral of 
the one-point function over the unit sphere,
$\int d^2 \vec n \,\langle  \cO_J (0, n) \rangle_q$,  yields the total charge of the source. 
It is easy to check that the formulas above satisfy the relation \eqref{eq:onepoincrt}. We remark that the $\omega$-deformed event shapes are not analytic  around  $\om =0$. This is very intuitive because, depending on the sign of $\omega$, the detectors either create or annihilate particles. We discuss this property further in Section~\ref{sec:nonanalytic}.

 The relations \re{f+} and \re{f-} are valid for a source of arbitrary scaling dimension  $\Delta_\phi$. In $\mathcal N=4$ SYM 
{we choose the} source {to} be the half-BPS scalar operator with dimension $\Delta_\phi=2$ (see 
Eq.~\re{halfBPS} below). In this case, the one-point functions  \re{f+} and \re{f-} simplify as
\begin{align}
f_\text{S}(\omega) = 1\,,\qqquad  f_\text{Q}(\omega) = 1+\frac{\omega}{2}\,,\qqquad f_\text{E}(\omega) = 1 +  {\omega} +\frac{\omega ^2}{6}\,.
\end{align}
We will derive these expressions in the next section where we compute the generalized event shapes using  the amplitude method. 

Likewise, for $\Delta_\phi=2$ the two-point function \re{eq:twopointgeneral} simplifies as
\begin{align}\label{2pt-simp}
\langle   \cO_{J_1} (\om_1, n_1)   \cO_{J_2} (\om_2, n_2)   \rangle_q =  {1\over 8\pi^2} {(q^2)^{J_1+J_2} \over (q  n_1)^{J_1 + 1}  (q  n_2)^{J_2+1}}  f_{\cO_{J_1} \cO_{J_2} }(\omega_1, \omega_2 , z)\,, 
\end{align}
where we replaced $ \Delta_{O_i}=2+J_i$. 
  
For the detectors described above we will follow the tradition in the literature and will call the two points functions \re{eq:twopointgeneral} {\it correlations}. For example, for the correlations of scalar, charge and energy we write
\begin{align}\label{eq:definingXXC}
&\text{SSC} \equiv f_{\cO \cO}(\omega_1, \omega_2 , z),  && \text{QQC} \equiv f_{\cQ \cQ}(\omega_1, \omega_2 , z) , \nn \\[2mm]
&\text{QSC} \equiv f_{\cQ \cO}(\omega_1, \omega_2 , z),  && \text{ESC} \equiv f_{\cE \cO}(\omega_1, \omega_2 , z) , \nn \\[2mm]
&\text{EQC} \equiv f_{\cE \cQ}(\omega_1, \omega_2 , z) ,  &&\text{EEC} \equiv f_{\cE \cE}(\omega_1, \omega_2 , z) \,.
\end{align}
Again in ${\cal N}=4$ SYM we would need to specify the choice of the R-symmetry components of the scalar and the current operators, this will be done below.

 The factor of $1/(8\pi^2)$ on the right-hand side of \re{2pt-simp} was inserted  to simplify the normalization conditions for the scaling function 
$f_{\cO_{J_1} \cO_{J_2} }(\omega_1, \omega_2 , z)$ at $\omega_i=0$. In the case of the energy-energy correlations these conditions take the form \cite{Korchemsky:2019nzm,Kologlu:2019mfz,Dixon:2019uzg} 
\begin{align}
\label{eq:WIstress}
\int_0^1 dz\, \text{EEC}(0,0,z) = \frac12\,,\qqqquad \int_0^1 dz\, z \,\text{EEC}(0,0,z) = \frac14\,.
\end{align}
These relations follow from the requirement that the total momentum of the particles produced in the rest frame of the source should be equal to $q^\mu=(q,\vec 0)$. Relations \eqref{eq:WIstress} follow from the  Ward identities satisfied by the stress-energy tensor.

Since the  $\omega$-deformed event shapes are not analytic around $\omega = 0$,  we should be more specific about the choice of the signs of $\omega$. Let us for example consider $\text{SSC}$ defined in \eqref{eq:definingXXC}. We have four choices $\text{SSC}_{++}$, $\text{SSC}_{+-}$, $\text{SSC}_{-+}$, $\text{SSC}_{--}$. As argued in the previous subsection,   these functions are real. Moreover, 
due to \eqref{eq:CRT2} only two of them are independent. 

More precisely, in the main case of interest  that we  consider in detail below, namely  $d=4$ and $\Delta_{\phi}=2$,  we get from \eqref{eq:CRT2}  
\be
\label{eq:relationused}
 \text{SSC}_{-+} (\omega_1, \omega_2 , z) &= {1 \over (1 + \omega_1 z) (1 + \omega_2 z)}  \text{SSC}_{+-} (\omega_1' , \omega_2' , z') , \nn \\
\text{SSC}_{--} (\omega_1, \omega_2 , z) &= {1 \over (1 + \omega_1 z) (1 + \omega_2 z)}  \text{SSC}_{++} (\omega_1' , \omega_2' , z') ,
\ee
where $\omega_i'$ and $z'$ were defined in \eqref{w'}. These relations allow us to restrict the discussion only to $\text{SSC}_{+-}$ and $\text{SSC}_{++}$. The same  applies to other event shapes:
\be
\label{eq:relationusedGen}
& \text{QSC} (\omega_1, \omega_2 , z)= {(1+\omega_1 +\omega_2 + \omega_1 \omega_2 z ) \over (1 + \omega_2 z)^2 (1 + \omega_1 z)}  \text{QSC}(\omega_1' , \omega_2' , z') \,, \nn \\
& \text{QQC} (\omega_1, \omega_2 , z) = {(1+\omega_1 +\omega_2 + \omega_1 \omega_2 z )^2 \over (1 + \omega_1 z)^2 (1 + \omega_2 z)^2}  \text{QQC}(\omega_1' , \omega_2' , z') \,, \nn \\
&\text{ESC} (\omega_1, \omega_2 , z) = {(1+\omega_1 +\omega_2 + \omega_1 \omega_2 z )^2 \over (1 + \omega_2 z)^3 (1 + \omega_1 z)}  \text{ESC}(\omega_1' , \omega_2' , z') \,, \nn \\
&\text{EQC} (\omega_1, \omega_2 , z) = {(1+\omega_1 +\omega_2 + \omega_1 \omega_2 z )^3 \over (1 + \omega_2 z)^3 (1 + \omega_1 z)^2}  \text{EQC}(\omega_1' , \omega_2' , z') \,, \nn \\
&\text{EEC} (\omega_1, \omega_2 , z) = {(1+\omega_1 +\omega_2 + \omega_1 \omega_2 z )^4 \over (1 + \omega_1 z)^3 (1 + \omega_2 z)^3}  \text{EEC}(\omega_1' , \omega_2' , z') \,. 
\ee

\subsection{Finiteness, commutativity and non-analyticity in $\omega$}
\label{sec:nonanalytic}

One might wonder to what extent the generalized event shapes \re{eq:eventshapesQ}
are well defined. To understand this issue it is convenient to go to the conformal frame \eqref{eq:nullplanerepr} {and to} consider the product of the detector operators inserted at the same null plane, 
\be
\label{eq:product}
\int_{- \infty}^{\infty} d u_1 e^{-i \hat{\omega}_1 u_1} O_{\underbrace{ {\scriptstyle u ... u} }_{J_1}} (u_1,v=0, \vec y_1) \int_{- \infty}^{\infty} d u_2 e^{-i \hat{\omega}_2 u_2} O_{\underbrace{\scriptstyle u ... u}_{J_2}} (u_2 ,v=0, \vec y_2) \,.
\ee
The conditions for the existence of this product for $\hat \omega_i =0$ were discussed in detail in \cite{Kologlu:2019bco}. The basic observation is that the product of operators \re{eq:product} is  potentially ill-defined because the two operators in \eqref{eq:product} become light-like separated when one of the detectors reaches infinity $u_i = \pm \infty$. An avatar of this in a collider experiment is that the corresponding observable is not IR safe. 

In the case   $\om_i=0$ the situation is discussed in Section~4.4 in \cite{Kologlu:2019bco}. An important condition that guarantees the existence and commutativity of the product \eqref{eq:product} for $\vec y_1 \neq \vec y_2$ is
\be
\label{eq:commutativecond}
J_1 + J_2 \geq 1 + J_0 \ ,
\ee
where $J_0$ is the Regge intercept of the theory and $J_1, J_2$ are the spins of the operators in \eqref{eq:product}.\footnote{For the precise conditions, see \cite{Kologlu:2019bco}.} 

The insertion of $\omega-$dependent phase factors in \eqref{eq:product} does not affect the analysis of \cite{Kologlu:2019bco} and therefore all the results apply  directly. This fact immediately implies that the $\omega$-deformed event shapes are finite and commutative for $\vec y_1 \neq \vec y_2$, or, equivalently, for $n_1 \neq n_2$ in the collider experiment frame, as long as the condition \eqref{eq:commutativecond} is satisfied.

There is a difference between the two cases in the situation when \eqref{eq:commutativecond} is violated. For $\om_i =0$ the product \re{eq:product} is ill-defined. The reason is that the integrals over $u_i$ diverge polynomially at large $|u_i| \gg 1$, see  \cite{Kravchuk:2020scc} for a recent discussion of this point. For $\om_i \neq 0$ the rapidly oscillating phases damp the contribution at $|u_i| \gg 1$ in  \eqref{eq:product}. As a consequence, the product of operators in \eqref{eq:product} is well defined (at least in the distributional sense) and commutative for $\vec y_1 \neq \vec y_2$,
\be
\Big[ \int_{- \infty}^{\infty} d u_1 e^{-i \hat{\omega}_1 u_1} \cO_{u ... u} (u_1,v=0, \vec y_1) , \int_{- \infty}^{\infty} d u_2 e^{-i \hat{\omega}_2 u_2}   \cO_{u ... u} (u_2 ,v=0, \vec y_2) \Big]  \stackrel{\vec y_1 \neq \vec y_2 }{=}0\,. 
\ee
One can also try to extend the $\omega$-deformed event shapes as distributions to include the case of coincident points $\vec y_1 = \vec y_2$. This turns out to be quite subtle and we will discuss it elsewhere \cite{contact}.  

For the same reason, we expect that when expanded around {arbitrary} $\om_i \neq 0$ the $\omega$-deformed event shapes \re{eq:eventshapesQ} are real analytic functions. This is indeed what we find in the examples considered below. One interesting aspect of this fact, when interpreted in terms of amplitudes, is that the $\omega$-deformed event shapes are IR finite to any order in perturbation theory for any choice of the detectors. The same argument, however, suggests that the $\omega$-deformed event shapes are not analytic around $\om_i=0$.

\section{Detector properties: amplitude definition}\label{sect:amp}
 
In this section, we explain how to compute the correlations of the flow operators \re{eq:eventshapesQ} using scattering amplitudes or more precisely matrix elements of local operators with respect to on-shell states.  In the context of conformal field theory this can be done if the weak coupling expansion around the free field point exists. In this case we expect that scattering amplitudes
can be defined order-by-order in perturbation theory. Computing event shapes using scattering amplitudes may involve IR divergences at the intermediate steps but they cancel in the final answer. In this section we work to leading order in the coupling constant in a weakly coupled CFT in $d=4$ dimensions. All the relevant scattering amplitudes, or more precisely form factors, are the tree-level ones and no IR divergences arise. 

Using the completeness condition $\sum_X \ket{X}\bra{X}=\m{I}$, we can rewrite the Wightman correlation function \re{eq:quantumeventshapes} in the following form
\begin{align}\label{XX'}
\sum_{X,X'} \vev{0| \phi^\dagger(\tilde q) |X'\rangle \langle X'| \mathcal O_{J_1}(\hat\omega_1,n_1) \dots \mathcal O_{J_k}(\hat\omega_k,n_k) |X\rangle \langle X|  \phi(q)|0},
\end{align}
where the sum runs over the asymptotic states $X$ and $X'$ containing an arbitrary number of on-shell particles. 
The relation \re{XX'} has a simple physical meaning. The source $\phi(q)$ excites the vacuum and creates the state $X$.
This state propagates through a collection of detectors located at points $n_1,\dots, n_k$  on the celestial sphere,  undergoes a transition to the state $X'$, which in turn is absorbed by the sink $\phi^\dagger(\tilde q)$. 

To apply \re{XX'} we have to provide expressions for the on-shell form factors $\vev{X|\phi(q)|0}$ and $ \vev{0|\phi^\dagger(\tilde q)|X'}$ describing the source and sink, respectively, and to evaluate the transition amplitude $\langle X'| \mathcal O_{J_1}(\hat\omega_1,n_1) \dots \mathcal O_{J_k}(\hat\omega_k,n_k) |X\rangle$. We recall that the flow operator $\mathcal O_{J_i}(\hat\omega_i,n_i)$ represents a detector that selects particles 
 propagating in the direction of the null vector $n_i$ and assigns a certain weight to  them. For $\hat\omega=0$ the flow operator \re{eq:definitiondet} acts on a single-particle state with  momentum $p^\mu=(p_0,\vec p)$ as follows \cite{Belitsky:2013bja}, 
\begin{align}\label{O-zero-w}
\mathcal O_{J}(\hat\omega=0,n)\ket{p} = \ket{p} \delta^{(2)}(\Omega_{\vec p}-\Omega_{\vec n}) w_{J}\,.
\end{align}
Here the delta function ensures that the three-momentum of the particle is aligned along the detector direction $n=(1,\vec n)$ and the function $w_{J}$ defines the weight. It depends on the type of detector as well as on the energy of the particle. For $\hat\omega\neq 0$ the detector transfers the momentum $n \hat\omega$ and modifies the particle state to $\ket{p+n\hat\omega}$. In addition, for $\hat\omega\neq 0$ the detector can either create (for $\hat\omega>0$) or annihilate (for $\hat\omega<0$) particles. This suggests that for $\hat\omega\neq 0$ the expression on the right-hand side of \re{O-zero-w} should contain additional terms describing an inelastic contribution. We compute such terms below, see Eq.~\re{eq:detectorsplit}.  

Because the detectors are located at different points on the celestial sphere, they are separated by space-time intervals and, therefore, they cannot interact with each other by exchanging on-shell particles with  nonzero energy.
One might therefore expect that the detectors should work independently. If this were the case, the transition amplitude would factorize into a product of 
transition amplitudes describing individual detectors, symbolically
\begin{align}\label{no-cross-talk}
\langle X',p_1',\dots,p_k'| \mathcal O_{ J_1}(\hat\omega_1,n_1) \dots \mathcal O_{ J_k}(\hat\omega_k,n_k) |X,p_1,\dots,p_k\rangle 
\stackrel{?}{=}  \vev{X'|X} \prod_i \vev{p_i'|\mathcal O_{ J_i}(\hat\omega_i,n_1) | p_i}\,.
\end{align}
It turns out that this relation is violated already at $\hat\omega_i=0$, see Ref.~\cite{Belitsky:2013bja}. The reason for this is that the transition amplitude in \re{no-cross-talk} receives a contribution from particles with zero energy. They generate  cross-talk between the detectors, thus invalidating the argument leading to \re{no-cross-talk}. We show below that the same phenomenon happens for $\hat\omega_i\neq 0$
and identify the missing contribution    in \re{no-cross-talk}. 

\subsection{Warm up example: free scalar field}\label{sect:warm}

In this subsection, we consider the properties of various detectors (scalar, charge and energy) defined in \re{eq:definitiondet}. 
The detectors are located on the celestial sphere where the interaction between the particles is switched off. This allows us to restrict our consideration to free fields.
  
To start with, we define the  conformal operators $O_J=O_{\mu_1\dots 
\mu_J}\bar n^{\mu_1} \dots \bar n^{\mu_J}$  (the same projection appears in  \p{eq:definitiondet}) with spin $J=0,1,2$  built out of a free complex scalar field,  
\begin{align}\notag\label{O-scal}
& O_{J=0}=\bar\varphi\varphi\,,\qquad
\\[2mm]\notag
& O_{J=1} = i(\bar\varphi \partial_+ \varphi -\partial_+  \bar\varphi  \varphi)\,,\qquad
\\
& O_{J=2}= \frac16\lr{  \partial_+^2 \bar\varphi  \varphi-4\partial_+ \bar\varphi  \partial_+\varphi+ \bar\varphi \partial_+^2  \varphi}\,,
\end{align}
where  $\partial_+=\bar n^\mu \partial_\mu$ denotes the projected space-time derivative. The operators $O_{J=1}$ and $O_{J=2}$ coincide with the $U(1)$ current and stress-energy tensor, respectively, projected on the auxiliary null vector $\bar n$.

A free scalar field can be expanded over  creation and annihilation operators, 
\be\notag\label{a}
\varphi(x) = \int_{- \infty}^\infty {d^4 p \over (2 \pi)^4} 2\pi \delta_+(p^2) ( a_p \e^{- i p \cdot x} + b^{\dagger}_p \e^{i p \cdot x})\,,
\\
\bar\varphi(x) = \int_{- \infty}^\infty {d^4 p \over (2 \pi)^4} 2\pi \delta_+(p^2) ( b_p \e^{- i p \cdot x} + a^{\dagger}_p \e^{i p \cdot x})\,,
\ee
where $a_p$ and $b_p$ are the annihilation operators of the scalars with  $U(1)$ charge $(+1)$ and $(-1)$, respectively. They satisfy 
the canonical commutation relations 
\be\label{a-com}
[a_{p} , a_{p'}^\dagger] = [b_{p} , b_{p'}^\dagger] = (2 \pi)^{3} 2 |\vec p| \delta^{(3)} (\vec p - \vec p')\,,
\ee
the remaining commutators vanish.
Substituting \re{a} and \re{O-scal} into the definition of the detector (or flow operator) \re{eq:definitiondet} we obtain after some algebra (see Appendix~\ref{app:free} for details)
\be
\label{eq:scalardetector}
\cO_{J=0}(\hat \omega,n) 
&= (n  \bar n) \int_{-\infty}^{\infty} d \alpha \,e^{-i  (n  \bar n) \hat \omega \alpha}  \lim_{r \to \infty} r^2 (\bar\varphi\varphi)(r n + \alpha \bar n)   \nn 
\\
&= \int_0^\infty {d s \over 4(2 \pi)^3} \left(  2 a^{\dagger}_{(s + \hat \omega)n} a_{s n}  - a^{\dagger}_{sn} b^\dagger_{(\hat \omega -s)n}  - a_{sn} b_{-(s+\hat \omega )n}  \right) + (a\leftrightarrow b) \,.
\ee
The last term is obtained by exchanging the operators with  opposite values of the $U(1)$ charge.
As expected, this relation does not depend on $\bar n^\mu$. 
The integral on the second line contains the annihilation and creation operators of the scalars with their on-shell momenta aligned along the null vector $n$.  {The integration variable $s$, like the deformation parameter $\ho$,  has the dimension of energy.}

The relation \re{eq:scalardetector} holds for arbitrary $\hat\omega$ and it is tacitly assumed that the annihilation/creation operators with the subscript $s'n$ vanish for $s'<0$. In particular, for $\hat\omega=0$ the last two terms inside the parentheses in \re{eq:scalardetector} vanish. The first term in \eqref{eq:scalardetector} defines the elastic contribution
\begin{align}
\cO^{\text{elastic}} (\hat \omega,n) =\int_0^\infty {d s \over 2(2 \pi)^3} 
\lr{a^{\dagger}_{(s + \hat \omega)n} a_{s n}+ b^{\dagger}_{(s + \hat \omega)n} b_{s n}}\,.
\end{align}
Its action on the single-particle state $\ket{p} = a^\dagger _p\ket{0}$ is given by
\begin{align}\label{O-el}
\cO^{\text{elastic}} (\hat \omega,n)\ket{p} =\ket{p+n\hat\omega} \delta^{(2)}(\Omega_{\vec p}-\Omega_{\vec n})  \theta(p^0+\hat\omega) (p^0)^{-1} \,.
\end{align}
For $\hat\omega=0$ this relation is in agreement with \re{O-zero-w}.

An interesting new effect due to $\hat \omega \neq 0$ is the inelastic character of the last two terms in the parentheses in \eqref{eq:scalardetector}. They create and annihilate a pair of particles, correspondingly. To make this explicit, we rewrite \re{eq:scalardetector} as follows
\be
\label{eq:detectorsplit}
\cO_{J=0} (\hat \omega,n) &=\cO^{\text{elastic}}(\hat \omega,n) - \theta(\hat \omega) \cO^{\text{creation}}(\hat \omega,n) - \theta(-\hat \omega) \cO^{\text{annihil.}}(-\hat \omega,n),
\ee
where we introduced the notation
\be\notag\label{O-inel}
\cO^{\text{creation}}(\hat \omega,n) &= \int_0^{\hat \omega} {d s \over 2(2 \pi)^3} a^{\dagger}_{sn} b^\dagger_{(\hat \omega -s)n} 
\,,  \\
\cO^{\text{annihil.}}(\hat \omega,n) &= \int_0^{\hat \omega} {d s \over 2(2 \pi)^3}  a_{sn} b_{(\hat \omega - s)n}  \, \,.
\ee 
Note that these two operators are conjugate to each other.
When acting on the vacuum, the operator $\cO^{\text{creation}}(\hat \omega,n)$ creates a two-particle state that consists of two scalars with opposite $U(1)$ charges.
Their on-shell momenta are aligned along the detector direction $n$ and the total momentum is $n\hat\omega$. Similarly, $\cO^{\text{annihil.}}(\hat \omega,n)$ annihilates a pair of scalars. 

It is straightforward to repeat the same analysis for the two remaining operators in \re{O-scal}.
For the charge $\mathcal Q(\hat\omega,n)=\cO_{J=1}$ and energy $\mathcal E(\hat\omega,n)=\cO_{J=2}$ detectors we find
\begin{align}\notag\label{Q-det}
\mathcal Q(\hat\omega,n) =\int_0^\infty {ds\over 2(2\pi)^3} 
& \Big[w^{(s)}_Q(s,s+\hat\omega)   (a^\dagger_{(s+\hat\omega)n}a_{sn}  - b^\dagger_{(s+\hat\omega)n}b_{sn})
\\
& +  w^{(s)}_Q(s,s-\hat\omega) b^\dagger_{sn}  a^\dagger_{-(s-\hat\omega)n} 
+  w^{(s)}_Q(s,s+\hat\omega) b_{sn} a_{-(s+\hat\omega)n}\Big],
\\[2mm]\notag\label{E-det}
 \mathcal E(\hat\omega,n) =
\int_0^\infty {ds\over 2(2\pi)^3} 
\Big[ &  w^{(s)}_E(s,s+\hat\omega)   \lr{a^\dagger_{(s+\hat\omega)n}a_{sn}+b^\dagger_{(s+\hat\omega)n}b_{sn}}
\\\notag
 +& \frac12 w^{(s)}_E( s,\hat\omega-s)\lr{ a^\dagger_{sn}  a^\dagger_{-(s-\hat\omega)n}  + b^\dagger_{sn}  b^\dagger_{-(s-\hat\omega)n} }
\\
 +&\frac12 w^{(s)}_E( s,-s-\hat\omega) \lr{ a_{sn} a_{-(s+\hat\omega)n}+b_{sn} b_{-(s+\hat\omega)n}} \Big] \,.
\end{align}
Here $w^{(s)}_Q$ and $w^{(s)}_E$ are the weights assigned to a scalar particle going through the charge and energy detectors, respectively, 
\begin{align}\label{weights}
w^{(s)}_Q(s,s') = s+s' \,,\qqqquad w^{(s)}_E(s,s') = \frac16(s^2 + s'{}^2 +4s s')\,.
\end{align}
For the scalar detector the analogous weight is $w^{(s)}_O=1$.

We recall that $a^\dagger_{sn}$ and $b^\dagger_{sn}$ are creation operators of scalars with opposite values of the $U(1)$ charge and on-shell momentum $p^\mu=sn^\mu$ with $s\ge 0$. The relations \re{Q-det} and \re{E-det} have a form similar to the scalar detector \re{eq:detectorsplit}. The first line in the expression for $\mathcal Q(\hat\omega,n)$ and $\mathcal E(\hat\omega,n)$ describes the elastic contribution and the remaining terms describe the creation and annihilation of a pair of scalars with aligned momenta. 

The weights \re{weights} are homogenous polynomials in the energies $s$ and $s'$ of degree equal to the  detector spin.  
We encounter similar polynomials by examining the expression of the twist-two operator $O_J(x)$. This operator is built out of two scalar fields and $J$ derivatives $i(\bar n\partial)$ and it takes the following form~\cite{Braun:2003rp} 
\begin{align}
O_J(x) = \bar\varphi(x)
P_J\left(i(\bar n \!\!\stackrel{\leftarrow}{\partial}),i(\bar n\!\!\stackrel{\rightarrow}{\partial})\right) \varphi(x)\,,
\end{align}
where $P_J$ is the polynomial defined in \re{Geg} and the arrows indicate the fields that the derivatives act upon.
The close examination shows that the two polynomials coincide up to a normalization factor,
\begin{align}
w^{(s)}_Q(s,s')=-P_{J=1}(s,-s') \,,\qqqquad w^{(s)}_E(s,s')=\frac13 P_{J=2}(s,-s') \,.
\end{align} 
This relation is not accidental and has the following explanation.

For an arbitrary spin $J$, the weight can be defined as
\begin{align}\label{wJ}
\bra{0} O_J(x=0) \ket{s_1,s_2} = w_J(s_1,-s_2) \,,
\end{align}
where $\ket{s_1,s_2}= a^\dagger_{s_1n} b^\dagger_{s_2n}\ket{0}$ is a two-particle state.  
Replacing the scalar fields with plane waves \re{a} and evaluating \re{wJ} we arrive at $w^{(s)}_J(s_1,s_2) \sim P_J(s_1,-s_2)$. This relation is powerful because it allows us to generalize the expressions  obtained in 
\re{Q-det} and \re{E-det} by including the contribution of the fermions $(\psi)$ and the gauge $(g)$ fields,
\begin{align}\notag\label{T}
& O^{(\psi)}_{J=1} = \bar\psi \gamma_+ \psi\,,
\\
& O^{(\psi)+(g)}_{J=2} = {i\over 2} \left(\bar\psi \gamma_+ \partial_+ \psi- \partial_+ \bar\psi \gamma_+ \psi\right) + F_{+\mu} F_{+\mu}\,.
\end{align}
Like in \re{wJ}, we act with these operators on  two-particle states built out of fermions and gauge fields and identify the corresponding weights as
\begin{align}\label{w-psi}
w_Q^ {(\psi)} (s,s') = (s s')^{1/2}
\,,\qquad
w_E^ {(\psi)} (s,s') = \frac12(s+ s')(s s')^{1/2}\,,\qquad w_E^{(g)}(s,s') = s s'\,,
\end{align}
where the factor $(s s')^{1/2}$ comes from the wave functions of the fermions. 

Comparing these expressions with \re{weights} we observe an important difference from the scalar case. The variables $s$ and $s'$ have the meaning of the energy of the particles entering or leaving the detectors. Going to the limit $s\to 0$ we find that the weights \re{w-psi} corresponding to the fermions and gauge fields vanish, whereas the weights for the scalars, Eq.~\re{weights}, approach a finite value. This implies that scalars with zero energy can give a non-vanishing contribution to the flow operators. As was mentioned above, this leads to cross-talk between the detectors and to the violation of \re{no-cross-talk}. 

\subsection{One-point functions from amplitudes}
 
To illustrate the amplitude approach, we apply \re{XX'} to computing  the one-point function  
\begin{align}\label{1pt-amp}
\vev{\mathcal O_{J} (\hat\omega,n)}_q & = \int d^4 x\e^{-i(qx)}  \sum_{X,X'}  \vev{0|\phi(0)|X'\rangle \langle X'| \mathcal O_{J} (\hat\omega,n) |X\rangle \langle  X|\phi(x)|0}\,.
\end{align}
For the sake of simplicity we chose the source and sink to be described by an  operator $\phi=\bar\varphi\varphi$ of dimension $\Delta=2$.
~\footnote{For operators with arbitrary scaling dimension $\Delta$, the corresponding expression for the one-point function 
can be found  in Appendix~\ref{app:1pt}.}
In a free theory the states 
$X$ and $X'$ contain two scalars with opposite $U(1)$ charges, e.g. $\ket{X} =a^\dagger_{p_1} b^\dagger_{p_2}\ket{0}\equiv \ket{p_1,p_2}$ with $p_i^2=0$. The form-factor takes the simple form $\vev{X|\phi(x)|0}=\e^{ix(p_1+p_2)}$ and its substitution into \re{1pt-amp} yields $(2\pi)^4 \delta^{(4)}(q-p_1-p_2)$ upon integration over $x$. In this way, we obtain
\begin{align}
\vev{\mathcal O_{J} (\hat\omega,n)}_q & =(2\pi)^4 \int \prod_{a=\{1,2,1',2'\}}{d^4 p_a\over (2\pi)^3} \delta_+(p_a^2) \delta^{(4)}(q-p_1-p_2)
 \vev{ p_1'p_2'| \mathcal O_{J} (\hat\omega,n) |p_1,p_2 }\,,
\end{align}
where $\delta_+(p^2) = \theta(p^0)\delta(p^2)$. 

As explained in the previous subsection, the flow operator $\mathcal O_{J} (\hat\omega,n)$ is  a sum of elastic and inelastic terms. The inelastic terms change the number of particles and do not contribute to the matrix element $\vev{ p_1'p_2'| \mathcal O_{J} (\hat\omega,n) |p_1,p_2}$. The elastic contribution factorizes into a product of single-particle matrix elements,
\begin{align}\label{O-fac}
\vev{ p_1'p_2'| \mathcal O_{J} (\hat\omega,n) |p_1,p_2} = 
\vev{ p_1'| \mathcal O^{\rm elastic}_{J} (\hat\omega,n) |p_1}
\vev{p_2'|p_2}+\vev{p_2'| \mathcal O^{\rm elastic}_{J} (\hat\omega,n) |p_2}\vev{p_1'|p_1}\,,
\end{align}
where $\vev{p_i'|p_i}=(2\pi)^3 p_i^0 \delta^{(3)}(\vec p_i-\vec p_i')$. The operator $\mathcal O^{\rm elastic}_{J} (\hat\omega,n)$ acts on the single-particle state as
\begin{align}\label{elastic}
\mathcal O^{\rm elastic}_{J} (\hat\omega,n)\ket{p} = \ket{p+n\hat\omega} \delta^{(2)}(\Omega_{\vec p_1}-\Omega_{\vec n})  \theta(p^0+\hat\omega) (p^0)^{-1}  w_J  \,,
\end{align}
where the weight $w_J= w_J(p^0,p^0+\hat\omega) $ depends on the energy of the particle entering and leaving the detector.
The two terms on the right-hand side of \re{O-fac} give the same contribution to \re{1pt-amp}:
\begin{align} 
\vev{\mathcal O_{J} (\hat\omega,n)}_q & \sim \int {d^4 p_1\over (2\pi)^3} \delta_+(p_1^2) \delta_+((q-p_1)^2)
 \delta^{(2)}(\Omega_{\vec p_1}-\Omega_{\vec n})  \theta(p_1^0+\hat\omega) w^{(s)}_J(p^0_1,p_1^0+\hat\omega)  (p_1^0)^{-1}\,.
\end{align}
Here the product of delta functions $\delta_+(p_1^2)\delta^{(2)}(\Omega_{\vec p_1}-\Omega_{\vec n})$ aligns the momentum of the scalar with the detector direction, $p_1=s n$ with $s>0$.
Applying the identity
\begin{align}\label{id}
\int_0^\infty ds\, f(s) \delta^{(4)}(p-n s) = 2f(p^0) (p^0)^{-1}\delta_+(p^2) \delta^{(2)}(\Omega_{\vec p}-\Omega_{\vec n}) \,,
\end{align}
with $f(s)$ being a test function, we obtain
\begin{align} \notag
\vev{\mathcal O_{J} (\hat\omega,n)}_q & \sim \int_0^\infty ds\,\delta_+((q-ns)^2) w^{(s)}_J(s,s+\hat\omega)  \theta(s+\hat\omega)
\\
& = {1\over 4\pi} {(q^2)^J\over (qn)^{J+1}} w^{(s)}_J\left(1,1+\omega\right)\theta\left(1+\omega\right)\,.
\end{align}
Here in the second relation we introduced the dimensionless variable $\om$ defined in \p{eq:rescaledfreq} and took into account that $w^{(s)}_J(s,s+\hat\omega)$ is a homogenous function of degree $J$.  
Replacing the weights $w_J$ by their explicit expressions \re{weights}, we verify that $\vev{\mathcal O_{J} (\hat\omega,n)}_q$ has the expected form \re{1pt-exp}, with the   function $ f_{\cO_J}$  given by \re{f+} and \re{f-} evaluated at $\Delta_\phi=2$.

\subsection{Cross-talk}\label{sect:talk}

To demonstrate the phenomenon of cross-talk between the detectors, we consider the two-point functions $\vev{\mathcal O_{J_1}(\hat\omega_1,n_1)\mathcal O_{ J_2}(\hat\omega_2,n_2)}_q$ with the source and sink given by a complex scalar field,
\begin{align}\label{OO-sing}\nn
\vev{ \mathcal O_{J_1}(\hat\omega_1,n_1)\mathcal O_{J}(\hat\omega_2,n_2)}_q 
& \equiv \vev{ \varphi^\dagger (\tilde q) \mathcal O_{ J_1}(\hat\omega_1,n_1)  \mathcal O_{ J}(\hat\omega_2,n_2) \varphi(q)}  
\\[2mm]
& =\vev{0|a_{\tilde q} \,\mathcal O_{ J_1}(\hat\omega_1,n_1) \mathcal O_{ J}(\hat\omega_2,n_2)a^\dagger_q|0}\,,
\end{align}
where $\tilde q=q+n_1\hat \omega_1+n _2\hat\omega_2$ and $q^2=\tilde q^2=0$. The expression on the right-hand side coincides with the transition amplitude on the left-hand side of \re{no-cross-talk} with single-particle \textit{in} and \textit{out} states. If the relation \re{no-cross-talk} were correct, the above expression would be proportional to the vacuum expectation value of the flow operator and, therefore, it would vanish.
Let us show that this is not the case.

\begin{figure}
\centering
\includegraphics[scale=0.8]{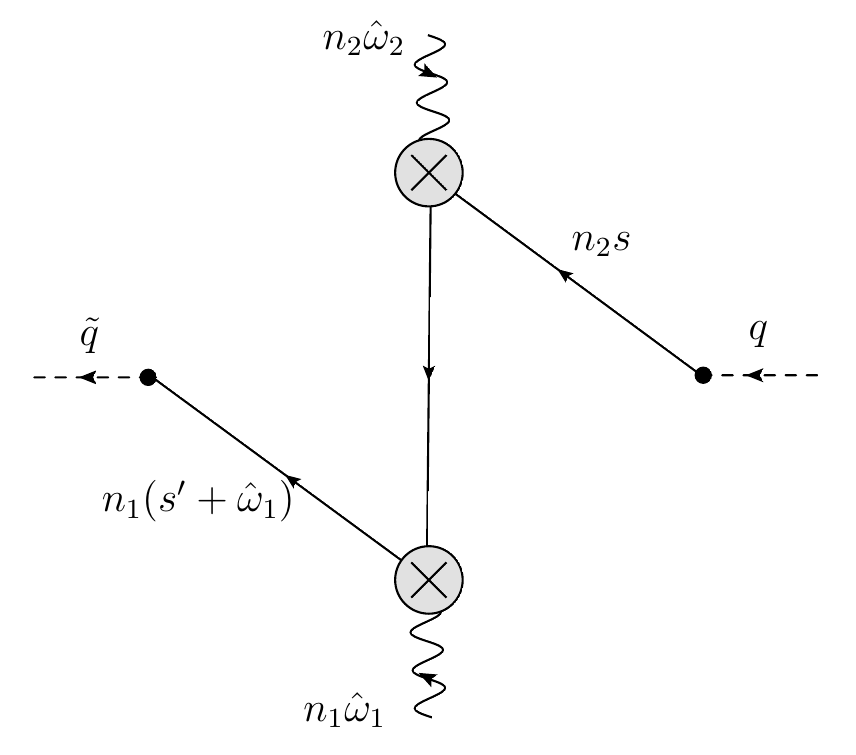} 
\caption{Cross-talk between the detectors. The grey blobs represent the detectors, the solid line depicts the propagator of a scalar particle, the arrows indicate the direction in which it propagates. }\label{Fig:x-talk} 
\end{figure}

As in the previous case, the inelastic terms in $\mathcal O_{J_1}(\hat\omega_1,n_1)$ and $\mathcal O_{J_2}(\hat\omega_2,n_2)$  do not contribute to \re{OO-sing}
for $n_1\neq n_2$. The contribution of the elastic terms can be found using \re{elastic}:
\begin{align}\notag\label{OO-el}
\vev{ \mathcal O_{J_1}(\hat\omega_1,n_1)\mathcal O_{J_2}(\hat\omega_2,n_2)}_q  = 
\int_0^\infty {ds ds'\over (2(2\pi)^3)^2} w_{J_2}(s,s+\hat\omega_2)w_{J_1}(s',s'+\hat\omega_1)
\\
  \times\vev{a_{\tilde q} a^\dagger _{(s'+\hat\omega_1)n_1}}
\vev{a_{s'n_1} a^\dagger_{(s+\hat\omega_2)n_2}}
\vev{a_{sn_2} a^\dagger_q}   + (\omega_1\leftrightarrow \omega_2,J_1\leftrightarrow J_2,n_1\leftrightarrow n_2)\,,
\end{align}
where the second term ensures the symmetry under the exchange of the detectors. The first term  describes the sequential propagation of a particle   through the two detectors,  as shown in Figure~\ref{Fig:x-talk}. Since the  detectors are located on the celestial sphere at two different directions, we expect that the energy of the particle exchanged between them should vanish. 

To show this we examine
\begin{align}\notag\label{limit}
\vev{a_{s'n_1} a^\dagger_{(s+\hat\omega_2)n_2}} & = (2\pi)^3 2 s \delta^{(3)}(s'\vec n_1-(s+\hat\omega_2)\vec n_2)
\\
& = (2\pi)^3 2 s \lim_{\epsilon\to 0}  {1\over (\pi\epsilon)^{3/2}}\e^{-(s'\vec n_1-(s+\hat\omega_2)\vec n_2)^2/\epsilon}
 =   {(2\pi)^2 \over (n_1n_2)}\delta(s')\delta(s+\hat\omega_2)\,,
\end{align}
where we used \re{a-com} and replaced $n_1=(1,\vec n_1)$ and $n_2=(1,\vec n_2)$ with $\vec n_1^2=\vec n_2 ^2=1$. To check the relation on the second line it is sufficient to integrate both sides against a test function. The expression on the right-hand side of \re{limit} is different from zero only for $\hat\omega_2<0$. Substituting \re{limit} into \re{OO-el} and taking into account \re{a-com} we obtain
\begin{align}\notag\label{OO-scal}
& \vev{ \mathcal O_{J_1}(\hat\omega_1,n_1)\mathcal O_{ J}(\hat\omega_2,n_2)}_q = {(2\pi)^2 \over (n_1n_2)}q^0 \tilde q^0& 
\\\nn
&\qquad \times 
\left[\delta^{(3)}(\vec q+\vec n_2\hat\omega_2)  \delta^{(3)}(\vec {\tilde q}-\vec n_1\hat\omega_1) 
w_{J_2}(-\hat\omega_2,0)w_{J_1}(0,\hat\omega_1)\theta(-\hat\omega_2)\theta(\hat\omega_1)\right.
\\
&\qquad \ \ + \left.
\delta^{(3)}(\vec q+\vec n_1\hat\omega_1)  \delta^{(3)}(\vec {\tilde q}-\vec n_2\hat\omega_2) 
w_{J_2}(0,\hat\omega_2)w_{J_1}(-\hat\omega_1,0)\theta(\hat\omega_2)\theta(-\hat\omega_1)\right].
\end{align}
Notice that this expression is different from zero for $\hat\omega_2$ and $\hat\omega_1$ of different signs. For instance, for $\ho_2<0$ and $\ho_1>0$, a particle with  on-shell momentum $q^\mu=(q^0,\vec q)$ enters the detector $\mathcal O_{J}(\hat\omega_2,n_2)$ and leaves it with  momentum $q+n_2\hat \omega_2=0$, or equivalently $s'=0$. The second detector $\mathcal O_{J_1}(\hat\omega_1,n_1)$ transfers the momentum $n_1\hat \omega_1$ and the particle reaches the sink with  momentum $\tilde q = n_1\hat \omega_1$. 

In the above analysis we chose the source and sink to be a scalar field. It is straightforward to show that the relation \re{OO-scal} also holds for fermion and gauge fields (up to an overall factor proportional to a power of $q^0 \tilde q{}^0$). However, comparing the relations \re{weights} and \re{w-psi}, we find 
that, unlike the scalars, the weights $w_J(0,s)$ and $w_J(s,0)$ vanish for fermions and gauge fields. Thus,  cross-talk can only be generated by the exchange of scalars with zero energy. As follows from \re{OO-scal}, its contribution is accompanied by a factor of $1/(n_1n_2)$ that becomes singular for $z\sim (n_1 n_2)\to 0$. Below we show  that an analogous phenomenon occurs in the interacting field theory. 

\subsection{Two-point functions in $\mathcal N=4$ SYM from amplitudes}\label{sub:cross}

In this subsection, we apply the technique described above to compute the two-point function \re{eq:twopointgeneral} in $\mathcal N=4$ SYM at weak coupling. 
This theory describes a gauge field coupled to four gauginos $\psi_i$ (with $i=1,\dots,4$) and six real scalars $\varphi_I$ ($I=1,\dots 6$), all in the adjoint representation of the gauge group $SU(N)$. The index $i$  belongs to the fundamental representation of the R-symmetry group  $SU(4)$, while $I$ is an index of the vector representation of $SO(6) \sim SU(4)$.

For the sake of simplicity, 
we choose the initial and final states, $\phi(q)\ket{0}$  and $\bra{0}\phi^\dagger (\tilde q)$, respectively, to be defined by the simplest  gauge invariant scalar operator of the form
\begin{align}\label{halfBPS}
\phi= Y^I Y^J \tr(\varphi_I \varphi_J)\,,\qqqquad  \phi^\dagger =\bar Y^I \bar Y^J \tr(\varphi_I \varphi_J)\,,
\end{align}
where $Y$ are auxiliary (complex) six-dimensional null vectors, $Y^IY^I =0$.  
The operator \re{halfBPS} is half-BPS and its scaling dimension $\Delta_\phi=2$ is protected from quantum corrections. 
It creates a pair of scalars out of the vacuum, which carries zero   total  color charge and nonzero R-charge corresponding to the representation $\bf 20'$ of the  $SU(4)$.
  
As mentioned above, we shall consider three different flow operators, scalar ($S$), charge $(Q)$ and energy $(E)$ operators. They are defined by local twist-two operators of spin $J=0,1,2$, respectively. The stress-energy tensor is given by the spin $J=2$ operators defined in \re{O-scal} and \re{T} with the only difference that the scalar and fermion fields are replaced by $\varphi_I$ and $\psi_i$, respectively. The scalar $J=0$ operator is defined as
\begin{align}\label{Os}
O_{J=0} (x,S) = S^{IJ} \tr(\varphi_I \varphi_J(x))\,,
\end{align}
where the symmetric traceless `polarization' tensor $S^{IJ}$ determines the orientation of the detector in the isotopic R-space. Like \re{halfBPS}, the operator \re{Os} is protected and lives in the same representation $\bf 20'$ of $SU(4)$. Finally, the spin $J=1$ operator is given by the R-current
\begin{align}\label{O1}
O_{J=1} (x,Q) = i Q^{IJ}  \tr[\varphi_I \partial_+ \varphi_J] + \dots\,,
\end{align}
where the charge polarization matrix $Q^{IJ}$ is an antisymmetric tensor in the adjoint representation $\bf 15$ of $SU(4)$. The dots denote terms involving gauge fields and gauginos, which do not contribute to the two-point correlations $\langle   \cO_{J_1} (\om_1, n_1)   \cO_{J_2} (\om_2, n_2)   \rangle_q$ to one-loop order at weak coupling. The operators \re{Os} and \re{O1} have  scaling dimension $\Delta=2+J$ and  twist $t=\Delta-J=2$.

According to \re{2pt-simp}, the two-point correlations are defined by the functions $f_{\cO_{J_1} \cO_{J_2} }$. At weak coupling, 
they admit an expansion in the powers of the 't Hooft coupling $a=g^2 N_c/(4\pi^2)$,
\begin{align}\label{weak-gen}
f_{\cO_{J_1} \cO_{J_2} }(\omega_1, \omega_2 , z) = f^{(0)}_{\cO_{J_1} \cO_{J_2} }(\omega_1, \omega_2 , z) + {a\over 4} f^{(1)}_{\cO_{J_1} \cO_{J_2} }(\omega_1, \omega_2 , z) + \dots\,,
\end{align}
where the first and the second terms describe the Born and the one-loop approximations, respectively. 
 
\begin{figure}[h!t]
\centering
\includegraphics[scale=0.7]{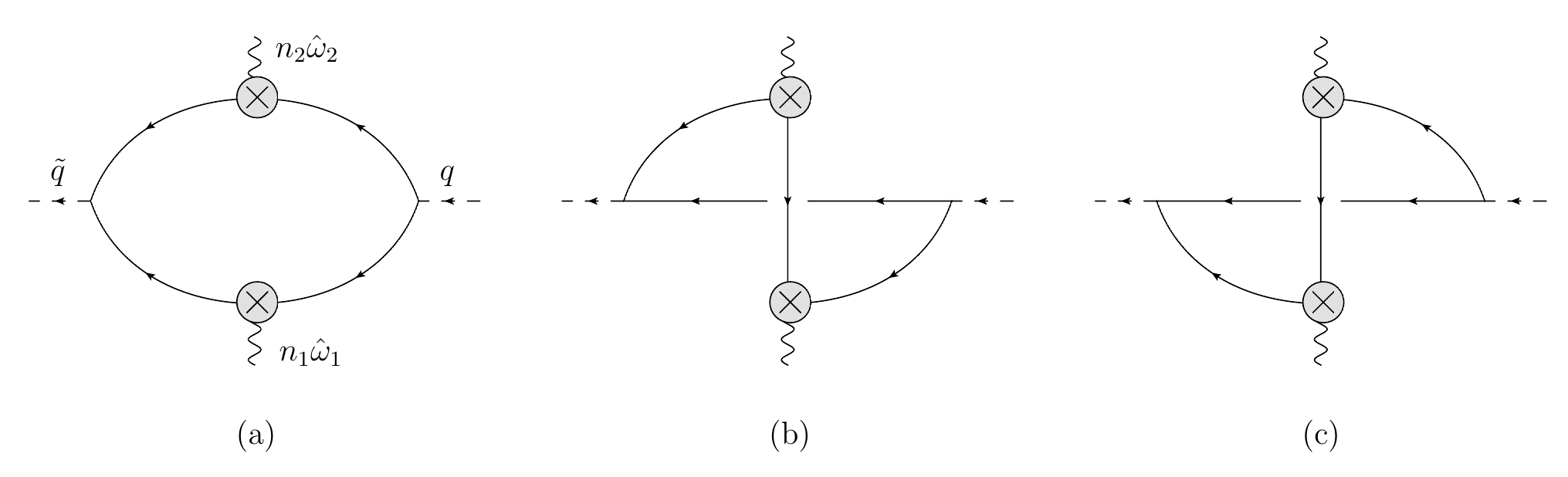} 
\caption{The two-point correlations $\vev{\mathcal O_{J_1}(\hat\omega_1,n_1)\mathcal O_{J_2}(\hat\omega_2,n_2)}_q$ in the Born approximation.}\label{Fig:born}
\end{figure}

In the Born approximation, the function $\vev{\mathcal O_{J_1}(\hat\omega_1,n_1)\mathcal O_{J_2}(\hat\omega_2,n_2)}_q ^{(0)}$ is given by a sum of two terms describing the two possible propagation channels of the two scalars to the final state: (i) each  scalar goes through one of the detectors (see Figure~\ref{Fig:born}(a)) and (ii) one of the scalars goes sequentially through the two detectors and  the other one remains undetected (see Figures~\ref{Fig:born}(b) and (c)).  The second channel generates  cross-talk between the detectors. 

Each of these contributions is accompanied by a R-symmetry factor arising from the contraction of the indices of the scalar fields in the source and sink with those of the detectors. In particular, for the scalar-scalar and scalar-charge detectors the first contribution comes with a factor of $(\bar Y S_1  Y)(\bar Y S_2  Y)$ and  $(\bar Y S_1  Y)(\bar Y Q_2  Y)$, respectively, where we used a shorthand notation for the contraction of the $SO(6)$ indices, $(\bar Y S_1  Y)=\bar Y^I S^{IJ} Y^J$ and similarly for the charge. For the energy detector, the corresponding isotopic polarization tensor is diagonal
in the $SO(6)$ indices,  $(E_i)^{IJ} =\delta^{IJ}$ and yields $(\bar Y E_i  Y)=(\bar Y Y)$. To simplify the formulae we normalize the $SO(6)$ tensors as follows,
\begin{align}
(\bar Y S  Y) = (\bar Y Q Y) = 1\,.
\end{align}
For the cross-talk contribution the R-symmetry factors look differently. For instance, for the scalar-scalar and scalar-charge correlations they take the form $(\bar Y Y) (\bar Y S_1 S_2 Y)$ and $(\bar Y Y) (\bar Y S_1 Q_2 Y)$, respectively, plus the same expressions with the detectors exchanged. As  mentioned above, the cross-talk between the detectors induces corrections that are
singular at $z=0$. We can eliminate them by imposing  additional conditions on the detector polarization tensors,
\begin{align}
S_1S_2=S_1S_2=0 \,,\qquad S_1Q_2=Q_2S_1=0\,.
\end{align}
Following Refs.~\cite{Belitsky:2013xxa,Belitsky:2013bja}, we choose 
\begin{align}\label{S1S2}
Y=(1,0,1,0,i,i)\,,\qquad S_1=\text{diag}(1,-1,0,0,0,0)\,,\qquad S_2=\text{diag}(0,0,1,-1,0,0)\,.
\end{align}
Notice that a similar condition cannot be imposed on the polarization tensor of the energy detector $(E_i)^{IJ} =\delta^{IJ}$. This means that the correlations involving the energy detector will necessarily get a  cross-talk contribution and, therefore, we expect them to contain   additional terms that are
singular for $z\to 0$.

The contribution of the first channel to the two-point correlation is
\begin{align}\notag\label{born}
&  \int {d^4 p_1 \over (2\pi)^3} \delta_+(p_1^2) \int {d^4 p_2 \over (2\pi)^3} \delta_+(p_2^2) (2\pi)^4 \delta^{(4)}(q-p_1-p_2) 
\\[1.5mm] 
&\times \prod_{i=1,2}  w^{(s)}_{J_i}(p_i^0,p_i^0+\hat\omega_i) 
\delta^{(2)}(\Omega_{\vec p_i}-\Omega_{\vec n_i})  
\theta(p_i^0+\hat\omega_i) (p_i^0)^{-1} \,,
\end{align}
where the first line contains an integral over the Lorentz invariant two-particle phase space and the second line 
contains the weights \re{weights} corresponding to a scalar going through the two detectors. 
The evaluation of \re{born} can be simplified by going to the rest frame of the source, $q^\mu = (Q,\vec 0)$. Then, $\vec p_1=-\vec p_2$ and $p_1^0=p_2^0=Q/2$, so that the integral does not vanish 
only if the detectors are located back-to-back, $\vec n_1=-\vec n_2$, or equivalently $z=1$. The calculation results in
\begin{align}\label{Born-1}
f^{(0)}_{\cO_{J_1} \cO_{J_2} }(\omega_1, \omega_2 , z)  =  \frac14\delta(1-z)  \prod_{i=1,2}  w^{(s)}_{J_i}(1,1+\omega_i)\,.   
\end{align}
This relation holds for all  detectors but the energy one. In the latter case, the   function  $f^{(0)}_{\cO_{J_1} \cO_{J_2} }$ receives an additional contribution due to the cross-talk 
between the detectors,
\begin{align}\notag\label{Born-2}
f^{(0)}_{\cO_{J_1} E}(\omega_1, \omega_2 , z)  &=  \frac14\delta(1-z)   w^{(s)}_{J_1}(1,1+\omega_1)w^{(s)}_E(1,1+\omega_2)
\\\notag
&+ {1\over z} w^{(s)}_{J_1}(1,0)w^{(s)}_E(1,0) \left[\delta(\omega_1-1)\delta(\omega_2+1) + \delta(\omega_1+1)\delta(\omega_2-1) \right]\,,
\\[2mm]
f^{(0)}_{E\cO_{J_1}}(\omega_1, \omega_2 , z)  &=f^{(0)}_{\cO_{J_1} E}(\omega_2, \omega_1 , z)  \,.
\end{align}
 {where the product of delta   functions originates from \re{OO-scal}. Namely, in the Born approximation the source creates a pair of scalar particles 
each having  energy $Q/2$ in the rest frame $q^\mu=(Q,\vec 0)$. The energy of one of the particles is absorbed by the detector and, therefore, the two-point correlation should be localized at $\omega_i=-1$. The second delta function arises because the other detector has to transfer the energy $Q/2$ to the same scalar particle before it reaches the sink. To avoid such contributions we  tacitly assumed elsewhere that $\omega_i>-1$.}

The relation \re{Born-2} holds for $0<z\le 1$ and it does not take into account the contact terms proportional to $\delta(z)$. Such contact terms have interesting properties and we shall discuss them elsewhere \cite{contact}. 
 
Turning on the interaction, we find that to order $O(a)$ the two-point correlation \re{XX'} receives contributions from the states $X$ and $X'$ containing up to four particles. Following the discussion in the previous subsection, we split $\vev{\mathcal O_{\hat\omega_1,J_1}(n_1)\mathcal O_{\hat\omega_2,J_2}(n_2)}_q$ into a sum of elastic and inelastic contributions. The former is given by \re{XX'} with $X$ and $X'$ containing the same number of particles. For $z\neq 1$, or equivalently $\vec n_1\neq -\vec n_2$ in the rest frame of the source, there are only three particles. The inelastic contribution takes into account the possibility for the detector to create or annihilate a pair of particles. As a result, the states $X$ and $X'$ can have $2$ or $4$ particles.

\subsubsection*{Elastic contribution}\label{sect:elastic}

In this case, the states $X$ and $X'$ in \re{XX'} contain three particles, either two scalars and a gluon, or a scalar and a pair of gauginos. Two of these particles with on-shell momenta $p_i$, $i=1,2$, go through the detectors which transfer to them the energy $\hat\omega_i$ and modify their momenta to $p_i'=p_i+n_i \hat\omega_i$.
The remaining undetected particle propagates from the source to the sink and has      momentum $p_3=q-p_1-p_2$. This results in the contribution
\begin{align}\notag\label{OO-amp-el}
\vev{\mathcal O_{J_1}(\omega_1,n_1)\mathcal O_{ J_2}(\omega_2,n_2)}_q^{\text{elastic}}
&= 2 \int {d^4 p_1 \over (2\pi)^3} \delta_+(p_1^2) \int {d^4 p_2 \over (2\pi)^3} \delta_+(p_2^2) \, 2\pi \delta_+((q-p_1-p_2)^2) 
\\[1.5mm] \notag
&\times \prod_{i=1,2}  w_{J_i}(p_i^0,p_i^0+\hat\omega_i) 
\delta^{(2)}(\Omega_{\vec p_i}-\Omega_{\vec n_i})  
\theta(p_i^0+\hat\omega_i) (p_i^0)^{-1} 
\\[1.5mm]
&\times A_{q\to p_1+p_2+p_3} \lr{ A_{\tilde q\to p_1'+p_2'+p_3} }^*\,,
\end{align}
 {where the first two lines contain the integral over the phase space of the three particles with on-shell momenta $p_1$, $p_2$ and $p_3=q-p_1-p_2$. The particles $p_1$ and $p_2$ enter the detectors located in the direction $\vec n_1$ and $\vec n_2$, respectively. }
The two factors in the last line of \re{OO-amp-el} describe the transition amplitudes $\ket{\text{source}} \to \ket{p_1,p_2,p_3}$ and
$\ket{p_1',p_2',p_3}\to \ket{\text{sink}} $. They coincide with the on-shell form factors of the operators \re{halfBPS},
\begin{align}\label{A}
A_{q\to p_1+p_2+p_3} = \vev{p_1,p_2,p_3|\phi(0)|0}\,,
\end{align}
and similarly for the sink.

The evaluation of \re{OO-amp-el} can be simplified by making use of the identity \re{id}, 
\begin{align}\notag\label{OO-amp-el1}
\vev{\mathcal O_{J_1}(\omega_1,n_1)\mathcal O_{ J_2}(\omega_2,n_2)}_q^{\text{elastic}}
= \frac12 \int_0^\infty ds_1 ds_2 \int {d^4 p_1 \over (2\pi)^3} \int {d^4 p_2 \over (2\pi)^3} 2\pi \delta_+((q-p_1-p_2)^2) 
\\[1.5mm]
\times  A_{q\to p_1+p_2+p_3} \lr{ A_{\tilde q\to p_1'+p_2'+p_3} }^* \prod_{i=1,2}  w_{J_i}(p_i^0,p_i^0+\hat\omega_i) 
\theta(p_i^0+\hat\omega_i) \delta^{(4)}(p_i-n_is_i)\,.
\end{align}
In this representation, it is manifest that the particle momenta $p_i=n_is_i$ are aligned with the null directions of the detectors and $s_i$ play the role of their energy. Switching to the dimensionless  integration variables $s_i\to s_iq^2/(2(qn_i))$ and   taking into account the homogeneity of the weights $ w_{J_i}$ we obtain from \re{OO-amp-el1}
\begin{align}\notag\label{OO-amp-el2}
& \vev{\mathcal O_{J_1}(\omega_1,n_1)\mathcal O_{ J_2}(\omega_2,n_2)}_q^{\text{elastic}}
\\[1.5mm]
& \sim \int_0^\infty ds_1 ds_2 \, \delta_+(1-s_1-s_1+s_1s_2 z) \theta(s_1+\omega_1)  \theta(s_2+\omega_2)  \mathcal M(s_1,s_2)\,.
\end{align}
Here the dimensionless parameters  $z$ and $\omega_i$ are defined in \re{z} and \re{eq:rescaledfreq}, respectively, and we denote
\begin{align}\label{M}
\mathcal M(s_1,s_2) =  A_{q\to p_1+p_2+p_3} \lr{ A_{\tilde q\to p_1'+p_2'+p_3} }^* \prod_{i=1,2}  w_{J_i}(s_i,s_i+\omega_i) \,.
\end{align}
The product of on-shell form factors is evaluated at 
$p_i^\mu =n_i^\mu s_i q^2/(2(qn_i))$ and $p_i' =p_i+n_i\omega_i$. 

As was already mentioned, to order $O(a)$ we have to consider two form factors \re{A} describing 
the transitions $\text{source} \to s+s+g$ and $\text{source}\to s+ \psi + \psi$, where $s$, $\psi$ and $g$ denote the scalar, gaugino and gluon on-shell states.
These form factors are given by (up to overall normalization factors)
\begin{align}\notag\label{ff}
& A(s(1)s(2)g^\lambda (3)) = \epsilon_\mu ^\lambda (p_3) \left[{p_1^\mu\over (p_1p_3)} - {p_2^\mu\over (p_2p_3)}\right] ,
\\
& A(\psi(1)\psi(2)s (3)) = \sqrt{2} {\vev{p_1 p_2}\over (p_1p_2)} = {2\sqrt{2}\over [p_2p_1]} \,,
\end{align}
where $s(i)$ denotes the scalar state with   momentum $k_i$ and similarly for the gaugino and the gluon. Here
$\epsilon_\mu ^\lambda (p_3)$ is the polarization vector of the gluon with helicity $\lambda=\pm 1$ and in the second relation we employed the spinor-helicity representation of the on-shell momenta, $2(p_1p_2) =\vev{p_1p_2}[p_2p_1]$.

Replacing the transition amplitudes $A_{q\to p_1+p_2+p_3}$ in \re{M} with their expressions \re{ff}, we have to specify which particles (scalars, gauginos or gluons) enter the detectors and replace the weights $w_{J_i}$ with the corresponding expressions \re{weights} and \re{w-psi}. 
As an example, consider the scalar-scalar correlation  
\begin{align}\label{SSC}
\vev{\mathcal O_{J_1=0}(\omega_1,n_1)\mathcal O_{J_2=0}(\omega_2,n_2)}_q = {1\over 8\pi^2  (q  n_1) (q  n_2)}  \text{SSC}(\omega_1, \omega_2 , z)
 \,.
\end{align}
In this case, the detector only selects the scalar particles and the second amplitude in \re{ff} does not contribute. The corresponding expression for the function \re{M} looks as
\begin{align}\notag\label{M-SSC}
\mathcal M_\text{SSC}(s_1,s_2) &= w^{(s)}_S(s_1,s_1') w^{(s)}_S(s_2,s_2')\sum_{\lambda=\pm 1} A(s(1)s(2)g^\lambda(3)) \widebar{ A(s(1')s(2')g^\lambda(3))}
\\
&= {4\over q^2} {s_1+s_2-1\over (1-s_1)(1-s_2)} \,.
\end{align}
Here we replaced $w^{(s)}_S=1$ and substituted the momenta $p_i=s_i n_i q^2/(2(qn_i))$ and $p_i'=s_i' n_i q^2/(2(qn_i))$  with $s_i'=s_i+\omega_i$. 

For the energy-scalar correlation
\begin{align}\label{ESC}
\vev{\mathcal O_{J_1=2}(\omega_1,n_1)\mathcal O_{J_2=0}(\omega_2,n_2)}_q={(q^2)^2 \over 8\pi^2(q  n_1)^3  (q  n_2)}  \text{ESC}(\omega_1, \omega_2 , z) \,,
\end{align}
the energy detector sees all the particles and the function \re{M} takes the form
\begin{align}\notag\label{M-ESC1}
\mathcal M_{\rm ESC} &= w^{(s)}_{E}(s_1,s'_1)  w^{(s)}_{S}(s_2,s'_2)  \sum_{\lambda=\pm 1} A(s(1)s(2)g^\lambda (3)) \widebar{A(s(1')s(2')g^\lambda (3)) }
\\\notag
&+w^{(g)}_{E}(s_1,s'_1)  w^{(s)}_{S}(s_2,s'_2)  \sum_{\lambda=\pm 1} A(s(3)s(2)g^\lambda (1)) \widebar{A(s(3)s(2')g^\lambda (1')) }
\\ 
 &+ w^{(\psi)}_{E}(s_1,s'_1)  w^{(s)}_{S}(s_2,s'_2)  \,A(\psi(1)\psi(3)s (2)) \widebar{A(\psi(1')\psi(3)s (2')) }\,,
\end{align}
where we changed the indices $i$ and $i'$ in the arguments of $A$ to specify the particles that go through the detectors.
Using the expressions for the energy weights, Eqs.~\re{weights} and \re{w-psi}, and going through the calculation we find from \re{M-ESC1}, term-for-term, 
\begin{align}\label{M-ESC}
\mathcal M_{\rm ESC} = \frac{z \left(6 s _1 \omega_1+6 s _1^2+\omega_1^2\right)}{6
   (1-z)}+\frac{\left(1-s _1 z\right){}^2}{(1-z) z}+\frac{\left(2 s _1+\omega
   _1\right) \left(1-s _1 z\right)}{1-z} \,,
\end{align}
where we used the delta function in \re{OO-amp-el2} to replace $s_2=(1-s_1)/(1-s_1z)$.   

Substituting \re{M-SSC} and \re{M-ESC} into \re{OO-amp-el2} we obtain the elastic contribution to the two-point functions \re{SSC} and \re{ESC},
\begin{align}\notag\label{2pt-elastic}
& \text{SSC}^{(1)}_{\rm elastic} =  {1\over 1-z} I_{\omega_1,\omega_2}(z)\,,
\\
& \text{ESC}^{(1)}_{\rm elastic} = {1+z  \omega_1 +\ft16 (z  \omega_1)^2 \over z^2(1-z)} I_{\omega_1,\omega_2}(z)\,,
\end{align}
where the superscript refers to the perturbative order in  \re{weak-gen}. The function
\begin{align}
I_{\omega_1,\omega_2}(z) = \int_0^1 {z ds_1\over 1-zs_1}
 \theta(s_1+ \omega_1)\theta\left({1+ \omega_2\over 1+z \omega_2}-s_1\right) 
\end{align}
depends on the angle variable $z$ and on the  energies $\omega_1$ and $\omega_2$ transferred by the detectors. 
It takes different forms depending on the signs of $\omega_i$:
\begin{align}\label{I-diff}
I_{\omega_1,\omega_2}(z) 
 =\left\{\begin{array}{ll}
 -\log(1-z)\,, &\ \   \omega_1>0\,, \omega_2>0  \,,
 \\[1.2mm]  -\log\left({1-z\over 1-z | \omega_2|}\right)\,, & \ \   \omega_1>0\,,  \omega_2< 0 \,,
 \\  -\log\left({1-z\over (1-z |  w_1|)(1-z |  \omega_2|)}\right)\,, & \ \  \omega_1<0\,,  \omega_2<0\,,
 \end{array}\right.
\end{align}
and satisfies the  relation $I_{\omega_1,\omega_2}(z)=I_{\omega_2,\omega_1}(z)$.
The following comments are in order.

It is straightforward to verify that the obtained expressions  \re{2pt-elastic} verify the crossing symmetry relations \re{eq:relationused} and \re{eq:relationusedGen}. 
For $\omega_1=\omega_2=0$ the inelastic contribution to the detectors vanishes and the relations \re{2pt-elastic} match the one-loop result for the two-point correlations previously found in Ref.~\cite{Belitsky:2013bja}. 

The elastic contributions to the correlations \re{2pt-elastic} are proportional to the same function $I_{\omega_1,\omega_2}(z)$ and they only differ by a rational prefactor. The same is true for all the remaining two-point correlations of flow operators that can be evaluated in a similar manner. To save space, we do not present their explicit expressions here. We shall encounter these expressions later in the paper when we compute the correlations using another approach based on  the correlation functions.  

We observed previously that the one-point correlations of the flow operators have different dependence on the detector energies for positive and negative $\omega$. It follows from \re{2pt-elastic} and \re{I-diff} that the same is true  for the two-point correlations.  

\subsubsection*{Inelastic contribution}\label{sect:inel}

In this subsection we compute the additional contribution to the correlations due to the possibility for the detectors $\mathcal O_{J}(\omega,n)$ to create (for $\omega>0$) and annihilate (for $\omega<0$) particles. 

We start with the scalar correlation \re{SSC}. As was already mentioned, 
depending on the signs of $\omega_1$ and $\omega_2$, we can distinguish between four different functions
$\text{SSC}_{\pm,\pm}$ and $\text{SSC}_{\pm,\mp}$, with $\text{SSC}_{++}$ corresponding to $\omega_1>0$ and $\omega_2>0$, etc. The
crossing symmetry relations \re{eq:relationused} allow us to limit the discussion to the two functions $\text{SSC}_{++}$ and $\text{SSC}_{+-}$.
The elastic contribution to these functions is given by \re{2pt-elastic} and \re{I-diff}. 

\begin{figure}[h!t]
\centering
\includegraphics[scale=0.8]{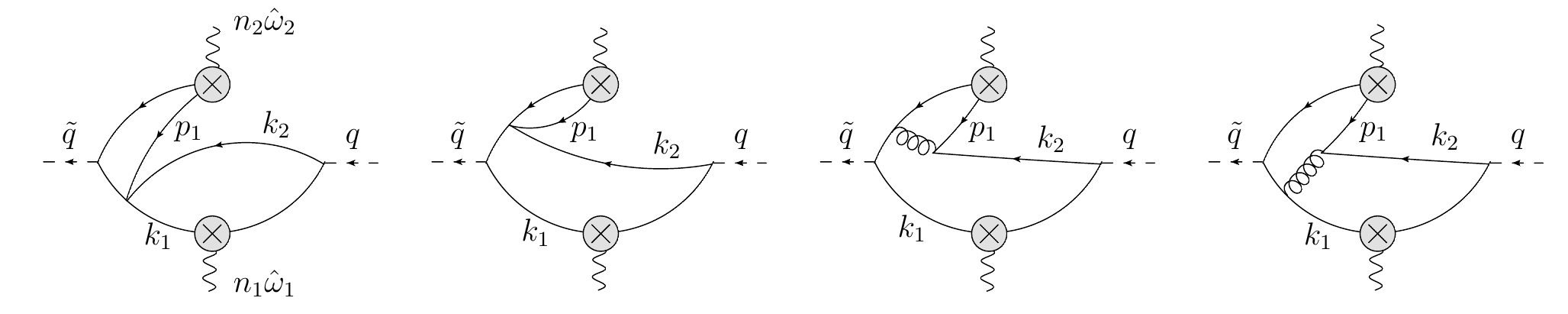} 
\caption{Creation of particles by the scalar detector $\mathcal O_{J=0}(\hat\omega_2,n_2)$.}\label{Fig:cr}
\end{figure}

To one-loop order, the creation of particles by the scalar detector $\mathcal O_{J=0}(\hat\omega_2,n_2)$ is described by the diagrams shown in Figure~\ref{Fig:cr}. 
The annihilation of particles is described by similar diagrams in which the initial and final states are swapped. These diagrams contain interaction vertices describing the quartic scalar coupling and the interaction of scalars with gluons in $\mathcal N=4$ SYM. The creation and annihilation of particles by the detector $\mathcal O_{J=0}(\hat\omega_1,n_1)$ is described by analogous diagrams. 

It is clear from Figure~\ref{Fig:cr} that to order $O(a)$ only one of the detectors can create/annihilate particles. This leads to the following expressions for the correlations: 
\begin{align}\notag\label{SSC-in1}
\text{SSC}_{++}^{\text{inelastic}} &=  \vev{\tilde q| ssss} \vev{ssss|\cO^{\text{elastic}}(\hat \omega_1,n_1)\cO^ {\text{creation}}(\hat \omega_2,n_2)|ss}\vev{ss|q} 
\\[2mm]
&+\vev{\tilde q| ssss} \vev{ssss|\cO^ {\text{creation}}(\hat \omega_1,n_1)\cO^ {\text{elastic}}(\hat \omega_2,n_2)|ss}\vev{ss|q} \,,
\\[2mm]\notag\label{SSC-in2}
\text{SSC}_{+-} ^{\text{inelastic}}&=  \vev{\tilde q| ss} \vev{ss|\cO^ {\text{elastic}}(\hat \omega_1,n_1)\cO^ {\text{annihil.}}(\hat \omega_2,n_2)|ssss}\vev{ssss|q} 
\\[2mm]
& + 
 \vev{\tilde q| ssss} \vev{ssss|\cO^ {\text{creation}}(\hat \omega_1,n_1)\cO^ {\text{elastic}}(\hat \omega_2,n_2)|ss}\vev{ss|q} \,, 
\end{align}
where $\vev{ssss|q}$ and $\vev{ss|q}$ are the transition amplitudes for the source going to two- and four-scalar states and  the detector operators were defined in Eqs.~\re{O-el} -- \re{O-inel}.

Let us consider the first term in the expression for $\text{SSC}_{++}$, Eq.~\re{SSC-in1}. According to \re{O-inel}, the operator $\cO^ {\text{creation}}(\hat \omega_2,n_2)$ creates a pair of scalars with   momenta $p_1=sn_2$ and $p_2=(\omega_2-s)n_2$, so that the four-particle state looks
as $\ket{s(p_1)s(p_2)s(k_1)s(k_2))}$. Here $k_1$ is the momentum of the scalar that leaves the detector $\cO^ {\text{elastic}}(\hat \omega_1,n_1)$. The incoming two-particle state is $\ket{s(k_1')s(k_2)}$ where $k_1'$ is the momentum of the scalar that enters the detector $\cO^ {\text{elastic}}(\hat \omega_1,n_1)$. The momenta $k_1$ and $k_1'$ should be aligned with the null vectors of the detectors, $k_1=E n_1$ and $k_1'=(E-\omega_1)n_1$. This leads to
\begin{align}\notag\label{1term}
& \vev{\tilde q| ssss} \vev{ssss|\cO^ {\text{elastic}}(\hat \omega_1,n_1)\cO^ {\text{creation}}(\hat \omega_2,n_2)|ss}\vev{ss|q} 
\\
& =   \int_0^{\omega_2} ds \int dE  \, \lr{A_{\tilde q\to s(p_1)s(p_2)s(k_1)s(k_2)}}^* A_{q\to s(k_1') s(k_2)} 2\pi \delta_+((q-k_1')^2)\,,
\end{align}
where the integration goes over the energy of the scalars $s(p_1)$ and $s(k_1)$.
Here we took into account that the scalar detector assigns a trivial weight to the scalar particles, $w^{(s)}_S=1$.
The delta function in \re{1term} comes from the on-shell propagator of the scalar $s(k_2)$ with momentum $k_2=q-k_1'$. 
Replacing $k_1'=(E-\omega_1)n_1$ we find that the on-shell condition $(q-k_1')^2=0$ leads to $E=(\omega_1+1) q^2/(2(qn_1))$ and fixes $k_1=En_1$ unambiguously.

The transition amplitudes entering \re{1term} are given by $A_{q\to s(k_1') s(k_2)}=1$ and 
\begin{align}\notag
A_{\tilde q\to s(p_1)s(p_2)s(k_1)s(k_2)}&= {1 \over (p_1+k_1+k_2)^2}  {-} {1\over (p_1+p_2+k_2)^2} 
\\
&+
{1\over 2(p_1k_2)}\left[{2(p_2(p_1-k_2))\over (p_1+p_2+k_2)^2 } - {2(k_1(p_1-k_2))\over (p_1+k_1+k_2)^2 }\right].
\end{align}
This relation takes into account the contribution of the diagrams shown in Figure~\ref{Fig:cr}. Substituting the particle momenta $p_1=sn_2$, $p_2=(\omega_2-s)n_2$, $k_1=n_1(\omega_1+1) q^2/(2(qn_1))$ and $k_2=q-k_1+n_1\hat\omega_1$, and performing the integration in \re{1term} we obtain
\begin{align}\notag\label{cre}
& \vev{\tilde q| ssss} \vev{ssss|\cO^{\text{elastic}}(\hat \omega_1,n_1)\cO^{\text{creation}}(\hat \omega_2,n_2)|ss}\vev{ss|q} 
\\
& \qquad
 \sim
  -\frac{ \left( \omega_1+1\right) z
  }{ (1-z) \left( \omega
   _1 z+1\right)}  \log \left(1+ \omega_1+ \omega_2+ \omega_2  \omega_1 z\over  \omega_1+1\right) \theta( \omega_2)\theta(1+\omega_1) \,.
\end{align}
The calculation of the first term in the expression for $\text{SSC}_{+-}$, Eq.~\re{SSC-in2}, goes along the same lines.
We have
\begin{align}\notag\label{2term}
& \vev{\tilde q| ss} \vev{ss|\cO^{\text{elastic}}(\hat \omega_1,n_1)\cO^{\text{annihil.}}(\hat \omega_2,n_2)|ssss}\vev{ssss|q} 
\\
& =  \int_0^{-\omega_2} ds\int d E \, \lr{A_{\tilde q\to s(k_1') s(k_2)}}^* A_{q\to s(p_1)s(p_2)s(k_1)s(k_2)}   2\pi \delta_+((\tilde q-k_1')^2)\,,
\end{align}
where $p_1=s n_1$ and $p_2=-\omega_2n_2-p_1$ are the momenta of the scalars absorbed by the detector $\cO^{\text{annihil.}}(\hat \omega_2,n_2)$. The momenta of the two remaining scalars are $k_1=E n_1$, $k_1'=n_1(E+\omega_1)$ and $k_2=\tilde q-k_1'$. As in the previous case, the relation $(\tilde q-k_1')^2=0$ fixes the energy $E=(1+\omega_2)/(1+z \omega_2) q^2/(2(qn_1))$. Going through the calculation of \re{2term} we find 
\begin{align}\notag\label{ann}
& \vev{\tilde q| ss} \vev{ss|\cO^{\text{elastic}}(\hat \omega_1,n_1)\cO^{\text{annihil.}}(\hat \omega_2,n_2)|ssss}\vev{ssss|q} 
\\
&\qquad \sim -\frac{ \left( \omega_2+1\right) z }{ (1-z) \left( \omega_2 z+1\right)} \log \left({1\over  \omega
   _2+1}\right)\theta( -\omega_2) \theta( 1+\omega_2) \,.
\end{align}
Note that this relation is independent of $\omega_1$.

Substituting the relations \re{cre} and \re{ann} into \re{SSC-in1} and \re{SSC-in2}, we  obtain the inelastic contribution to the scalar-scalar correlations.
Combining it with the elastic contribution \re{2pt-elastic} we finally arrive at
\begin{align}\label{SSC++}\notag
\text{SSC}^{(1)}_{++}(\omega_1,\omega_2,z)
&= \frac{1}{1-z}\log
   \left(\frac{1}{1-z}\right)
\\&\notag
+\frac{(\omega_1+1) z  }{2(1-z) (\omega_1 z+1)}
   \log \lr{\omega_1+\omega_2+\omega_1 \omega_2
   z+1\over \omega_1+1}
\\ &   
   +\frac{(\omega_2+1) z }{2(1-z) (\omega_2 z+1)}
    \log \lr{\omega_1+\omega_2+\omega_1 \omega_2
   z+1\over \omega_2+1}\,,
   \\[2mm]\label{SSC+-}
   \text{SSC}^{(1)}_{+-}(\omega_1,\omega_2,z) &=\frac{\log \left(\frac{1+\omega_2 z}{1-z}\right)}{1-z}+\frac{\left(\omega_2+1\right) z
   }{2 (1-z) \left(\omega_2 z+1\right)}\log \left(\frac{\omega_1+\omega_2+\omega_2 \omega_1 z+1}{\left(\omega
   _2+1\right)^2}\right)\,.
\end{align}
Here the first terms in both relations correspond to the elastic contribution and the remaining terms describe the inelastic contribution. The relations \re{SSC++} and \re{SSC+-} are valid for $0< z<1$ and they do not take into account contact terms of the types $\delta(z)$ and $\delta(1-z)$. 

For $\omega_1=\omega_2=0$ the inelastic contribution to \re{SSC++} and \re{SSC+-} vanishes and one recovers the known result for the SSC derived in Ref.~\cite{Belitsky:2013bja}.
We observe that the above relations are given by a linear combination of logarithms with rational coefficients. To make this property manifest, it is convenient to assign weight $1$ to the logarithm and weight $0$ to the rational functions. Then, all the terms in the expressions for the correlations have the same weight $1$. This  is yet another manifestation of the uniform weight property previously observed for various quantities in $\mathcal N=4$ SYM. 

Applying the crossing symmetry relations \re{eq:relationused} we can find from \re{SSC++} and  \re{SSC+-} the two remaining scalar-scalar correlations,
\begin{align}\label{SSC--}
\text{SSC}^{(1)}_{--}(\omega_1,\omega_2,z)
&= \frac{1}{1-z}\log
   \left(\frac{(1+\omega_1 z)(1+\omega_2 z)}{1-z}\right)
\\&\notag
+\frac{(\omega_1+1) z  }{2(1-z) (\omega_1 z+1)}
   \log \lr{1\over \omega_1+1} 
   +\frac{(\omega_2+1) z }{2(1-z) (\omega_2 z+1)}
    \log \lr{ 1\over \omega_2+1}\,,
   \\[2mm]\label{SSC-+}
   \text{SSC}^{(1)}_{-+}(\omega_1,\omega_2,z) &=\text{SSC}^{(1)}_{+-}(\omega_2,\omega_1,z)\,.
\end{align}
Here the last relation can be obtained by taking into account the symmetry of the SSC under the exchange of the detectors. 

Comparing the above expressions we observe an interesting relation,
\begin{align}\label{mag}
\text{SSC}^{(1)}_{++}(\omega_1,\omega_2,z)+\text{SSC}^{(1)}_{--}(\omega_1,\omega_2,z)=\text{SSC}^{(1)}_{+-}(\omega_1,\omega_2,z)+\text{SSC}^{(1)}_{-+}(\omega_1,\omega_2,z)\,,
\end{align}
where it is tacitly assumed that the functions are continued from the domain of their validity (positive or negative $\omega$'s)  to arbitrary real $\omega_1$ and $\omega_2$. We elucidate the origin of this relation in Section~\ref{sect:id} below. We will show that 
it is rather general and holds for the various correlations both at weak and strong coupling.

It is straightforward to extend the above analysis to the correlations involving charge and energy detectors. These detectors receive contributions from all the particles (scalars, gauginos and gluons). Due to the growing number of relevant diagrams, the calculation is more involved. To save space we do not present it here. In the next section, we introduce a more efficient approach to computing the various correlations based on correlation functions. We have checked that the two approaches yield the same expressions for the correlations.

\section{Generalized event shapes with scalar detectors}
\label{sec:SSCcorrelator}

In this section, we consider the four-point Wightman function of scalar primaries. We introduce the Mellin representation
for the connected part of this function and use it  
to compute the $\omega$-deformed
event shapes \re{eq:eventshapesQ}. This section generalizes the analysis of 
\cite{Belitsky:2013xxa,Belitsky:2013bja} to $\omega_i \neq 0$.

Following \cite{Belitsky:2013bja}, we consider the  four-point function of ${\bf 20}'$ operators in ${\cal N}=4$ SYM 
\be
\label{eq:fourpoint}
&\langle \phi^\dagger (x_4) O(x_1,S_1) O(x_2,S_2) \phi(x_3) \rangle
= \frac{G^{\bf (105)}(u,v) }{ x_{12}^4 x_{34}^4} \,,  
\ee
where the explicit expressions for the operators can be found in 
\eqref{halfBPS}, \re{Os} and \eqref{S1S2}, and the two conformal cross-ratios are defined as
\begin{align}
 u = {x_{12}^2 x_{34}^2 \over x_{13}^2 x_{24}^2}\,,\qqqquad v = {x_{14}^2 x_{23}^2 \over x_{13}^2 x_{24}^2} \ \,.
\end{align}
In \p{eq:fourpoint} the notation  $G^{\bf (105)}(u,v)$ emphasizes that with the choice of R-symmetry polarizations 
 \eqref{S1S2}  only the ${\bf 105}$ representation of $SU(4)$ appears in the OPE of the detector
operators $ O(x_1,S_1) O(x_2,S_2)$. This particular representation played a privileged role in \cite{Belitsky:2013xxa} because for $\ho_i=0$ it is very simply related to the energy-energy correlation,  which is the observable of our prime interest. 

It is convenient to write $G^{\bf (105)}(u,v)$ as follows
\be
\label{eq:fourpointneq4g105}
G^{\bf (105)}(u,v)&=
 \frac{c_T}{2(2\pi)^4}
 \Big( u^2 + \frac{u^2 }{ v^2} \Big)  + \frac{1}{(2\pi)^4} \frac{u^2 }{v} \Big( \frac 1 2 + u \Phi(u,v) \Big),
\ee
where the central charge $c_T$ is given by
\be
c_T &= \frac{N_c^2-1}{4}.
\ee
Let us briefly comment on the structure of \eqref{eq:fourpointneq4g105}. The rational term $\frac{c_T}{2(2\pi)^4} \Big( u^2 + \frac{u^2 }{ v^2} \Big)$ is the disconnected part of the correlation function.  {It does not contribute to the generalized event shapes at separated points ($z \neq 0$) and can be safely discarded.}
 The other rational  term $\frac{1}{(2\pi)^4} \frac{u^2 }{2v}$ is the connected correlator at zero coupling (i.e. Born level or free theory). Finally, the most important for our purposes part of  \p{eq:fourpointneq4g105} is the function $\Phi(u,v)$. At weak coupling it encodes the perturbative corrections, i.e.  it is proportional  the 't Hooft coupling constant  $a \equiv  {g_{\mathrm{YM}}^2 N_c}/{(4\pi^2)}$. This function  satisfies the crossing symmetry relations
\be
\label{eq:crossingphi}
\Phi(u,v)= \Phi(v,u) = \frac{1 }{ v}\Phi\left( \frac{u }{ v} , \frac{1 }{ v} \right) \,.
\ee

The starting point of our discussion is the Mellin representation of the  function $\Phi(u,v)$ in the form used in 
\cite{Belitsky:2013xxa,Belitsky:2013bja} 
\be
\label{eq:MellinformB}
\Phi(u,v) &=\int_{{\cal C}_0} {d j_1 d j_2 \over (2 \pi i)^2} \left[ \Gamma(1-j_1) \Gamma(1-j_2) \Gamma(j_1+j_2)\right]^2 M(j_1, j_2) u^{j_1} v^{j_2} \ ,
\ee
where the integration contour ${\cal C}_0$ runs parallel to the imaginary axis and satisfies $ {\rm Re}(j_1 + j_2) >-1$ and ${\rm Re} (j_1),{\rm Re} (j_2) <0$. 
The crossing symmetry \p{eq:crossingphi} implies that
\be
M(j_1, j_2) = M(j_2 , j_1) = {(j_1+ j_2)^2 (j_1+ j_2+1)^2 \over j_1^2 (1+j_1)^2} M(-1-j_1-j_2, j_2) \,.
\ee
Equivalently, if we write 
\be
\label{eq:tildeM}
M(j_1, j_2) = (j_1+ j_2)^2 (j_1+ j_2+1)^2 \tilde M(j_1, j_2)
\ee
then $\tilde M(j_1 , j_2)$ is
fully crossing-symmetric $\tilde M(j_1, j_2) = \tilde M(j_2 , j_1) = \tilde M(-1-j_1-j_2, j_2)$. 

The leading weak and strong coupling results take the form
\be
\label{eq:Mellin amplitudes at weak and strong coupling}
M^{\text{weak}}(j_1, j_2) &= - {a \over 4} {(j_1 + j_2)^2 \over j_1^2 j_2^2} , \nn \\
M^{\text{strong}}(j_1, j_2) &= - {1 \over 2}{(j_1 + j_2)^2 (1+j_1 +j_2) \over j_1 j_2} \ \,.
\ee

Using the $\cN=4$ supersymmetry Ward identities, similar but more complicated expressions can be written for other four-point functions, in particular those with the insertions of the R-symmetry current and the stress-energy tensor at the detector  points 1 and 2 (see Section~\ref{sN4WI}). 

Our next task is to go from the Mellin representation of the correlator to the Mellin representation of generalized event shapes. This amounts to three steps:
\begin{enumerate}
\item Take the detector limit \eqref{eq:definitiondet}. 
\item Fourier transform with respect to the detector working time. 
\item Fourier transform with respect to the position of the sink.
\end{enumerate}
In the Mellin representation of the correlator  \eqref{eq:MellinformB} all the dependence on the positions of the operators is encoded in the factor $u^{j_1} v^{j_2}$,  to which we apply the steps above. All the dynamical information (i.e., the coupling dependence)  on the other hand is contained in the Mellin amplitude $M(j_1, j_2)$ which factors out. This computation is a straightforward generalization of the one done in \cite{Belitsky:2013xxa} and we refer the reader to appendix \ref{sec:derivationSSC} for the details. The result takes the following form
\be\label{4.10}
\langle  \cO ( \omega_1, n_1) \cO( \omega_2 , n_2) \rangle_q &= {1 \over 8 \pi^2}  {\text{SSC} (\omega_1, \omega_2, z) \over (q n_1) (q n_2)} ,  
\ee
where $\text{SSC} (\omega_1, \omega_2 , z)$ is the convolution of the Mellin amplitude of the four-point correlation function $M(j_1, j_2)$ and the detector kernel $K_{\text{SS}}$
\begin{align} \label{M*K}
\text{SSC} (\omega_1, \omega_2 , z) &=  \int_{{\cal C}_0} {d j_1 d j_2 \over (2 \pi i)^2}  M(j_1, j_2) K_{\text{SS}}(j_1, j_2 | \omega_1, \omega_2,z) \,.
\end{align}
In the undeformed case $\omega_i=0$  the detector kernel $K_{\text{SS}}$  is very simple  \cite{Belitsky:2013bja}:
\be
\label{eq:zeroores}
K_{\text{S} \text{S}}(j_1, j_2 |0, 0, z)=\norm   {2\pi \over \sin \pi (j_1+j_2)} z^{-j_1 - j_2}\left(1 - z  \right)^{j_1 + j_2 - 1}\ \,.
\ee
Let us now see how this simple result changes when $\omega_i \neq 0$.

We start by quoting the result for the detector kernel in the most compact form. As discussed earlier we consider separately the cases where the detector frequencies have the same sign $\om_i \geq 0$ (we denote this case by $++$), and the case when they have  opposite signs $\om_1 \geq 0$, $\om_2 \leq 0$ (we denote this case by $+-$).  The result takes the form (see Appendix~\ref{sec:derivationSSC} for details)
\be
\label{eq:kernelSSCpm} 
K_{\text{S} \text{S}}^{+-}(j_1, j_2 | \om_1, \om_2, z) &= \norm  {1 \over 2(1 -z)} \Big( -{z \om_1 \om_2 \over 1 + \om_2} \Big)^{-\Jone} \Big( {z(1+ \om_2) \over 1- z} \Big)^{-\Jtwo} {\Gamma(1-\Jtwo)^2 \Gamma(\Jtwo+\Jone) \over \Gamma(1+\Jone - \Jtwo)}\nn \\
&\times \ _2 F_1 \left(\Jone, 1-\Jtwo , 1+\Jone - \Jtwo, {(1+\om_1)(1 + \om_2)  \over \om_1 \om_2 (1-z)}\right) \, ,
\\[2mm]
\label{eq:kernelSSCpp}
K_{\text{S} \text{S}}^{++}(j_1, j_2 | \om_1, \om_2, z) &=\norm   {1 \over 2(1 -z)} \Big( {z \om_1 \om_2 \over 1 + \om_2} \Big)^{-\Jone} \Big( {z( 1+ \om_2) \over 1- z} \Big)^{-\Jtwo} \Gamma(1-\Jtwo) \Gamma(1-\Jone) \Gamma(\Jtwo+\Jone) \nn \\
&\times \ _2 F_1 \left(\Jone, 1-\Jtwo , 1, 1 - {(1+\om_1)(1 + \om_2)  \over \om_1 \om_2 (1-z)}\right) \, \,.
\ee
If  one of  the frequencies vanishes, the kernels  become  simply related to the undeformed one in \p{eq:zeroores}, e.g.
\be
\label{eq:onezeroKSCC}
K_{\text{S} \text{S}}^{+-}(j_1, j_2 | 0, \om_2, z) =K_{\text{S} \text{S}}^{++}(j_1, j_2 | 0, \om_2, z) = K_{\text{S} \text{S}}(j_1, j_2 | 0, 0, z) (1+ \om_2)^{-j_1}.
\ee
While the above formulas are rather compact, they obscure some of the  properties of the kernels which are useful in the subsequent computation of the Mellin integral in \p{M*K}. Below we present another representation of the kernels which makes these properties manifest. 

\subsection{SSC kernel: opposite signs}\label{s4.1}

\begin{figure}[t!]
 \centering
\begin{tikzpicture}[scale=0.40]
	\begin{pgfonlayer}{nodelayer}
		\node [style=none] (3) at (2, 7) {};
		\node [style=none] (5) at (2, 7.5) {};
		\node [style=none] (43) at (3, 9) {};
		\node [style=none] (44) at (3, 5.5) {};
		\node [style=none] (45) at (3, 12.5) {};
		\node [style=none] (46) at (3, 9) {};
		\node [style=none] (47) at (3, 19) {};
		\node [style=none] (48) at (3, -1) {};
		\node [style=none] (49) at (3, 1.5) {};
		\node [style=none] (50) at (3, 3.5) {};
		\node [style=none] (51) at (3, 17) {};
		\node [style=none] (52) at (3, 17) {};
		\node [style=none] (53) at (3, 17) {};
		\node [style=none] (77) at (2.5, 7.5) {};
		\node [style=none] (78) at (2.5, 7) {};
		\node [style=none] (79) at (3.5, 7) {};
		\node [style=none] (80) at (3.5, 7.5) {};
		\node [style=none] (81) at (4, 7.5) {};
		\node [style=none] (82) at (4, 7) {};
		\node [style=none] (83) at (5, 7) {};
		\node [style=none] (84) at (5, 7.5) {};
		\node [style=none] (85) at (5.5, 7.5) {};
		\node [style=none] (86) at (5.5, 7) {};
		\node [style=none] (87) at (6.5, 7) {};
		\node [style=none] (88) at (6.5, 7.5) {};
		\node [style=none] (89) at (7, 7.5) {};
		\node [style=none] (90) at (7, 7) {};
		\node [style=none] (91) at (8, 7) {};
		\node [style=none] (92) at (8, 7.5) {};
		\node [style=none] (93) at (8.5, 7.5) {};
		\node [style=none] (94) at (8.5, 7) {};
		\node [style=none] (95) at (4.5, 14.5) {};
		\node [style=none] (96) at (4.5, 15) {};
		\node [style=none] (97) at (5, 15) {};
		\node [style=none] (98) at (5, 14.5) {};
		\node [style=none] (99) at (6, 14.5) {};
		\node [style=none] (100) at (6, 15) {};
		\node [style=none] (101) at (6.5, 15) {};
		\node [style=none] (102) at (6.5, 14.5) {};
		\node [style=none] (103) at (7.5, 14.5) {};
		\node [style=none] (104) at (7.5, 15) {};
		\node [style=none] (105) at (8, 15) {};
		\node [style=none] (106) at (8, 14.5) {};
		\node [style=none] (107) at (9, 14.5) {};
		\node [style=none] (108) at (9, 15) {};
		\node [style=none] (109) at (9.5, 15) {};
		\node [style=none] (110) at (9.5, 14.5) {};
		\node [style=none] (111) at (10.5, 14.5) {};
		\node [style=none] (112) at (10.5, 15) {};
		\node [style=none] (113) at (11, 15) {};
		\node [style=none] (114) at (11, 14.5) {};
		\node [style=none] (115) at (-3.25, 11) {};
		\node [style=none] (116) at (-3.25, 11.5) {};
		\node [style=none] (117) at (-2.75, 11.5) {};
		\node [style=none] (118) at (-2.75, 11) {};
		\node [style=none] (119) at (-1.75, 11) {};
		\node [style=none] (120) at (-1.75, 11.5) {};
		\node [style=none] (121) at (-1.25, 11.5) {};
		\node [style=none] (122) at (-1.25, 11) {};
		\node [style=none] (123) at (-0.25, 11) {};
		\node [style=none] (124) at (-0.25, 11.5) {};
		\node [style=none] (125) at (0.25, 11.5) {};
		\node [style=none] (126) at (0.25, 11) {};
		\node [style=none] (127) at (1.25, 11) {};
		\node [style=none] (128) at (1.25, 11.5) {};
		\node [style=none] (129) at (1.75, 11.5) {};
		\node [style=none] (130) at (1.75, 11) {};
		\node [style=none] (131) at (2.75, 11) {};
		\node [style=none] (132) at (2.75, 11.5) {};
		\node [style=none] (133) at (3.25, 11.5) {};
		\node [style=none] (134) at (3.25, 11) {};
		\node [style=textdot] (135) at (2.5, 6.5) {$0$};
		\node [style=textdot] (136) at (5.5, 6.5) {$2$};
		\node [style=textdot] (137) at (4, 6.5) {$1$};
		\node [style=textdot] (138) at (7, 6.5) {3};
		\node [style=textdot] (139) at (8.5, 6.5) {$4$};
		\node [style=textdot2] (140) at (1.5, 10.5) {$-1-j_2$};
		\node [style=textdot2] (141) at (3.5, 10.5) {$-j_2$};
		\node [style=textdot2] (142) at (0, 12) {$-2-j_2$};
		\node [style=textdot2] (144) at (-1.5, 10.5) {$-3-j_2$};
		\node [style=textdot2] (145) at (-3, 12) {$-4-j_2$};
		\node [style=textdot2] (146) at (4.75, 14) {$1-j_1-j_2$};
		\node [style=textdot2] (147) at (10.75, 14) {$5-j_1-j_2$};
		\node [style=textdot2] (148) at (9.25, 15.5) {$4-j_1-j_2$};
		\node [style=textdot2] (149) at (7.75, 14) {$3-j_1-j_2$};
		\node [style=textdot2] (150) at (6.25, 15.5) {$2-j_1-j_2$};
		\node [style=textdot] (151) at (14, 18) {};
		\node [style=textdot] (152) at (14, 18) {$s$};
		\node [style=none] (153) at (13.5, 18.25) {};
		\node [style=none] (154) at (13.5, 17.5) {};
		\node [style=none] (155) at (14.25, 17.5) {};
	\end{pgfonlayer}
	\begin{pgfonlayer}{edgelayer}
		\draw [bend left=90, looseness=2.25] (44.center) to (43.center);
		\draw [bend right=90, looseness=2.25] (46.center) to (45.center);
		\draw (45.center) to (47.center);
		\draw (44.center) to (48.center);
		\draw [style=arrow end] (49.center) to (50.center);
		\draw [style=arrow end] (45.center) to (53.center);
		\draw (3.center) to (77.center);
		\draw (5.center) to (78.center);
		\draw (79.center) to (81.center);
		\draw (80.center) to (82.center);
		\draw (83.center) to (85.center);
		\draw (84.center) to (86.center);
		\draw (87.center) to (89.center);
		\draw (88.center) to (90.center);
		\draw (91.center) to (93.center);
		\draw (92.center) to (94.center);
		\draw (95.center) to (97.center);
		\draw (96.center) to (98.center);
		\draw (99.center) to (101.center);
		\draw (100.center) to (102.center);
		\draw (103.center) to (105.center);
		\draw (104.center) to (106.center);
		\draw (107.center) to (109.center);
		\draw (108.center) to (110.center);
		\draw (111.center) to (113.center);
		\draw (112.center) to (114.center);
		\draw (115.center) to (117.center);
		\draw (116.center) to (118.center);
		\draw (119.center) to (121.center);
		\draw (120.center) to (122.center);
		\draw (123.center) to (125.center);
		\draw (124.center) to (126.center);
		\draw (127.center) to (129.center);
		\draw (128.center) to (130.center);
		\draw (131.center) to (133.center);
		\draw (132.center) to (134.center);
		\draw (153.center) to (154.center);
		\draw (154.center) to (155.center);
	\end{pgfonlayer}
\end{tikzpicture}
 \caption{Integration contour ${\cal C}_{K}$ in the $s$-plane for the kernel in \eqref{eq:kernelpm}. The poles come from $\Gamma(-s) \Gamma(1-\Jtwo-\Jone - s)   \Gamma(\Jone + s) $. In drawing the plot we used the fact that ${\rm Re}[j_2]<0$ and $-1<{\rm Re}[j_1+j_2]<0$ as in  \eqref{eq:MellinformB}.}
 \label{fig:contourkernel}
 \end{figure}
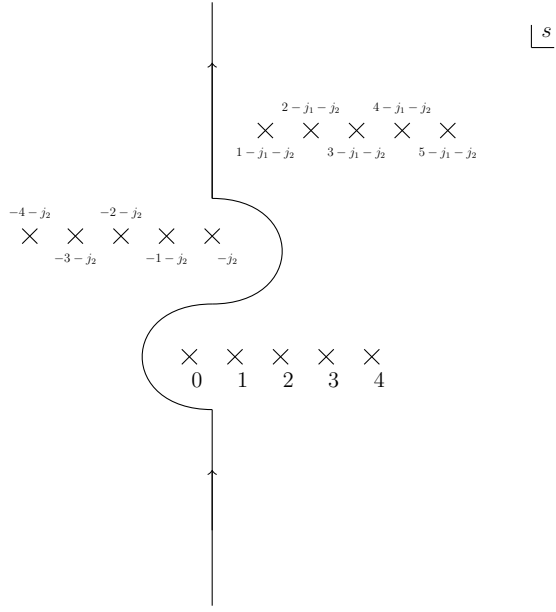

Let us first consider the case where the frequencies of the detectors have opposite signs, namely $\omega_1 \omega_2 < 0$. This covers the cases $+-$ and $-+$. 
The kernel \eqref{eq:kernelSSCpm} can be recast  in the following convenient form (see appendix \ref{sec:derivationSSC} for the derivation)
\be
\label{eq:kernelpm} 
&{K^{+-}_{\text{SS}}(j_1, j_2 | \omega_1, \omega_2, z) \over K_{\text{S} \text{S}}(j_1, j_2 | 0, 0, z)} = (1+\omega_1)^{-\Jone} (1+\omega_2)^{-\Jtwo}  \nn \\
&\times\int_{{\cal C}_{K}} {d s \over 2 \pi i}  \Gamma(-s) {   \Gamma(1-\Jtwo-\Jone - s)   \Gamma(\Jone + s)  \Gamma(1-\Jtwo)\over  \Gamma(1-\Jtwo-\Jone)    \Gamma(\Jone) \Gamma(1 - \Jtwo - s)}  \Big(- {(1-z) \omega_1 \omega_2 \over (1+ \omega_1) (1 + \omega_2)}  \Big)^s \,.  
\ee
The  integration contour ${\cal C}_{K}$ runs parallel to the imaginary axis and separates the poles generated by the product of gamma functions in the numerator, i.e. ascending poles coming from $\Gamma(\dots -s)$ from the left and descending poles coming from $\Gamma(\dots+s)$ from the right, see Figure~\ref{fig:contourkernel}.  
 
For $\text{SSC}_{-+}$ we define
\be
K^{-+}_{\text{SS}}(j_1, j_2 | \omega_1, \omega_2, z) \equiv K^{+-}_{\text{SS}}(j_2 , j_1 | \omega_2, \omega_1,z) ,
\ee
where on the right-hand side we permuted both $\om_1 \leftrightarrow \om_2$ and $j_1 \leftrightarrow j_2$ using the symmetry of the Mellin amplitude $M(j_1, j_2)=M(j_2, j_1)$.

The Mellin integral in \eqref{eq:kernelpm} can be done explicitly resulting in \re{eq:kernelSSCpm}. The representation \eqref{eq:kernelpm} makes many of the properties of the kernel manifest, so we can equally use it instead. Let us discuss these properties one by one. 

Firstly, consider the small $\om_i \to 0$ limit of the kernel. This corresponds to closing the $s$-contour in $\eqref{eq:kernelpm}$ to the right with the contribution from $s=0$ producing $1$ and thus correctly reproducing \eqref{eq:zeroores}. Similarly, if we are interested in  the small $\omega_i$ expansion of our event shapes, we can consistently do it by keeping more and more residues from the poles at $s = n$ and $s = 1 -\Jtwo-\Jone + n$, $n \in \mathbb{Z}_+$. The poles at $s=n$ correspond to the analytic part $(\om_1 \om_2)^n$ of the expansion of the kernel around $\om_i = 0$. The poles at $s = 1 -\Jtwo-\Jone + n$ result in the non-analytic part  $(\om_1 \om_2)^{1-\Jtwo-\Jone+n}$.

Secondly, the ratio of  kernels \eqref{eq:kernelpm} is analytic for ${\rm Re}(\Jone)>0$, ${\rm Re}(\Jtwo)<0$  and fixed $\Jtwo+\Jone$. This is clear from the fact that 
no pinch in the $s$ integration contour arises in this case. This is particularly useful if we take into account that $K_{\text{S} \text{S}}(\Jone, \Jtwo | z, 0, 0)$ is analytic for fixed $\Jtwo+\Jone$ as well.  We find this property very helpful  in the perturbative computations.

For $\om_1 > 0$, $\om_2 < 0$ and $\Jtwo+\Jone$ fixed the kernel decays if we deform the $\Jtwo$ contour to the left. The fact that the kernel is analytic means that we only pick the contribution from the poles of $M(j_1, j_2)$. As can be seen from \eqref{eq:Mellin amplitudes at weak and strong coupling}, both at weak and at strong coupling only a single pole  at $j_2=0$ contributes. This dramatically simplifies the calculation.

Thirdly, note that $\text{SSC}_{+-}$ should go into itself under the combination of the crossing symmetry transformation \eqref{eq:relationused} and the permutation of the detectors. This is indeed reflected in the following property of the kernel,
\be
\label{eq:kernelidentity}
K^{+-}_{\text{SS}}(\Jone, \Jtwo | \omega_2, \omega_1, z) = {1 \over (1+ \omega_1 z)(1+ \omega_2 z)} K^{+-}_{\text{SS}}(\Jone, \Jtwo | \omega_1', \omega_2', z')\,.  
\ee
To check this property note that the argument of the $s$-integral in \eqref{eq:kernelpm} is invariant under this transformation, namely
\be
 - {\omega_1 \omega_2 (1-z) \over 1+ \omega_1 + \omega_2 + \omega_1 \omega_2 z} =- {\omega_1' \omega_2' (1-z') \over 1+ \omega_1' + \omega_2' + \omega_1' \omega_2' z'} \ ,
\ee
which makes checking \eqref{eq:kernelidentity} trivial.

Using the properties of the kernel described above  and the explicit expressions  \eqref{eq:Mellin amplitudes at weak and strong coupling} for the Mellin amplitudes it is easy to compute 
\be
\label{eq:SSCpm}
\text{SSC}_{+-}^{\text{weak}}&= {a\over 4}\bigg[ \frac{   \log \left(\frac{1+\omega_2 z}{1-z}\right) }{ 1-z } +{(1+\omega_2) z \over 2(1-z)(1+ z\omega_2)}
\log \left({\omega_1+\omega_2+\omega_1 \omega_2 z+1\over (\omega_2+1)^2 }\right)\bigg] , 
\\
\text{SSC}_{+-}^{\text{strong}}&={z^2\over 2} \frac{ \left(\omega_2+1\right)^2}{\left(\omega_2 z+1\right)^3}\, \,. \label{4.21}
\ee
Let us reiterate that only the $j_2=0$ pole of the Mellin amplitude contributes to  the above computations.  
The expressions for $\text{SSC}_{-+}$ can be obtained from $\text{SSC}_{+-}$ by permuting $\omega_1 \leftrightarrow \omega_2$.
We have also checked our calculation by performing the Mellin integrals numerically and found perfect agreement with the formulas above.

\subsection{SSC kernel:  same sign}

Consider next the case when the frequencies of the detectors have the same sign,   $\omega_1 \omega_2 > 0$. This covers both cases $++$ and $--$. The convenient representation of the kernel \eqref{eq:kernelSSCpp} takes the form (see appendix  \ref{sec:derivationSSC} for the derivation)
\be
\label{eq:kernelpp}
&{ K^{\text{++}}_{\text{SS}}(j_1, j_2 | \omega_1, \omega_2, z) \over K_{\text{S} \text{S}}(j_1, j_2 | 0, 0, z)} =(1+\omega_1)^{- \Jone} (1+\omega_2)^{- \Jtwo}\nn \\
&\times \int_{{\cal C}_{K}} {d s \over 2 \pi i} \Gamma(-s) {\Gamma(1-\Jtwo-\Jone-s)\Gamma(\Jone+s) \Gamma(\Jtwo+s) \over \Gamma(1-\Jtwo-\Jone)\Gamma(\Jone) \Gamma(\Jtwo) } \Big({\omega_1 \omega_2 (1-z) \over (1+ \omega_1)(1 + \omega_2)} \Big)^{s} ,
\ee
where the contour of integration ${\cal C}_{K}$ is defined in the same way as in \re{eq:kernelpm}. 

  Let us comment on the relevant properties of \eqref{eq:kernelpp}.
Firstly, as expected, the kernel is invariant under the permutation of the detectors $\omega_{1} \leftrightarrow \omega_{2}$ combined with $j_{1} \leftrightarrow j_{2}$. The latter exchange does not affect the event shape since $M(j_1, j_2) = M(j_2, j_1)$. Due to the relation (which one can explicitly check  using \re{eq:kernelpp})
\be
\label{eq:relpp}
K^{\text{++}}_{\text{SS}}(j_1, j_2 | \omega_1, \omega_2, z) ={1 \over (1 + \omega_1 z) (1+ \omega_2 z)} K^{\text{++}}_{\text{SS}}(j_2, j_1 | \omega_1', \omega_2', z') \,,
\ee
where $\omega_i'<0$, and the crossing symmetry relations \eqref{eq:relationused},  the right-hand side of \eqref{eq:kernelpp} defines the kernel for the $--$ case  as well.

Secondly, as before it is trivial to extract the  $\om_i\to 0$   limit from \eqref{eq:kernelpp}  by closing the Mellin contour in \eqref{eq:kernelpp} to the right and picking the contributions form the poles at $s = n$ and $s = 1 -\Jtwo-\Jone + n$, $n \in \mathbb{Z}_+$. For example, the case $\omega_i = 0$ is simply given by the residue at $s=0$, which produces $1$ for the ratio \eqref{eq:kernelpp}. These two series of poles can be resummed into a pair of hypergeometric functions \eqref{eq:kernelSSCppTwoHypers}.

Thirdly, there is an important difference in the analytic properties of the kernel \re{eq:kernelpp} compared to the $+-$ case.
 Namely, for fixed $\Jtwo+\Jone$ the kernel has an infinite number of singularities both to the left and to the right of the $\Jtwo$ contour. This means that in evaluating the Mellin integrals we have to deal with infinitely many poles in both Mellin variables, even though the Mellin amplitudes of interest \eqref{eq:Mellin amplitudes at weak and strong coupling}  are relatively simple. We will see that the result can still be computed analytically.  

Because of the complicated properties of the Mellin kernel,
the computation of $\text{SSC}_{++}$ is  more difficult than in the $+-$ case. The result agrees with the amplitude computation and takes the following form 
(the details of the computation  can be found in Appendix \ref{sec:SSCppcomputationMellin})
\be
\label{eq:SSCpp}\nn 
\text{SSC}_{++}^{\text{weak}}&={a\over 4}\bigg[ \frac{1}{1-z}\log
   \left(\frac{1}{1-z}\right)
\\&\notag
+\frac{(\omega_1+1) z  }{2(1-z) (\omega_1 z+1)}
   \log \lr{\omega_1+\omega_2+\omega_1 \omega_2
   z+1\over \omega_1+1}
\\ &   
   +\frac{(\omega_2+1) z }{2(1-z) (\omega_2 z+1)}
    \log \lr{\omega_1+\omega_2+\omega_1 \omega_2
   z+1\over \omega_2+1} \bigg], 
   \\\label{SSC-st} \nn
\text{SSC}_{++}^{\text{strong}} &= {z^2\over 2} \bigg[ \frac{6 \omega_1^2  \omega
   _2^2\left(\omega_1+\omega_2\right)}{\left(\omega_1+\omega_2+\omega_2 \omega_1 z\right)^5}   
  +\frac{3 \omega_1
   \left(2 \omega_2 \omega_1^2+2 \omega_2^2 \omega_1+\omega_1^2+\omega_2^2\right)
   \omega_2}{\left(\omega_1+\omega_2+\omega_2 \omega_1 z\right)^4}
 \\   &  
 +\frac{\omega
   _2^2 \omega_1^3+\omega_2^3 \omega_1^2+2 \omega_2 \omega_1^3+2 \omega
   _2^3 \omega_1+\omega_1^3+\omega_2^3}{\left(\omega_1+\omega_2+\omega_2 \omega_1
   z\right)^3}\bigg] \,.
\ee
As a consistency check, we verify that the relations \eqref{eq:SSCpm} and \eqref{eq:SSCpp} coincide for $\omega_2=0$. 
The singularities at $\omega_1+\omega_2+\omega_2 \omega_1 z= 0$ are related to the bulk point singularities of the correlator. They only appear at strong coupling and we discuss them in more details in Section~\ref{sect:bulk}.

For the $--$ case, we use the transformation \eqref{eq:relationused}  to get
\be\notag
\text{SSC}_{--}^{\text{weak}}&= {a\over 4}\bigg[\frac{1}{1-z}\log
   \left(\frac{(\omega_1 z+1) (\omega_2 z+1)}{1-z}\right)
\\&
+\frac{(\omega_1+1) z  }{2(1-z) (\omega_1 z+1)}
   \log \lr{1\over \omega_1+1}   
   +\frac{(\omega_2+1) z }{2(1-z) (\omega_2 z+1)}
    \log \lr{1 \over \omega_2+1}\bigg] ,  \\
\text{SSC}_{--}^{\text{strong}} &= {z^2\over 2} \bigg[  \frac{\left(\omega_1+1\right)^2}{\left(\omega_1
   z+1\right)^3}+\frac{ \left(\omega_2+1\right)^2}{\left(\omega_2
   z+1\right)^3}  
\nn \\ &\notag
- \frac{6 \omega_1^2  \omega
   _2^2\left(\omega_1+\omega_2\right)}{\left(\omega_1+\omega_2+\omega_2 \omega_1 z\right)^5}   
  -\frac{3 \omega_1
   \left(2 \omega_2 \omega_1^2+2 \omega_2^2 \omega_1+\omega_1^2+\omega_2^2\right)
   \omega_2}{\left(\omega_1+\omega_2+\omega_2 \omega_1 z\right)^4}
 \\ &  
 -\frac{\omega
   _2^2 \omega_1^3+\omega_2^3 \omega_1^2+2 \omega_2 \omega_1^3+2 \omega
   _2^3 \omega_1+\omega_1^3+\omega_2^3}{\left(\omega_1+\omega_2+\omega_2 \omega_1
   z\right)^3}\bigg] \,.
\ee
We recall that the above relations are valid for $0<z<1$ and $\omega_i>-1$.

\subsection{Closing the contour}

The reader might have noted something puzzling about our result above: if the kernel does not depend on the choice $++$ versus $--$, as explained around \eqref{eq:relpp}, 
how is it possible that the expressions for $\text{SSC}_{++}$ and $\text{SSC}_{--}$ are different? 

To understand the  reason,  
 it is instructive to consider the following simple integral
\be
\label{eq:simpleintegral}
\int_{- \epsilon - i \infty}^{- \epsilon + i \infty} {d j \over 2 \pi i} {(1+\omega)^{-j} \over j } =  - \theta(\omega) \,. 
\ee
For negative $\omega$ we evaluate the integral by closing the contour to the left and get zero. For positive $\omega$  we are allowed to close the contour to the right. Picking up the residue of the pole at $j=0$ we arrive at \re{eq:simpleintegral}. 

The same mechanism is at work for our observables. Consider first the simpler case $\om_1=0$ with the kernel given by \eqref{eq:onezeroKSCC}. Switching to the integration variables $j_1$ and $j=j_1+j_2$  in the Mellin integral, we see that the dependence on $j_1$ enters only through the factor $(1+\om_2)^{-j_1}$. Therefore, as in the simple example \eqref{eq:simpleintegral}  above,  the allowed direction of closing the $j_1$ contour depends on the sign of $\om_2$. For $\om_2 \geq 0$ we can close the contour to the right, whereas for $\om_2 \leq 0$ we can close the contour to the left. This can lead to different results and it explains the difference between $\text{SSC}(0,\om_2,z)$ for positive and negative $\om_2$, even though the Mellin kernel does not depend on the sign of $\om_2$. 

For $\om_1, \om_2 \neq 0$, the kernels become much more complicated. Nevertheless, one can identify by  inspection the allowed directions of closing the $j_1,j_2$ contours, depending on the parameters of the problem. We found that the representations of the kernel \eqref{eq:kernelSSCppTwoHypers} and \eqref{eq:kernelSSCpmTwoHypers} are particularly useful for this purpose. The result is, as expected, that the generalized event shapes are not analytic  around $\om_1,\om_2=0$.

\subsection{Detector identity}\label{sect:id}

We observe that both at weak and strong coupling the scalar-scalar correlations satisfy the identity (cf. \p{mag})
\be
\label{eq:relationSSCcont}
\text{SSC}_{--} + \text{SSC}_{++} -  \text{SSC}_{+-}  - \text{SSC}_{-+} = 0\,.
\ee
It  is understood as an {\it analytic continuation} from the region where each function has been originally computed to some
common values of $\omega_i$. In Section~\ref{sec:summaryofresults} we check that this relation also holds for the  other  detectors 
except for the energy one. In the latter case, the identity is violated at weak coupling due to the contribution from the cross-talk between the detectors.
Here we derive a general formula for the left-hand side of \eqref{eq:relationSSCcont} which is valid at finite coupling and from which its vanishing at weak and strong coupling follows immediately.

In order to proceed we have to specify the analytic continuation in $\omega$ of the event shapes in \eqref{eq:relationSSCcont}. Consider for example $\text{SSC}_{++}$ written for the detector operators defined in the null plane conformal frame \re{eq:nullplanerepr}. We have schematically 
\be
\label{eq:PP}
\text{SSC}_{++} = \int_{- \infty}^{\infty} du_1 d u_2\, e^{-i \sum_{k=1}^2\hat \om_k u_k}  \langle O_4 O_1(u_1) O_2(u_2) O_3 \rangle \,,
\ee
where we suppressed the dependence on the positions of the source and the sink, as well as the transverse coordinates of the detector operators. Such details will not be relevant for what follows.

Since the light-ray operators with $\hat \om_i \geq 0$ annihilate the left vacuum (see Eq.~\re{eq:generalizedannihil}), we can rewrite \p{eq:PP} in the following equivalent way
\be
\label{eq:commutatorPP}
\text{SSC}_{++} = \int_{- \infty}^{\infty}  du_1 d u_2\, e^{-i \sum_{k=1}^2\hat \om_k u_k}   \langle [ [ O_4 , O_1(u_1) ] , O_2(u_2) ] O_3 \rangle \,.
\ee
Now we can `analytically continue' $\text{SSC}_{++}$ to any sign of $\hat \omega_i$ by using Eq.~\eqref{eq:commutatorPP} instead of \eqref{eq:PP} for any signs of $\hat \omega_i$. Let us emphasize again that \eqref{eq:commutatorPP} is equivalent to \eqref{eq:PP} only  for $\hat \omega_i >0$ but they differ otherwise. Similarly, we get the following relations for other choices of the signs of $\om_i$
\be
\label{eq:commutatorRest}
\text{SSC}_{--} &= \int_{- \infty}^{\infty}  du_1 d u_2\, e^{-i \sum_{k=1}^2\hat \om_k u_k}   \langle  O_4  [ O_1(u_1) , [ O_2(u_2) , O_3 ] \rangle , \nn \\
\text{SSC}_{+-} &= \int_{- \infty}^{\infty}  du_1 d u_2\, e^{-i \sum_{k=1}^2\hat \om_k u_k}  \langle  [O_4 , O_1(u_1) ]  [ O_2(u_2) , O_3 ] \rangle , \nn \\
\text{SSC}_{-+} &= \int_{- \infty}^{\infty}  du_1 d u_2\, e^{-i \sum_{k=1}^2\hat \om_k u_k}   \langle  [O_4 , O_2(u_2) ]  [ O_1(u_1) , O_3 ] \rangle ,
\ee
where we used  again Eq.~\re{eq:generalizedannihil}. Switching to commutators does not produce any effect in the region where each of the event shapes was originally defined. However, as we go away from this region (by flipping some of the signs of $\om_i$), formulas \eqref{eq:commutatorPP} and \eqref{eq:commutatorRest} define the analytic continuation needed to check the identity \eqref{eq:relationSSCcont}.

Now we can evaluate the formulas above for any sign of $\hat \omega_i$. We get the following result (where we also use the fact that the detectors commute with each other for $z\neq 0$)  
\be
&\text{SSC}_{--} + \text{SSC}_{++} -   \text{SSC}_{+-} - \text{SSC}_{-+} = \int_{- \infty}^{\infty}  du_1 d u_2\, e^{-i \sum_{k=1}^2\hat \om_k u_k} 
 \Big[   \langle  O_1(u_1) O_2(u_2)  O_4   O_3  \rangle  \nn \\
&   +  \langle   O_4   O_3  O_1(u_1) O_2(u_2)  \rangle -   \langle  O_1(u_1)  O_4   O_3 O_2(u_2) \rangle -   \langle  O_2(u_2)  O_4   O_3 O_1(u_1) \rangle \Big] \,.
\ee
Let us now specify the signs of $\hat \omega_i$ and choose, for example,  $\hat \omega_i > 0$. Since the positive frequency detectors annihilate the left vacuum (see Eq.~\re{eq:generalizedannihil}), we find that only one term survives,
\be
\label{eq:combinationcomm}
 {\text{SSC}_{--} + \text{SSC}_{++} -  \text{SSC}_{+-} - \text{SSC}_{-+}  \stackrel{\hat \om_i > 0}{=} \int du_1 d u_2\, e^{-i \sum_{k=1}^2\hat \om_k u_k}     \langle   O_4   O_3  O(u_1) O(u_2)  \rangle \,. }
\ee
This formula is completely general and it should hold for any choice of the detectors.

Compared to \re{eq:PP}, the Wightman function on the right-hand side of \re{eq:combinationcomm} has different operator ordering. 
In a close analogy with \re{M*K}, the integral over $u_i$ in \re{eq:combinationcomm}  {can be expressed as} the convolution of the Mellin amplitude $M(j_1,j_2)$ and the kernel
$\tilde K_{SS}(j_1, j_2 | \omega_1, \omega_2, z)$. Going through the same steps as in Appendix \ref{sec:derivationSSC}, we find that this kernel takes the form
\be
\label{eq:strange}
\tilde K_{SS} (j_1, j_2 | & \omega_1, \omega_2, z) = {(j_1+j_2 - 1) (q n_1) (q n_2)  \over 2^{j_1+j_2-1} (n_1 n_2)^{j_1+j_2}}  \int_0^{\hat \omega_1} ds_1 \int_0^{\hat \om_2} d s_2 \nn \\
& \times\big[ (\hat \omega_1 - s_1) s_2 \big]^{-j_1}\big[ s_1 (\hat \omega_2 - s_2) \big]^{-j_2} \big[ (q + n_1 s_1 + n_2 s_2)^2 \big]^{j_1+j_2 - 2} \,.
\ee
When integrated against the Mellin amplitudes \eqref{eq:Mellin amplitudes at weak and strong coupling} it gives zero both at  weak  and strong coupling.
 Indeed,  from the integral representation \re{eq:strange} it follows that $\tilde K_{SS} $ is an analytic function of $j_2$ for ${\rm Re}\, j_2<0$. The same is true for the Mellin amplitudes \eqref{eq:Mellin amplitudes at weak and strong coupling}. Closing the integration contour for ${\rm Re}\, j_2<0$ we get that the integral on the right-hand side of \re{eq:combinationcomm} is zero.
For more complicated Mellin amplitudes with extra poles at negative $j_2$ the integral in \eqref{eq:combinationcomm} will not vanish. 
This happens in particular for the correlations involving the energy detector.
We checked  that \eqref{eq:combinationcomm} correctly reproduces our results in  these cases as well.

\subsection{Bulk point singularity}\label{sect:bulk}

The reader might have noticed something peculiar about the expressions for SSC at strong coupling, Eqs.~\eqref{4.21} and \eqref{SSC-st}. We see that there is a new type of singularity that appears in $\text{SSC}_{++}^{\text{strong}}$ at
\be
\label{eq:bulkpointlocus}
\omega_1+\omega_2+\omega_2 \omega_1 z = 0 \,,
\ee
but is absent in $\text{SSC}_{+-}^{\text{strong}}$. 
This locus cannot be reached for $\om_{1,2}>0$ and $0< z <1$ but a singularity appears if we analytically continue $\om_2$ to negative values. Using \eqref{eq:commutatorPP} and \re{eq:commutatorRest} we get
\be
\label{eq:bulkpointsingCorr}
\text{SSC}_{++}  \stackrel{\hat \om_2 < 0}{=} \text{SSC}_{+-} -  \int du_1 d u_2\, e^{-i \sum_{k=1}^2\hat \om_k u_k}  \langle  O_2(u_2) O_4 O_1(u_1) O_3 \rangle \,.
\ee
Given that $\text{SSC}_{+-}^{\text{strong}}$ is regular at \eqref{eq:bulkpointlocus}, we deduce that the singular contribution is generated by the second term in \eqref{eq:bulkpointsingCorr}.
 
Coming back to the  four-point correlation function \re{eq:fourpoint},  it is a well-known fact that there are certain Lorentzian singularities which are present at strong coupling but are absent at weak coupling \cite{Gary:2009ae}, namely the bulk point limit singularities \cite{Maldacena:2015iua}. It is therefore natural to conjecture that the singularity at \eqref{eq:bulkpointlocus} is nothing but an avatar of the bulk point singularity in the $\omega$-deformed event shape. We will now demonstrate that this is indeed the case.

Recall that the emergence of the bulk point singularity in Mellin space is related to the behavior of the Mellin amplitude \re{eq:MellinformB} in the limit $j_i \to \infty$ with ${j_1 / j_2}$  fixed, see \cite{Penedones:2010ue}. We see from   \re{eq:Mellin amplitudes at weak and strong coupling}   that
\be
M^{\text{weak}}(\lambda j_1, \lambda j_2)  = O(\lambda^{-2})\,,   \qqqquad M^{\text{strong}}(\lambda j_1, \lambda j_2) = O(\lambda)\,,
\ee
and indeed at strong coupling the Mellin amplitude grows as $\lambda \to \infty$. 

To make the connection between the bulk point singularity and the structure of the $\omega$-deformed event shapes more direct,  let us consider the following toy model
\be
\label{eq:toymellin}
M_p (j_1 , j_2) =(j_1+j_2)^2 (1+j_1+j_2)^2 \big[ j_1 j_2 (1 + j_1 + j_2)   \big]^{p-2}\,,
\ee
 {where $p$ is positive integer.}  
This Mellin amplitude is polynomial for $p\ge 2$ and thus corresponds to a contact quartic interaction in the AdS dual theory \cite{Penedones:2010ue}.
In the large $j$ limit we have $M_p (\lambda j_1 , \lambda j_2) = O(\lambda^{3p-2})$.

We now want to compute the $\omega$-deformed event shapes \re{M*K} for the toy Mellin amplitude \eqref{eq:toymellin}. First, one can readily see that 
\be
\text{SSC}_{p,+-} = 0 \,. 
\ee
This immediately follows from the analytic properties of the kernel described in the previous sections. A more interesting case is $\text{SSC}_{p,++}$ for which we get
\be\label{fp}
\text{SSC}_{p,++} = z^2 {f_p(\om_1 , \om_2) \over (\omega_1 + \omega_2 + \omega_1 \omega_2 z)^{6p - 1}} + \dots \ \,.
\ee
The details of the computation  are given in Appendix \ref{app:bulk}. For the first few values of $p$ we find
\be
f_1 (\om_1 , \om_2)  &= - 12 (\om_1 \om_2)^2 (\om_1 + \om_2) \,, \nn \\
f_2 (\om_1 , \om_2)  &=6 \times 7! \times (\omega_1 \omega_2)^3 (\omega_1 + \omega_2)^2 \,. 
\ee
As expected, the maximal power of the singularity in \re{fp} is linked to the growth of the amplitude in the large $\lambda$ limit. 

Let us also write down the complete answer for $p=2$
\be
\label{eq:ambiguityp2}
\text{SSC}_{p=2,++} =  {12\om_1^3 \om_2^3 \over (\omega_1 + \omega_2 + \omega_1 \omega_2 z)^{11} } \sum_{k=0}^7 c_k(\omega_1, \omega_2) (\omega_1 + \omega_2 + \omega_1 \omega_2 z)^k \, ,
\ee
where the explicit form of the coefficients $c_k(\omega_1, \omega_2)$ can be found in appendix \ref{app:bulk}.  

The function
$\text{SSC}_{p=2,++}$ has interesting properties in the zero frequency limit $\om_i \to 0$. If we consider $\om_1 \to 0$ with $\om_2$ fixed the result vanishes as $O(\om_1^3)$. If  on the other hand we take both $\om_i \to 0$ at the same rate, the result diverges as $\text{SSC}_{p=2,++}(z, \lambda \om_1, \lambda \om_2) \sim {1 / \lambda}$. 

{Another interesting subtlety related to polynomial Mellin amplitudes of the type \eqref{eq:toymellin} is that they lead to nontrivial distributional terms in the 
generalized event shapes which are
supported at $\om_i=0$. Let us consider for simplicity  $\om_1=0$ and $p=2$ (this example will also be relevant in section \ref{sec:QGsection} when we discuss one-loop supergravity and stringy corrections). 
In this case we apply \re{M*K}, \re{eq:zeroores} and \re{eq:onezeroKSCC} to get the following expression for the scalar-scalar correlation}
\be\label{bulk-delta}
\text{SSC}_{p=2}(0,\om_2,z) &= \int {dj d j_1 \over (2 \pi i)^2}  {\pi \over 2\sin \pi j} z^{-j}\left(1 - z  \right)^{j - 1} j^2 (1+j)^2 (1+\om_2)^{-j_1} \nn \\
&=2 z^2 (1-6 z+6 z^2) \delta(\om_2) \,.
\ee
{Here the $j_1$ integral was obtained by differentiating \eqref{eq:simpleintegral} with respect to $\om$.}

 \section{$\cN=4$ supersymmetry Ward identities for the kernels} \label{sN4WI}

In the previous sections we used the known results for the four-point correlation functions of scalar half-BPS operators in $\mathcal N=4$ SYM to compute the correlation between the scalar detectors SSC. In this section, we extend the consideration to more complicated correlations that involve charge and energy flow operators. According to their definition \re{eq:definitiondet}, they are light-ray operators built out of the $R-$current $J^R_\mu$ and the stress-energy tensor $T_{\mu\nu}$. As a consequence, their correlations can be expressed in terms of the four-point correlation functions of these operators. 

We start in Section~\ref{sN41} with a brief review  of the $\cN=4$ supersymmetry Ward identities for the four-point correlation functions and for the {undeformed} flow correlations, i.e. for $\om_i=0$. Then in Section~\ref{sN42} we give a summary of the new relationships between the Mellin kernels of the various flow correlations with the frequencies $\om_i$ turned on. The detailed derivation is presented   in Appendix~\ref{AppWI}.

\subsection{Stress-energy tensor multiplet in ${\cal N}=4$ SYM}\label{sN41}

In what follows, we assume that the source and sink are given by scalar operators $\phi(3)\equiv O_{\bf 20'}(x_3,Y)$ and $\phi^\dagger(4)\equiv O_{\bf 20'}(x_4,\bar Y)$  (see \p{halfBPS}) and consider the following correlation functions\footnote{For definiteness we always assume that $J_1 \geq J_2$.} 
\begin{align}\label{G-SS}
G_{J_1 J_2} = \vev{\phi^\dagger(4) O_{J_1}(1) O_{J_2}(2) \phi(3)}\,,
\end{align}
where $O_J(i) \equiv O_J(x_i)$ denotes an operator with spin $J$ whose Lorentz indices are projected on the light cone,  
\begin{align}\label{J-ex}
 O_{J=1}(i)=\bar n_i^\mu J_\mu^R(x_i)\,,\qqqquad O_{J=2}(i)=\bar n_i^\mu \bar n_i^\nu T_{\mu\nu}(x_i)\,,
\end{align}
and $O_{J=0}$ was defined in \re{Os}.
In a generic CFT, the correlation functions \re{G-SS} are very complicated and are not related to each other. 

A unique feature of $\mathcal N=4$ SYM is that the operators \re{J-ex} belong to the same stress-tensor supermultiplet. As a result,  their correlation functions are related to each other by $\mathcal N=4$  superconformal Ward identities. Solving them one can 
express \re{G-SS} in terms of the simplest four-point  function involving scalar operators only \cite{Belitsky:2014zha,Korchemsky:2015ssa}.   
The correlation functions \p{G-SS} have a rather complicated R-symmetry structure. It corresponds to the overlap of the tensor product of  two real $SO(6) \sim SU(4)$ representations $\mathbf{20'} \times \mathbf{20'} = \mathbf{1} + \mathbf{15} + \mathbf{20'} + \mathbf{84} + \mathbf{105} + \mathbf{175}$ for the scalar source ($\phi$) and sink ($\phi^\dagger$), and that for the detectors $O_J$, for example for two R-symmetry currents $\mathbf{15} \times \mathbf{15} =  \mathbf{1} + \mathbf{15}_s +\mathbf{15}_a + \mathbf{20'} + \mathbf{45} + \mathbf{\overline{45}}  + \mathbf{84}$. We are not going to study this rich structure in full generality. Instead, we will concentrate on a single R-symmetry channel in each correlation function, namely the one corresponding to the top  value of the $SU(4)$ quadratic Casimir. In the example above this is the channel $ \mathbf{84}$. The reason for this choice is the remarkably simple relation between the various flow correlations in this channel  for vanishing $\om_i$  discovered in \cite{Belitsky:2013bja,Belitsky:2014zha} (see \p{e1.13} below). 

We wish to find out to what extent these relations are preserved in the case of interest $\om_i\neq0$. So, we concentrate on the preferred  channel and show only the space-time dependence of the correlation functions, dropping the isotopic variables  $Y$ and $S$ of Section~\ref{sub:cross}. For instance, the all-scalar correlation function in the top Casimir channel $\mathbf{105}$  (denoted by $G^{\mathbf{(105)}}$ in  \p{eq:fourpoint}) is
\begin{align}
G_{00} =  (x_{12}^2 x_{34}^2)^2 F(x)\,.
\end{align}
Here the dependence on the coupling resides in the function $ F(x)$ given by
\begin{align}\label{1.4}
 F(x) = {1 \over (2 \pi)^4}{\Phi(u,v) \over u v (x_{12}^2 x_{34}^2)^4 } \qq  u={x_{12}^2 x_{34}^2\over x_{13}^2 x_{24}^2}
\qq  v={x_{23}^2 x_{14}^2\over x_{13}^2 x_{24}^2}\,,
\end{align}
where $\Phi(u,v)$ is a function of the conformal cross-ratios $u,v$ that admits the Mellin representation \p{eq:MellinformB}.\footnote{Note that in this section  we do not consider the explicit rational (Born-level) terms in \p{eq:fourpointneq4g105}.}

Following \cite{Korchemsky:2015ssa}, in order to present explicit expressions for the higher-spin correlator components $G_{J_1 J_2}$, it is convenient to  introduce a pair of  auxiliary commuting Weyl spinors, $\lambda_{i,\alpha}$ and $\bar\lambda_{i,\dot\alpha}$, for each spinning operator in \re{G-SS}. They form the null vectors  $\bar n_i$ in \p{J-ex},
\begin{align}\label{la}
\bar n_{i,\mu} \sigma^\mu_{\alpha\dot\alpha} = \lambda_{i,\alpha}  \bar \lambda_{i\dot\alpha} \qq i=1,2\,,
\end{align}
where $\sigma^\mu=(1,\vec \sigma)$ is the four-vector of Pauli matrices. The  general form of the correlation function \re{G-SS} is (up to normalization and only for the top Casimir channel), see Ref.~\cite{Korchemsky:2015ssa},
\begin{align}\label{e1.6}
G_{J_1 J_2} =\cD_1^{J_1} \cD_2^{J_2}\Big[\cM_{J_1,J_2} F(x)\Big]\ ,
\end{align}
where $\cM_{J_1,J_2}$ is a homogenous polynomial in $\lambda_1^\alpha$ and $\lambda_2^\alpha$ of degree $2J_1$ and $2J_2$, respectively. $\cD_i$ is a differential operator acting both on the coordinates of the operator $x_i$ and the spinors $\lambda_i$,
\begin{align}\label{Di}
 \cD_i = \bar\lambda_{i\dot\alpha} \lr{\sigma^\mu}^{\dot\alpha\alpha}
 {\partial\over\partial\lambda_{i}^{\alpha} }
 {\partial\over\partial x_i^\mu}\,.
\end{align}
The functions $\cM_{J_1,J_2}$ are expressed in terms of three building blocks,
\begin{align} \label{e1.8}
& \cX_{1[234]} =\lambda_1^{\alpha} (x_{12})_{\alpha\dot\alpha}(\tx_{23})^{\dot\alpha\beta}(x_{34})_{\beta\dot\g}(\tx_{41})^{\dot\g\delta}\lambda_{1,\delta} \equiv \vev{1|x_{12} \tx_{23}x_{34}\tx_{41}|1} \,, \qqqquad \nt[2mm]
& \cX_{[12]i}=\lambda_1^{\alpha} (x_{1i})_{\alpha\dot\alpha}(\tx_{i2})^{\dot\alpha\beta}\lambda_{2,\beta} \equiv \vev{1|x_{1i}\tx_{i2}|2}\qq (i=3,4)\,,
\end{align}
as follows:
\begin{align}\notag\label{Xs}
& \cM_{0,0} = 1\,,&& \cM_{1,0} = x_{12}^2x_{34}^2 \cX_{1[234]} \,, 
\\[2mm]\notag
& \cM_{2,0} =(\cX_{1[234]})^2\,, && \cM_{1,1} =x_{12}^2 x_{34}^2\cX_{[12]3} \cX_{[12]4}\,,&&
\\[2mm]
& \cM_{2,1} =\cX_{1[234]}  \cX_{[12]3} \cX_{[12]4} \,,&& \cM_{2,2} =(\cX_{[12]3} \cX_{[12]4})^2 \,.
\end{align}
 
As explained in Section~\ref{sec:SSCcorrelator}, we can use the correlation functions $G_{J_1 J_2}$ to obtain the various flow correlations. To this end, we have to apply the transformations specified in \re{eq:definitiondet} to  the operators at points 1 and 2. This amounts to sending both operators to the celestial sphere in the direction of the null vectors $n_i$,  that is $x_i=r_i n_i + \alpha_i \bar n_i$ with $r_i\to\infty$, and integrating along the light-rays $\bar n_i$ with the exponential weight $\e^{-i(n_i\bar n_i)\hat\omega_i \alpha_i}$. As a result the operators at points 1 and 2 in the four-point function \p{G-SS} are converted into flow operators and the correlator becomes a function of the two sets of detector  variables $(n_i ,\ho_i)$ and of the separation between the source and sink $x_{34}$.   This is followed by a Fourier transform to momentum space, $x_{34}\to q$. In the simplest case of SSC , i.e. $J_1=J_2=0$, the result is the convolution of the Mellin amplitude $M(j_1,j_2)$ and the detector kernel $K_{\text{SS}}(j_1,j_2|z,\omega_1,\omega_2)$ {(see Eq.~\re{M*K})}. It is important to remember that the dynamical information (i.e. the coupling dependence) resides in the Mellin amplitude while the kernel is universal, so it applies to both weak and strong coupling. 

Because the various correlation functions \re{e1.6} are determined by the same function $\Phi(u,v)$, we expect that the same Mellin representation  \re{M*K} should hold for the two-point correlations of flow operators with arbitrary spin, schematically
\begin{align}\label{XYC}
\text{XYC} (\omega_1, \omega_2,z) = \int{d j_1 d j_2 \over (2 \pi i)^2} \,   K_{\text{XY}}(j_1,j_2|z,\omega_1,\omega_2) \,M(j_1, j_2)\, \,.
\end{align}
Here $X$ and $Y$ stand for the scalar $(S)$, charge $(Q)$ or energy $(E)$ flow operators. The kernels $K_{\text{XY}}$ are coupling-independent  and have a fixed form for each pair of detectors.  They are functions of the scaling variable $z$ (the angle between the detectors)  and the energies $\omega_i$ transferred by the detectors.  

The special form of the correlation functions \re{e1.6} is dictated by $\mathcal N=4$ superconformal symmetry. It is natural to ask whether analogous relations exist between the correlations \re{XYC}. Indeed, in Ref.~\cite{Belitsky:2014zha} it was shown  that  for $\omega_i=0$ the correlation  of two flow operators of spins $J_1,J_2$ in the top $SU(4)$ Casimir channel,  before the final Fourier transform $x \to q$,\footnote{Without loss of generality we can set $x_3\equiv x$ and $x_4=0$.}   is a {\it total space-time derivative} of the fundamental scalar-scalar one (for $J_1 \geq J_2$): 
\begin{align}\label{2.5} 
{(\Box_{x})^{J_2}(n_2\pa_{x})^{J_1-J_2} \over   2^{J_1-1}(n_1 n_2)^{J_1+1} } \left( \frac{\cG(\g)}{x^2} \right)  
  = {2^{J_2+1} (xn_2)^{J_1-J_2} \over (n_1 n_2)^{J_1+1}(x^2)^{J_1+1}}{d^{J_1} \over d\g^{J_1}}  \left[(1-\g)^{J_1} \g^{J_2} 
 \cG^{(J_2)}(\g)\right]  \,.
 \end{align}
 Here  $\g=  {2(x n_1) (x n_2) /( x^2  (n_1 n_2)) }$ is the Lorentz-invariant and dimensionless  angle variable. The function $\cG(\g)$ contains the coupling constant dependence.  After the Fourier transform $x \to q$ the left-hand side of \p{2.5} gives rise to a very simple relation between the corresponding  kernels $K_{\text{XY}}$ in \p{XYC}.  Stripping off the kinematic prefactor $\sim (q^2)^{J_2} (qn_2)^{J_1-J_2}/(n_1 n_2)^{J_1+1}$  (i.e., the Fourier transform of the derivatives on the left-hand side of \p{2.5}), and replacing it with the factor  $\sim{(q^2)^{J_1+J_2} /( (q  n_1)^{J_1 +1}  (q  n_2)^{J_2+1})}$ from  the  definition  \p{2pt-simp},   one finds  
\begin{align}\label{e1.13} 
K_{\text{XY}}(j_1,j_2|z,0,0) = z^{-J_{X}} K_{\text{SS}}(j_1,j_2|z,0,0)\,,  
\end{align}
where $J_X\ge J_Y$. 
In application to \re{XYC} this implies that for $\omega_i=0$ all the flow correlations are given by the same universal function of $z$, up to an overall power of $z$.

\subsection{Summary of the relationships between the kernels for $\om_i\neq0$}\label{sN42}

Below we show that for $\omega_i\neq 0$ the relation \p{e1.13} is generalized in the form of differential operators expressing each  $K_{\text{XY}}$ in terms of the basic scalar one $K_{\text{SS}}$:
\begin{align}\label{e1.14}
K_{\text{XY}}(j_1,j_2|z,\om_1,\om_2) =z^{-J_X} \sum_{p,q=0}^2 C_{\text{XY}}^{p,q}(j_1,j_2|z,\om_1,\om_2)  \pa^p_{\om_1} \pa^q_{\om_2} \, K_{\text{SS}}(j_1,j_2|z,\om_1,\om_2) \,,
\end{align}
where $J_X \ge J_Y$ and  the coefficients $C_{\text{XY}}$ are polynomials in $z$ and $\om_i$ and  rational functions of the Mellin parameters $j_1$ and $j_2$. The order of the differential operator is at most 2 in each variable, depending on the choice of the  detectors. 

Here we list some of these non-vanishing coefficients:

(i)  Charge-scalar correlation:
\begin{align}\label{coQS}
&C_{\text{QS}}^{0,0} =1+\frac{j_2 \omega_1 z}{j_1+j_2} \qq C_{\text{QS}}^{1,0} = \frac{\omega_1 \left(\omega_1 z+1\right)}{2 \left(j_1+j_2\right)} \qq C_{\text{QS}}^{0,1} =-\frac{\omega_1 \left(\omega_2 z+1\right)}{2 \left(j_1+j_2\right)}\,;
\end{align}

(ii) Charge-charge correlation:
\begin{align}\label{coQQ}
&C_{\text{QQ}}^{0,0} = 1+\frac{ \left(j_1+j_2\right)^2 ( \omega_1+\omega_2   +  z \omega_1 \omega_2)+  j_1 j_2( \omega_1+\omega_2) }{2 \left(j_1+j_2\right){}^2}\,,\nt
&C_{\text{QQ}}^{1,0} = \frac{\omega_1 \left(2 j_1 \left(\omega_1+1\right)+\left(j_1+j_2\right) \omega_2 \left(\omega_1 z+1\right)\right)}{4 \left(j_1+j_2\right){}^2}\,, \nt
&C_{\text{QQ}}^{0,1} = \frac{\omega_2 \left(2 j_2 \left(\omega_2+1\right)+\left(j_1+j_2\right) \omega_1 \left(\omega_2 z+1\right)\right)}{4 \left(j_1+j_2\right){}^2}\,;
\end{align}

(i)  Energy-scalar correlation:
\begin{align}  
&C_{\text{ES}}^{0,0} = 1+\frac{z\om_1 \left[  j_2 \left((j_1+j_2)^2 +j_1-1\right)   \omega_1 z +\left((j_1+j_2)^2-1\right)  \left(j_1+2 j_2\right)\right]}{\left(j_1+j_2-1\right) \left(j_1+j_2\right) \left(j_1+j_2+1\right)}\,,\nt
&C_{\text{ES}}^{1,0} = \frac{\om_1  \left(\omega_1 z+1\right) \left[ \left(\omega_1 z+1\right)((j_1+j_2)^2 -1)    +2 j_1 j_2 \omega_1 z \right] }{ \left(j_1+j_2-1\right) \left(j_1+j_2\right) \left(j_1+j_2+1\right)}\,, \nt
&C_{\text{ES}}^{0,1} = \frac{\om_1 \left[j_2 \omega_1 z \left(\omega_2 z+1\right)+((j_1+j_2)^2-1) (z-1) \right]}{ \left(j_1+j_2-1\right) \left(j_1+j_2\right) \left(j_1+j_2+1\right)}\,, \nt
&C_{\text{ES}}^{1,1} =\frac{\om_1^2 \Big[ \left(j_1+j_2+2\right)(1+z  \om_1+ z  \om_2 +z^2 \om_1 \om_2)   +(j_1  + j_2)(2 z -1)+z +1\Big]}{3 \left(j_1+j_2-1\right) \left(j_1+j_2\right) \left(j_1+j_2+1\right)}\,, \nt
&C_{\text{ES}}^{2,0} = \frac{\om_1^2   \left(z  \om_1+1\right){}^2}{6  \left(j_1+j_2\right) \left(j_1+j_2+1\right)}\,,  \qquad  C_{\text{ES}}^{0,2} = \frac{\om_1^2   \left(z  \om_2+1\right){}^2}{6  \left(j_1+j_2\right) \left(j_1+j_2+1\right)}\,. \label{coES}
\end{align}
We see that setting $\om_i=0$ reproduces the relations  \p{e1.13}. 
The differential operators \re{e1.14} for EQC and EEC are presented in the ancillary file. The detailed derivation of the various differential operators from the $\cN=4$ supersymmetry Ward identity \p{e1.6} is explained  in Appendix~\ref{AppWI}.

Another interesting result of the above analysis is the observation that the basic kernel $K_{\text{SS}}$ satisfies a simple homogeneous partial differential equation:
\begin{align}\label{5.1}
&\Big[j_1 j_2 \left(j_1 \left(\om_1+\om_2+\om_2 \om_1 z\right)+j_2 \left(\om_1+\om_2+\om_2 \om_1 z\right)-\om_1-\om_2\right) \nt
& + j_1 \om_1 \left(\left(j_1+j_2-1\right) \left(\om_1+1\right)+\left(j_1+j_2\right) \om_2 \left(\om_1 z+1\right)\right)\, \pa_{\om_1} \nt
&+ j_2 \om_2 \left(\left(j_1+j_2-1\right) \left(\om_2+1\right)+\left(j_1+j_2\right) \om_1 \left(\om_2 z+1\right)\right)\, \pa_{\om_2}  \nt
&+
\left(j_1+j_2\right) \om_1 \om_2 \left(\om_1+\om_2+\om_2 \om_1 z+1\right)\, \pa_{\om_1} \pa_{\om_2}
\Big] K_{\text{SS}}(j_1,j_2|z,\om_1,\om_2) =0 \,.
\end{align}
Its solutions are discussed in Appendix~\ref{s5}.

We remark that the differential operators that relate the various  kernels depend on the Mellin parameters $j_1,j_2$, so they do not commute with the Mellin integral \p{XYC}. Consequently, the event shapes \re{XYC} are not related to each other in a simple way, except for two special cases discussed in the next subsection.  

\subsection{Relationship between generalized event shapes in special cases} \label{special case}
 
We might ask the question what happens if we switch off one of the frequencies? It turns out that the relations between the kernels described above simplify dramatically. The differential operators \p{e1.14} are reduced to simple factors {\it independent of the Mellin parameters} $j_1,j_2$. Consequently, we can establish very simple relations between the Mellin integrals \p{XYC}, i.e. between the flow correlations, and not only between the kernels. 

\subsubsection*{Vanishing frequency  $\om_1=0$}  

Let us start with the asymmetric correlations QSC, ESC and EQC. From \p{coQS} and \p{coES} it is obvious that setting $\om_1=0$ trivializes the differential operators, reducing them to 1.  The situation is similar for EQC. Denoting the ratio
\begin{align}\label{}
r_{\text{XY}}(z,\om_2) =  {K_{\text{XY}}(j_1,j_2| z,0, \om_2)\over  K_{\text{SS}}(j_1,j_2| z,0, \om_2)} \,,
\end{align} 
we find
\begin{align}\label{e1.19}
& r_{\text{QS}}=  \frac1{z} \qq    r_{\text{ES}}=  \frac1{z^2} \qq   r_{\text{EQ}}=  \frac1{z^2}\left(1+\frac{\om_2}{2}\right) \,.
\end{align}
At the same time, setting $\om_2=0$ in  \p{coQS} leaves the derivative $\left. \pa_{\om_2} K_{\text{SS}} \right\vert_{\om_2=0}$ in place, and we do not observe any substantial simplification  of the differential operator. The same applies to the other  two  asymmetric correlations ESC and EQC.  

The two symmetric correlations QQC and EEC also undergo drastic simplifications if we set $\om_1=0$ (or similarly $\om_2=0$, due to the symmetry). Take for example  QQC: putting $\om_1=0$ in  \p{coQQ} we get  
\begin{align}\label{}
 C_{{QQ}}^{0,0} = 1 + \frac{\left(j_1^2+3 j_2 j_1+j_2^2\right){\om_2}}{2 \left(j_1+j_2\right)^2} \,, \qquad  C_{{QQ}}^{0,1} = \frac{j_2{\om_2} ({\om_2}+1)}{2 \left(j_1+j_2\right)^2}   \,, \qquad C_{{QQ}}^{1,0} = 0 \,.
\end{align}
This is not all, we still have to explore the consequences of the differential equation \p{5.1}, which now reads simply
\begin{align}\label{}
\big[ (\om_2+1) \pa_{\om_2}  +j_1\big] K_{\text{SS}}(z,\om_1=0, \om_2) = 0\,.
\end{align}
This equation allows us to eliminate the term with $\pa_{\om_2}$  {in  \p{e1.14}},
\begin{align}\label{}
C_{{QQ}}^{0,0} - \frac{j_1}{\om_2+1} C_{{QQ}}^{0,1}= 1+\frac{\om_2}{2}\,,
\end{align}
so that the differential operator \p{e1.14} collapses to (compare with the third relation in  \p{e1.19})
\begin{align}\label{e1.23}
 r_{\text{QQ}}=  \frac1{z}\left(1+\frac{\om_2}{2}\right) \,.
\end{align}
The analysis of the differential operator for EEC yields an equally simple relation,
\begin{align}\label{e1.24}
 r_{\text{EE}} =   \frac1{z^2}\left(1+\om_2 +\frac{\om_2^2}{6}\right) \,.
\end{align}
As mentioned
 earlier, the remarkable feature of the relations \p{e1.19}, \p{e1.23} and \p{e1.24} is that they do not depend on the Mellin parameters $j_i$. Inserting them in \re{XYC} yields equivalent relations between the various event shapes,
\begin{align}\label{rXY}
{\text{XYC}(0,\omega_2,z)\over \text{SSC}(0,\omega_2,z)} =  r_{\text{XY}}\qq J_X\ge J_Y\,. 
\end{align}
Most importantly, these relations hold both perturbatively (at weak and at strong coupling) and nonperturbatively (at finite coupling).

\subsubsection*{Event shapes ${\rm XYC}_{+-}$ at strong coupling} 

As explained in Section~\ref{s4.1}, the computation of the event shape ${\rm SSC}_{+-}^{\rm strong}$ is drastically simplified by using the integral representation \p{eq:kernelpm}  of the kernel $K^{+-}_{\rm SS}$. The expression in \p{eq:kernelpm} is analytic in $j_1,j_2$, while the  Mellin amplitude $M^{\text{strong}}$ from \p{eq:Mellin amplitudes at weak and strong coupling} has simple poles. Consequently, the Mellin integral over $j_2$ with $j_1+j_2$ fixed amounts to evaluating \p{eq:kernelpm} at $j_2=0$. To this end we have to analyze the behavior of the factors $\Gamma(-s) \Gamma(j_2+s)/\Gamma(j_2)$ under the $s-$integral, when $s,j_2\to0$. The pole in $\Gamma(j_2+s)$ at $s=-j_2 \to0$ lies outside the contour whereas the pole in $\Gamma(-s)$ at $s=0$ has the trivial residue $-1$. After setting $j_2=0$, the only remaining $\om$ dependence is in the factor $(1+\om_2)^{-j_1}$ in front of the integral. The Mellin integral over $j_1$ is easily done by residues resulting   in the simple expression  \p{4.21}.

The same scenario applies when we act with  the differential operators \p{e1.14}. We observe another drastic simplification, namely the Mellin integral over $j_2$ turns the differential operators into simple factors {\it independent of $j_1$}. They go through the $j_1-$integral and the resulting event shapes  ${\rm XYC}_{+-}^{\rm strong}$ are obtained by   multiplying  ${\rm SSC}_{+-}^{\rm strong}$ by these factors.

Let us illustrate the mechanism on the simplest example of  ${\rm QSC}_{+-}^{\rm strong}$. When the differential operator \p{e1.14} with coefficients from \p{coQS} acts under the integral in \p{eq:kernelpm}, it produces terms $\sim s \to 0$. Similarly, its action on the factor $(1+\om_1)^{-j_2}$  in front of the integral  produces terms $\sim j_2 \to 0$. So, we just need to evaluate
\begin{align}\label{5.26}
&\frac1{(1+\om_2)^{-j_1}}\left[\frac1{z}  \sum_{q=0}^2 C^{0,q}_{\rm QS}(j_1,0|\om_1,\om_2) \pa^q_{\om_2}\, (1+\om_2)^{-j_1}\right]  \nt
&=\frac1{(1+\om_2)^{-j_1}}\left[\frac1{z}  \left( 1 -\frac1{2 j_1} \om_1 (\om_2 z+1) \pa_{\om_2}\right)\, (1+\om_2)^{-j_1}\right]    = \frac{  \omega_1+2 \omega
   _2+\omega_2 \omega_1 z+2 }{2z\left(\omega_2+1\right)}\,.  
\end{align}
As explained above,  this result is independent of $j_1$, so it immediately gives the ratio of the event shapes  ${\rm QSC}_{+-}^{\rm strong}/{\rm SSC}_{+-}^{\rm strong}$ listed in  \p{6.22} and  \p{6.20}, respectively. The calculation with all the other differential operators \p{e1.14} is equally simple and we  obtain all the event shapes  ${\rm XYC}_{+-}^{\rm strong}$ from Section~\ref{s6.2}. The non-trivial feature of the operators \p{e1.14} is  that their action on $(1+\om_2)^{-j_1}$ gives a result independent of $j_1$. 

Note that the situation is different if we employ the  Mellin amplitude $M^{\text{weak}}$ from \p{eq:Mellin amplitudes at weak and strong coupling}. In order to evaluate the residue at the double pole  $j_2=0$, we need to consider  the complete sequence of poles of the integrand in \p{eq:kernelpm} at $s=0,1,2,\ldots$. This results in an infinite sum which, together with that originating from the poles at $j_1=-1,-2,-3,\ldots$ gives rise to the logs in   \p{eq:SSCpm}. Further, the action of the differential operators \p{e1.14} on the kernel \p{eq:kernelpm}  modifies the residue at the double pole $j_2=0$, so the result is not as simple as in \p{5.26}.

\section{Summary of the generalized event shapes in ${\cal N}=4$ SYM}
\label{sec:summaryofresults}

In this section, we present the explicit expressions for the two-point correlations \re{eq:twopointgeneral} in $\mathcal N=4$ SYM at weak and strong coupling.~\footnote{For the reader's convenience, we presented the expressions for all  two-point correlations in a Mathematica readable format in the ancillary file included with an arXiv submission.}
 Following \re{eq:definingXXC} we distinguish six different functions $\rm XYC$ with $X,Y=\{S,Q,E\}$. 

\subsection{Weak coupling}

At weak coupling, the correlations take the form
\begin{align}
{\rm XYC} (\omega_1,\omega_2,z)= {\rm XYC}^{(0)} +{ a\over 4} \, {\rm XYC}^{(1)} + O(a^2)\,,
\end{align} 
where $a=g^2N/(4\pi^2)$. 
In the Born approximation, the correlations are given by \re{Born-1} and \re{Born-2}. The functions ${\rm XYC}^{(0)}$ vanish for $0<z<1$ 
and $\omega_i>-1$.
To one-loop order, depending on the signs of $\omega_1$ and $\omega_2$ we find the following expressions:

\subsection*{Scalar-scalar correlation}

\begin{align}
\text{SSC}^{(1)}_{+-} &=\frac{\log \left(\frac{\omega_2 z+1}{1-z}\right)}{1-z}+\frac{\left(\omega_2+1\right) z
   \log \left(\frac{\omega_1+\omega_2+\omega_2 \omega_1 z+1}{\left(\omega
   _2+1\right)^2}\right)}{2 (1-z) \left(\omega_2 z+1\right)}
\\ 
\text{SSC}^{(1)}_{++}
&= \frac{\log
   \left(\frac{1}{1-z}\right)}{1-z} 
+\frac{(\omega_1+1) z \log \lr{\omega_1+\omega_2+\omega_1 \omega_2
   z+1\over \omega_1+1} }{2(1-z) (\omega_1 z+1)}  
   +\frac{(\omega_2+1) z   \log \lr{\omega_1+\omega_2+\omega_1 \omega_2
   z+1\over \omega_2+1}  }{2(1-z) (\omega_2 z+1)}  
\end{align}  

\subsection*{Charge-scalar correlation}

\begin{align}
 \text{QSC}^{(1)}_{+-} &=  \frac{\left(\frac{\omega_1 z}{2}+1\right)
   \log \left(\frac{\omega_2 z+1}{1-z}\right)}{(1-z) z}+\frac{\left(\omega_1+2 \omega
   _2+\omega_2 \omega_1 z+2\right) \log \left(\frac{\omega_1+\omega_2+\omega_2
   \omega_1 z+1}{\left(\omega_2+1\right)^2}\right)}{4 (1-z) \left(\omega_2
   z+1\right)}
 \\
  \text{QSC}^{(1)}_{-+} &=  
\frac{\left(\frac{\omega_1 z}{2}+1\right) \log \left(\frac{\omega_1
   z+1}{1-z}\right)}{(1-z) z} +\frac{\left(\omega_1+2\right) \log \left(\frac{\omega_1+\omega
   _2+\omega_2 \omega_1 z+1}{\left(\omega_1+1\right)^2}\right)}{4
   (1-z)}
 \\\notag
 \text{QSC}^{(1)}_{++}&=\frac{\left(1+\frac{\omega_1 z}{2}\right) \log
   \left(\frac{1}{1-z}\right)}{(1-z) z} 
   +
\frac{\left(\omega_1+2\right) \log \left(\frac{\omega_1+\omega_2+\omega_2 \omega_1
   z+1}{\omega_1+1}\right)}{4 (1-z)}
\\    
&   +\frac{\left(\omega_1(1+\omega_2 z) +2 (\omega_2+1)\right) \log \left(\frac{\omega_1+\omega_2+\omega_2 \omega_1
   z+1}{\omega_2+1}\right)}{4 (1-z) \left(\omega_2 z+1\right)}  
\end{align}

\subsection*{Charge-charge correlation}

\begin{align}\notag
\text{QQC}^{(1)}_{+-} &=
\frac{( \omega_1 \omega_2 z+2 (\omega_1+\omega_2+2)) \log
   \left(\frac{\omega_2 z+1}{1-z}\right)}{4(1-z) z}
\\   
  & +\frac{(\omega_2+2) (\omega_2
   (\omega_1 z+2)+\omega_1+2) \log \left(\frac{\omega_1 \omega_2
   z+\omega_1+\omega_2+1}{(\omega_2+1)^2}\right)}{8(1-z) (\omega_2 z+1)}
\\
\notag
\text{QQC}^{(1)}_{++}&=
\frac{\left(2 \omega_1+2 \omega
   _2+\omega_2 \omega_1 z+4\right)\log\left(\frac{1}{1-z}\right)}{4 (1-z) z} 
\\   \notag
& +\frac{\left(\omega
   _1+2\right) \left(\omega_2(1+\omega_1 z)+2(\omega_1+1)\right) \log
   \left(\frac{\omega_1+\omega_2+\omega_2 \omega_1 z+1}{\omega_1+1}\right)}{8 (1-z)
   \left(\omega_1 z+1\right)}
\\   
&+ \frac{\left(\omega_2+2\right) \left(\omega_1 (1+\omega_2 z)+2 (\omega_2+1) \right) \log \left(\frac{\omega_1+\omega_2+\omega_2 \omega_1 z+1}{\omega
   _2+1}\right)}{8 (1-z) \left(\omega_2 z+1\right)}
\end{align}

\subsection*{Energy-scalar correlation}

\begin{align}
 \text{ESC}^{(1)}_{+-} &=
 \frac{ \left(\omega_1^2 z^2+6 \omega_1 z+6\right)
   \log \left(\frac{\omega_2 z+1}{1-z}\right)}{6(1-z) z^2}-\frac{ \omega_1^2 \log
   \left(\frac{\left(\omega_2+1\right) z}{\omega_2 z+1}\right)}{6(\omega_2 +1)z}
\\ \notag
&   
   +\frac{
   \left(6 \left(\omega_2+1\right)^2+6 \omega_1 \left(\omega_2+1\right) \left(\omega
   _2 z+1\right)+\left(\omega_1+\omega_2 \omega_1 z\right)^2\right) \log
   \left(\frac{\omega_1+\omega_2+\omega_2 \omega_1 z+1}{\left(\omega
   _2+1\right)^2}\right)}{12\left(\omega_2+1\right) (1-z) z \left(\omega_2
   z+1\right)}
\\\notag
  \text{ESC}^{(1)}_{-+} &=
\frac{ \left(\omega_1^2 z^2+6 \omega_1 z+6\right) \log \left(\frac{\omega
   _1 z+1}{1-z}\right)}{6(1-z) z^2}-\frac{\omega_1^2 \log \left(\frac{\left(\omega
   _1+1\right) z}{\omega_1 z+1}\right)}{6\left(\omega_1+1\right) z}
 \\
 &  
   +\frac{ \left(\omega
   _1^2+6 \omega_1+6\right) \left(\omega_1 z+1\right) \log \left(\frac{\omega_1+\omega
   _2+\omega_2 \omega_1 z+1}{\left(\omega_1+1\right)^2}\right)}{12\left(\omega
   _1+1\right) (1-z) z}
 \\\notag
 \text{ESC}^{(1)}_{++} & =\frac{ \left(\omega_1^2 z^2+6 \omega_1 z+6\right) \log\lr{1\over 1-z}}{6(1-z) z^2}
\\ &   \notag
   +\frac{ \left(\omega_1^2+6 \omega_1+6\right) \left(\omega_1 z+1\right) \log \left(\frac{\omega_1+\omega
   _2+\omega_2 \omega_1 z+1}{\omega_1+1}\right)}{12\left(\omega_1+1\right) (1-z) z}
\\ &   \notag
   +\frac{ \left(6 \left(\omega
   _2+1\right)^2+6 \omega_1 \left(\omega_2+1\right) \left(\omega_2 z+1\right)+\omega_1^2\left(\omega_2
   z+1\right)^2\right) \log \left(\frac{\omega_1+\omega_2+\omega_2 \omega_1 z+1}{\omega_2+1}\right)}{12\left(\omega
   _2+1\right) (1-z) z \left(\omega_2 z+1\right)}
\\ &
-\omega_1^2 \left(\frac{ \log
   \left(\frac{\left(\omega_1+1\right)^2 \omega_2 z}{\omega_1 \left(\omega_1+\omega_2+\omega_2 \omega_1
   z+1\right)}\right)}{12\left(\omega_1+1\right) z}+\frac{ \log \left(\frac{\omega_1 \left(\omega_2+1\right)^2
   z}{\omega_2 \left(\omega_1+\omega_2+\omega_2 \omega_1 z+1\right)}\right)}{12\left(\omega_2+1\right)
   z}\right)    
\end{align}

\subsection*{Energy-charge correlation}

\begin{align}\notag
\text{EQC}^{(1)}_{+-}&=\frac{ \left(6 \left(\omega_2+2\right)+\omega_1 \left(z \left(2 \omega_1+6 \omega
   _2+\omega_2 \omega_1 z+6\right)+6\right)\right) \log \left(\frac{\omega_2
   z+1}{1-z}\right)}{12(1-z) z^2}
\\ \notag &   
   +\frac{ \left(\omega
   _2+2\right) \left(\omega_2+1\right)\left(6  +6 {\omega_1
   \left(\omega_2 z+1\right)\over \left(\omega_2+1\right) }+{\omega_1^2 \left(1+\omega_2 z\right)^2\over \left(\omega_2+1\right)^2}\right)  \log \left(\frac{\omega_1+\omega_2+\omega_2 \omega_1 z+1}{\left(\omega
   _2+1\right)^2}\right)
}{24(1-z) z \left(\omega_2 z+1\right)}
 \\  
&+\frac{\omega_2 \omega_1^2 \log \left(\frac{z\left(\omega
   _2+1\right)}{\omega_2 z+1}\right)}{12 z(\omega_2 +1)}  
\\\notag
\text{EQC}^{(1)}_{-+} &=\frac{\left(6 \left(\omega_2+2\right)+\omega_1 \left(z \left(2 \omega_1+6 \omega
   _2+\omega_2 \omega_1 z+6\right)+6\right)\right) \log \left(\frac{\omega_1
   z+1}{1-z}\right)}{12(1-z) z^2}
\\ \notag &   
   +\frac{
   \left(\omega_1^2+6\omega_1+6\right) \left(\omega_2+\omega_1
   \left(\omega_2 z+2\right)+2\right) \log \left(\frac{\omega_1+\omega_2+\omega_2
   \omega_1 z+1}{\left(\omega_1+1\right)^2}\right) }{24\left(\omega_1+1\right) (1-z)
   z}
\\ &   
   -\frac{\omega_2 \omega_1^2 \log \left(\frac{z\left(\omega
   _1+1\right)}{\omega_1 z+1}\right)}{12 z\left(\omega_1+1\right)}   
\\
\notag
\text{EQC}^{(1)}_{++} & =\frac{ \left(6 \left(\omega_2+2\right)+\omega_1 \left(z \left(2 \omega_1+6 \omega_2+\omega_2 \omega_1
   z+6\right)+6\right)\right) \log \lr{1\over 1-z}}{12(1-z) z^2}
\\ &   \notag
   +\frac{ \left(\omega_1^2+6\omega
   _1+6\right) \left(\omega_2+\omega_1 \left(\omega_2 z+2\right)+2\right) \log \left(\frac{\omega_1+\omega
   _2+\omega_2 \omega_1 z+1}{\omega_1+1}\right)}{24\left(\omega_1+1\right) (1-z) z}
\\ &   \notag
   +\frac{ \left(\omega_2+2\right)
   \left(6 \left(\omega_2+1\right)^2+6 \omega_1 \left(\omega_2+1\right) \left(\omega_2 z+1\right)+ \omega_1^2\left(\omega_2  z+1\right)^2\right) \log \left(\frac{\omega_1+\omega_2+\omega_2 \omega_1 z+1}{\omega
   _2+1}\right)}{24\left(\omega_2+1\right) (1-z) z \left(\omega_2 z+1\right)}
\\ &   
   - \omega_2 \omega_1^2 \left(\frac{ \log \left(\frac{\left(\omega
   _1+1\right)^2 \omega_2 z}{\omega_1 \left(\omega_1+\omega_2+\omega_2 \omega_1 z+1\right)}\right)}{24\left(\omega
   _1+1\right) z}-\frac{ \log \left(\frac{\omega_1 \left(\omega_2+1\right)^2 z}{\omega_2 \left(\omega_1+\omega
   _2+\omega_2 \omega_1 z+1\right)}\right)}{24\left(\omega_2+1\right) z}\right)   
\end{align}

\subsection*{Energy-energy correlation}
 
\begin{align}\notag\label{EECpm-weak}
 \text{EEC}^{(1)}_{+-} & =
 \frac{ \left(\omega_1^2 (\omega_2 z
   (\omega_2 z+6)+6)+6 \omega_1 (\omega_2 ((\omega_2+3) z+3)+6)+6 (\omega_2 (\omega_2+6)+6)\right) \log\lr{ \omega_2 z+1\over 1-z}}{36(1-z)
   z^2} 
\\ &   \notag
   + \frac{(\omega_2+1) (\omega_2^2+6\omega_2+6) 
   \left( {\omega_1^2(  \omega_2 z+1)^2\over (\omega_2+1)^2}+{6 \omega_1  (\omega_2 z+1)\over (\omega_2+1)}
   +6 
  \right)\log \left(\frac{\omega_1 \omega_2 z+\omega_1+\omega_2+1}{(\omega_2+1)^2}\right)}{72(1-z) z (\omega_2 z+1)} 
\\ & -\frac{\omega_1^2 \omega_2^2}{36(\omega_2 +1)z}
\log\lr{z(\omega_2+1)\over \omega_2 z+1}
\end{align} 
\begin{align}\notag\label{EEC-weak}
\text{EEC}^{(1)}_{++} & =\frac{\left(6 \left(\omega_1^2+6 \omega_1+6\right)+\omega_2^2 \left(\omega_1^2
   z^2+6 \omega_1 z+6\right)+6 \omega_2 \left(\omega_1^2 z+3 \omega_1
   (z+1)+6\right)\right) \log \lr{1\over 1-z}}{36(1-z) z^2}
\\&      \notag
   +\frac{ \left(\omega
   _1+1\right) \left(\omega_1^2+6 \omega_1+6\right) \left(\frac{\omega_2^2
   \left(\omega_1 z+1\right)^2}{\left(\omega_1+1\right)^2}+\frac{6 \omega_2
   \left(\omega_1 z+1\right)}{\omega_1+1}+6\right) \log \left(\frac{\omega_1+\omega
   _2+\omega_2 \omega_1 z+1}{\omega_1+1}\right)}{72(1-z) z \left(\omega_1
   z+1\right)}
\\&      \notag
   +\frac{\left(\omega_2+1\right) \left(\omega_2^2+6 \omega_2+6\right)
   \left(\frac{\omega_1^2 \left(\omega_2 z+1\right)^2}{\left(\omega
   _2+1\right)^2}+\frac{6 \omega_1 \left(\omega_2 z+1\right)}{\omega_2+1}+6\right)
   \log \left(\frac{\omega_1+\omega_2+\omega_2 \omega_1 z+1}{\omega
   _2+1}\right)}{72(1-z) z \left(\omega_2 z+1\right)}
 \\&   \notag
   - \omega_1^2 \omega_2^2
   \left(\frac{\log \left(\frac{\left(\omega_1+1\right)^2 \omega_2 z}{\omega_1
   \left(\omega_1+\omega_2+\omega_2 \omega_1 z+1\right)}\right)}{72\left(\omega
   _1+1\right) z}+\frac{\log \left(\frac{ \left(\omega_2+1\right)^2\omega_1 z}{\omega
   _2 \left(\omega_1+\omega_2+\omega_2 \omega_1 z+1\right)}\right)}{72\left(\omega
   _2+1\right) z}\right)
\\&      
-  \omega_1^2 \omega_2^2\frac{ \log \left(\frac{z \left(\omega_1 \omega_2
   z+\frac{1}{2} \left(\omega_1+\omega_2+\sqrt{\left(\omega_1+\omega_2\right)^2-4
   \omega_1 \omega_2 z}\right)\right)^2}{\omega_1 \omega_2 \left(\omega_1+\omega
   _2+\omega_2 \omega_1 z+1\right)}\right)}{12 z \sqrt{\left(\omega_1+\omega
   _2\right)^2-4 \omega_1 \omega_2 z}}
\end{align}
 
The following comments are in order. 

The above relations are valid for $0<z<1$ and $\omega_i>-1$.
We verify that for $\omega_1=\omega_2=0$ they simplify to
\begin{align}
\text{XYC}^{(1)} (0,0,z) = {1\over (1-z) z^{J_{X}}} \log\lr{1\over 1-z}\,,
\end{align}
where $J_X\ge J_Y$. This relation is in agreement with the findings of Ref.~\cite{Belitsky:2013bja}.

For $\omega_1=0$ and nonzero $\omega_2$ it is possible to show that the above expressions verify the relation \re{rXY}.
For nonzero $\omega_i$, the first term in the above expressions for the correlations describes the elastic contribution discussed in Section~\ref{sect:elastic}.
For the correlations SSC, QSC and QQC, the remaining terms describe the annihilation and production of particles by the detectors, depending on the sign of $\omega_i$ (see Section~\ref{sect:inel}). In addition to such terms, the correlations ESC, EQC and EEC also contain terms that scale at small $z$ as $(\log z)/z$. They are accompanied by powers of $\omega_1$ and $\omega_2$ and take the form
\begin{align}
\text{EXC}^{(1)}(\omega_1,\omega_2,z) \stackrel{z\to 0}{\sim} \omega_1^2 \omega_2^{J_X}\times  {\log z\over z} \,,
\end{align}
where $J_X$ is the spin of the detector $X=\{S,Q,E\}$. These additional terms come from the cross-talk between the detectors. 
{For the energy-energy correlation \re{EEC-weak}, such terms are accompanied by a square-root $\sqrt{\left(\omega_1+\omega
   _2\right)^2-4 \omega_1 \omega_2 z}$. Its origin and physical interpretation are discussed in Appendix~\ref{app:sing}.} 

We did not display the expressions for $\text{XYC}_{--}$ and $\text{XXC}_{-+}$. They can be obtained by applying the crossing  relation \re{eq:relationusedGen} to $\text{XYC}_{++}$ and $\text{XXC}_{+-}$, respectively. Doing so, we examined the  relation  
\begin{align}\label{mag1}
\text{XYC}_{--}+\text{XYC}_{++} \stackrel{?}{=} \text{XXC}_{+-}+\text{XXC}_{-+}\,,
\end{align}
which was previously observed to hold for the scalar-scalar correlation \re{mag}.
We found that it is satisfied for the correlations  SSC, QSC and QQC, which do not receive contributions from  the cross-talk between the detectors. At the same time, it is violated for the correlations involving the energy detector.
As explained in Section~\ref{sect:id}, 
{this can be understood using the general formula \eqref{eq:combinationcomm} and the properties of the corresponding kernels.}

 \subsection{Strong coupling}\label{s6.2}

At strong coupling, in the leading supergravity approximation, we find the correlations:
 
\subsection*{Scalar-scalar correlation}

\begin{align}
\text{SSC}_{+-} = & \frac{z^2 \left(\omega_2+1\right)^2 }{2\left(\omega_2 z+1\right)^3} \label{6.20}
\\
\notag
\text{SSC}_{++}  =&  {z^2\over 2} \bigg[  \frac{\left(\omega_1+1\right)^2}{\left(\omega_1
   z+1\right)^3}+\frac{ \left(\omega_2+1\right)^2}{\left(\omega_2
   z+1\right)^3}  
\\ &\notag
- \frac{6 \omega_1^2  \omega
   _2^2\left(\omega_1+\omega_2\right)}{\left(\omega_1+\omega_2+\omega_2 \omega_1 z\right)^5}   
  -\frac{3 \omega_1
   \left(2 \omega_2 \omega_1^2+2 \omega_2^2 \omega_1+\omega_1^2+\omega_2^2\right)
   \omega_2}{\left(\omega_1+\omega_2+\omega_2 \omega_1 z\right)^4}
 \\ &  
 -\frac{\omega
   _2^2 \omega_1^3+\omega_2^3 \omega_1^2+2 \omega_2 \omega_1^3+2 \omega
   _2^3 \omega_1+\omega_1^3+\omega_2^3}{\left(\omega_1+\omega_2+\omega_2 \omega_1
   z\right)^3}\bigg] 
\end{align}

\subsection*{Charge-scalar correlation}

 \begin{align}
 \text{QSC}_{+-}  &=\frac{ z\left(\omega_2+1\right)  \left(\omega_1+2 \omega
   _2+\omega_2 \omega_1 z+2\right)}{4\left(\omega_2 z+1\right)^3}  \label{6.22}
 \\
  \text{QSC}_{-+} &=\frac{z\left(\omega_1+1\right) \left(\omega_1+2\right) }{4\left(\omega_1
   z+1\right)^2}
\\
\notag
\text{QSC}_{++}  &={z\over 4}  \bigg[\frac{ \left(\omega_1+1\right) \left(\omega_1+2\right) }{\left(\omega_1
   z+1\right)^2}+\frac{\left(\omega_2+1\right)  \left(\omega_1+2 \omega
   _2+\omega_2 \omega_1 z+2\right)}{\left(\omega_2 z+1\right)^3}
\\ &   \notag
-\frac{6  \omega_1^2\omega_2 \left(\omega_1+\omega_2\right)}{\left(\omega
   _1+\omega_2+\omega_2 \omega_1 z\right)^4}-\frac{2 \left(3 \omega_2 \omega
   _1^2+3 \omega_2^2 \omega_1+\omega_1^2+2 \omega_2^2\right) \omega_1}{\left(\omega
   _1+\omega_2+\omega_2 \omega_1 z\right)^3}
\\ &     
   -\frac{\omega_2 \omega_1^3+\omega_2^2 \omega_1^2+3 \omega_2^2 \omega_1+\omega
   _1^3+2 \omega_2^2}{\left(\omega
   _1+\omega_2+\omega_2 \omega_1 z\right)^2}\bigg]
\end{align}

\subsection*{Charge-charge correlation}

\begin{align}
\text{QQC}_{+-}=& \frac{z (\omega_2+1) (\omega_2+2)  (\omega_1 \omega_2 z+\omega_1+2
   \omega_2+2)}{8(\omega_2 z+1)^3}
\\
\notag
\text{QQC}_{++} = & {z\over 16} \bigg[ \frac{2 \left(\omega_1+1\right) \left(\omega_1+2\right)   \left(2 \omega_1+\omega
   _2+\omega_2 \omega_1 z+2\right)}{\left(\omega_1 z+1\right)^3}
\\&   \notag
   +\frac{2
   \left(\omega_2+1\right) \left(\omega_2+2\right)   \left(\omega_1+2 \omega_2+\omega
   _2 \omega_1 z+2\right)}{\left(\omega_2 z+1\right)^3}
   \\&\notag
   -\frac{2 \left(\omega_2 \omega_1+\omega_1+\omega_2\right)
   \left(\omega_2 \omega_1^2+\omega_2^2 \omega_1-2 \omega_2 \omega
   _1+2 \omega_1^2+2 \omega_2^2\right)}{\left(\omega_1+\omega_2+\omega_2 \omega_1
   z\right)^2}
\\& \notag 
   -\frac{4 \left(\omega_2^3 \omega_1^3+7 \omega_2^2 \omega_1^3+7 \omega
   _2 \omega_1^3+7 \omega_2^3 \omega_1^2+7 \omega_2^3 \omega_1+2 \omega_1^3+2
   \omega_2^3\right)}{\left(\omega_1+\omega_2+\omega_2 \omega_1 z\right)^3} 
\\& \notag
   -\frac{24 \omega_1 \omega_2\left(\omega_2^2
   \omega_1^2+3 \omega_2 \omega_1^2+3 \omega_2^2 \omega_1+\omega_1^2+\omega
   _2^2\right)  }{\left(\omega_1+\omega_2+\omega_2 \omega_1
   z\right)^4}
\\&
     -\frac{24 \omega_1^2\omega_2^2
   \left(\omega_1 \omega_2+2 (\omega_1+\omega_2)\right)  }{\left(\omega
   _1+\omega_2+\omega_2 \omega_1 z\right)^5}\bigg]
\end{align}
 
\subsection*{Energy-scalar correlation}
 
\begin{align}
 \text{ESC}_{+-} &= \frac{ \left(\omega_2+1\right)^2 \left(\frac{\omega_1^2 \left(\omega_2
   z+1\right)^2}{\left(\omega_2+1\right)^2}+\frac{6 \omega_1 \left(\omega_2
   z+1\right)}{\omega_2+1}+6\right)}{12\left(\omega_2 z+1\right)^3}
 \\
  \text{ESC}_{-+} &=  
\frac{ \left(\omega_1^2+6 \omega_1+6\right)}{12(\omega_1 z+1)}
\\
\notag
\text{ESC}_{++} &=  {1\over 12}\bigg[
   \frac{ \left(\omega_1^2+6 \omega
   _1+6\right)}{(\omega_1 z+1)}      
   +\frac{
   \left(\omega_2+1\right)^2 }{\left(\omega_2 z+1\right)^3}\left(\frac{\omega_1^2 \left(\omega_2
   z+1\right)^2}{\left(\omega_2+1\right)^2}
   +\frac{6 \omega_1 \left(\omega_2
   z+1\right)}{\omega_2+1}+6\right)
\\ &   \notag
- \frac{ \omega_1^2}{z \left(\omega_1+\omega_2+\omega_2 \omega_1 z\right)}
  -\frac{ \left(\omega_1^3+(\omega_1^2+6\omega_1+6) \omega_2\right)}{(\omega_1+\omega_2+\omega_2 \omega_1 z)}
\\[2mm] & 
   -\frac{\left(6
   \omega_1^2+(\omega_1^2+6  \omega_1+6) \omega_2\right) \omega_1}{\left(\omega_1+\omega
   _2+\omega_2 \omega_1 z\right)^2} 
   -\frac{2
   \left(\omega_2 \omega_1+3 (\omega_1+\omega_2)\right) \omega_1^2}{\left(\omega
   _1+\omega_2+\omega_2 \omega_1 z\right)^3}\bigg]
\end{align}

\subsection*{Energy-charge correlation}
 
\begin{align}
\text{EQC}_{+-} &=  
   \frac{
   \left(\omega_2+2\right) \left(\omega_2+1\right)^2 \left( { \omega_1^2 \left(\omega_2
   z+1\right)^2 \over (\omega_2+1)^2} +{6 \omega_1  \left(\omega_2 z+1\right)\over \left(\omega
   _2+1\right) }+ 6\right)}{24\left(\omega_2 z+1\right)^3} 
\\
\text{EQC}_{-+} &=
 \frac{\left(\omega_1^2+6\omega_1+6\right) \left(2+2\omega_1+\omega_2+
\omega_1 \omega_2 z\right)}{24\left(\omega_1 z+1\right)^2}
\\
\notag
\text{EQC}_{++} &= {1\over 24}\bigg[ \frac{4
   \left(\omega_2+2\right) \left(\omega_2+1\right)^2 }{\left(\omega_2 z+1\right)^3} \left( { \omega_1^2 \left(\omega_2
   z+1\right)^2 \over (\omega_2+1)^2} +{6 \omega_1  \left(\omega_2 z+1\right)\over \left(\omega
   _2+1\right) }+ 6\right)
\\&\notag
+ \frac{  \left(\omega_1^2+6\omega_1+6\right) \left(2+2\omega_1+\omega_2+
\omega_1 \omega_2 z\right)}{\left(\omega_1 z+1\right)^2}      
\\&\notag
-\frac{ \omega_1^2\omega_2(\omega_1-\omega_2)(2+\omega_1+\omega_2)}{z(\omega_1+\omega_2+\omega_1 \omega_2
   z)^3}
  -\frac{12 \omega_1^2\omega_2 \left(2\omega_1  \omega_2 +3( \omega_1+\omega_2)\right) }{\left(\omega
   _1+\omega_2+\omega_2 \omega_1 z\right)^4}
\\&   \notag
   +\frac{\left(\omega_2^2 \omega_1^3-\left(\omega_2 \left(\omega
   _2 \left(\omega_2+24\right)+54\right)+12\right) \omega_1^2-54 \omega_2^2 \omega_1-24 \omega_2^2\right)
   \omega_1}{\left(\omega_1+\omega_2+\omega_2 \omega_1 z\right)^3}
\\&   \notag
   -\frac{20 \omega_2 \omega_1^3+12 \omega
   _1^3+2 \left(\omega_1+2\right) \left(2 \omega_1 \left(\omega_1+3\right)+3\right) \omega_2^2}{\left(\omega
   _1+\omega_2+\omega_2 \omega_1 z\right)^2}
\\&   
   -\frac{(\omega_2+2) \omega_1^3+\left(\omega_1^2 + 6 
 \omega_1+6\right) \omega_2^2}{\omega_1+\omega_2+\omega_2 \omega_1 z}\bigg]
\end{align}

\subsection*{Energy-energy correlation}

\begin{align}
\text{EEC}_{+-} &= \frac{ (\omega_2+1)^2 \left(\omega_2^2+6 \omega_2+6\right) \left(\frac{\omega_1^2
   (\omega_2 z+1)^2}{(\omega_2+1)^2}+\frac{6 \omega_1 (\omega_2
   z+1)}{\omega_2+1}+6\right)}{72(\omega_2 z+1)^3}
\\
\notag
\text{EEC}_{++} &= {1\over 72}\bigg[\frac{ (\omega_2+1)^2 }{(\omega_2 z+1)^3}\left(\omega_2^2+6 \omega_2+6\right) \left(\frac{\omega_1^2
   (\omega_2 z+1)^2}{(\omega_2+1)^2}+\frac{6 \omega_1 (\omega_2
   z+1)}{\omega_2+1}+6\right)
\\&\notag
+\frac{ (\omega_1+1)^2}{(\omega_1 z+1)^3}  \left(\omega_1^2+6 \omega_1+6\right)   \left(\frac{\omega_2^2
   (\omega_1z+1)^2}{(\omega_1+1)^2}+\frac{6 \omega_2 (\omega_1
   z+1)}{\omega_1+1}+6\right)
\\&\notag
-\frac{\omega_1^2\omega_2^2(\omega_1^2-4\omega_1\omega_2+\omega_2^2)}{z(\omega_1+\omega_2+\omega_1 \omega_2
   z)^5} ((\omega_1+\omega_2)^2+6 (\omega_1+\omega_2)+6)
\\\notag     
&-{\omega_1^3(\omega_2^2+6 \omega_2+6)+\omega_2^3(\omega_1^2+6 \omega_1+6)\over (\omega_1+\omega_2+\omega_1 \omega_2
   z)}   
\\\notag
&- {2(4 \omega_1^3 \omega_2^3+24 \omega_1^2\omega_2^2 \left(\omega_1+\omega_2\right) +39 \omega_1 \omega_2
   \left(\omega_1^2+\omega_2^2\right)+18 \left(\omega_1+\omega_2\right) \left(\omega_1^2-\omega_2
   \omega_1+\omega_2^2\right) )\over (\omega_1+\omega_2+\omega_1 \omega_2
   z)^2}
\\\notag   
&-{116 \omega_1^3 \omega_2^3+\omega_1^3\omega_2^3 \left(\omega_1+\omega_2\right) +264 \omega_1^2\omega_2^2 \left(\omega
   _1+\omega_2\right) +180 \omega_1\omega_2 \left(\omega_1^2+\omega_2^2\right)  \over (\omega_1+\omega_2+\omega_1 \omega_2
   z)^3}
\\\notag   
&-{  36\left(\omega
   _1+\omega_2\right) \left(\omega_1^2-\omega_2 \omega_1+\omega_2^2\right) \over  (\omega_1+\omega_2+\omega_1 \omega_2
   z)^3}   
 -{\omega_1^3\omega_2^3(312+6  \left(\omega_1+\omega_2\right)  ) \over (\omega_1+\omega_2+\omega_1 \omega_2
   z)^4}  
\\   \notag
&-{\omega_1\omega_2( \omega_1^2 \omega
   _2^2\left(\omega_1^2-4 \omega_2 \omega_1+\omega_2^2\right) +432 \omega_1\omega_2 \left(\omega_1+\omega_2\right)+108 \left(\omega
   _1^2+\omega_2^2\right)) \over (\omega_1+\omega_2+\omega_1 \omega_2
   z)^4}
\\   
&-{\omega_1^2\omega_2^2(6 \omega_1\omega_2 \left(\omega_1+\omega_2\right) + \omega_1\omega_2 
   \left(6+\omega_1+\omega_2\right) \left(\omega_1^2-4 \omega_2 \omega_1+\omega
   _2^2\right)  +216 (\omega_1 \omega_2+\omega_1+\omega_2))\over (\omega_1+\omega_2+\omega_1 \omega_2
   z)^5}   \bigg]
\end{align}

Let us summarize the properties of the above results.

We verify that for $\omega_1=\omega_2=0$ they are in agreement with the findings of Ref.~\cite{Belitsky:2013bja,Belitsky:2014zha},
\begin{align}
\text{XYC} (0,0,z) = \frac12 z^{2-J_{X}}\,,
\end{align}
where $J_X\ge J_Y$. For $\omega_1=0$ and $\omega_2\neq 0$ they satisfy the relation \re{rXY}.

We observe that the expressions for the correlations $\text{XYC}_{++}$ are more complicated  compared to $\text{XYC}_{+-}$ and $\text{XYC}_{-+}$. They contain  additional terms involving inverse powers of 
$\tilde q^2/q^2-1=\omega_1+\omega_2+\omega_1 \omega_2z$. As we explained in Section~\ref{sect:bulk}, such terms are generated by a bulk point singularity.

As at weak coupling, the remaining correlations $\text{XYC}_{--}$ can be found using the crossing relation \re{eq:relationusedGen}. They have a form  similar to $\text{XYC}_{++}$. Then, examining the sum 
$\text{XYC}_{--}+\text{XYC}_{++}$ we observe that all the terms with the bulk singularity induced scale $\omega_1+\omega_2+\omega_1 \omega_2z$ cancel against each other and the resulting expression exactly matches the sum $\text{XYC}_{-+}+\text{XYC}_{+-}$. We therefore conclude that the relation 
\re{mag1} holds at strong coupling. We recall that at weak coupling this relation is broken for the correlations that receive a contribution from the detector cross-talk. In terms of the OPE, the cross-talk is associated with the contribution of the leading, twist-two operators in the product of the detector operators. At weak coupling, these operators develop anomalous dimensions that manifest themselves through $\log z$ terms in the small $z$ limit. At strong coupling, these operators become infinitely heavy and do not contribute. This explains why the cross-talk contribution is not present at strong coupling.

\section{Quantum gravity corrections to event shapes}
\label{sec:QGsection}

In this section we analyze  ${1/ c_T}\sim 1/N_c^2$ corrections to the generalized energy-energy correlations at strong coupling. Such corrections {come from} gravitational loops in the bulk theory. Here we only consider the case $\om_1=0$ which simplifies the analysis dramatically. We use the Mellin amplitude recently computed in \cite{Alday:2018kkw} to study the structure of the generalized event shapes at one loop in supergravity. As expected on general grounds \cite{Kologlu:2019bco}, we find that the result is divergent when $\om_i \to 0$. To understand the nature of this divergence  we derive a dispersive representation of the generalized event shapes using CFT dispersion relations in Mellin space \cite{Penedones:2019tng}. We then discuss the expected structure of the undeformed ($\om_i =0$) event shapes at finite $\lambda$ and $c_T$.

\subsection{One-loop supergravity}

Conventional event shapes, like the energy-energy correlation, cannot be computed perturbatively in ${1 / c_T}$ at strong coupling. The reason is that the correlation function becomes more and more singular in the Regge limit at higher loops which renders the event shape of interest divergent.
This divergence is an artifact of perturbation theory and, based on general grounds, it is absent at finite $c_T$, see \cite{Kologlu:2019bco}.
Since the correlation functions at finite $c_T$ are not accessible, we can alternatively deform the observable in a way that makes it finite order by order in ${1 / c_T}$. We can then try to understand what happens as we take the deformation parameter to zero. Introducing non-zero $\om_i$ is one natural way to do it.

The divergence of the undeformed event shape manifests itself in the appearance of divergences in the $\omega_i \to 0$ limit. In other words, the $c_T \to \infty$ and $\omega_i \to 0$ limits do not commute. Here we compute the $\omega$-deformed event shapes using the one-loop Mellin amplitude in supergravity. The relevant one-loop supergravity correlation function was recently found in \cite{Alday:2017xua,Aprile:2017bgs,Alday:2017vkk}.

The starting point of our analysis is the one-loop Mellin amplitude \cite{Alday:2018kkw} 
\be
\label{eq:1loopsugra}
M^{\text{1-loop sugra}} (j_1 , j_2) &= {1 \over 4 c_T} (j_1 + j_2)^2 (j_1 + j_2 + 1)^2 \sum_{m,n=0}^\infty c_{m,n} \Big[ {1 \over (j_1 - m- 1)(j_2 - n - 1)} \nn \\
&- {1 \over (j_1 - m - 1) (j_1 + j_2 + 2 +n)} - {1 \over (j_2 - m - 1) (j_1 + j_2 + 2 +n)}   \Big] \,.
\ee 
The expression inside the brackets is fully crossing symmetric. The expansion coefficients are given by
\be
c_{m,n} &= c_{n,m} = {1 \over 5 (m+n-1)_5} \Big( 15 m^4 n^2+25 m^4 n+12 m^4+30 m^3 n^3+120 m^3 n^2+114 m^3 n \nn \\
&+36 m^3+15 m^2 n^4+120 m^2 n^3+216 m^2 n^2+77 m^2 n-8 m^2+25 m n^4+114 m n^3\nn \\
&+77 m
   n^2 -76 m n-40 m+12 n^4+36 n^3-8 n^2-40 n \Big) \,.
\ee
The double sum in \eqref{eq:1loopsugra} is divergent. This is a reflection of the UV divergence in the bulk theory at one loop which needs to be renormalized. The divergence does not depend on $j_i$ and leads to an ambiguity in the renormalized one-loop supergravity amplitude which we denote by $c_\infty$. 

 We introduce a crossing-symmetric regulator in the sum, $\sum_{m,n} c_{m,n} \to \sum_{m,n} c_{m,n} {x^m+x^n \over 2}$. To recover the original sum we take $x \to 1$. The divergent part of the sum is given by $\propto {1 /(1- x)}$
 and is indeed $j_i$ independent. We then arrive at 
\be
\label{eq:oneloopsugramellin}
M^{\text{1-loop sugra}} (j_1 , j_2) &= (j_1 + j_2)^2 (j_1 + j_2 + 1)^2 \left( c_\infty  + \hat M^{\text{reg}} (j_1, j_2) \right) ,
\ee
where $\hat M^{\text{reg}}(j_1, j_2)$ can be easily computed explicitly using the regulator above and dropping the divergent ${1 /(1- x)}$ part. It takes a relatively simple form of a sum of various polygamma functions with rational coefficients, and we do not present it here to save space.\footnote{It can be found in the ancillary file that we provided together with the submission. In fact, the explicit expression for the one-loop Mellin amplitude was recently found in \cite{Aprile:2020luw}. We checked that our results agree. We thank Francesco Aprile for pointing this out to us.} After this step, the Mellin integral representation for the the generalized event shapes becomes well defined and it can be studied both analytically and numerically.
The result takes the following form
\be\label{c0}
\text{SSC}_{++}^\text{1-loop sugra} (\omega_1,\omega_2,z) = c_\infty \, \text{SSC}_{++}^{p=2} (\omega_1,\omega_2,z) + \text{SSC}_{++}^{\text{reg}}(\omega_1,\omega_2,z) \, ,
\ee
where $\text{SSC}_{++}^{p=2}$ is given in \eqref{eq:ambiguityp2} and $c_\infty$ is fixed in the nonperturbative theory. It is known explicitly using the power of supersymmetric locailzation \cite{Chester:2019pvm}
\be
\label{eq:cinfty}
c_\infty = {1 \over c_T} \Big( {5 \sqrt{\lambda} \over 16} + 18 \zeta(3) - {3 \pi^2 \over 2} - {2141 \over 192} + O(\lambda^{-{3 \over 2}}) \Big) \,.
\ee

Instead of keeping $\om_i$ general, here we focus our attention on understanding the small $\om_i$ limit. 
Setting $\omega_1 = 0$ does not spoil the finiteness of the result but dramatically simplifies the kernel, see  \eqref{eq:onezeroKSCC}. 
According to \re{bulk-delta}, $\text{SSC}_{++}^{p=2}(0, \om_2,z) \sim \delta(\om_2)$ and, therefore, the $\text{SSC}_{++}^\text{1-loop sugra} (0,\omega_2,z)$  does not depend on the coefficient $c_\infty$ for $\om_2 \neq 0$. In fact, any four-point contact interaction in AdS does not contribute to the generalized event shape for $\om_1 = 0$ and $\om_2 \neq 0$.~\footnote{Microscopically this is related to the extra suppression of the double trace operators familiar from studies of undeformed event shapes.}

The result for the generalized energy-energy correlation at one loop in supergravity takes the form
\be
\label{eq:resoneloopsugra}
&\text{EEC}^\text{1-loop sugra}(0, \omega_2,z) ={5 \over 4 c_T} \bigg[  {1 - 6 z + 6 z^2 \over \omega_2^2} + {3 \over 2} {2 - 15 z + 24 z^2 - 10 z^3 \over \omega_2} \nn \\
&+{12 \over 25} (1-36 z +216 z^2 - 400 z^3 + 225 z^4) \log |\omega_2| \log (z^2 |\om_2|) \nn \\
&+ \Big({716 - 20496 z + 111396 z^2 \over 125} - 1560 z^3 + 846 z^4  \Big)  \log |\omega_2|  + O(\om_2^0) \bigg] \ ,
\ee
where we only presented the terms that are singular in the $\om_2 \to 0$ limit {and did not include the contact term $\delta(\om_2)$}.\footnote{{The leading contact term contribution can be read off from \eqref{bulk-delta} and \eqref{eq:cinfty} and it is given by ${5 \sqrt{\lambda} \over 8 c_T} \delta(\om_2) (1 - 6 z + 6 z^2)$.}} 
One can also check that the expression above correctly implements the transformation property \eqref{eq:relationusedGen} which relates the cases $\om_2 >0$ and $\om_2 < 0$.
 
As an immediate consistency check, we verified that the stress-tensor Ward identities \eqref{eq:WIstress} continue to hold when we include the one-loop correction, because \eqref{eq:resoneloopsugra} satisfies
\be
\label{eq:WIst}
& \int_0^1 d z\, \text{EEC}^\text{1-loop sugra}(z,0, \omega_2) = \int_0^1 d z \, z \, \text{EEC}^\text{1-loop sugra}(z,0, \omega_2) = 0 \,. 
\ee

\subsection{Dispersive representation of the energy-energy correlator}
\label{sec:dispersionrelations}

To elucidate the origin of the ${1 / \om_2}$ divergences in \re{eq:resoneloopsugra} and to understand what happens to them at finite coupling, it is useful to consider the dispersive representation for the Mellin amplitude \cite{Penedones:2019tng}.\footnote{It would be very interesting to generalize this analysis to the coordinate space as well using the coordinate space CFT dispersion relations \cite{Carmi:2019cub,Mazac:2019shk,Caron-Huot:2020adz}.}
 
To make as many symmetries as possible manifest, we find it convenient to write down dispersion relations for the
Mellin amplitude $\tilde M(j_1, j_2) \equiv { M(j_1, j_2) /( (j_1+ j_2)^2 (j_1+ j_2+1)^2)}$ introduced in \eqref{eq:tildeM}.
The advantage of using $\tilde M(j_1, j_2) $ is that it is fully crossing-symmetric function of $j_1$ and $j_2$
\be
\label{eq:tildeMcrossing}
\tilde M(j_1, j_2)  = \tilde M(j_2, j_1)  
= \tilde M(j_1,-1-j_1-j_2) \ \,. 
\ee
 
Consistency with the OPE implies that $M(j_1,j_2)$ has simple poles at the positions dictated by the quantum numbers of the exchanged operators \cite{Mack:2009mi,Mack:2009gy}.
Let us start by recalling the structure of these poles in variable $j$ defined as 
\be
j = j_1 + j_2 \,.
\ee
Poles in $j$ correspond to the  $\text{detector} \times \text{detector} $ OPE channel which due to our choice of detectors is the same as the ${\bf 105}$  R-symmetry channel.  The residues at the poles are fixed in terms of the three-point functions of the exchanged operator, see e.g. \cite{Costa:2012cb} for the detailed derivation,
\be
\label{eq:OPEMpole}
M(j-j_2,j_2) = - {1 \over 2}{ C_{\tau,J}^2 {\cal Q}_{J,m}^{\tau}(2(1-j_2))  \over j  + {\tau \over 2} + m - 2} +O\left(j  + {\tau \over 2} + m - 2\right).
\ee
Here $\tau$ and $J \in 2 \mathbb{Z}_+$ are the twist and spin of the exchanged operator correspondingly. {At finite $c_T$ and $\lambda$ these include all operators present in the theory. Below, when analyzing the one-loop supergravity, only double trace operators will contribute. }
 Descendants are labeled by integer $m \geq 0$, and ${\cal Q}_{J,m}^{\tau}(2(1-j_2))$ are the so-called Mack polynomials known explicitly, see e.g. \cite{Gopakumar:2021dvg} for an explicit formula. $C_{\tau,J}^2 $ is the square of the three-point function of the external operators and the operator that appears in the OPE.
Only even spin operators appear in the formula above due to the crossing symmetry $M(j_1,j_2) = M(j_2,j_1)$. 

Let us now translate the {relation} \eqref{eq:OPEMpole} to the statement about the singularities of the crossing-symmetric modified Mellin amplitude $\tilde M(j_1, j_2)$.
The {relation} \eqref{eq:OPEMpole} implies that it has simple poles in $j=j_1+j_2$ with the following residues
\be
\label{eq:tildeMpoles}
\tilde M(j-j_2, j_2) = - {1 \over 2}{ C_{\tau,J}^2 {\cal Q}_{J,m}^{\tau}(2(1-j_2))  \over j  + {\tau \over 2} + m - 2} {1 \over ({\tau \over 2} + m-2)^2 ({\tau \over 2} + m-3)^2} +O\left(j  + {\tau \over 2} + m - 2\right), 
\ee
where $J \in 2 \mathbb{Z}_+$.
Of course, in addition it could have singularities at $j=0$ and $j=-1$ since {$\tilde M(j-j_2, j_2) \equiv { M(j-j_2, j_2) /( j^2 (j+1)^2)}$.}

Since $\tilde M(j_1, j_2)$ is fully crossing-symmetric, see \eqref{eq:tildeMcrossing}, by replacing $j_2 \to -(1 + j)$ and $j\to -(1+j_2)$ 
 {in \re{eq:tildeMpoles} we find that $\tilde M(j-j_2, j_2)$ has additional poles in $j_2$. They correspond to the OPE channel detector $\times$ source.}
{The residues at these poles can be found by applying the crossing to \eqref{eq:tildeMpoles} }
\be
\tilde M(j-j_2,j_2) = {1 \over 2}{ C_{\tau,J}^2 {\cal Q}_{J,m}^{\tau}( -2(2+j)  )  \over j_2  - {\tau \over 2} - m + 3} {1 \over ({\tau \over 2} + m-2)^2 ({\tau \over 2} + m-3)^2} +O(j_2  - {\tau \over 2} - m + 3) \,.
\ee
The additional singularities at $j=-1,0$ are mapped to $j_2=0,-1$ under the crossing. 

We now write down the dispersive representation for $\hat M(j_1,j_2)$ for fixed $j=j_1+j_2$
\be
\label{eq:dispersivemellinB}
\tilde M(j-j_2,j_2) &= \oint {d j_2' \over 2 \pi i} {\tilde M(j-j_2',j_2') \over j_2' - j_2} = - \sum_{\alpha = \{0,-1,j,j-1 \}}{\rm Res}_{j_2'=\alpha} {\tilde M(j-j_2',j_2') \over j_2' - j_2}   \nn \\
&- {1 \over 2} \sum_{\tau,J - \text{even}, m}  { C_{\tau,J}^2 {\cal Q}_{J,m}^{\tau}(- 2(2+j)  ) \over({\tau \over 2} + m-2)^2 ({\tau \over 2} + m-3)^2} \Big(  {1 \over  {\tau \over 2} + m - 3 - j_2 } + {1 \over  {\tau \over 2} + m - 3 - (j - j_2) } \Big) \,.
\ee
In deriving {this relation} we used the non-perturbative bound on the Mellin amplitude $|M(j_1 , j_2) | \leq | j_2 |$ as $|j_2| \to \infty$, see \cite{Penedones:2019tng}, which translates to  $|\tilde M(j_1,j_2)| \leq |j_2|^{-3}$ as $|j_2| \to \infty$. Using the crossing symmetry this also implies that $|\tilde M(j-j_2,j_2)| \leq |j_2|^{-3}$ for fixed $j$, therefore the arc at infinity does not contribute to \eqref{eq:dispersivemellinB}. 
 {The sum on the first line of \eqref{eq:dispersivemellinB} comes from the contribution of protected twist-two operators and it is fixed by the OPE.}\footnote{Here we assume that the coupling is finite and there are no higher spin conserved currents in the spectrum.} 

As a result we get the following representation for the nonperturbative Mellin amplitude
\be
\label{eq:dispersivesuper}
M(j_1, j_2)&= - {1 \over 2}{(j_1 + j_2)^2 (1+j_1 +j_2) \over j_1 j_2} - {(j_1+ j_2)^2 (j_1+ j_2+1)^2 \over 2}  \   \nn \\
& \times \sum_{\tau, J - \text{even}, m}  { C_{\tau,J}^2 {\cal Q}_{J,m}^{\tau}(- 2(2+j_1 + j_2) ) \over({\tau \over 2} + m-2)^2 ({\tau \over 2} + m-3)^2} \Big(  {1 \over  {\tau \over 2} + m - 3 - j_2 } + {1 \over  {\tau \over 2} + m - 3 - j_1 } \Big) \,.
\ee
Let us emphasize that this formula is exact. The first term comes from the protected twist two operators and it coincides with the leading result at strong coupling \re{eq:Mellin amplitudes at weak and strong coupling}. The infinite sum encodes corrections to this result.

For the one loop supergravity Mellin amplitude, the only operators that appear in the sum in \eqref{eq:dispersivesuper} are the double trace operators which have twist $\tau_{n,J} = 8+2n+ \gamma(n,J)$ with $\gamma(n,J) \sim {1 / c_T}$.\footnote{The shift of the twist $\tau$ by $(+4)$ is a consequence of superconformal symmetry, see e.g. Eqs.~(2.5) and (2.6) in \cite{Alday:2017vkk}. In the language of \cite{Aprile:2017bgs}, we have $\tau = 4 + 2 t$, where $t \geq 2$.} The
Mack polynomials ${\cal Q}_{J,m}^{\tau}$ that enter \eqref{eq:dispersivesuper} are importantly proportional to
\be
{\cal Q}_{J,m}^{\tau} \sim \sin^2 \lr{\pi \tau \over 2} \,.
\ee
For the double trace operators this becomes
\be
\label{eq:scalingDT}
\sin^2 \lr{\pi \tau_{n,J} \over 2} = \sin^2 \lr{\pi \gamma_{n,J} \over 2} \sim {1 \over c_T^2}
\ee
Double trace operators have three-point functions that scale as $C_{\tau,J}^2 \sim c_T$ which can be seen from form of the disconnected part of the correlator in \eqref{eq:fourpointneq4g105}. Combining this fact with \eqref{eq:scalingDT} we correctly recover the ${1 / c_T}$ scaling of the leading correction {to \re{eq:dispersivesuper}} in supergravity.

Next we can use the dispersive representation \eqref{eq:dispersivesuper} to compute the $\omega$-deformed event shapes at finite coupling in terms of the OPE data. For simplicity we consider $\om_1 = 0$ and $\om_2 > 0$. Using relations \eqref{e1.24}, \eqref{rXY}, and \eqref{eq:onezeroKSCC}, we get the following representation for the event shape
\be
\label{eq:dispersiveB}
\text{EEC}_{+}(0, \omega_2,z) &=\left(1+\om_2 +\frac{\om_2^2}{6}\right)  \Big[\norm {1 \over 2} + \norm \sum_{\tau,J - \text{even} } 2 \sin^2 \lr{\pi \tau \over 2} C_{\tau,J}^2  (1+\om_2)^{- ({\tau \over 2} - 1)}  \widetilde{\text{EEC}}_{\tau,J} (z, 0, \om_2) \Big] ,  
\ee
where we have defined a function
\be
\widetilde{ \text{EEC}}_{\tau,J}(z,0, \omega_2) &\equiv - {1 \over 2}  \int {d j \over 2 \pi i}  {\pi \over 2 \sin \pi  j } z^{-j-2}\left(1 - z  \right)^{j - 1} j^2 (1+j)^2 \nn \\
&\times \sum_{m=0}^\infty  (1+\om_2)^{-m}   {{\cal Q}_{J,m}^{\tau}(-2 (j+2)) \over 2 \sin^2 \lr{\pi \tau \over 2}}  {1 \over({\tau \over 2} + m-2)^2 ({\tau \over 2} + m-3)^2}  \,.
\ee
It encodes the contribution {to the energy-energy correlation} of a set of poles in \eqref{eq:dispersivesuper} associated to a given primary operator.~\footnote{For correlation functions these are known as Polyakov blocks, see e.g.  \cite{Carmi:2019cub,Mazac:2019shk,Caron-Huot:2020adz} for details. Therefore $\widetilde{ \text{EEC}}_{\tau,J}(z,0, \omega_2)$ encodes the contribution of a given Polyakov block to the energy-energy correlation.} In writing the formula above we used the fact that for $\om_2 > 0$ we can close the $j_1$ contour to the right half-plane since this is the direction in which $(1+ \om_2)^{-j_1}$ decays. 

It is  {possible} to compute $\widetilde{ \text{EEC}}_{\tau,J}(z,0, \omega_2)$ explicitly for the first few values of $J$.  For $J=0$ we get
\be\label{tEEC}
\widetilde{ \text{EEC}}_{\tau,J=0}(z,0,\om_2) &= - (1 - 6 z + 6  z^2) \frac{\Gamma \left(\frac{\tau -6}{2}\right)^2 \Gamma (\tau) }{\pi
   ^2 \Gamma \left(\frac{\tau }{2}\right)^4}  \ _2 F_1 \left({\tau - 6 \over 2}, {\tau - 6 \over 2} , \tau - 1 , {1 \over 1 + \om_2}\right) \,. 
\ee
For $J>0$ we similarly get polynomials in $z$, $\widetilde{ \text{EEC}}_{\tau,J}(z,0,\om_2) = \sum_{i=0}^{J+2} c_i(\tau, \om_2) z^i$. Curiously, we also observed  {the relation}
\be
\label{eq:zdependenceleading}
\lim_{\tau \to \infty} \widetilde{ \text{EEC}}_{\tau,J}(z,0,\om_2) \sim (1 - 6 z + 6  z^2) \Big( 1 + O( \tau^{-1} ) \Big) ,
\ee
which will be useful for our discussion below. Note that the dependence  of \eqref{eq:zdependenceleading}  on $z$ matches the leading ${1 / \om_2^2}$ term in \eqref{eq:resoneloopsugra}.

It is now easy to understand how the terms singular {for $\om_2\to 0$} are generated in \eqref{eq:resoneloopsugra}. The  {underlying} mechanism has already been  discussed in the literature, see for example \cite{Maldacena:2015iua}, and it is related to the contribution of double trace operators to the sum \eqref{eq:dispersiveB}.  
In this case the factor $\sin^2 \lr{\pi \tau / 2}$ becomes $\sin^2 \lr{\pi \gamma_{n,J,i} / 2}$ where $n$ is related to the twist, $J$ is spin, and $i$ accounts for the degeneracies of the double trace operators in question. We now see that
\be
\label{eq:sinanom}
\sin^2 \lr{\pi \gamma_{n,J,i} \over 2} \leq 1 \qq c_T - \text{finite}.
\ee
If on the other hand we first expand at large $c_T$ using the fact that $\gamma_{n,J,i} \sim {1 / c_T} \ll 1$ 
we get
\be
\sin^2 \lr{\pi \gamma_{n,J,i} \over 2} = {\pi^2\over 4} \gamma_{n,J,i}^2 + O(1/c_T^4)  \,. 
\ee
To compute the contribution of double trace operators to \eqref{eq:dispersiveB} we need to sum both over spin $J$ and twists of the operators labeled by $n$. Because the anomalous dimensions $\gamma_{n,J,i}$ are growing functions of $n$, it is the sum over twists of the operators in \re{eq:dispersiveB} that produces terms that diverge in the limit $\om_2 \to 0$. Non-perturbatively we know that an infinite series of ${1 / c_T}$ corrections turn $\gamma_{n,J,i}^2$ into \eqref{eq:sinanom} in this way regulating the sum.
This mechanism of regulating the divergences in perturbation theory is sometimes called ``eikonalization'' \cite{Fitzpatrick:2015qma}. To summarize, we see that the large twist operators can generate divergences in ${1 / c_T}$ perturbation theory but they are absent at finite $c_T$. 

Let us next work out the form of the ${1 / \om_2^2}$ term in \eqref{eq:resoneloopsugra} in more detail. In order to do it {we replace $C_{\tau,J}^2$ in \re{eq:dispersiveB} with its leading asymptotic behaviour in mean field theory at tree level}
\be
\lim_{\tau \to \infty} \langle C_{\tau,J}^2 \rangle = c_T 2^{5} \pi (J+1) 4^{-J} 4^{- \tau} \tau^2 \,.
\ee
For the anomalous dimensions we use the results of \cite{Aprile:2017bgs} to get
\be
\label{eq:oneloopssugraasym}
\lim_{c_T \to \infty}{ \langle 2 \sin^2 \lr{\pi \tau \over 2} C_{\tau,J}^2  \rangle \over \langle C_{\tau,J}^2 \rangle} \sim  {\pi^2 \over 2} { \langle \gamma_{\tau,J}^2 C_{\tau,J}^2  \rangle \over \langle C_{\tau,J}^2 \rangle}  \sim {\tau^{11} \over c_T^2} \,.
\ee
{Substituting these relations into \re{eq:dispersiveB} and taking into account \re{tEEC} we find that the contribution of the operators with large twist and $J=0$ involves the following integral}
\be
&\int_{\tau_0}^\infty d \tau \ \langle C_{\tau,J}^2 \rangle { \langle 2 \sin^2 \lr{\pi \tau \over 2} C_{\tau,J}^2  \rangle \over \langle C_{\tau,J}^2 \rangle} \widetilde{ \text{EEC}}_{\tau,J=0}(z,0,\om_2) (1+\om_2)^{-{\tau \over 2} }  \nn \\
&\sim {1 \over c_T} \int_{\tau_0}^\infty d \tau \, 2^{- \tau} \tau^{{15 \over 2}}  \ _2 F_1 \left({\tau - 6 \over 2}, {\tau - 6 \over 2} , \tau - 1 , {1 \over 1 + \om_2}\right) (1+\om_2)^{-{\tau \over 2} } \,,
\ee
{where we replaced the sum over twist by an integral over $\tau> \tau_0$.}
To understand  {its behaviour for $\om_2 \to 0$} it is convenient to use the integral representation of the hypergeometric function. The result then takes the following form at large $\tau$
\be
{1 \over c_T}\int_0^1 d t \int_{\tau_0}^\infty d \tau \tau^8 {(1-t) (1-t+\om_2)^3 \over t^4} \Big( {1 - t + \om_2 \over t (1-t)} \Big)^{- {\tau \over 2}} \sim {1 \over c_T \om_2^2}\,. 
\ee
The integral over $\tau$ is trivial to do. The integral over $t$ can be easily analyzed numerically or by the saddle point method, with the expected result that it behaves as ${1 / \om_2^2}$. In this way we have also correctly reproduced the dependence on $\om_2$ of the first term in \eqref{eq:resoneloopsugra}.  We have not tried to reproduce the ${1 / \om_2}$ term using the formulas above. It would be interesting to do it by analyzing the ${1 / \tau}$ corrections  {as well as to generalize the above analysis} to higher loops.

Let us also comment on how contact terms $\delta(\om_2)$ originate from the finite coupling formulas above in the perturbative expansion. Consider an exchange by a heavy operator ($\tau \gg 1$), which could be for example a stringy mode, in the dispersion representation \eqref{eq:dispersivemellinB} of the Mellin amplitude $\tilde M(j-j_2,j_2)$. Expanding the right-hand side of \eqref{eq:dispersivemellinB}
in ${1 / \tau}$, we get polynomial in $j$ and $j_2$ corrections to the Mellin amplitude. As we explained, these yield contact terms in the generalized event shape $\text{EEC}(0, \omega_2,z)$  localized at $\om_2 =0$. Let us now see how this comes about directly from \eqref{eq:dispersiveB}. In this case we can write the following distributional identity
\be
\lim_{\tau \to \infty}  (1+\om_2)^{-{\tau \over 2}} =  {2 \over \tau} \delta(\om_2) + {4 \over \tau^2} \Big( \delta(\om_2) + \delta'(\om_2) \Big) + O(1/\tau^3) \,.
\ee
These are terms that will appear in the generalized event shapes in  string perturbation theory, with the 't Hooft coupling $\lambda$   controlling the twist of the stringy modes. On the other hand, by going back to the finite coupling we  {get}
\be
\lim_{\tau \to \infty} \lim_{\om_2 \to 0} (1+\om_2)^{-{\tau \over 2}} = 1 ,
\ee
{so} that the limits $\tau \to \infty$ and $\om_2 \to 0$ do not commute. {To summarize, the} contact terms in the perturbative expansion of generalized event shapes signal the  presence of  enhanced terms in the undeformed event shapes. 
{In the example above, $2\delta(\om_2)/\tau$ gets replaced with $1$ after resummation of corrections in $1/\tau$.}
 
 \subsection{Small $\om$ and the Regge limit}

Let us show that the leading behavior of the EEC as $\om_2 \to 0$ is related to the behavior of the Mellin amplitude in the Regge limit.  
We use \eqref{eq:zeroores}, \eqref{eq:onezeroKSCC} and \eqref{e1.24} to write down the expression for $\text{EEC}(0, \omega_2,z) $ in the Mellin space
\be\notag\label{eec-int}
\text{EEC}(0, \omega_2,z) = \left(1+\om_2 +\frac{\om_2^2}{6}\right) \int {d j \over 2 \pi i} {1 \over 2} {\pi \over \sin \pi j} {z^{-j-2}\left(1 - z  \right)^{j - 1} \over j^2 (j+1)^2}
\\
\times \int {d j_2 \over 2 \pi i} (1 +\om_2)^{- j_2} \tilde M(j - j_2, j_2)\, \,.
\ee
In order to understand the small $\om_2$ expansion, the relevant integral is thus
\be
\label{eq:relevantintegral}
\int_{- \delta - i \infty}^{- \delta + i \infty} d j_2\, (1 +\om_2)^{- j_2} \tilde M(j - j_2, j_2)  = \int_0^\infty d j_2  (1 +\om_2)^{- j_2} {\rm Disc}_{j_2} \tilde M(j - j_2, j_2)  \, , 
\ee
where we evaluated the integral by closing the contour to the right for $\om_2>0$ and picking the contribution from the poles in $j_2$, or, equivalently, the discontinuity of the Mellin amplitude. Imagine that  the discontinuity of the Mellin amplitude has the following Regge behavior  {as $j_2\to\infty$} \footnote{Since $\tilde M(j - j_2, j_2)$  is a meromorphic function of $j_2$, this formula should be understood as a statement about averaged discontinuity.}  
\be
{\rm Disc}_{j_2}  \tilde M(j - j_2, j_2) \sim j_2^\alpha \,.
\ee
Via the integral \eqref{eq:relevantintegral} it will produce terms as singular as ${1 / \om_2^{1+\alpha}}$ in the Regge limit.
At one loop we get ${\rm Disc}_{j_2} \tilde M(j - j_1, j_2) \sim j_2$ as $j_2 \to \infty $, which via \eqref{eq:relevantintegral} produces ${1 / \om_2^2}$. 
It would be interesting to explore the limit $\om_2 \to 0$ and correspondingly the Regge limit of the Mellin amplitude at two and higher loops using the results of \cite{Bissi:2020woe}.

We however did check the relevant scalings by looking at the discontinuity ${\rm Disc}_s M(s,t)$ of the corresponding flat space supergravity amplitude $M(s,t)$, where $s$ and $t$ are the standard Mandelstam invariants in flat space.
At one loop the supergravity amplitude takes the following form in the Regge limit, see for example formula (39) in \cite{Alday:2018kkw},
\be
{\rm Disc}_s M^{1-\text{loop}}(s,t) \sim s \,.
\ee
The leading term being analytic in $t$ corresponds to scattering at zero impact parameters, which is consistent from the analysis above where it emerged from intermediate operators of fixed spin and large twist. A more familiar eikonal phase in ten dimensions behaves instead as ${t^2 \log (-t) / s}$. It controls the leading Regge behavior of scattering at non-zero impact parameters which manifests itself in the fact that it is not analytic in $t$. We observe the same phenomenon for the expansion of ${\rm Disc}_{j_2}  \tilde M(j - j_2, j_2)$ in one loop supergravity. The term ${1 / j_2}$ comes with the $j$-dependent pre-factor which is non-analytic. Via \eqref{eq:relevantintegral} it will contribute to $\log \om_2$ term in \eqref{eq:resoneloopsugra}. As opposed to ${1 /\om_2^2}$ and ${1 / \om_2}$ terms, it is non-analytic in $z$ (due to the presence of $\log z$). This can be traced to non-analyticity in $j$ of the ${1 / j_2}$ term in the large $j_2$ expansion of ${\rm Disc}_{j_2}  \tilde M(j - j_2, j_2)$. This in turn can be traced back to the universal high energy limit of the eikonal phase in gravity at non-zero impact parameters, see e.g. \cite{Amati:1990xe}. It would be interesting to make this connection more explicit using the eikonal phase in the AdS analysis of \cite{Cornalba:2007zb}.

\section{Ordinary event shapes at finite $\lambda$ and finite $c_T$}
\label{sec:finiteN}

At weak coupling,  given the  perturbative expansion of the correlation function one can systematically compute the perturbative corrections to the event shapes. The situation is very different at strong coupling. In this case there is no simple way of translating the perturbative expansion of the correlation function to the perturbative expansion of the event shape. At the technical level, this is related to the fact that both the stringy ${1 / \lambda}$ and the ${1 / c_T}$ gravitational loop corrections yield expressions for the correlation functions that grow too fast in the Regge limit for the event shape of interest to be well defined.  In the collider physics setup studied in this paper, the Regge limit divergence is translated to a large detector time effect. It arises because the correlation function describes the energy correlations with the property that the energy flux does not decay with time and generates a divergence when the working time of the detector goes to infinity.

The divergences are absent at finite $\lambda$ and finite $c_T$ where, based on general arguments, we know that the Regge limit is such that, say, the energy-energy correlation is well defined. Computing the corrections to the event shapes thus requires resummation of the naive perturbative expansion. Let us illustrate this phenomenon with a few examples.

\subsection{Stringy corrections}

The simplest example of this phenomenon arises when we consider stringy corrections. The leading order correction to the Mellin amplitude takes the form 
\cite{Goncalves:2014ffa,Binder:2019jwn}
\be
\label{eq:stringymellin}
\tilde M(j_1, j_2) = \tilde M^{\text{strong}}(j_1, j_2) + {60 \zeta(3) \over \lambda^{3/2}} + O(\lambda^{-5/2}) \ \,.
\ee
Substituting this relation into \re{eec-int} and taking into account \re{bulk-delta}, we can immediately write down the result for the generalized energy-energy correlations with non-zero $\om_2$,
\be
\label{eq:stringygenevent}
\text{EEC}(0,\om_2, z) &= \left(1 + \om_2 + {\om_2^2 \over 6}\right) \nn \\  
&\left[ {1 \over 2} \left(\theta(\om_2) + \theta(-\om_2){(1+\om_2)^2 \over (1+ \om_2 z)^3} \right) +{120 \zeta(3) \over \lambda^{3/2}} \delta(\om_2) (1 - 6 z + 6 z^2) + \dots \right] \,.
\ee 
On the other hand, the direct computation of the $\om_2=0$ event shape in string theory gives~\cite{Hofman:2008ar}
\be
\label{eq:hofmanmaldacenastringy}
\text{EEC}(z) \equiv \text{EEC}(0,0,z) = {1 \over 2} \left( 1 + {4 \pi^2 \over \lambda} (1 - 6 z + 6 z^2) + \dots \right) \,,
\ee
where dots in this and the previous expression denote corrections suppressed by powers of $\lambda^{-1/2}$.

Let us discuss how the three formulas above are related to each other. The relation \eqref{eq:hofmanmaldacenastringy} can be understood 
by taking into account that the leading correction to the Mellin amplitude \eqref{eq:stringymellin} scales as 
$\tilde M(j_1, j_2) |_{\lambda^{-3/2}} \sim 1$ in the Regge limit $j_2 \to \infty$. For this reason the integral $\int d j_2 \tilde M (j - j_2, j_2)$ that appears in the computation of the energy-energy correlation \re{eec-int} is divergent {at $\om_2=0$}.  
The stringy effects effectively cut the Mellin integral off at $j_2 \sim \sqrt{\lambda}$ leading to \eqref{eq:hofmanmaldacenastringy}. The evaluation of the precise coefficient in \eqref{eq:hofmanmaldacenastringy} is a nontrivial task,  requiring the resummation of the string perturbative corrections. Yet it can    be done explicitly, see \cite{Goncalves:2014ffa} for details. In terms of the dispersion relations the mechanism is precisely the same as described above in Section \ref{sec:dispersionrelations}, which also explains the dependence on~$z$.

Comparing the relations \re{eq:stringygenevent} and \re{eq:hofmanmaldacenastringy}, we observe that the leading correction to both expressions is proportional to the same function of $z$ but has different dependence on $\lambda$. The presence of the $\delta(\om_2)$ term in \re{eq:stringygenevent} signifies that the corresponding detector measures a time-independent energy flux. If we imagine a detector with a finite working time $T$ this would produce an effect of order ${T /\lambda^{3/2}}$. Letting the working time $T \to \infty$ produces then a divergent result. What happens at finite $\lambda$ however is that at times $T_{\text{str}} \sim \lambda^{1/2}$  string perturbation theory breaks down and the effect of the resummation is that the energy flux effectively becomes zero for times $T> T_{\text{str}}$. 

\subsection{Gravitational loops}

Let us now try to generalize this discussion to include gravitational loops. It is natural to expect that the same mechanism is at play: the parts of the one-loop result \eqref{eq:resoneloopsugra} that diverge in the $\om_2 \to 0$ limit get enhanced by $\lambda$ or $c_T$ dependent factors coming from the long time effects in the detector. 
The presence of such terms, which are enhanced compared to the naive expectation coming from the perturbative expansion of the correlation function, is a characteristic feature of strongly coupled conformal field theories with semi-classical gravity duals.

Based on the OPE data of double trace operators and the mechanism discussed in section \ref{sec:dispersionrelations}, we associate the terms ${1 / \om_2^2}$ and ${1 / \om_2}$ with high energy scattering at zero impact parameters. In the flat space limit this manifests itself through the fact that the corresponding terms in the Regge limit of ${\rm Disc}_s M(s,t)$ are analytic in $t$. In string theory, a fixed-angle scattering is very soft in the high energy limit \cite{Gross:1987ar}. We therefore expect that the  divergences {for $\om_2\to 0$} get regularized in the same way as in the example with string corrections above: namely the $j_2$ integral is cut off at  $j_2 \sim \sqrt{\lambda}$. Equivalently, we expect that the divergences to leading order get substituted by ${1 / \om_2} \to \sqrt{\lambda}$. From this argument we expect the leading stringy correction to be\footnote{The same term originates from the correction $c_\infty \delta(\om_2)$ discussed in the previous section.}
\be
\label{eq:stringyQG}
\text{EEC}^{\text{QG}}(z) \propto {\lambda \over c_T} (1 - 6 z + 6 z^2) + \dots \ \,.
\ee
Computing  the precise coefficient requires a more detailed analysis. The correction \eqref{eq:stringyQG} originates from two different sources: $T{\sqrt{\lambda} / c_T}$ and $T^2/c_T$ which grow with time and both get saturated at $T \sim T_{\text{str}}\sim \sqrt{\lambda}$.

We expect a different mechanism to be at play for the terms that diverge in the $\om_i \to 0$ limit which originate from scattering at non-zero impact parameters. Such divergences will be present in the ${1 / c_T}$ expansion of the event shape even at finite string tension, or, equivalently, $\lambda$. These terms get unitarized at finite $c_T$ via the eikonalization of the gravitational loops. Because of this we also expect to have $c_T$-enhanced terms in the expansion of the event shapes. Effects coming from non-zero impact parameters originates from operators of arbitrary large spin, see e.g. 
\cite{Cornalba:2006xk,Cornalba:2006xm}. Given that $ \widetilde{ \text{EEC}}_{\tau,J}(z,0,\om_2) $ are polynomials in $z$ of the maximal power $z^{J}$, any non-analyticity in $z$ can only come from the large spin operators. As an example at one loop,  the divergent and non-analytic in $z$ piece $\sim {1 \over c_T} \log |\om_2| \log z$ should originate from the expansion of the eikonal phase in AdS \cite{Cornalba:2007zb}. Indeed, the flat space limit of the part of the one-loop supergravity Mellin amplitude that generates ${1 \over c_T} \log |\om_2 | \log z$ term matches the expected expansion of the eikonal phase 
\cite{DiVecchia:2019myk,DiVecchia:2019kta}. 

The $\log |\om_2| $ divergence in \re{eq:resoneloopsugra} corresponds to the long-time tail which grows  as $\log T$ as a function of the detector working time $T$. We thus see that at finite but large working times the leading contribution to the energy correlation takes the form
\be
\label{eq:leadingQGT}
\text{EEC}^{\text{QG}}(z,T) \propto {\log T \over c_T} \left(1-36 z+216 z^2 -400 z^3 +225 z^4\right) \log z \,. 
\ee
This correction is universal (it does not depend on the details of the spectrum of the bulk theory) and originates from the fact that the dual theory is gravitational, with gravity being the leading interaction at high energies and large impact parameters. As a nontrivial consistency check, we verified that \eqref{eq:leadingQG} satisfies the Ward identities \eqref{eq:WIst}.

To recover the energy-energy correlation $\text{EEC}(z)$ we have to take the limit $T \to \infty$. Doing it rigorously requires the resummation of the $1/c_T$ perturbation theory. This might be possible using the eikonalization of the gravitational amplitudes in AdS but we do not attempt it in the present paper. Still, let us discuss the simplest possible scenario in which the energy flux that grows as $\log T$ gets saturated at some time $T_{\text{QG}}$. Here $T_{\text{QG}}$ is some characteristic time that goes to infinity as $c_T \to \infty$. From the eikonalization of the gravitational amplitudes we expect that $\log T_{\text{QG}} \sim \log c_T$. In this simple scenario, we get that the leading quantum gravity correction to the energy-energy correlation takes the form
\be
\label{eq:leadingQG}
\text{EEC}^{\text{QG}}(z) &\stackrel{?}{\propto} {\log c_T \over c_T} \left(1-36 z+216 z^2 -400 z^3 +225 z^4\right) \log z \,. 
\ee
It would be very interesting to derive the leading quantum gravity correction to the energy-energy correlation \eqref{eq:leadingQG} rigorously. 

\section{Conclusions and future directions}\label{sect:conc}

In this paper we have introduced a new class of observables in a collider physics setting which we dubbed generalized event shapes. They are defined as the  matrix elements of light-ray operators  \eqref{eq:energydetom} with non-zero momentum, \eqref{eq:momentumdef}. These light-ray operators have a finite resolution in time, given by ${1 / \hat \om}$, and thus allow us to probe the longitudinal structure of the state. By setting $\hat \om =0$ they smoothly transit over to the familiar event shapes. 

The generalized event shapes are {\it not} the standard differential cross sections, since they are sensitive to the relative phases of the various scattering amplitudes. This makes them harder to measure, but from the theoretical point of view they are very natural observables. They are IR safe and can be evaluated both in terms of scattering amplitudes and correlation functions.
If defined in terms of integrated Wightman functions, for $\hat \om \neq 0$ the new observables retain all the information contained in the former. We therefore expect that they form a {convenient} basis of {collider-type} observables (at least in a CFT).  Indeed, many popular  jet observables can be restated in terms of energy correlations, see e.g. \cite{Chen:2020vvp}.

The generalized event shapes exhibit new features compared to the usual event shapes. One interesting example is that they develop a bulk point singularity at strong coupling. Approaching the singularity requires analytic continuation of the physical parameters and in this way it is reminiscent of the flat space singularity in the de Sitter correlators, see e.g.  \cite{Arkani-Hamed:2015bza}.  Another characteristic feature of the generalized event shapes is that they are non-analytic around $\hat \om = 0$. This is related to the fact that, depending on the sign of $\hat \om$, the detector operators can either create or annihilate particles. Nonperturbatively this fact manifests itself in Eq.~\eqref{eq:annihilationE}.

We studied the generalized event shapes perturbatively in planar ${\cal N}=4$ SYM  both at weak and strong coupling. The results are summarized in Section~\ref{sec:summaryofresults}. First, we considered the case where  the source, the sink and the detectors are built out of local scalar operators. In this case, at weak coupling  we obtained the  leading-order result using both scattering amplitudes and correlation functions. At strong coupling we have only the correlation functions approach at our disposal. Second, we used  the $\cN=4$ supersymmetry Ward identities to compute  generalized event shapes with detectors  defined by conserved currents, namely charge and energy correlations. In the special case where only one of the detectors has $\hat \om \neq 0$,  in Section~\ref{special case} we found remarkably simple relations between the various observables. When both detectors carry non-zero momentum we found that relatively simple relations exist not among the observables themselves but among their Mellin kernels, see Eq.~\p{e1.14}.

Unlike the ordinary event shapes, the generalized event shapes can be computed perturbatively order by order in the ${1 / c_T}$ and ${1 / \lambda}$   {at strong coupling}.   A characteristic feature of the stringy and quantum-gravitational corrections to the generalized event shapes is that they are divergent in the  limit  $\hat \om \to 0$. These divergences signify a longitudinal broadening of the state and the presence of long-time tails in the measured radiation. For ordinary event shapes, which perform measurements at arbitrarily late times, this leads to corrections which are enhanced compared to those coming from the perturbative expansion of the correlation function. 
We studied the one-loop supergravity corrections to the generalized event shapes in Section~\ref{sec:QGsection}. We also derived a dispersive representation of the generalized event shape and used it to analyze the result. The structure of the ordinary event shapes at finite $\lambda$ and finite $c_T$ is discussed in Section~\ref{sec:finiteN}.

Let us list some  interesting open questions that we did not address in the present paper:

\begin{itemize}
\item Probably the most important question is to understand whether the generalized event shapes, or some of their cousins, can be measured in an experiment. We commented on the special case of low energy elastic scattering, where this should be possible  indirectly. It would be interesting to find out if the same or a similar method  generalizes to higher energies.
\item It would be very interesting to study generalized event shapes in QCD. The computation is analogous to the scattering amplitude one in Section~\ref{sect:amp}.  {We expect that the final result will be qualitatively similar to that in  ${\cal N}=4$ SYM, though the details will differ. This is what happens to the ordinary event shapes and we expect this relation between ${\cal N}=4$ SYM and QCD to hold for the generalized event shapes as well. } 
\item We motivated our study by thinking about energy fluxes in real time, but for technical reasons it was more convenient to work with detectors that carry a definite momentum, rather than detectors that work for a given period of time. We demonstrated the transition between the two pictures on the simple example of the one-point generalized event shapes in Appendix~\ref{sec:timeprofiledet}.  {We observed the effect of the longitudinal broadening at strong coupling which manifests itself through the divergences in the $\hat \om \to 0$ limit, or, equivalently, the long-time tails of the radiation measured by the detectors.} It would be interesting to connect our results to the broadening picture explored in \cite{Hatta:2012kn}.
\item We emphasized that for $\hat \om \neq 0$ we can render a broader class of correlations well defined. A prominent example is the charge-charge correlation which is not IR safe for $\hat \om = 0 $ beyond one-loop order. It would be interesting to understand the physics of the $\hat \om \to 0$ limit of such observables and see if there is a way of constructing an observable that stays finite in that limit.
\item We did not discuss the OPE of light-ray operators with non-zero momentum. It would be useful to generalize the analysis of \cite{Kologlu:2019bco,Kologlu:2019mfz,Chang:2020qpj} to this case. As we have seen, one obvious difference is that the spin selection rule $J = J_1 + J_2 -1$ (here $J$ is the spin of a light-ray operator that appears in the OPE of light-ray operators of spin $J_1$ and $J_2$) does not hold anymore since $\hat \om$ carries non-zero spin. The light-ray OPE can also shed interesting light on the $\hat \om \to 0$ limit of the observables discussed in this paper.
\item We found that thinking about generalized event shapes is helpful for understanding the  stringy and quantum gravity corrections to the ordinary event shapes. The basic phenomenon is that the ${1 / \hat \om}$ divergences  {present in the generalized event shapes} turn into $\lambda$ or $c_T$ enhanced terms in the ordinary event shapes. To establish this fact  it is convenient to use a dispersive representation of the event shapes (we did it for the leading stringy corrections). For gravitational loop corrections that unitarize the scattering amplitudes at large impact parameters, a more thorough analysis should be possible.
\item The way how we introduced time dependence in the measurement at infinity is not unique. Our  choice of generalized event shapes was dictated by the symmetries of the problem: given a source that carries a definite momentum, it is natural to have detectors that carry definite momenta as well. Another advantage of these detectors is that they naturally suppress the potential divergences at large working times $u$, see Eq.~\eqref{eq:nullplanerepr}. This makes the corresponding matrix elements well defined, even if the original event shape at $\hat \om =0$ was not. Other bases of time-dependent light-ray operators have been discussed in the literature, see for instance \cite{Besken:2020snx}. 
\item It is known that the light transform of the stress-energy tensor, also called  ANEC operator, is a non-negative operator in any unitary QFT 
\cite{Faulkner:2016mzt,Hartman:2016lgu}. If instead we restrict the time integration to some finite interval we get what is known as the quantum null energy condition (QNEC) 
\cite{Bousso:2015mna,Bousso:2015wca,Balakrishnan:2017bjg}. It bounds the energy flux  measured over a time interval $T$ from below by the shape variations of the entanglement entropy. It would be interesting to explore the implications of QNEC to the  generalized event shapes.
\item In this paper, we only studied one- and two-point correlations. It would be interesting to extend the consideration to multi-point correlations, see e.g. \cite{Chen:2019bpb,Chen:2021gdk}. The computation of the corrections to the lowest order in the coupling within the amplitude approach would require incorporating  tree-level form factors with an arbitrary number of on-shell particles. Results for such form factors have been obtained   in $\mathcal N=4$ SYM~\cite{Bork:2014eqa,Bianchi:2018peu}. In the correlation function approach one would need expressions for the mutli-point correlation functions involving conserved currents. Such correlation functions can be computed following the approach developed in Refs. \cite{Chicherin:2015bza,Chicherin:2014uca,Fleury:2019ydf}.
\item By placing the detector operators with non-zero null momentum both at past and future null infinity we can naturally define transition amplitudes in CFT. It would be interesting to explore the properties of such objects using   CFT methods. It would be also very interesting to understand   the connection between such objects and the transitions studied in \cite{Korchemsky:2018hnb,Gillioz:2020mdd}. 
\item Generalized event shapes can be defined and studied in gravity  (see \cite{Gonzo:2020xza} for a recent discussion of the ordinary event shapes). Indeed, they are IR safe observables and can be computed using the standard scattering amplitudes techniques. Moreover, in the case of gravitational waves for obvious reasons we cannot place the detectors around the collision point and essentially all the information acquired in the observation comes from the longitudinal, or time-dependent, details of the signal.
\end{itemize}

\section*{Acknowledgments}

We are grateful to Alex Belin, Andrei Belitsky, Agnese Bissi, Cyuan-Han Chang, Dima Chicherin, Miguel Correia, Murat Kologlu, Petr Kravchuk, David Simmons-Duffin, Gabriele Veneziano, Matthew Walters for useful discussions. AZ is grateful to the participants of the SwissMap workshop ``Advances in Quantum Gravity'' for stimulating discussions on related topics. This project has
received funding from the European Research Council (ERC) under the
European Union's Horizon 2020 research and innovation programme (grant
agreement number 949077).  The work of GK was supported by the French National Agency for Research grant ANR-17-CE31-0001-01.

\appendix

\section{Conformal properties of the deformed light-ray operators}\label{AppXZ}

Here we recall some key points about the embedding formalism  and index-free notation used in Section~\ref{sec:definition}. A more detailed introduction to it can be found, e.g., in \cite{Costa:2011mg}. We then apply the formalism to give an alternative derivation of the conformal \p{eq:specialconformal} and dilatation  \p{eq:dilatationFIN} transformations of the deformed light-ray operators. 

\subsection*{Embedding formalism and index-free notation}

The embedding formalism provides a redundant description of $d-$dimensional conformal tensors by raising them to $(d+2)-$dimensional ones. For example, a vector $V_A(X)$ with index $A=(+,-,\mu)$  has $d+2$ components instead of the expected  $d$ for a vector $V_\mu(x)$. The  $(d+2)-$dimensional null  vector $X^A$ (with $X\cdot X \equiv X^+ X^- + X^\mu X_\mu=0$) is also redundant, having $d+1$ 
{independent coordinates} instead of $d$.  

This redundancy is removed in three steps:

1) Require {\it scale invariance},
\begin{align}\label{eqq1.1}
\Phi(\la X)= \la^{-\Delta}\Phi(X)
\end{align}
with some weight which will later on be identified with the conformal weight. Assuming that $X^+\neq0$ this allows us to define the scale invariant field
\begin{align}\label{eq2.4}
\phi(x)= (X^+)^\Delta \Phi(X)\,,
\end{align}
where the $d-$dimensional  vector $x^\mu$ is identified with
\begin{align}\label{eq2.5}
{X^A\over X^+} = \left( 1,\ {X^- \over X^+} \equiv -x^2,\ {X^\mu \over X^+} \equiv x^\mu\right) \ \Rightarrow \ X\cdot X=0 \,.
\end{align}

2) Impose {\it transversality} of the tensor fields, e.g. for the vector
\begin{align}\label{eq2.1}
X\cdot V \equiv X^A V_A(X) = \frac1{2}(X^+ V^- + X^- V^+) + X^\mu V_\mu = 0\,.
\end{align}
This condition removes one degree of freedom. 

3) Consider the tensor field modulo {\it gauge transformations}, e.g. for the vector
\begin{align}\label{eq2.2}
V_A(X) \ \to \ V_A(X) + X_A\, s(X)\,,
\end{align}
where $s(X)$ is an arbitrary scalar field that scales as $s(\la X)= \la^{-\Delta-1} s(X)$.  This is the second redundant degree of freedom of the vector field that needs to be removed.  Clearly, the transversality condition \p{eq2.1} is invariant under such gauge transformations. 

The index-free notation consists in projecting the index $A$ of the vector field with an auxiliary null vector $Z^A$ (with $Z\cdot Z=0$),
\begin{align}\label{eqe2.3}
V(X,Z)= Z^AV_A(X) \,.
\end{align}
To respect the gauge invariance \p{eq2.2}  the vectors $X$ and $Z$ are assumed orthogonal,
\begin{align}\label{eq2.3}
Z\cdot X=0 \,. 
\end{align}
Then the transversality \p{eq2.1}  takes the form of a {\it gauge invariance} condition  (recall \p{eq:primaryB}), 
\begin{align}\label{eqeq2.7}
V(X,Z+\b X)= V(X,Z)\,.
\end{align}

The auxiliary vector $Z^A$ undergoes independent scale transformations reflecting the homogeneous polynomial dependence on $Z$ of the projected tensor. In the example of the vector field  \p{eqe2.3} we obviously have $
V(X,\rho Z) = \rho V(X,Z)$. This scaling, together with \p{eqq1.1},  is generalized to  the case of arbitrary positive integer spin $J$, see \p{eq:primaryA}, or even to non-integer and negative spins by analytic continuation. 

One can reduce the  $(d+2)-$dimensional null vector $Z^A$, satisfying the orthogonality condition \p{eq2.3},  to a $d-$dimensional null vector $z^\mu$  by exploiting the { gauge invariance}  \p{eqeq2.7}. Indeed, consider the linear combination $Z^A+ \b X^A$ and choose $\b=- Z^+/X^+$: 
\begin{align}\label{eq2.8}
Z^A+ \b X^A   \ \stackrel{\b=- {Z^+ \over X^+}}{=}\ \left( 0 ,\ Z^-  - {Z^+  \over X^+} X^-  \equiv -2z\cdot x ,\ Z^\mu  - { Z^+ \over X^+} X^\mu  \equiv z^\mu  \right)   \Rightarrow  \ z^2=0\,.
\end{align}
The vector $z^\mu$ is traditionally used for describing symmetric traceless tensors, see e.g. \cite{Dobrev:1977qv}.

 \subsection*{Conformal transformations}
 
 The main point of the embedding formalism is that the $d-$dimensional conformal group acts linearly on the $(d+2)-$dimensional vectors,    e.g. $X^A \to \Lambda^A{}_B X^B$ with $\Lambda \in SO(2,d)$. The benefit is that one can  easily build invariants of such  linear transformations, see e.g.  \cite{Costa:2011mg}. The price to pay is the redundancy discussed above and the necessity to fix various gauges, if we want to recover the familiar Lorentz tensors, e.g.  the vector field $V(X,Z) \to V_\mu(x)$. 
 
 According to this concept, a $Z-$projected tensor field is {\it invariant} under $SO(2,d)$,
 \begin{align}\label{eq2.10}
O(\Lambda X, \Lambda Z) = O(X,Z) \,,
\end{align} 
and in addition has the scaling properties \p{eq:primaryA}. 

The conformal group $SO(2,d)$ is extended to $O(2,d)$ by adding the discrete operation of {\it inversion}, $X^{d+2} \to - X^{d+2}$ or equivalently,  $X^+ \leftrightarrow -X^-$. In the fixed frame \p{eq2.5} we find 
\begin{align}\label{}
x^\mu = {X^\mu \over X^+} \ \stackrel{I}{\rightarrow} \ -{X^\mu \over X^-} = -{X^+ \over X^-} {X^\mu \over X^+}= {x^\mu \over x^2}\,.
\end{align}
The inversion of the vector $Z^A$ in  the fixed frame \p{eq2.8}  is accompanied by a compensating gauge transformation \p{eqeq2.7} with parameter $\b=Z^-/X^- - Z^+/ X^+$. This implies \cite{Dobrev:1977qv}
\begin{align}\label{inversion}
z^\mu   \ \stackrel{I}{\rightarrow} \  z^\mu - \frac{2z\cdot x}{x^2} x^\mu \qq (I[z^\mu])^2=0 \qq  I^2[z^\mu]=z^\mu\,.
\end{align}

\subsubsection*{Infinitesimal translations and conformal boosts}

The  conformal algebra  is generated by $d-$dimensional translations $P^{+\mu}$ and conformal boosts $K^{-\mu}$. In the embedding space they are realized by $SO(2,d)$ matrices of the block form
\begin{align}\label{}
\Lambda^A{}_B=\left(
    \begin{array}{c|c}
     \frac1{2}\delta^\pm{}_\pm  & 2b^{\pm}{}_{ \nu}\\
      \hline
      -b^{ \mu}{}_{\pm} & \delta^\mu{}_\nu
    \end{array}
    \right)  
\end{align}
with infinitesimal parameters $b^\pm \equiv b_\mp$. We find 
\begin{align}\label{eq2.12}
\delta X^\pm =2 b^{\pm \mu} X_\mu\qq \delta X^\mu =  -b^{-\mu} X^+ - b^{+\mu} X^- \quad \Rightarrow\quad \delta(X\cdot X)=0\,,
\end{align}
so that
\begin{align}\label{eq2.15}
\delta x^\mu &= \delta \left( {X^\mu \over X^+} \right) 
  = -b^{-\mu} +  b^{+\mu}\, x^2 - 2 (b^{+} x)\,  x^\mu \,,
\end{align}
where $ b^{-\mu}$ is the  parameter of infinitesimal translations and $b^{+\mu}$ of conformal boosts.

The auxiliary null vector $Z^A$ transforms in the same way,
\begin{align}\label{eq2.12Z}
\delta Z^\pm =2 b^{\pm \mu} Z_\mu\qq \delta Z^\mu =  -b^{-\mu} Z^+ - b^{+\mu} Z^- \quad \Rightarrow\quad \delta(Z\cdot Z)=0\,.
\end{align}
Applied to the fixed frame variable $z^\mu$ in \p{eq2.8} this transformation becomes\footnote{The transformation  \p{eqe2.16} can also be obtained from the inversion   \p{inversion} and the  relation $K=IPI$.} 
\begin{align}\label{eqe2.16}
&\delta z^\mu = \delta\left( Z^\mu  - { Z^+ \over X^+} X^\mu \right) =    2 b^{+\mu} (x  z)  - 2 x^\mu (z b^{+})  \quad \Rightarrow\  \delta(z^2)=0 \,.
\end{align}
We see that $z^\mu$ transforms only under conformal boosts and   $\delta z^\mu \vert_{x=0}=   0$. 

\subsubsection*{Transformations of  tensor fields} 

As we explained, the $Z-$projected fields $O(X,Z)$ are inert under conformal transformations, see \p{eq2.10}. To recover the familiar tensor field transformations we have to `undress' $O(X,Z)$ to an ordinary field with  Lorentz indices. Let us see how this works in the example of the vector field \p{eqe2.3}. First, we use the gauge freedom  \p{eq2.2} and \p{eqeq2.7}  to fix the gauges
\begin{align}\label{eq2.17}
V^-(X)=Z^+ = 0\,.
\end{align}
A conformal boost with parameter $b^+$ should be accompanied by  compensating gauge transformations to maintain the gauges \p{eq2.17}. Then we use the fact that the projected vector field $V(X,Z)$ is  invariant under the conformal transformations  to write down
\begin{align}\label{vector}
0&=\delta V(X,Z)\nt
& = z^\mu \left( \delta V_\mu(x) +\delta x^\nu \pa_\nu V_\mu(x) -2\Delta (b^+ x) \, V_\mu(x) + 2b^+_\mu x^\nu V_\nu(x)     - 2x_\mu b^{+\nu} V_\nu(x)   \right)\,.
\end{align}
Here $\delta x^\nu$ is the coordinate transformation \p{eq2.15} and the weight factor comes from the definition  \p{eq2.4} of the ordinary field and from \p{eq2.12}. We have also used the transformation \p{eqe2.16} of   $z^\mu$.  Stripping off  $z^\mu$, we find the familiar conformal transformation of a vector field. The generalization to  tensors of arbitrary integer spin is straightforward. 

\subsection*{Transformations of the deformed light-ray operators} 
 
{Comparing \p{eq:transformationB} and  \p{eq:primaryB}, we observe that the light-ray operators acquire a phase under the gauge transformation. }
When we fix the frame \p{eq2.8}, the gauge parameter $\b=- Z^+/X^+$ appears in the exponential factor
\begin{align}\label{eqe2.20} 
{\bf L}_{\hat{\omega}}[O]\left(X, Z- {Z^+ \over X^+} X\right)  =  e^{i \hat{\omega} Z^+/X^+} {\bf L}_{\hat{\omega}}[O]( X,  Z) \,.
\end{align}
According to \p{eq2.15} and \p{eqe2.16} both arguments of $O\left(x - {z \over \alpha}, z\right) $ in \p{eq:definitionpoincare} are inert  under conformal boosts at the origin $x=0$.      However, the exponential factor in \p{eqe2.20} transforms: 
\begin{align}\label{}
  \delta_{b^+} \left( {Z^+ \over X^+} \right) =   {2 b^+\cdot Z \over X^+} - {2 b^+\cdot X\, Z^+ \over (X^+)^2} \  \ \underset{X^+=1, \ Z^\mu=z^\mu}{\overset{x^\mu = X^\mu/X^+=0}{\xrightarrow{\hspace*{20mm}}}}
    \  \  2 (b^{+} z)   \,.
\end{align}  
This leads to the relation \p{eq:specialconformal} but its derivation  is different from that in \p{eee1.13}. There we implicitly `undressed' the tensors from  the projection variables $z^\mu$ and then applied the standard tensor transformation rules like  \p{vector}. Here we treated both $x$ and $z$ on the same footing and we saw that the effect  \p{eq:specialconformal} originates from the compensating $\b-$gauge transformations needed to restore the special frame \p{eq2.8}. 

The dilatation transformation {can be obtained from} the commutator $ [P^{+\mu} , K^{-\nu} ] = \eta^{\mu\nu} D + L^{\mu\nu}$. Commuting two variations with parameters $b^+$ and $b^-$, we find 
\begin{align}\label{}
&\delta_D x^\mu = \tau\, x^\mu \,, \quad   \delta_D z^\mu =0\,, \quad  \delta_{D}X^+ =  -\tau  X^+  \,, \quad    \delta_{D}Z^+ = -\tau  Z^+ \,,
\end{align}
where  $\tau = -2 b^+\cdot b^-$ is the infinitesimal dilatation parameter. The  variation of $x$ in  the combination $x-z/\a$ on the right-hand side  of \p{eq:definitionpoincare} is  accompanied by a compensating transformation of the integration parameter $\a$, $\delta_D \a =  -\tau \, \a $.
The change of variable in the integral in \p{eq:definitionpoincare} produces the weight $\la^{1-J}$ (with $\la=1+\tau $)  in \p{eq:dilatationFIN}. The extra weight $\la^{-\Delta}$  is absorbed by the defining factor $(X^+)^{-\Delta}$ in \p{eq:embpoinc} (see also \p{eq2.4}), which is implicit in \p{eq:definitionpoincare}. 
This reproduces \p{eq:dilatationFIN}, without setting $x=0$ in the dilatation transformations. 

\section{Reality property of generalized event shapes}
\label{app:CRT}
 
{In this appendix we}
 derive the reality condition for the one-point generalized event shapes \eqref{eq:CRT1}. 
We have 
\be
 \langle  \cO(\hat \omega,n)  \rangle_q &=
{(-1)^{J} \over \eta_{\cO}} \langle \phi^\dagger(\tilde q) \Big( (\mathsf{CRT}) \cO(\hat \omega, n) (\mathsf{CRT})^{-1} \Big)^\dagger \phi(q) \rangle \nn \\
&={(-1)^{J} \over \eta_{\cO}} \langle (\mathsf{CRT}) \cO(\hat \omega, n) (\mathsf{CRT})^{-1} \phi(\tilde q) |  (\mathsf{CRT}) (\mathsf{CRT})^{-1} \phi(q) \rangle \nn \\
&={(-1)^{J} \over \eta_{\cO}}\langle (\mathsf{CRT})^{-1} \phi(q) |  \cO(\hat \omega, n)(\mathsf{CRT})^{-1} \phi(\tilde q)   \rangle \nn \\
&={(-1)^{J} \over \eta_{\cO}} \langle \phi^\dagger(q) (\mathsf{CRT}) \cO(\hat \omega, n) (\mathsf{CRT})^{-1}  \phi(\tilde q) \rangle \nn \\
&=\langle \phi(q)^\dagger \cO(-\hat \omega,n)  \phi(\tilde q) \rangle = \langle  \cO(-\hat \omega,n) \rangle_{q + \hat \omega_1 n_1 + \hat \omega_2 n_2} ,
\ee
Let us set, for example,  $\Delta_\phi = 2$, $d=4$. We then get for the detectors of interest
\be
\label{eq:hermiticity}
f_{\text{S}}( \omega) &= f_{\text{S}}\left(-{ \omega \over  \omega + 1}\right) \ , \nn \\
f_{\text{Q}}( \omega) &= (1 +  \omega) f_{\text{Q}}\left(-{ \omega \over  \omega + 1}\right)  \ , \nn \\
f_{\text{E}}( \omega) &= (1 +  \omega)^2 f_{\text{E}}\left(- { \omega \over  \omega + 1}\right) \ \,.
\ee
The explicit {expressions} \eqref{f+} and \eqref{f-} satisfy \eqref{eq:hermiticity}  indeed.

{For the one-point generalized event shapes \eqref{eq:hermitiantwop} we have}
\be
 \langle  \cO_1(\hat \omega_1,   n) \cO_2(\hat \omega_2, n_2)  \rangle_q &=
{(-1)^{J_1 + J_2} \over \eta_{\cO_1} \eta_{\cO_2}} \langle \phi^\dagger (\tilde q)\Big( (\mathsf{CRT}) \cO_1(\hat \omega_1, n_1) \cO_2(\hat \omega_2, n_2) (\mathsf{CRT})^{-1} \Big)^\dagger \phi(q) \rangle \nn \\
&={(-1)^{J_1 + J_2} \over \eta_{\cO_1} \eta_{\cO_2}} \langle (\mathsf{CRT}) \cO_1(\hat \omega_1, n_1) \cO_2(\hat \omega_2, n_2) (\mathsf{CRT})^{-1} \phi(\tilde q) |  (\mathsf{CRT}) (\mathsf{CRT})^{-1} \phi(q) \rangle \nn \\
&={(-1)^{J_1 + J_2} \over \eta_{\cO_1} \eta_{\cO_2}}  \langle (\mathsf{CRT})^{-1} \phi(q) |  \cO_1(\hat \omega_1, n_1) \cO_2(\hat \omega_2, n_2) (\mathsf{CRT})^{-1} \phi(\tilde q)   \rangle \nn \\
&={(-1)^{J_1 + J_2} \over \eta_{\cO_1} \eta_{\cO_2}}\langle \phi^\dagger(q) (\mathsf{CRT}) \cO_1(\hat \omega_1, n_1) \cO_2(\hat \omega_2, n_2) (\mathsf{CRT})^{-1}  \phi(\tilde q) \rangle \nn \\
&=\langle \phi^\dagger(q) \cO_1(-\hat \omega_1,n_1) \cO_2(-\hat \omega_2,n_2)  \phi(\tilde q) \rangle \nn \\[2mm]
&= \langle  \cO_1(-\hat \omega_1, n_1)  \cO_2(-\hat \omega_2, n_2)  \rangle_{q + \hat \omega_1 n_1 + \hat \omega_2 n_2}\ \,.
\ee
Going from the second to the third line we used the anti-unitary property of $\mathsf{CRT}$. Together with \re{eq:twopointgeneral} this results in the relation \eqref{eq:CRT2}.

\section{One-point function}\label{app:1pt}

In this appendix we consider the simplest correlation function  involving a single flow operator  
\begin{align}\label{1pt}
\vev{ \phi(\tilde q)  \mathcal O_{J}(\ho,n)  \phi(q)} = \int d^d x_1 e^{-i\tilde q x_1} \int d^d x_2 e^{-iqx_2} \vev{\phi(x_1)  \mathcal O_{J}(\ho,n) \phi(x_2)} \,,
\end{align}
where the scalar operators $\phi(x_2)$ and $\phi(x_1)$ define the source and sink, respectively, and the flow operator is given by \re{eq:definitiondet}. Its calculation   
will serve as an illustration of the general formalism for computing more complicated observables from correlation functions.

For  simplicity we work in $d=4$ dimensions, for arbitrary $d$ the formulae become more complicated. For the lowest values of the spin $J=0,1,2$, the flow operator in \re{1pt} is identified as the scalar, charge or energy detector, respectively.
After inserting the definition  \re{eq:definitiondet} of $\mathcal O_{J}(\ho,n)$ in \re{1pt}, we encounter a three-point Wightman correlation function involving
two scalar operators $\phi$ with  scaling dimension $\Delta$ and the spinning twist-two operator $O_J(x) \equiv \mathcal O_{\mu_1\dots\mu_J}(x)\bar n^{\mu_1} \dots \bar n^{\mu_J}$. Its form is fixed by  conformal symmetry up to an overall normalization factor,
\begin{align}\label{phiOphi}
\vev{\phi(x_1) \mathcal O_J (x_3) \phi(x_2)} =   {\mathcal N_J\over x_{13}^2 x_{32}^2 x_{12}^{2\Delta-2} } \lr{{2i(x_{13}\bar n)\over x_{13}^2} + {2i(x_{32}\bar n)\over x_{32}^2}}^J\,, 
\end{align}
where we used a shorthand notation for $x_{ij}=x_i-x_j$. The poles of the correlation function come with an `$\pm i0$' prescription whose sign depends on the time separation of the operators:
\begin{align}
x_{12}^2 \to x_{12}^2 - i0 x_{12}^0\,,\qquad x_{13}^2 \to x_{13}^2 - i0 x_{13}^0\,,\qquad x_{32}^2 \to x_{32}^2 - i0 x_{32}^0 \,. 
\end{align} 
In \re{phiOphi} we  have tacitly assumed that the operator $O_J(x)$ has the canonical scaling dimension $2+J$. This is certainly true for $J=1$ and $J=2$ since the corresponding operators are the conserved $U(1)$ current and the stress-energy tensor. In a generic interacting conformal theory, the operators with $J=0$ and $J\ge 3$ are not protected and their scaling dimension differs from the canonical dimension $2+J$, unless there exists an additional symmetry that protects them.~\footnote{This happens in particular in $\cN=4$ SYM  where the scalar operator with $J=0$ is half-BPS.} Since we are mostly interested in the case $J\le 2$, we shall assume that the operators $O_J(x)$ are protected.

To obtain $\vev{\phi(x_1)  \mathcal O_{J}(\ho,n) \phi(x_2)}$ from \re{phiOphi}, we have to replace $x_3=rn+\alpha \bar n$ and apply the operations specified in \re{eq:definitiondet}
(take the limit $r\to\infty$ and integrate over $\alpha$).  This can be done using the following identity 
\begin{align}\label{P}
{1\over x_{13}^2 x_{32}^2} \lr{{2i(x_{13}\bar n)\over x_{13}^2} + {2i(x_{32}\bar n)\over x_{32}^2}}^J = P_J\left(i(\bar n \partial_{x_1}),i(\bar n \partial_{x_2})\right)  {1\over x_{13}^2 x_{32}^2} \,,
\end{align}
where $P_J(s_1,s_2)$ is a homogenous polynomial in $s_i$ of degree $J$. Its form can be found by comparing the coefficients in front of the powers of $(x_{13}\bar n)$ on both sides of \re{P}\,,
\begin{align}\label{Geg}
P_J(s_1,s_2) =\sum_{k=0}^J  \lr{J\atop k} {(-s_1)^k\over k!} {s_2^{J-k}\over (J-k)!} \,,
\end{align}
and can be identified as   Gegenbauer polynomials. Replacing $x_3=rn+\alpha \bar n$ we find for $r\gg 1$
\begin{align}\notag\label{Sch}
{1\over x_{13}^2 x_{32}^2} &\to {1\over 4r^2  (n(x_1-\alpha\bar n)-i0)((n(\alpha\bar n-x_2)-i0) } 
\\[2mm]
&=-{1\over 4r^2}\int_0^\infty ds_1 ds_2  \, e^{-i s_1 (n(x_1-\alpha\bar n))-is_2(n(\alpha\bar n-x_2)) }\,,
\end{align}
where we only kept the leading term.
Here the minus sign in the exponent in the second relation is dictated by the `$-i0$' prescription of the propagator in the first line. 

The rational for using the integral representation \re{Sch} is that, upon substituting \re{Sch} into \re{P},  the action of the differential operator amounts to inserting the polynomial $P_J(s_1(n\bar n), -s_2(n\bar n))=(n\bar n)^J P_J(s_1,-s_2)$ into the integral over $s_1$ and $s_2$.
Putting together the above relations and carying out the $\a-$integration in  \re{eq:definitiondet} we find
\begin{align}\label{1pt-x}
 \vev{\phi(x_1)  \mathcal O_{J}(\ho,n) \phi(x_2)} = - {\pi \mathcal N_J\over 2x_{12}^{2\Delta-2} } \int_0^\infty ds_1 ds_2  \, e^{-i s_1 (n x_1)+is_2(nx_2) }
 \delta(s_1 -s_2-\hat \omega) P_J(s_1,-s_2)\,,
\end{align}
where the delta function arises from the integration over $\alpha$ in \re{eq:definitiondet} and the factor of $1/r^2$ in \re{Sch} cancels against the analogous factor in \re{eq:definitiondet}. Note that the dependence on the auxiliary null vector $\bar n$ dropped out of this relation, as it should be. The dependence on $J$ resides in the normalization factor and in the polynomial 
\re{Geg}. 

To obtain the correlation \re{1pt} we have to Fourier transform \re{1pt-x} with respect to $x_1$ and $x_2$. This can be done using the following relation 
\begin{align}\label{sigma}
\sigma_\Delta(q) = \int {d^4 x \e^{-i(qx)}\over (-x^2+i0 x^0)^\Delta} =\theta(q)  {2\pi^3 (q^2/4)^{\Delta-2}\over \Gamma(\Delta)\Gamma(\Delta-1)}\,,
\end{align}
where $\theta(q)\equiv\theta(q^0)\theta(q^2)$ ensures that $\sigma_\Delta(q) \neq 0$ only  for time-like $q$'s. In this way we find 
\begin{align}\notag\label{O-q}
\vev{ \mathcal O_{J}(\ho,n)}_q & \equiv \int d^4 x_2\e^{-iqx_2}  \vev{\phi(0)  \mathcal O_{J}(\ho,n) \phi(x_2)} 
\\
& =\mathcal N'_J  \int_0^\infty ds_1 ds_2 \,  \delta(s_1 -s_2-\hat \omega) P_J(s_1,-s_2) \sigma_{\Delta-1}(q-ns_2) \,,
\end{align}
where the proportionality factor $\mathcal N'_J$ does not depend on $\omega$ and, therefore, it cancels in the ratio $\vev{ \mathcal O_{J}(\ho,n)}_q/\vev{\mathcal O_{J} (\omega=0,n)}_q$. The expression for $\vev{\mathcal O_{J} (\omega=0,n)}_q$ was previously derived in \cite{Belitsky:2013xxa,Belitsky:2013bja}.

\begin{figure}
\centering
\includegraphics{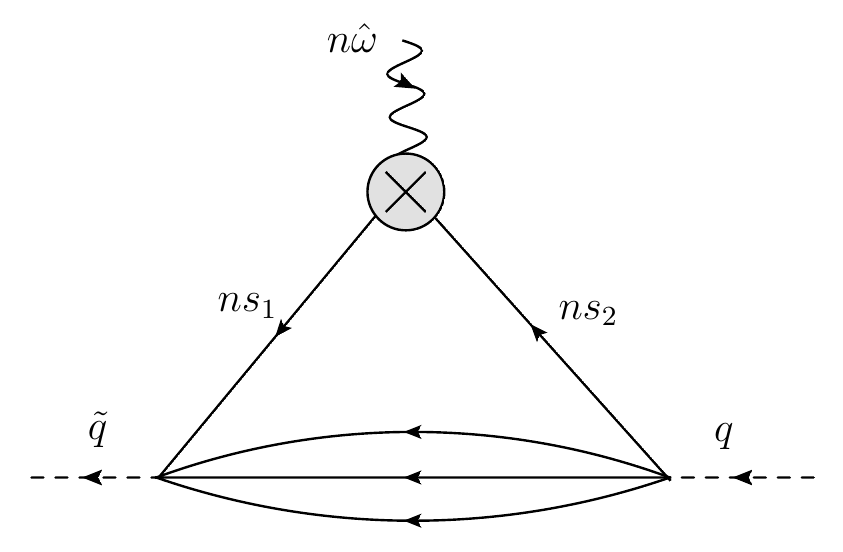} 
\caption{Diagrammatic interpretation of the one-point correlation \re{O-q}. The grey blob represents  the detector and the arrows indicate the propagation direction of  the particles.}\label{Fig:1pt}
\end{figure}

The relation \re{O-q} has a simple interpretation in a weakly coupled CFT, as shown in Figure~\ref{Fig:1pt}. The operator $\phi(x_2)$ excites the vacuum and creates the state $\int d^4 x_2 \e^{-iqx_2}\phi(x_2) $ containing an arbitrary number of particles with the total momentum $q$. These particles propagate to the final state and are absorbed by the sink $\phi(x_1)$. The flow operator $ \mathcal O_{J}(\ho,n)$ plays the role of a detector located on the celestial sphere in the direction $n$. It selects a particle carrying momentum $ns_2$ and transfers the light-like momentum $n\hat\omega$  to it, so that the momentum of the outgoing particle is  $n s_1=ns_2+n\hat\omega$. Thus, the integration variables $s_i$ in \re{O-q} have the meaning of  particle energies  and 
the polynomial $P_J(s_1,-s_2)$ defines the transition amplitude. Finally, $\sigma_{\Delta-1}(q-ns_2)$ gives the probability for the remaining particles with  total momentum $q-ns_2$ to be absorbed by the sink.  

Changing the integration variable to  $s_2=q^2/(2(qn)) s$ {and replacing $\hat\omega = q^2/(2(qn)) \omega$}, we find after some algebra  
\begin{align}
\vev{ \mathcal O_{J}(\ho,n)}_q\sim {(q^2)^{J+\Delta-2} \over (qn)^{J+1} }
 \int^1_{\max(0,- \omega)} ds\,  P_J(s+\omega,-s) (1-s)^{\Delta-3} \,.
\end{align}
Here the upper integration bound follows from the condition $(q-n s_2)^2\ge 0$ 
and the lower bound is due to the condition $s_1=s_2+\hat\omega\ge 0$. It is easy to check that this relation is in agreement with \re{eq:onepoint}. 
 We find that up to a normalization factor the corresponding scaling function is given by  
\begin{align}\label{fO}
f_{\mathcal O}(\omega,J)= c_J \int^1_{\max(0,- \omega)} ds\,  P_J(s+\omega,-s) (1-s)^{\Delta-3} \,,
\end{align}
where the coefficient  $c_J= {(-1)^{J} \Gamma (J+\Delta -1)}(J!)^2 /\lr{\Gamma (\Delta -2)  (2 J)!}$ is chosen in such a way that $f_{\mathcal O}(0,J)=1$.
Notice that the lower integration bound depends on the sign of $\omega$. As we show in a moment, this makes the function $ f_{\mathcal O}(\omega,J)$  non-analytic  around $\om=0$.

Replacing the polynomial $P_J$ in \re{fO} with its explicit expression \re{Geg} we find that for the lowest values of the spin $J=0,1,2$
 the function $f_{\mathcal O}(\omega,J)$ is given by \re{f+} and \re{f-}.
It is straightforward to verify that these expressions satisfy the crossing symmetry relation \re{eq:onepoincrt} (upon replacing $\Delta_\phi=\Delta$, $\Delta_O=J+2$ and $d=4$).

The relations \re{f+} and \re{f-} define the function $f_\mathcal{O}(\omega,J)$ on the positive and negative semi-axes of $\omega$, respectively. Although the functions
\re{f+} and \re{f-} are continuous at the origin, their derivatives $f'_S$,  $f^{'''}_Q$  and $f^{(5)}_E$ have a jump at $\omega=0$. 
For an arbitrary spin $J$, this happens for the derivative of order $2J+1$.

\subsection*{Fixing the one-point function from symmetries}\label{secOnept}

Let us now ask if we can fix the result for the one-point function using symmetries. A similar discussion for the case $\hat \om=0$ can be found in Section 5.1.3 in \cite{Kologlu:2019bco}.  From the formulas above, it is clear that the conditions \eqref{eq:transformationA}, \eqref{eq:transformationB}, and \eqref{eq:zresccaling} are not enough to fix the result. Indeed, consider the following ansatz for the one-point function
\be
\label{eq:ansatz1}
&\langle \phi(X_1) {\bf L}_{\hat{\omega}}[O]( X_3,  Z_3)  \phi(X_2)   \rangle = {1 \over (-2 X_1 \cdot X_3)^{{\Delta - J \over 2}} (-2 X_2 \cdot X_3)^{{\Delta - J \over 2}}} {1 \over (-2 X_1 \cdot X_2)^{\Delta_{\phi} +{J-\Delta \over 2}}} \nn \\
&\times \int_0^\infty d s_1 d s_2 \, s_1^{\Delta - 2} f\left({s_2 \over s_1} \right) \delta(s_1 - s_2 - \hat \omega) e^{{i \over 2} \Big( s_2 {Z_3 \cdot X_2\over X_2 \cdot X_3} - s_1 {Z_3 \cdot X_1 \over X_1 \cdot X_3} \Big) } ,
\ee
where $f\left({s_2 / s_1} \right)$ is an unknown function. We note that \eqref{eq:ansatz1} has the correct transformation properties for any $f\left({s_2 / s_1} \right)$. Moreover, assuming that $f(1) \neq 0$, for $\hat \omega =0$ it yields the known expression for the three-point function.

We can fix the function $f(s_2/s_1)$ in \re{eq:ansatz1} by noting that the light-ray operator in \re{eq:ansatz1} is polynomial 
in the polarization vector $Z$. To implement this condition, we observe that a local operator satisfies the shortening condition 
\be
\label{eq:shortening}
( w \cdot \partial_z)^{J+1} O(x,z) = 0 ,
\ee
where we introduced an extra polarization vector which is null and transverse to $z^\mu$, see section 3.1 in \cite{Chang:2020qpj} for more details, 
\be
z^2 = z \cdot w = w^2 =0 \,.
\ee 
In writing \eqref{eq:shortening} we assumed that $O(x,z)$ is a local operator with  spin $J$.  The relation \re{eq:shortening}
is just an economic way of saying that $O(x,z)$ is a polynomial in the polarization tensor $z$ of degree $J$.   We can now pull $( w \cdot \partial_z)^{J+1}$ through the $\omega$-deformed light transform to get a shortening condition when acting on \eqref{eq:ansatz1}. This fixes the unknown function $f\left({s_2 / s_1} \right)$. 

Let us do this exercise explicitly for the simplest case $J=0$. In this case we get the relation
\be
\Big( (z \cdot \partial_x)(w \cdot \partial_z) - (z \cdot \partial_z)(w \cdot \partial_x) \Big) {\bf L}_{\hat \omega}[O](x,z) = 0,  \qquad \text{(scalar $O$)} ,
\ee
which was also discussed in \cite{Chang:2020qpj}. It leads to the following equation for the three-point function
\be
\label{eq:shorteningonepoint}
\Big( (z \cdot \partial_x)(w \cdot \partial_z) - (z \cdot \partial_z)(w \cdot \partial_x) \Big) \langle \phi(x_1) {\bf L}_{\hat{\omega}}[O](x,z)  \phi(x_2)   \rangle = 0 \,. 
\ee
Plugging \eqref{eq:ansatz1} into \eqref{eq:shorteningonepoint} we find  
\be
f(x) = c_0 \, x^{{\Delta-2 \over 2}} ,
\ee
which is indeed the correct expression. Analogous computations can be done for $J>0$ as well.

\section{Free scalar field at null infinity}\label{app:free}

Let us consider a free massless real scalar field in $d=4$ dimensions. It can be represented in terms of creation and annihilation operators,  
\be
\varphi(x) = \int_{- \infty}^\infty {d^4 p \over (2 \pi)^3} \delta_+(p^2) ( a_p e^{- i (p  x)} + a^{\dagger}_p e^{i (p  x)}) \,,
\ee
where $\delta_+(p^2) =\theta(p^0)\delta(p^2)$.
 
Following the discussion in Section~\ref{sect:warm}, we send the scalar field to null infinity by replacing $x=r n + \alpha \bar n$ and taking the limit $r\to\infty$ 
\be
\label{eq:freefield}
\lim_{r \to \infty} r \varphi(r n + \alpha \bar n) = \lim_{r \to \infty} r  \int_{- \infty}^\infty {d^4 p \over (2 \pi)^3} \delta_+(p^2) \Big[ a_p e^{-i r (p n)  - i \alpha (p \bar n)} + a^{\dagger}_p e^{i r (p  n)+ i \alpha (p  \bar n)} \Big] \,.
\ee
It is convenient to apply Sudakov's decomposition of the momenta
\be
p = p_+ n + p_- \bar n + p_\perp \ ,
\ee
where $n^2 = \bar n^2 = 0$ and $(p_\perp n) =(p_\perp \bar n) = 0$. Switching to the variables $(p_+, p_-, p_\perp)$ in \eqref{eq:freefield} we get
\be
\lim_{r \to \infty} r (n \bar n) & \int_{- \infty}^\infty {d p_+ d p_- d^2 p_{\perp} \over (2 \pi)^3} \theta(p_+) \theta(p_-) \delta(2 p_+ p_- (n \bar n) - p_\perp^2) \nn \\
& \times \Big[ a_{p_+,p_-,  p_\perp} e^{-i (n \bar n) ( p_-  r + p_+ \alpha )} + a^{\dagger}_{p_+,p_-,  p_\perp}  e^{i (n \bar n) ( p_- r + p_+ \alpha )} \Big] \,.
\ee
Next we rescale $p_- \to {p_- /( (n \bar n) r)}$ and take the limit $r \to \infty$ to get
\be
& \int_{- \infty}^\infty {d p_+ d p_- d^2 p_{\perp} \over (2 \pi)^3} \theta(p_+) \theta(p_-) \delta(p_\perp^2) \left[ a_{p_+,0_-  p_\perp} e^{-i  (p_-  + (n \bar n) p_+ \alpha)} + a^{\dagger}_{p_+,  0_-,p_\perp}  e^{i ( p_-  + (n \bar n)  p_+ \alpha)}\right] \,.
\ee
Integrating over $p_-$ and using the identity $\int d^2 p_\perp  \delta(p_\perp^2) = \pi \int d p_\perp^2  \delta(p_\perp^2)  = \pi$ we  arrive at
\be
\label{eq:freescalarlim'}
 {\lim_{r \to \infty} r  \varphi(r n + \alpha \bar n) =i \int_0^\infty {d p_+ \over 8 \pi^2}  \left[a^{\dagger}_{p_+,0_-,0_\perp}  e^{i (n \bar n)  p_+ \alpha} - a_{p_+, 0_-,0_\perp} e^{-i (n \bar n) p_+ \alpha } \right]}\,.
\ee
Here the creation and annihilation operators are defined for the momentum $(p_+,0_-,0_\perp)$ that is aligned along the null vector $n$. Denoting this momentum by $sn^\mu$ we can rewrite the relation \re{eq:freescalarlim'} in a covariant form,
\be
\label{eq:freescalarlim}
 {\lim_{r \to \infty} r  \varphi(r n + \alpha \bar n) =i \int_0^\infty {d s \over 8 \pi^2}  \left[a^{\dagger}_{sn}  e^{i (n \bar n) s \alpha} - a_{sn} e^{-i (n \bar n) s \alpha } \right]}\,.
\ee
The complex scalar field satisfies a similar relation with the only difference that  the operator $a^\dagger_{sn}$ is replaced by the annihilation operator $b^\dagger_{sn}$ with  
 opposite charge. Then, substituting \re{eq:freescalarlim} into \re{eq:scalardetector} and performing the integration over $\alpha$ we obtain a representation of the scalar flow operator that coincides with the second line of \re{eq:scalardetector}. 
  
\section{Derivation of the SSC kernels}
\label{sec:derivationSSC}

Here we present the details of the derivation of the scalar-scalar kernels, Eqs.~\re{eq:kernelSSCpm} and \re{eq:kernelSSCpp},  in Mellin space. The derivation is a direct generalization of the analysis in \cite{Belitsky:2013xxa}. We start with the four-point correlation function \eqref{eq:fourpointneq4g105}, in which we retain  only the interaction part. It is described by the function $\Phi(u,v)$ which has the Mellin representation \eqref{eq:MellinformB}. 
Using the symmetry properties of the Mellin amplitude $M(j_1, j_2)$ and doing  the crossing transformation $j_1 \to -1 -j_1 - j_2$, we can recast  \eqref{eq:MellinformB} as follows:
\be
\label{eq:MellinformApp}
\Phi(u,v) &=  \int_{{\cal C}_0} {d j_1 d j_2 \over (2 \pi i)^2} [\Gamma(1-j_1) \Gamma(1-j_2)\Gamma(j_1+j_2)]^2 M(j_1, j_2) u^{-1-j_1-j_2} v^{j_2}\,,
\ee
where the integration contour satisfies ${\rm Re}(j_1)$, ${\rm Re}(j_2)<0$ and ${\rm Re}(j_1 + j_2)>-1$.

The relevant integral takes the form\footnote{The relation between \eqref{eq:MellinformApp} and the variables used in \cite{Chang:2020qpj} is  $\gamma_{12} =j_1 + j_2 , ~~~\gamma_{14} =1- j_2$ \,.}
\be
\label{eq:startingpoint}
&\int_{{\cal C}_0} {d j_1 d j_2 \over (2 \pi i)^2}  [\Gamma(1-j_1) \Gamma(1-j_2) \Gamma(j_1+j_2)]^2 M(j_1, j_2) \nn \\
&\times {1 \over (2 \pi)^4} \int d^4 x_3 d^4 x_4\, e^{i (\tilde q x_4) - i (q x_3)} \int_{-\infty}^{\infty} d \alpha_1 d \alpha_2\, e^{- i \alpha_1 \hat \om_1 (n_1 \bar n_1) - i \alpha_2 \hat \om_2 (n_2 \bar n_2)} \nn \\
&\times {1 \over x_{12}^4 x_{34}^4} \Big( {x_{14}^2 x_{23}^2 \over x_{13}^2 x_{24}^2 } \Big)^{j_2 - 1} \Big( {x_{12}^2 x_{34}^2 \over x_{13}^2 x_{24}^2 } \Big)^{2 - j_1 - j_2} ,
\ee
where we kept the ordering of the operators implicit. Next we set $x_i = r_i n_i + \alpha_i \bar n_i$ (for $i = 1, 2$) at the detector points and take the  limit $r_i \to \infty$ according to \eqref{eq:definitiondet}.  We also rescale $\alpha_i \to {\alpha_i /( n_i  \bar n_i)}$ to get (skipping the first line of \eqref{eq:startingpoint})
\be
&{1 \over (2 \pi)^4} \int d^4 x_3 d^4 x_4 e^{i (\tilde q x_4) - i (q x_3)} \int_{-\infty}^{\infty} d \alpha_1 d \alpha_2 e^{- i \alpha_1 \hat \om_1 - i \alpha_2 \hat \om_2} {2^{\Jtwo+\Jone-4} \over (n_1 n_2)^2 x_{34}^4}  \nn \\
&\times \left[ {x_{34}^2 (n_1 n_2) \over (\alpha_1 - (x_4 n_1) - i \epsilon)( (x_3 n_2) - \alpha_2 - i \epsilon ) } \right]^{1-\Jone}
 \left[  {x_{34}^2 (n_1 n_2) \over (\alpha_2 - (x_4 n_2) - i \epsilon)( (x_3 n_1) - \alpha_1 - i \epsilon ) }\right]^{1-\Jtwo}\, \,.
\ee
Here we introduced the $i \epsilon$ prescription that captures the ordering of the operators according to \eqref{eq:fourpoint}. Note that the dependence on $\bar n_i$ has dropped out, as expected. The identity 
\be
\label{eq:schwinger}
&\int_{-\infty}^{\infty} d \alpha  {e^{-i \hat \omega \alpha} \over  (\alpha - n \cdot x_3 - i \epsilon)^{a} ( n \cdot x_4 - \alpha - i \epsilon)^{b}}\nn \\
& = {i^{a+b} \over \Gamma(a) \Gamma(b)}  \int_0^\infty d s_1 d  s_2 \  s_1^{a-1} s_2^{b-1} 2 \pi \delta(s_1 - s_2 +  \hat \omega ) e^{i s_1  (n  x_3)  -i s_2  (nx_4) } ,  
\ee
where ${\rm Re}(a)$,  ${\rm Re }(b)>0$,
allows us to perform the integral over the detector times, or equivalently  the light transform.

As a result we get the following expression
\be
&{1 \over (2 \pi)^2} \int d^4 x_3 d^4 x_4 \, e^{i (\tilde q x_4) - i (q x_3) + i s_1 (x_3 n_1) + i s_1'  (x_3 n_2) -  i s_2 (x_4 n_1) - i s_2'  (x_4 n_2) }  { 2^{\Jtwo+\Jone-4} \over \Gamma(1-\Jone)^2 \Gamma(1-\Jtwo)^2}  \nn \\
&
\times \int_0^\infty ds_1 d s_1' ds_2 ds_2' \delta(s_1 - s_2 + \hat \om_1) \delta(s_1' -s_2' +\hat \om_2) {1 \over (n_1 n_2)^{\Jtwo+\Jone}}{1 \over (- x_{34}^2 + i \epsilon x_{43}^0 )^{\Jtwo+\Jone}} \,.
\ee
One of the space-time integrations produces the momentum conserving delta function $(2 \pi)^4 \delta^4(\tilde q- q - \hat \om_1 n_1 - \hat \om_2 n_2)$ as in \eqref{eq:quantumeventshapes}. The remaining integral over $x_{34}$ can be done using the  formula  
\be
\label{eq:sigmatot}
\int d^d x {e^{-i (q x)} \over (- x^2 + i \epsilon x^0)^\Delta } = \theta(q) {2 \pi^{d/2+ 1} \over \Gamma(\Delta)\Gamma(\Delta+1 - {d \over 2})} \lr{q^2/4 }^{\Delta - d/2} \,. 
\ee
In this way we arrive at the desired formula for the scalar-scalar  correlation
\be
\langle  \cO ( \omega_1, n_1) \cO( \omega_2 , n_2) \rangle_q = {1\over 8\pi^2}  {\text{SSC} (z, \omega_1, \omega_2 ) \over (q n_1) (q n_2)}  ,
\ee
where
\be
\label{eq:SSCkerndef}
 \text{SSC} (z, \omega_1, \omega_2 ) = \int_{{\cal C}_0} {d \Jone d \Jtwo \over (2 \pi i)^2}  K_{\text{SS}}(z, \omega_1 , \omega_2) M(j_1, j_2) ,
\ee
and the kernel admits the following  representation
\be 
\label{eq:schwingerkernelSS}
K_{\text{SS}} &= {(\Jtwo+\Jone - 1) (q n_1) (q n_2)  \over 2^{\Jtwo+\Jone-1} (n_1 n_2)^{\Jtwo+\Jone}}  \int_0^\infty ds_1 d s_2 \theta(s_1 + \hat \omega_1)  \theta(s_1 + \hat \omega_2) \theta(q - n_1 s_1 - n_2 s_2)
\nn \\
&\times\left[ (s_1 + \hat \omega_1) s_2 \right]^{-\Jone}\left[ s_1 (s_2 + \hat \omega_2) \right]^{-\Jtwo} \left[ (q - n_1 s_1 - n_2 s_2)^2 \right]^{\Jtwo+\Jone - 2} \,.
\ee
We  rescale the integration variables $s_i \to {2 (q n_i)s_i/q^2}$ and  introduce the dimensionless variables \re{eq:rescaledfreq} and the cross ratio \re{z} to get
\be
\label{eq:kernelSS}
 K_{\text{SS}}(z, \omega_1 , \omega_2) &= {1 \over 2} (\Jtwo+\Jone - 1) z^{- \Jtwo -\Jone} \int_0^\infty ds_1 d s_2 \theta(s_1 + \omega_1)  \theta(s_2 + \omega_2) \theta(1 - s_1 -s_2 +s_1 s_2 z)\nn \\
&\left[(s_1 +  \omega_1) s_2 \right]^{-\Jone}\left[ s_1 (s_2 +  \omega_2) \right]^{-\Jtwo} \left(1 - s_1 -s_2 +s_1 s_2 z \right)^{\Jtwo+\Jone - 2}  ,
\ee
As explained in the bulk of the paper, it is sufficient to consider two choices of the signs of $\omega_{1}, \omega_2$, namely $+-$ and $++$. The two other choices are obtained from the relations \re{eq:relationused}. Here we restrict ourselves to the case $\omega_1 >0$, so that the  integration range of $s_1$ in \eqref{eq:kernelSS} is 
\be
0 \leq s_1 \leq {1- s_2 \over 1 - s_2 z} \,. 
\ee
The result takes the following form
\be\label{K-ssc}
& K_{\text{SS}}(z, \omega_1 , \omega_2)= {\Gamma(\Jtwo+\Jone) \Gamma(1-\Jtwo)\over 2 \Gamma(\Jone)}  z^{-\Jtwo-\Jone} \omega_1^{1-\Jtwo-\Jone}  \nn \\
&\hskip10mm \times \int_{{\rm max}[0, - \omega_2]}^1 d s_2 s_2^{-\Jtwo} (1-s_2)^{\Jone - 1} (s_2 + \omega_2)^{-\Jtwo} \Big(1 +\omega_1 - s_2(1 +  \omega_1 z) \Big)^{\Jtwo - 1} \,.
\ee
Note that we first compute the kernel in the region where the integrals converge, namely  ${\rm Re}(\Jtwo+\Jone)>1$,  and then define it in the region where  we use the Mellin amplitude via analytic continuation.

Next we switch to the more convenient integration variable $s_2 = (1+{1+\omega_2 \over | \omega_2|} X)^{-1}$ to get
\be\label{last}
K_{\text{SS}} &= {\Gamma(\Jtwo+\Jone) \Gamma(1-\Jtwo)\over 2 \Gamma(\Jone)}  z^{-\Jtwo-\Jone} \omega_1^{1-\Jtwo-\Jone} |\omega_2|^{1-\Jtwo-\Jone} (1+\omega_1)^{\Jtwo-1} (1+\omega_2)^{\Jone-1}  \nn \\
&\times\int_0^{\infty} d X\, \theta(1+{\rm sign}(\omega_2) X)  X^{\Jone - 1} (1+{\rm sign}(\omega_2) X)^{-\Jtwo} \Big(X+{(1-z) \omega_1 |\omega_2| \over (1+ \omega_1) (1+\omega_2)} \Big)^{\Jtwo - 1} \,.
\ee
At this point we can in principle simply evaluate the last integral in terms of hypergeometric functions. The convergence of the integral requires first considering the region ${\rm Re}(\Jone)>0$ and then analytically continuing to the region of interest, ${\rm Re}(\Jone)<0$. 

To do this more explicitly  it is convenient to perform the integration in a slightly different manner. We first represent the last factor on the right-hand side of \re{last} in Mellin form:
\be
 \Big(X+{(1-z) \omega_1 |\omega_2| \over (1+ \omega_1) (1+ \omega_2)} \Big)^{\Jtwo - 1}  =  \int {d s \over 2 \pi i} { \Gamma(1-\Jtwo + s) \Gamma(-s) \over  \Gamma(1- \Jtwo)  }X^{\Jtwo -1 -s } \Big( {(1-z) \omega_1 |\omega_2| \over (1+ \omega_1) (1+ \omega_2)}  \Big)^s \,.
\ee
After that we integrate over $X$,
\be
\omega_2 > 0:& ~~~ \int_0^\infty d X  X^{\Jtwo+\Jone - s - 2} (1+X)^{-\Jtwo} = {\Gamma(1-\Jone + s) \Gamma(\Jtwo+\Jone - 1 - s) \over \Gamma(\Jtwo)} , \nn \\
\omega_2 < 0:& ~~~ \int_0^1 d X  X^{\Jtwo+\Jone - s - 2} (1-X)^{-\Jtwo} = {\Gamma(1-\Jtwo) \Gamma(\Jtwo+\Jone-1-s) \over \Gamma(\Jone - s)} \,.
\ee
Finally, we shift  the Mellin variable $s \to \Jtwo+\Jone-1 + s$ to get
\be
\label{eq:appendixKernel}
K_{\text{SS}}^{++} &= {1 \over 2} \Gamma(\Jtwo+\Jone)   \Gamma(1-\Jtwo-\Jone)   z^{-\Jtwo-\Jone} (1-z)^{1-\Jtwo-\Jone} (1+\omega_1)^{-\Jone} (1+\omega_2)^{-\Jtwo} \nn \\
&\times\int {d s \over 2 \pi i}    {\Gamma(-s)  \Gamma(\Jone + s)  \Gamma(\Jtwo + s)  \Gamma(1-\Jtwo-\Jone - s) \over  \Gamma(1-\Jtwo-\Jone) \Gamma(\Jone)  \Gamma(\Jtwo)} \Big( {(1-z) \omega_1 \omega_2 \over (1+ \omega_1) (1+ \omega_2)}  \Big)^s \,.
\ee
This is precisely the formula \re{eq:kernelpp} we quoted in the bulk of the paper. The integration contour in \eqref{eq:appendixKernel} is fixed as follows: we first consider a straight line for ${\rm Re}(\Jone)>0$ that runs parallel to the imaginary axis and separates the poles generated by the product of gamma functions in the numerator, i.e. the poles coming from $\Gamma(\dots -s)$ from the left and the poles coming from $\Gamma(\dots+s)$ from the right. We then deform the contour to the region of interest ${\rm Re}(\Jone)<0$. 
 
Let us also write down the explicit result of the integration   
\be
\label{eq:kernelSSCppTwoHypers}
&K_{\text{SS}}^{++} =  {\pi \over 2 \sin \pi (\Jtwo+\Jone)} (1+\om_1)^{-\Jone}(1+\om_2)^{-\Jtwo} z^{-\Jtwo-\Jone} (1-z)^{\Jtwo+\Jone-1}\nt
&\hskip20mm \times \ _2 F_1 \left(\Jone, \Jtwo, \Jtwo+\Jone,{(1-z) \omega_1 \omega_2 \over (1+ \omega_1) (1+ \omega_2)}  \right) \nn \\
&+  {\Gamma(1-\Jone)\Gamma(1-\Jtwo) \Gamma(\Jtwo+\Jone)\Gamma(\Jtwo+\Jone-1) \over 2\Gamma(\Jone) \Gamma(\Jtwo)} z^{-\Jtwo-\Jone}(\om_1 \om_2)^{1-\Jtwo-\Jone} (1+\om_1)^{\Jtwo-1}(1+\om_2)^{\Jone - 1} \nn \\
&\hskip20mm \times  \ _2 F_1 \left(1-\Jone, 1-\Jtwo, 2 -\Jtwo-\Jone,{(1-z) \omega_1 \omega_2 \over (1+ \omega_1) (1+ \omega_2)} \right ) \,.
 \ee
 
In the case $\omega_2 < 0$ we get instead 
\be
K_{\text{SS}}^{+-} &= {\pi \over 2 \sin \pi (\Jtwo+\Jone)}   z^{-\Jtwo-\Jone} (1-z)^{\Jtwo+\Jone-1} (1+\omega_1)^{-\Jone} (1+\omega_2)^{-\Jtwo}  \nn \\
&\ \ \ \times \int {d s \over 2 \pi i}  \Gamma(-s) {   \Gamma(1-\Jtwo) \Gamma(\Jone + s)  \Gamma(1-\Jtwo-\Jone - s) \over \Gamma(1 - \Jtwo - s)  \Gamma(\Jone) \Gamma(1-\Jtwo-\Jone) }  \Big(- {(1-z) \omega_1 \omega_2 \over (1+ \omega_1) (1 + \omega_2)}  \Big)^s ,
\ee
which is the formula \p{eq:kernelpm} quoted in the bulk of the paper. Doing the integral we get
\be
K_{\text{SS}}^{+-} &= {1 \over 2} z^{-\Jtwo-\Jone}  (1-z)^{\Jtwo - 1} (- \om_1 \om_2)^{-\Jone} (1+\omega_2)^{\Jone - \Jtwo} {\Gamma(1-\Jtwo)^2 \Gamma(\Jtwo+\Jone) \over \Gamma(1+\Jone - \Jtwo)} \nn \\
&\hskip20mm\times \ _2 F_1 \left(\Jone, 1-\Jtwo , 1+\Jone - \Jtwo ,  { (1+ \omega_1) (1 + \omega_2) \over (1-z) \omega_1 \omega_2}\right). 
\ee
This form is not particularly convenient for taking the  limit $\om_i \to 0$. Using a standard identity for the hypergeometric functions we can instead write it as follows
\be
\label{eq:kernelSSCpmTwoHypers}
&K_{\text{SS}}^{+-} =  { \pi \over 2\sin \pi (j_1+j_2)} (1+\om_1)^{-j_2}(1+\om_2)^{-j_1} z^{-j_1-j_2} (1-z)^{j_1+j_2-1}\nt
&\hskip20mm \times \ _2 F_1 \left(j_2, j_1, j_1+j_2, {(1-z) \omega_1 \omega_2 \over (1+ \omega_1) (1+ \omega_2)}  \right) \nn \\
&+ {\Gamma(1-j_1)^2 \Gamma(j_1+j_2)\Gamma(j_1+j_2-1) \over 2\Gamma(j_2)^2} z^{-j_1-j_2}(-\om_1 \om_2)^{1-j_1-j_2} (1+\om_1)^{j_1-1}(1+\om_2)^{j_2 - 1} \nn \\
&\hskip20mm\times  \ _2 F_1 \left(1-j_2, 1-j_1, 2 -j_1-j_2,{(1-z) \omega_1 \omega_2 \over (1+ \omega_1) (1+ \omega_2)}  \right) \,.
 \ee
 
 \subsection*{Null limit of the kernel}
  
Here we comment on the existence of a special simplifying limit in which the momentum of the sink becomes null. The representation \eqref{eq:kernelpp} suggests to consider the limit
\be
\label{eq:specialkin}
{\omega_1 \omega_2 (1-z) \over 1+ \omega_1 + \omega_2 + \omega_1 \omega_2} = 1 \, ,
\ee
which corresponds to 
\be
z = - {1 + \om_1 + \om_2 \over \om_1 \om_2}\, \,. 
\ee
Physically, this condition means that either the source or the sink  carries  null momentum,
\be
\eqref{eq:specialkin}: ~~~ q^2 = 0 ~~\text{or}~~(q+ \hat \omega_1 n_1 + \hat \omega_2 n_2)^2 = 0 \,. 
\ee
Keeping $q^2$ timelike and requiring that $z$ takes the physical values $0<z<1$ yields the constraint  $\om_i < 0$ and $\om_1+\om_2 < - 1$.

In this kinematics the Mellin integral in \eqref{eq:kernelpp} simplifies dramatically and can be immediatelly computed using Barnes' lemma with the following result
\be
\label{eq:kernelppnull}
&K^{\text{null}}_{\text{SS}}(j_1, j_2 | \omega_1, \omega_2)\nt
&\ \    =\norm  { \omega_1 \omega_2 \over 2(1+\omega_1) (1+\omega_2)} \left(- {1 + \omega_1 + \omega_2 \over (1+\omega_1) ( 1 +  \omega_2) } \right)^{-j_1-j_2}\Gamma(1-\Jtwo)  \Gamma(1-\Jone)  \Gamma(\Jtwo + \Jone)\, ,
\ee
where we have set $z = - {1 + \omega_1 + \omega_2 \over \omega_1 \omega_2}$ according to  \eqref{eq:specialkin}. Note that on this locus two-point generalized event shape becomes one-dimensional since the kernel depends nontrivially only on the combination ${z \over 1 - z} = - {1 + \omega_1 + \omega_2 \over (1+\omega_1) ( 1 +  \omega_2) }$.

 \section{Computing $\text{SSC}_{++}$ using correlation functions}
 \label{sec:SSCppcomputationMellin}
 
Here we show the details of the computation of the correlation  $\text{SSC}_{++}$ in \eqref{eq:SSCpp} and \eqref{SSC-st}. The starting point  is the following representation of the relevant Mellin kernel
 \be
 \label{eq:samekernelApp}
 &K^{++}_{\text{SS}}(j_1, j_2| z, \omega_1, \omega_2) = - \norm {1 \over 2} { j_1 j_2 \over \Jtwo+\Jone} \int_0^\infty dX \int_0^1 d t  {(1-t)^2 \over t^2 z^2 (1 + \omega_1 \omega_2 X)}Y_1^{\Jone - 1} Y_2^{\Jtwo-1}\,,
 \ee
 where 
 \begin{align}\label{F.2}
Y_1 = {(1-t)  (1- z + (1+\om_1)(1+\om_2) X)\over t z (1 + \om_2) (1+ \om_1 \om_2 X)} \geq 0\,,\qqquad Y_2 =  {(1-t) (1 + \om_2 ) X  \over z} \geq 0 \,.
\end{align}
The advantage of this representation is that for simple Mellin amplitudes like \eqref{eq:Mellin amplitudes at weak and strong coupling} the  integrals over $j_{i}$   are now very easy to take both at weak and at strong coupling. The basic integrals that we need are  
 \be
 \label{eq:basicintegral}
& \int_{- \ep - i \infty}^{- \ep + i \infty} {d j \over 2 \pi i} Y^{j-1} = \delta(1-Y) - \delta(Y) \,, \nn \\
&  \int_{- \ep - i \infty}^{- \ep + i \infty} {d j \over 2 \pi i} {1 \over j} Y^{j-1} =- {\theta(1-Y) \over Y_+}\,,
 \ee
 where  $\ep>0$ and  ${\theta(1-Y) \over Y_+}$ is the distribution that acts on test functions as $\int_0^1 {d Y \over Y} (f(Y) - f(0))$.  
 
 At weak coupling we get the distribution
 \be
& \int {d \Jone \Jtwo \over (2 \pi i)^2} \Big( {1 \over \Jone}  + {1 \over \Jtwo} \Big) Y_1^{\Jone - 1} Y_2^{\Jtwo - 1}\nt
& = - {\theta(1-Y_1) \over (Y_1)_+} \Big( \delta(1-Y_2) - \delta(Y_2) \Big) -  {\theta(1-Y_2) \over (Y_2)_+} \Big( \delta(1-Y_1) - \delta(Y_1) \Big) \,. 
 \ee
 We then act with this distribution on the test function
 \be
  g(Y_1 , Y_2) &= - {\pa (t,X) \over \pa (Y_1 , Y_2)}  \left.   {(1-t)^2 \over 2 t^2 z^2 (1 + \omega_1 \omega_2 X)} \right|_{t = t(Y_1 , Y_2) , X = X(Y_1, Y_2) }  
 \ee
 and obtain the result   \eqref{eq:SSCpp}.
 
 At strong coupling we get instead
 \be
 \label{eq:strongcouplingPP}
&\int_{- \ep - i \infty}^{- \ep + i \infty} {d \Jone \Jtwo \over (2 \pi i)^2} (\Jtwo+\Jone) (1 + \Jtwo+\Jone) Y_1^{\Jone - 1} Y_2^{\Jtwo - 1} \nn \\
&= (\hat j_1 + \hat j_2) (1 + \hat j_1 + \hat j_2) \int_{- \ep - i \infty}^{- \ep + i \infty} {d \Jone \Jtwo \over (2 \pi i)^2} Y_1^{\Jone - 1} Y_2^{\Jtwo - 1} ,
 \ee
 where $\hat \Jone = Y_1 \partial_{Y_1} + 1$ and $\hat \Jtwo = Y_2 \partial_{Y_2} + 1$. Using \eqref{eq:basicintegral} and \eqref{eq:strongcouplingPP} one can check that the result takes the form
 \be
 \text{SSC}_{++}^{\text{strong}} &= {1 \over 2} ( \pa_{Y_1}^2 g(1,0) + \pa_{Y_2}^2 g(0,1)  -  \pa_{Y_1}^2 g(1,1) - \pa_{Y_2}^2 g(1,1) - 2  \pa_{Y_1} \pa_{Y_2} g(1,1) ) \,  
 \ee
yielding  \eqref{SSC-st}.

\section{Bulk point singularity}\label{app:bulk}

According to \re{SSC-st},  the expressions for the correlations at strong coupling with $\omega_1$ and $\omega_2$ of the same sign contain a new scale $\omega_1+\omega_2 + z \omega_1\omega_2$. To elucidate its origin, we revisit the calculation of the SSC with the Mellin amplitude of the form \re{eq:toymellin}.
For $p=0$ and $p=1$ it coincides with the   expression at weak and strong coupling. For $p\ge 1$ 
the corresponding expression for the four-point correlation function has a bulk singularity. 

Applying \re{K-ssc} and \re{eq:SSCkerndef}, we change the integration variable as $s_2=1/(1+\omega_1 s)$ and obtain for $\omega_1>0$  
\begin{align}\notag \label{exa}
\text{SSC}_p(\omega_1,\omega_2)  &= \int_{{\rm Re}\, j_i=-\delta} {dj_1 dj_2\over (2\pi i)^2} 
 [j_1 j_2(1+j_1+j_2)]^p
{(j_1+j_2) \over j_1  j_2}{\Gamma(-j_2)\Gamma(1+j_1+j_2)\over \Gamma(1+j_1)}
z^{-j_1-j_2} 
\\
&\times 
\int_0^{s_{\rm max}} ds \, s^{j_1-1}(1+\omega_2(1+\omega_1s))^{-j_2}(s(1+\omega_1) + (1-z))^{j_2-1} \,,
\end{align}
where $s_{\rm max}$ is given by $\infty $ and $-(1+\omega_2)/(\omega_1\omega_2)$ for positive and negative $\omega_2$, respectively.
The integration over small $s$ produces a pole $1/j_1$. For $p>1$ it is compensated by the prefactor and, therefore, does not contribute. This implies that 
$\text{SSC}_p(\omega_1,\omega_2)=0$ for negative $\omega_2$ whereas for positive $\omega_2$ it does not contains  powers of $1/(1+\omega_1 z)$ and $1/(1+\omega_2 z)$.

For $p\ge 2$ we rewrite the previous relation as
\begin{align}\label{SSCp}
\text{SSC}_p(\omega_1,\omega_2) = &(\partial_1 \partial_2)^{p-1}(1+\partial_1+\partial_2)]^p(\partial_1+\partial_2) I(\epsilon_1,\epsilon_2)\Big|_{\epsilon_i=0}\,,
\end{align}
where  $\partial_i = \partial_{\epsilon_i}$ and the notation was introduced for
\begin{align}\notag
I(\epsilon_1,\epsilon_2) &=\int_{{\rm Re}\, j_i=-\delta} {dj_1 dj_2\over (2\pi i)^2} {\Gamma(-j_2)\Gamma(1+j_1+j_2)\over \Gamma(1+j_1)}
z^{-j_1-j_2} \e^{\epsilon_1 j_1+\epsilon_2 j_2}
\\
&\times 
\int_0^{s_{\rm max}} ds \, s^{j_1-1}(1+\omega_2(1+\omega_1s))^{-j_2}(s(1+\omega_1) + (1-z))^{j_2-1}\,.
\end{align}
Using the identity
\begin{align}
\int_{{\rm Re}\, j_i=-\delta} {dj_1 dj_2\over (2\pi i)^2} {\Gamma(-j_2)\Gamma(1+j_1+j_2)\over \Gamma(1+j_1)} X_1^{j_1} X_2^{j_2} = \delta(1+X_2-X_1) \,,
\end{align}
we obtain
\begin{align}
I(\epsilon_1,\epsilon_2) &= \int_0^{s_{\rm max}} {ds\over s(s(1+\omega_1)+1-z)}\delta\lr{1+{s(1+\omega_1)+1-z\over z(1+\omega_2(1+\omega_1 s))}\e^{\epsilon_1} -{s\over z} \e^{\epsilon_2}}\,.
\end{align}
The integral is localized on the solution of the equation
\begin{align}
{1+{s_\star(1+\omega_1)+1-z\over z(1+\omega_2(1+\omega_1 s_\star))}\e^{\epsilon_1} -{s_\star\over z} \e^{\epsilon_2}}=0
\end{align}
satisfying the condition  $0\le s_\star\le s_{\rm max}$ and it is given by
\begin{align}
I(\epsilon_1,\epsilon_2) &= {z(1+\omega_2(1+\omega_1 s_\star))\over s_\star(s_\star(1+\omega_1)+1-z) (\omega_1\omega_2 z - \e^{\epsilon_1}(1+ \omega_2(1+2s_\star\omega_1))+\e^{\epsilon_2}(1+ \omega_1))} \,.
\end{align}
At small $\epsilon_i$ we have
\begin{align}\label{s-star}
s_\star =  {1\over\omega_2} + z + O(\epsilon_i)\,,\qquad
I(\epsilon_1,\epsilon_2) &= -{\omega_2^2 z\over (1+\omega_2 z)(\omega_1+\omega_2+\omega_1\omega_2 z)}(1+ O(\epsilon_i)) \,.
\end{align}
Notice that $s_\star<0$ for $-1< \omega_2<0$ and, therefore, $I(\epsilon_1,\epsilon_2)=0$ in this case. For $\omega_2>0$ we observe that $I(\epsilon_1,\epsilon_2)$ contains two factors in the denominator. As mentioned above, the first one $(1+\omega_2 z)$ disappears in the 
final expression for $\text{SSC}_p(\omega_1,\omega_2)$. The derivatives in \re{SSCp} increase the power of $\omega_1+\omega_2+\omega_1\omega_2 z$ in the denominator. The close examination shows that each power of $\epsilon_i$ in the expansion of $I(\epsilon_1,\epsilon_2)$ is accompanied by a factor of 
$1/(\omega_1+\omega_2+\omega_1\omega_2 z)^2$. This leads
to \re{fp}. For $p=2$ we obtain from \re{SSCp} the relation \re{eq:ambiguityp2} with the coefficient functions given by
\begin{align}\nn
& c_0=2520 \left(\omega_1+\omega_2\right){}^2,
\\\nn
& c_1=
1260 \left(\omega_1+\omega_2\right)
   \left(5 \omega_1+5 \omega_2-4\right),
   \\   \nn
& c_2= 280 \left(20 \omega_1^2+40 \omega_2 \omega
   _1-43 \omega_1+20 \omega_2^2-43 \omega_2+9\right),
\\   \nn
& c_3=  140 \left(15 \omega_1^2+30
   \omega_2 \omega_1-72 \omega_1+15 \omega_2^2-72 \omega_2+41\right),
\\   \nn
& c_4=    60 \left(5
   \omega_1^2+10 \omega_2 \omega_1-58 \omega_1+5 \omega_2^2-58 \omega
   _2+75\right),
   \\   \nn
& c_5= 10 \left(\omega_1^2+2 \omega_2 \omega_1-44 \omega_1+\omega_2^2-44
   \omega_2+141\right),
\\   \nn
& c_6=    
   -4 \left(3 \omega_1+3 \omega_2-38\right),
   \\   
& c_7= 3 \,.
\end{align}

\section{Details of the derivation in Section \ref{sN4WI}}\label{AppWI}

\subsection*{Summary of the SSC correlation}

In this subsection we recall some of the key steps in obtaining the SSC correlation as an integrated correlation function.

 It proves very efficient to go to new coordinates $x^\mu \to y^\mu$ by performing a conformal transformation \cite{Cornalba:2007zb, Hofman:2008ar}  (see Section~\ref{sec:nullplane}), 
\begin{align}\label{y-var}
y^+ = - {1\over x^+}\,,\qquad y^-=x^- - {\vec x^2 \over x^+}\,,\qquad \vec y={\vec x\over  x^+}\,,\qquad dy^2 = {dx^2\over (x^+)^2} \,,
\end{align}
where the notation was introduced for the light-like coordinates $x^\pm$ and $\vec x= (x,\bar x)$,
\begin{align}\label{LC}
x_{\alpha\dot\alpha} = x_\mu (\sigma^\mu)_{\alpha\dot\alpha}
= \bigg[\begin{array}{ll} \ x^+ & \bar x \\ \ x & x^- \end{array}\bigg]\,,\qquad dx^2 = dx^+ dx^- - dx \,d \bar x\,.
\end{align} 
Then sending the detectors at points 1 and 2 to null infinity is equivalent to setting  $y^+_{1,2}=0$. The calculations are greatly simplified by using conformal invariance to `gauge away' some of the coordinates at the source points 3 and 4. So, we work in the special frame 
\begin{align}\label{gauge}
& y^+_1=y^+_2=0\,, \qquad  y^+_3=1 \qq \vec y_3=\vec y_4=y_4^-=0\,,\qquad  y_4^+\to\infty \,.
\end{align}
At the end the covariant form can be easily restored by undoing the gauges and switching back to the coordinates $x_{3,4}$.

The next step is the Fourier transform from  the detector times to frequencies, $y^-_{1,2} \to \ho_{1,2}$. Applied to the correlation function $G_{00}$ in \p{G-SS}, this procedure results in   
\begin{align}\label{G-int}
\m{G}_{00}(y;n_i, \ho_i) {}& = \int_{-\infty}^\infty dy_1^- dy_2^-  e^{-i (y^-_1 \ho_1 +  y^-_2 \ho_2)}\vev{\phi^\dagger(y_4) O(0^+,y_1^-,\vec y_1) O(0^+,y_2^-,\vec y_2)  \phi(y_3)}  \nt
&=\int_{-\infty}^\infty dy_1^- dy_2^-  e^{-i (y^-_1 \ho_1 +  y^-_2 \ho_2)}\, F(y) \,, 
\end{align}
where  the function $F$ was defined in \re{1.4}. After integrating over $y_1^-, y_2^-$  we restore the covariant notation and obtain 
\begin{align}\label{cal-G}
 \m{G}_{00}(x;\g,\Om_i)= {1\over x^2  } \cG(\g,\Om_i)  \,,
\end{align}
where $x\equiv x_{34}$. The function $\mathcal G$ generalizes that in \p{2.5}. It depends on the dimensionless variables
\begin{align}\label{eq1.3}
&\g=  {2(x n_1) (x n_2) \over x^2  (n_1 n_2) } \qq \Om_i =\ho_i  (xn_i) \,.
\end{align}

The  Mellin kernel  for the dimensionless function $\cG(\g,\Om_i)$ in the coordinate representation is given by the double Schwinger integral (see Appendix~\ref{sec:derivationSSC}) 
\begin{align}\label{eq1.2}
&K_{\cG}(j_i|x)= (-\g)^{j_1+j_2-1}  \int_0^\infty dS_1 dS_2 \,  e^{-i\left(S_1+S_2\right)} \left({S_2 \left(S_1+\Om _1\right)} \right)_+^{-j_2} \left( {S_1 \left(S_2+\Om _2\right)}\right)_+^{-j_1}  \,,
\end{align}
where the parameters $S_i =s_i (xn_i) $  are dimensionless (see \p{eq1.3}).  The $S-$factors  are understood as the $+$ distribution $1/S_+^j$ and   $1/(S+\Om)_+^j$. This makes differentiating with respect to $S$ or $\Om$ safe, as long as $j\neq -1,-2, \ldots\ $. 

After the Fourier transform to momentum space $x\to q$ we get (see \p{eq:schwingerkernelSS})
\begin{align} 
&K_{\text{SS}}(j_i|q)= ((n_1 n_2)/2)^{-j_1-j_2} \, 
{\mathbb{I}}_{j_1,j_2}\,  \Big\{ \s_{j_1+j_2}(q-n_1s_1-n_2s_2)\Big\} \,, \label{1.2}
\end{align}
where
\begin{align}\label{1.5'}
&{\mathbb{I}}_{j_1,j_2} \big\{ \sigma \big\} =  \int_0^\infty ds_1 ds_2 \,  \left(s_2 \left(s_1+\ho_1\right)\right)_+^{-j_2} \left(s_1 \left(s_2+\ho_2\right)\right)_+^{-j_1}\, \sigma\,, \\
&\s_j(\hat q) = \int {d^4 x \e^{-i\hat qx}\over (-x^2+i0 x^0)^j} =C_j\, \theta(\hat q^0)\theta(\hat q^2)  (\hat q^2)^{j-2}    \,,\quad \hat q=q-n_1s_1-n_2s_2   \label{1.5}
\end{align}
and $ C_j={\pi^3 /( 2^{2j-3}\Gamma(j)\Gamma(j-1))} $.
The variables $s_i$ have the same dimension and Lorentz charge as the frequencies  $\hat\om_i$. 
It is convenient to introduce new dimensionless variables $z$ and $\om_i$  (cf. \p{z} and \p{eq:rescaledfreq}), as well as to rescale the integration variables $ s_i  \to  {(q^2 /2(qn_i)}) s_i$. After this the scalar kernel becomes
\begin{align}\label{1.7}
&  K_{\text{SS}}  =C_{j_1+j_2} \, z^{1-j_1-j_2} \,  \mathbb{I}_{j_1,j_2} \Big\{  \q(q^0-n^0_i s_i)\, \q(1-s_1-s_2+z s_1 s_2) \, (1-s_1-s_2+z s_1 s_2) ^{j_1+j_2-2} \Big\},  
\end{align}
{where the integral operator $\mathbb{I}_{j_1,j_2}$ is  defined in \re{1.5'} but now with $\ho \to \om$.}
In the rest frame $q^\mu=(q^0, \vec 0)$ with $q^0>0$ and $n^\mu_i=(1,\vec n_i)$ with $\vec n_i^2=1$ we have $q^2=(q^0)^2$, $(qn_i)=q^0$. The condition $\hat q^0 \geq0$ becomes $2-s_1-s_2\geq0$, thus  reducing the integration domain to $0\leq s_i \leq 1$.  The conditions $s_i + \om_i \geq 0$ restrict the domain further, if one or both $\om_i <0$. In this section most of the time we will consider the simplest case $\om_i \geq 0$ and will comment on the choice $\om_1<0, \ \om_2>0$ in Section~\ref{s4.2}. The last condition $\hat q^2\geq0 \ \Rightarrow \ 1-s_1-s_2+z s_1 s_2\geq 0$ can be resolved by the following change of variables:  
\begin{align}\label{eq1.7}
s_1\to u \qq s_2\to \frac{(1-u) v}{1-z  u}\,; \qquad u\in [\max(0,-\om_1),1]\,, \quad v \in [0,1]\,.
\end{align}

 \subsection*{Relations QSC/SSC and ESC/SSC}
 
 In this subsection we work out the simplest example  QSC/SSC in detail and explain the main steps. The principal difference compared to the standard case is that instead of integrating over the detector time $x^-$, we now Fourier transform it into the energy $\ho$. Thus, if earlier we could neglect the total time derivative $\pa_{x^-}$ contained in the operator \p{Di}, now it contributes to the Fourier transform terms $\sim \ho$. This calls for a more efficient formalism.
 
 \subsubsection*{Spinor formalism}
 
We introduce the following notation for the $\la$-projections consistent with their Lorenty charges. The auxiliary spinors  $\la_\pm \equiv \bra{\pm}$ and  $\tl_\pm \equiv [\pm|$  satisfy completeness conditions
\begin{align}\label{}
&\ket{+}^\a \bra{-}_\b - \ket{-}^\a \bra{+}_\b = \delta^\a_\b \ \Rightarrow \ \bra{+}^\a \ket{-}_\a \equiv \vev{+|-}=1 \,,\nt
&|+]^\da [-|_\db - |-]^\da [+|_\db =\delta^\da_\db \ \Rightarrow \  [+|^\da |-]_\da \equiv [+|-]=1 \,.
\end{align} 
The four-vectors have the projections 
\begin{align}\label{1.11}
& y^+  = \bra{-}^\a y_{\a\da} |-]^\da  = [-|_\da \ty^{\da\a} \ket{-}_\a \equiv\bra{-}y|-] = [-|\ty \ket{-}\,, \   \nt
&y^- = \bra{+}y|+] = [+|\ty\ket{+} \qq   y = \bra{+}y|-] = [-|\ty\ket{+}\qq  \bar y = \bra{-}y|+] = [+|\ty\ket{-} \,.
\end{align}
In terms of projected variables and in the frame \p{gauge}  the matrices \p{e1.8} become
\  \begin{align}\label{3.6}
 \cX_{1[234]}&\equiv\left(
\begin{array}{ccc}
 \bra{-} \cX_{1[234]}\ket{-} & &  \bra{-} \cX_{1[234]}\ket{+} \\
\bra{+} \cX_{1[234]}\ket{-}& &  \bra{+} \cX_{1[234]}\ket{+} \\
\end{array}
\right) \nt
&=  \left(
\begin{array}{cccc}
\yb_{12} (y_2 \yb_1-y^-_3) & & & y_2\yb_{12} y^-_1 \\
 y_2 (\yb_1 y^-_{13}+\yb_2 y^-_3)-y_1 (\yb_1 y^-_{23}+\yb_2 y^-_3)+y^-_{21} y^-_3 & & & y^-_1 (y_2 y^-_{13}+y_1 y^-_{32})  \\
\end{array}
\right) ,\nt
 {\cX}_{[12]3} &=\left(
\begin{array}{ccc}
 \yb_{12} & & -y_2 \yb_1-y^-_{23} \\
 y_1 \yb_2+y^-_{13} & & y_1 y^-_{23}+y_2 y^-_{31} \\
\end{array}
\right)  \qq  \cX_{[12]4} =\left(
\begin{array}{ccc}
 \yb_{12}&  & -y^-_2 \\
 y^-_1 & & 0 \\
\end{array}
\right) \,.
\end{align}

 The operator \p{Di} is decomposed into two projections:
\begin{align}\label{115}
&\cD= \tl_{-\da} \tilde\pa^{\da\a}_y \frac\pa{\pa\la_-^\a} = \pa_{y^-} \pa_{+}  + \pa_{\yb}  \pa_0 \,, \nt
&  \cD^2=\pa_{\bar y}^2 \pa_0(\pa_0-1)  +   2 \pa_{y^-}  \pa_{\bar y} \pa_{+}(\pa_0  -1)  +\pa_{y^-}^2  \pa_{+}^2 \,,
\end{align}
where the projected  spinor derivatives act as follows:
\begin{align}\label{112}
&\pa_0 \ket{\pm} =\mp \ket{\pm}  \,, \quad   \pa_{+}  \ket{-} = \ket{+} \,, \quad   \pa_{+}  \ket{+} = 0 \,,\nt
&\pa_0 y^\pm = \pm y^\pm \,, \quad  \pa_0 \yb=\yb \,, \quad    \pa_0 y=-y \,, \quad  \pa_+ y^+ = y \,, \quad  \pa_+ \yb=y^- \,, \quad  \pa_+ y^- = \pa_+ y=0 \,,\nt
& \pa_{y^-} \equiv  [-|\tilde\pa_y\ket{-}     \ \Rightarrow \  \pa_{y^-} y^-=1 \,, \quad  \pa_{\yb} \equiv  - [-|\tilde\pa_y\ket{+}     \ \Rightarrow \  \pa_{\yb} \yb=1\,, \ \text{otherwise 0}     \,.
\end{align}

 These rules allow us  to expand out the differential operators  in \p{e1.6} and are easily automatized. Below we show in detail how this is done for the simplest relation QSC/SSC. 
 
 \subsubsection*{QSC differential operator in coordinate space}

The QSC  correlation is obtained from the relations \p{e1.6}--\p{Xs} and  is given by
 \begin{align}\label{e3.9}
&\text{QSC}=   ( \pa_{y^-_1} \pa_{+} + \pa_{\bar y_1} \pa_0) \left[\bra{-} \cX_{1[234]}\ket{-} {\cG(\g,\Om_i) \over (y_{12} \yb_{12} y^-_3 )^2 } \right]\nt
&=\pa_{y^-_1}  \left[\Big( \bra{+} \cX_{1[234]}\ket{-} + \bra{-} \cX_{1[234]}\ket{+}\Big) {\cG \over (y_{12} \yb_{12} y^-_3 )^2 } \right] +\pa_{\bar y_1} \left[2\bra{-} \cX_{1[234]}\ket{-} {\cG \over (y_{12} \yb_{12} y^-_3 )^2 } \right]  \,,
\end{align}
 where (see \p{eq1.3})
 \begin{align}\label{e3.10}
&\Om_i = (y^-_3-y_i\yb_i)\ho_i   \qq  \g=-{(y_3^--y_1\bar y_1)(y_3^- - y_2\bar y_2) \over  y_{12} \bar y_{12}  y_3^-}   \,.
\end{align}

The next step is the Fourier transform $y^- \to \ho$  (recall \p{G-int}). The first term  in \p{e3.9} with the total detector time derivative $\pa_{y^-_1}$ gives rise to terms proportional to $\Om_1$. This is in  contrast with the standard case (i.e. $\ho_i=0$) where the time derivatives were suppressed by the time integral. We distribute the derivatives in \p{e3.9} and  restore the covariant form to obtain  the generalization of the first relation in \p{2.5},
\begin{align}\label{3.12}
\text{QSC}(x)= \frac{ (x n_2)}{(n_1 n_2)^2  x^4}\left[  \frac{\Om_1 }{2 \g }\Big((\g -2) \cG_{\Om_1}-\g  \cG_{\Om_2}+i\cG\Big)+(\g -1) \cG_{\g }+\cG\right]   \,.
\end{align}
 On the other hand, we have the identity 
\begin{align}\label{3.13}
 \frac{(n_2\pa_x)}{2(n_1 n_2)^2 } \left[ \frac{\cG(\g,\Om)}{x^2}\right]  = \frac{ (x n_2)}{(n_1 n_2)^2  x^4}  \, \left[ (1-\g)\cG_\g + \frac{\Om_1}{\g}   \cG_{\Om_1} -\cG \right]   \,.
\end{align}
Recalling that $\text{SSC}(x) = \cG(x)/x^2$ (see \p{cal-G}), we can combine \p{3.12} and \p{3.13} into
\begin{align}\label{3.14}
\text{QSC}(x) = -\frac1{(n_1 n_2)}\,  (n_2\pa_x) \text{SSC}(x)  + \frac{ (x n_2)}{(n_1 n_2)^2  x^4} \,  \Om_1\left[ \cG_{\Om_1 }  - \cG_{\Om_2} + \frac{ i }{\g}\cG  \right]   \,.
\end{align}

 \subsubsection*{QSC differential operator in momentum space}\label{s3.3} 
 
 Our final goal is to express the QSC kernel as a differential operator in momentum space acting on the SSC kernel. We start by inserting the Mellin kernel $K_{\cG}(x) $ defined in   \p{eq1.2} into the right-hand side of \p{3.12} and then we switch  back to the dimensionfull variables $s_i$ and $\ho_i$: 
\begin{align}\label{3.15}
K_{\text{QS}}(x)&= \frac1{(n_1 n_2)^2}(-x^2 (n_1 n_2)/2)^{-j_1-j_2} {\mathbb{I}}_{j_1,j_2}\,  \Bigg\{   e^{-is_k (xn_k)} \Bigg[ \frac{i \ho_1}{2} +\left(1-j_1-\frac{j_2 s_1}{s_1+\ho_1}\right) \frac{1}{ (xn_1)}\nt
 &+ \frac{j_1 \ho_1}{ (n_1 n_2) x^2(s_2+\ho_2)} (xn_1)
  +\frac{2 j_1 \left(s_1+\ho_1\right)+j_2 \left(2 s_1+\ho_1\right)}{ (n_1 n_2) x^2\left(s_1+\ho_1\right)}(xn_2)  \Bigg]\Bigg\}  \,.
\end{align} 
When Fourier transforming the right-hand side of \p{3.15}, $
K_{\text{QS}} = \int d^4x e^{-ixq}\,  K_{\text{QS}}(x)$,
we encounter the basic Fourier integral \p{1.5} and two new ones: 
\begin{align} 
& \int d^4x e^{-ix\hat q} \,  (xn_k) (-x^2)^{-j-1} = -i\pa_{s_k} \s_{j+1}= \frac{i}{j}[(qn_k)-(n_k n_i)s_i] \, \s_{j} \,,\label{3.17}\\
& \int d^4x e^{-ix\hat q} \,  \frac{(-x^2)^{-j}}{(xn_1)}   \quad  \stackrel{- i\pa/\pa{s_1}}{\longrightarrow} \quad  \s_{j} \,.\label{3.18}
\end{align}
We use the relation \p{3.18}  in  \p{3.15} by pulling out a total derivative $\pa_{s_1}$ from the pole $1/ (xn_1)$,
\begin{align}\label{3.19}
 \left(s_1+\ho_1\right)^{-j_2}  s_1^{-j_1}\left(1-j_1-\frac{j_2 s_1}{s_1+\ho_1}\right)    =\pa_{s_1} \Big[  \left(s_1+\ho_1\right)^{-j_2}  s_1^{-j_1}\Big]
\end{align}
 and integrating it by parts.\footnote{The boundary terms vanish under suitable conditions on the parameters $j_i$. } Thus, all the terms  in  \p{3.15} are reduced to the basic Fourier integral \p{1.5}. Switching over to the dimensionless variables $ s_i  \to  {(q^2 /2(qn_i)}) s_i$, we obtain the following Schwinger integral with the measure defined in \p{1.7}:
 \begin{align}\label{3.21}
&K_{\text{QS}} = \nt
& \frac{i (qn_2)}{2(n_1 n_2) \left(j_1+j_2\right)}\mathbb{I}_{j_1,j_2} \Bigg\{ \Bigg[2 \left(j_1+j_2 + j_2 z  \om_1 \right) -\frac{j_2 \om_1 \left(z  \om_1+1\right)}{s_1+\om_1}  +\frac{j_1 \om_1 \left(z  \om_2+1\right)}{s_2+\om_2}\Bigg] \s_{j_1+j_2}\Bigg\} \,.
\end{align}
The terms with poles $1/(s_i+\om_i)$ can be interpreted as derivatives of $K_{\text{SS}}(q) \sim  \mathbb{I}_{j_1,j_2}\{ \s_{j_1+j_2}\}$  with respect to $\om_i$. In this way we arrive at the final result
\begin{align}\label{3.22}
K_{\text{QS}}  =    \frac{i (qn_2)}{(n_1 n_2) } \Bigg\{1+ \frac{\om_1}{2\left(j_1+j_2\right)}\Big[2  j_2 z+ (1+z\om_1) \pa_{\om_1} -(1+z\om_2) \pa_{\om_2} \Big] \Bigg\} K_{\text{SS}}   \,.
\end{align}
This is our first example of a differential operator \p{e1.14} relating two   correlations. 
We remark that Eq.~\p{3.22} contains the kinematic prefactor $i (qn_2)/(n_1 n_2) $ which has been stripped off the kernels in \p{XYC}, \p{e1.13} and \p{e1.14}. Setting $\om_i=0$ reproduces the standard relation \p{e1.13}.

 \subsubsection*{ESC differential operator }
 
The case ESC/SSC is treated in a similar way. The analog of \p{e3.9}, after doing the spinor variables algebra is (recall \p{3.6}--\p{112})
\begin{align}\label{}
 &\text{ESC}=12\pa_{\bar z_1}^2 \left[ \bra{-} \cX_{1[234]}\ket{-}^2 \, G\right] + 12 \pa_{z_1^-}\pa_{\bar z_1}  \left[\Big(\bra{+} \cX_{1[234]}\ket{-} + \bra{-} \cX_{1[234]}\ket{+}\Big)\bra{-} \cX_{1[234]}\ket{-} \, G  \right]\nt
&+2\pa_{z_1^-}^2 \left[\Bigg(2 \bra{+} \cX_{1[234]}\ket{+}\bra{-} \cX_{1[234]}\ket{-} +  \Big( \bra{+} \cX_{1[234]}\ket{-}+ \bra{-} \cX_{1[234]}\ket{+}\Big)^2 \Bigg)\, G\right], \end{align}
with $G=\cG/(z_{12} \zb_{12} z^-_3 )^3$. After distributing the derivatives and restoring the covariant notation, the analog of \p{3.15}  contains  a pole $1/(xn_1)$ (the measure $ \hat {\mathbb{I}}_{j_1,j_2}$ is not shown):
\begin{align}\label{}
&\frac{ \ho_1}{4 (xn_1)} \left[ \frac{j_2 (xn_2) \left(\left(j_1+j_2\right) s_1+\left(j_1-1\right) \ho_1\right)}{z x^2 \left(s_1+\ho_1\right){}^2}  -\frac{i \left(\left(j_1+j_2-1\right) s_1+\left(j_1-1\right) \ho_1\right)}{ \left(s_1+\ho_1\right)}\right]\nt
&\stackrel{\int ds_1\,   \pa_{s_1}}{\longrightarrow} \quad  -\frac{1}{4} s_1 \ho_1 \left(1+\frac{i j_2 (xn_2)}{z  x^2 ( s_1  +\ho_1)}\right)   \,.
\end{align}
Here  we pulled  $\pa_{s_1}$ out and  integrated  it by parts (recall \p{3.19}). The remaining terms are polynomial of the type  $(xn_1)^p  (xn_2)^q$. Their Fourier transforms  are done by repeated use  of  the relations \p{1.5}, \p{3.17} and \p{3.18}. The sum of all the terms contains poles in $s_i+\om_i$ (but no positive powers!) which are converted   to derivatives with respect to $\om_i$. The result for the coefficients of the differential operator \p{e1.14} is  shown in \p{coES}. 
Setting $\om_i=0$  we recover the standard relation  \p{e1.13}. 

 \subsubsection*{Remarks}

We conclude this subsection by a couple of remarks. The first term in \p{3.14} is a total $x-$space derivative with a simple Fourier transform, so we might have saved some work by  using the shorter  from \p{3.14} of the relation \p{3.12}.  However, the Fourier transform of the total derivative term is not so obvious in the cases QQC, EQC and EEC, as we explain in the next subsection. 

A related issue is a  key feature of the  right-hand side of \p{3.21}, namely the absence of positive powers of $s_i+\om_i$. They would have been  an obstruction for the interpretation of the right-hand side  as a differential operator, as we did for example   in \p{3.22}. This property is automatic in the cases QSC and ESC. However, in the cases QQC, EQC and EEC positive powers do appear, already in the total derivative term written as a  Schwinger integral. Since we know the expected result of its Fourier transform, we can derive  highly non-trivial  identities relating the obstruction terms to simpler ones. It turns out that these identities suffice for eliminating the obstruction terms also in the main (non-derivative) part. In this way we derive a differential operator for the relation to the SSC kernel. We illustrate this phenomenon in the next subsection on the simplest case QQC. 

  \subsection*{Relation QQC/SSC}
  
   \subsubsection*{Differential operator in coordinate space}

The QQC  correlation is obtained from   \p{e1.6}--\p{Xs} (see \p{3.6} and \p{e3.10}):
 \begin{align}\label{4.1}
&\text{QQC} =   ( \pa_{y^-_1} \pa_{1_+} + \pa_{\bar y_1} \pa_{1_0})  ( \pa_{y^-_2} \pa_{2_+} + \pa_{\bar y_2} \pa_{2_0})  \left[ \bra{1_-} {\cX}_{[12]3}\ket{2_-} \bra{1_-} {\cX}_{[12]4}\ket{2_-} {\cG(\g,\Om_i) \over (y_{12} \yb_{12} y^-_3 )^2 }\right] \,.
 \end{align}
Notice that the spinor derivatives $\pa_+, \pa_0$ act separately at points 1 and 2. Doing the algebra and restoring the covariant form we obtain a differential operator that consists of two parts:

(i) total derivative applied to a function of $\g$ and $\Om$ (the generalization of  \p{2.5}):
\begin{align}\label{4.2}
\text{QQC}_{\rm tot.der.}&=   \frac1{(n_1 n_2)^2}\Box_{x} \, \left(\frac{\cG(\g,\Om)}{x^2}\right)  = {1 \over  (n_1n_2) ^2x^4} \Big\{\pa_{\g }[(1-\g) \g  \pa_{\g }]  \nt
& - \left((\g -1) \pa_{\g }+1\right)\Om _1  \pa_{\Om _1} - \left((\g -1) \pa_{\g }+1\right)\Om _2  \pa_{\Om _2} +\frac1{\g} \Om_1 \Om_2  \pa_{\Om _1,\Om _2}
\Big\}{\cG}\,;
\end{align}

(ii) deviation from the total derivative by terms proportional to $\Om$:
\begin{align}\label{4.3}
\text{QQC}_{\rm dev.}&= {1 \over  (n_1n_2) ^2x^4}\, {i\Om_1 \over  4\g} \nt
& \times\left\{ \frac{i}{2} \Om_2 + 2 \g     +2 \g \left(\g   - 1 \right)  \pa_{\g }+\left(\g  -2 \right)   \Om_2 \pa_{\Om_1} +  i \g  \Om_2  \pa_{\Om_1,\Om_2}\right\}{\cG}+(1\leftrightarrow 2) \,.
\end{align}

   \subsubsection*{Differential operator in momentum space. Total derivative identity} \label{s4.2}
 
The  deviation term \p{4.3} is treated as in Section~\ref{s3.3}. The analog of \p{3.21} contains no positive powers of $s_i+\om_i$, so it can be  turned into a differential operator. The latter is simplified with the help of  the differential equation for the scalar kernel \p{5.1} with the result shown in \p{coQQ}. 

The total derivative term \p{4.2} requires special care. Although its Fourier transform is expected to be very simple, this is by no means obvious. Indeed, in order to obtain this Fourier transform we need to prove the following highly non-trivial identity:
\begin{align}\label{4.5}
 \mathbb{I}_{j_1,j_2}& \Bigg\{\Bigg[
 j_1 \left(z  \om_1 \om_2+\om_1+2 \om_2+2\right)
 -\om_1 -1   -\frac{j_2 \om_1\left( z  \om_1\om_2+\om_1+\om_2+1\right)}{s_1+\om_1}\nt
&+(s_2+\om_2) \Bigg( \frac{j_2 \om_1 \left(z  \om_1+1\right)}{s_1+\om_1}
- \left(z j_2  \om_1+j_1+j_2-1\right) \Bigg)  \Bigg] \s_{j_1+j_2}+(1\leftrightarrow2) \Bigg\}=0\,.
\end{align}
Notice the presence of positive power of $s_i+\om_i$ in the second line. This is an example of `obstruction' terms being replaced by terms of the differential operator type.  

The proof of \p{4.5} makes use of the variables \p{eq1.7} spanning  the interval  $u,v \in [0,1]$ if $\om_i \geq 0$. We can rewrite  the integrand in \p{4.5} is a sum of total derivatives:\footnote{ES is grateful to Dima Chicherin for a discussion on this point.}
\begin{align}
&\int_0^1 du dv \   \Bigg\{ \partial_v  \Big[  v^{1-j_2} (1-v)^{j_1+j_2-1}  (\om_2(1-uz)  +v(1-u))^{-j_1}\Big] \nt
&\hskip15mm \times  u^{-j_1}  (1-u)^{j_1-1} (1-z  u)^{j_1+j_2-2} \left(z  u^2-2 z  u+1\right)(\om_1+u)^{-j_2} \label{4.6}\\
&  + \partial_u \Big[ u^{1-j_1} (1-u)^{j_1}(1-z  u)^{j_1+j_2-1}  (\om_1+u)^{-j_2}(\om_2(1-u z)  +v(1-u))^{-j_1}\Big]v^{-j_2}  (1-v)^{j_1+j_2-1}  \Bigg\} \,.  \nonumber
\end{align}
The boundary terms in the two integrals vanish if $ 0<{\rm Re}(j_1)<1$, ${\rm Re}(j_2)<1$ and ${\rm Re}(j_1+j_2)>1$.
If $\om_1<0$ and $\om_2 \geq 0$, the variable $u$  is in the interval $u\in[-\om_1,1]$. In this case the factor $ (\om_1+u)^{-j_2}$ in  \p{4.6} vanishes at the boundary $u=-\om_1$ provided that ${\rm Re}(j_2)<0$. The  application domain of this identity can be extended by analytic continuation  in $j_1,j_2$.

\subsection*{Relations EQC/SSC and EEC/SSC}\label{s5}

The new feature in the cases EQC/SSC and EEC/SSC is that obstruction terms with positive powers of $s_i+\om_i$ appear not only in the Fourier transform of the total derivative part but also in the deviation terms.  Requiring that the Fourier transforms of the total derivative terms  equal the expressions originating from \p{2.5}, we find two new identities of the type \p{4.5}. More identities are generated by differentiating the existing ones with respect to $\om_i$, thus lowering the  powers of $s_i+\om_i$. For example, the  derivative $\pa_{\om_1,\om_2}$ of \p{4.5} removes all the positive powers, resulting in a differential equation for $K_{\text{SS}}(q)$.  We find altogether 11 linearly independent  identities. Adding the appropriate linear combinations of identities to the deviation terms, we succeed in removing all the obstruction terms. The resulting differential operators for EQC and EEC are listed in the ancillary file.

An interesting byproduct of the  procedure is a particular linear combination of identities that amounts to a  simple homogeneous partial differential equation for the scalar kernel $K_{\text{SS}}$, see \p{5.1}.
It arises as a consistency condition for the Fourier transform of the Schwinger integral in \p{eq1.2} and its space-time derivatives. 

Equation \p{5.1}  can be solved by separation  of variables. For example, we can  substitute  
\begin{align}\label{5.2}
K_{\text{SS}}(z,\om_i)=({\om_2}+1)^{j_2-j_1} ({\om_1} {\om_2})^{-j_2} F\left(X\right) \qq X=\frac{z{\om_1} {\om_2} +{\om_1}+{\om_2}+1}{{\om_1} {\om_2} (z-1)}\,.
\end{align}
We obtain an ordinary differential equation for the function $F(X)$:
\begin{align}\label{}
(1-X) X F''(X)+\left((j_1 -j_2-2) X+1\right) F'(X)+\left(j_1-1\right) j_2 F(X)=0 \,.
\end{align}
The relevant solution 
\begin{align}\label{5.4}
F(X)= c_1 \ _2F_1\left(1-j_1,j_2;1;X\right) 
\end{align}
is selected by the requirement that it be regular at $X=0$.  This boundary condition  amounts to setting $z{\om_1} {\om_2} +{\om_1}+{\om_2}+1=0$ in the integral \p{1.7}.  The  constant $c_1$ is  fixed by taking the limit $\om_i\to0$ and comparing with the value of the integral \p{1.2}.

An alternative substitution is 
\begin{align}\label{5.5}
K_{\text{SS}}(z,\om_i)=({\om_1}+1)^{j_1-j_2} (-{\om_1} {\om_2})^{-j_1} G(Y) \qq Y=\frac{({\om_1}+1) ({\om_2}+1)}{{\om_1} {\om_2} (1-z)} \,,
\end{align}
yielding the differential equation
\begin{align}\label{5.6}
&(Y-1) Y G''(Y)+\left((j_1-j_2+2)  Y-j_1+j_2-1\right) G'(Y) - j_1 \left(j_2-1\right) G(Y)=0 \ \Rightarrow \nt
&G(Y)=c_2 \ _2F_1\left(j_1,1-j_2;1+j_1-j_2;Y\right)  \,.  
\end{align}
Again, the relevant solution is selected by the boundary condition of regularity at $Y=0$. This condition amounts to setting $\om_1=-1$. Both the solution and the integral \p{1.7} vanish at this point. The integration constant $c_2$ can be fixed as before.

We remark that using well-known identities,  these solutions can be rewritten in an alternative form involving two  hypergeometric functions each. For example, the solution \p{5.2},  \p{5.4}  becomes (cf. \p{eq:kernelSSCppTwoHypers})
\begin{align}\notag
K_{\text{SS}}(z,\om_i) =&  (1-z)^{j_1+j_2-1} z^{-j_1-j_2} \frac{\left(\omega_1+1\right)^{-j_2} \left(\omega_2+1\right)^{-j_1}\Gamma\left(1-j_1-j_2\right) }{ [\Gamma\left(1-j_1\right) \Gamma \left(1-j_2\right) ]^2\Gamma \left(j_1+j_2\right)}
\\ \notag &
\times   
    \,
   _2F_1\left(j_1,j_2;j_1+j_2;\frac{\omega_1 \omega_2(1-z)}{({\om_1}+1) ({\om_2}+1)}\right)
\\[2mm] \notag
+{}&{} z^{-j_1-j_2}\frac{(\omega_1\omega_2)^{-j_1-j_2+1} \left(\omega_1+1\right)^{j_1-1} \left(\omega_2+1\right)^{j_2-1}  \sin \left(\pi
    j_1\right) \sin \left(\pi  j_2\right) }{\pi ^2 \left(j_1+j_2-1\right)}
\\
&    
\times    \, _2F_1\left(1-j_1,1-j_2;2-j_1-j_2;\frac{\omega_1 \omega_2(1-z)}{({\om_1}+1) ({\om_2}+1)}\right)\,.
\end{align}
Both terms in this expression are particular solutions of the differential equation \p{5.1}. The boundary conditions are less obvious in this form. The first term admits a power series  expansion  $\sum_{m,n=0}^\infty c_{m,n} \om_1^m \om_2^n$ while the second terms has a non-analytic behavior near the origin. It can be shown that the regular solution  of  \p{5.1} is unique up to normalization. So, the scalar kernels share the same regular part but differ by their non-analytic parts.

\section{Singular terms in the energy-energy correlation} \label{app:sing}

In this appendix we explain the origin of the last two terms in the expression \re{EEC-weak} for the energy-energy correlation. A feature  of these terms is that they scale at small $z$ as $\log z/z$,  
\begin{align} \label{sing} 
\text{EEC}^{(1)}_{++} & \sim 
   -  {\log z\over z}\omega_1^2 \omega_2^2
   \left(\frac{1}{72\left(\omega
   _1+1\right)}+\frac{1}{72\left(\omega
   _2+1\right) }+\frac{ 1}{12\sqrt{\left(\omega_1+\omega
   _2\right)^2-4 \omega_1 \omega_2 z}}\right)\,.
\end{align}
As explained in Section~\ref{sect:talk}, such singular terms come from the detector cross talk. 

Applying \re{XYC} and \re{eq:Mellin amplitudes at weak and strong coupling} we get 
\begin{align}
\text{EEC}^{(1)}_{++} (\omega_1, \omega_2 ,z) =
- 
 \int{d j_1 d j_2 \over (2 \pi i)^2} \,   K_{\text{EE}}(j_1,j_2|z,\omega_1,\omega_2) \,{(j_1 + j_2)^2 \over j_1^2 j_2^2}\,,
\end{align}
where the kernel $K_{\text{EE}}$ can be obtained from the analogous kernel for the scalar detectors $K_\text{SS}$ using the relation \re{e1.14}.
Replacing $K_\text{SS}$ with its integral representation \re{eq:kernelSS} and going to the limit $z\to 0$, we find after some algebra

\begin{align}\notag\label{ints}
 \text{EEC}^{(1)}_{++} (\omega_1, \omega_2 ,z) \sim &- \int{d j_1 d j  \over (2 \pi i)^2}
  {z^{-2-j }  \over (1+j)}  \int_0^\infty ds_1 ds_2 \,  f(s_1,s_2|\om_1,\om_2)
 \\\notag
&\times 
 (s_1(s_2+\omega_2))_+^{-j_1-2} (s_2(s_1+\omega_1))_+^{-j+j_1-2}\,  
\\[2mm]
&\times 
 (1-s_1-s_2+ z s_1 s_2)^{j } \theta(1-s_1-s_2+ z s_1 s_2) \,,
\end{align}
where we changed the integration variable $j=j_1+j_2$ and introduced  the function 
\begin{align}
 f(s_1,s_2|\om_1,\om_2)=  {{\omega_1^2 \omega_2^2\over 72} \left(s_1^2 \omega_2^2+s_2^2 \omega
   _1^2+4 s_2 s_1 \omega_1 \omega_2+6
   s_2 s_1^2 \omega_2+6 s_2^2 s_1 \omega_1+6 s_2^2 s_1^2\right)} \,.
\end{align}
The integration contour in the Mellin integral in \re{ints} satisfies ${\rm Re}(j_1)<0$ and ${\rm Re} \, j >-1$. The two factors in the second line of \re{ints} are defined with the `$+$' prescription, see \re{plus}. This is needed to make the integral over small $s_1$ and $s_2$ in \re{ints} well defined.

Closing the integration contour over $j$ in \re{ints} to the left half-plane, we find that the leading contribution for $z\to 0$ comes from the pole at $j =-1$. An additional pole arises after the integration over small $w\equiv 1-s_1-s_2+ z s_1 s_2$. Applying the identify
\begin{align}\label{plus}
{1\over w^{-j}} = -{1\over 1+j } \delta(w) +\left[1\over w\right]_+ + O(1+j )
\end{align}
and taking the residue at the double pole, we obtain from \re{ints}
\begin{align}\notag\label{ints2}
\text{EEC}^{(1)}_{++} (\omega_1, \omega_2 ,z) \stackrel{z\to 0}{\sim} & -{\log z\over z} \int{d j_1 \over 2 \pi i}
   \int_0^\infty ds_1 ds_2 \,  f(s_1,s_2|\om_1,\om_2)
 \\
&\times 
 (s_1(s_2+\omega_2))_+^{-j_1-2} (s_2(s_1+\omega_1))_+^{j_1-1}\,  
\delta(1-s_1-s_2+ z s_1 s_2) \,.
\end{align}
As in the previous case, the integration over small $s_1$ and $s_2$ yields additional poles at $j_1=-1$ and $j_1=0$, respectively. Their contribution to the right-hand side of \re{ints2} is
\begin{align}
- {\log z\over z}\times {\omega_1^2\omega_2^2\over 72}   \lr{ {1 \over 1+\omega_1}+{1 \over 1+\omega_2}} \,.
\end{align}
The remaining contribution from nonzero $s_i$ looks as
\begin{align}\notag\label{for}
&-24{\log z\over z} \omega_2^4  \int_0^\infty ds_1 ds_2 \, {(s_1+\omega_1)^2\over (s_2+\omega_2)^2}  \delta(s_1\omega_2-s_2\omega_1)\delta(1-s_1-s_2+z s_1 s_2)
\\\notag
&=-{\log z\over z}  {\omega_1^2\omega_2^2\over 12} \int_0^\infty ds_1 ds_2 \,  \delta(s_1\omega_2-s_2\omega_1)\delta(1-s_1-s_2+z s_1 s_2)
\\
&=-{\log z\over z}  {\omega_1^2\omega_2^2\over 12} {1\over \sqrt{(\omega_1+\omega_2)^2 -4z \omega_1\omega_2}}\,,
\end{align}
where $\delta(s_1\omega_2-s_2\omega_1)$ results from the integration over $j_1$ in \re{ints2}. Putting together the last two relations we recover \re{sing}.
Notice that the integral \re{for} vanishes for $\omega_1$ and $\omega_2$ of opposite signs. This explains why the expression \re{EECpm-weak} for $\text{EEC}_{+-}$ does not contains terms like \re{for}.

\section{Time profile in the detector}
\label{sec:timeprofiledet}

In this appendix we  address the question 
what is the longitudinal profile observed by a calorimeter in the physical state 
 \be
 | \Psi \rangle \simeq \int d^4 x \ e^{i P \cdot x} e^{- {t^2+\vec x^2 \over 4 \sigma^2}} J(x) | \Omega \rangle \,. 
 \ee
 We only consider here the simplest case of the one-point energy correlation. Consider the energy calorimeter centered around the retarded time $t_0$ and working in the interval $t_0 - T \leq t \leq t_0+T$
  \be
 \cE_{t_0, T}(n) = \int_{-T}^T d u \ T_{uu} (u-t_0, \vec n) = \int_{-\infty}^\infty {d \hat \omega \over \pi} \ {\sin \hat \omega T  \over \hat \omega } e^{i \hat \omega t_0} \cE(\ho, n) \,. 
 \ee
 We can easily relate the one-point function of $ \cE_{t_0, T}(n)$ to the one-point function studied in the bulk of the paper
  \be
 \label{eq:observable}
 \langle \Psi | \cE_{t_0, T}(n) | \Psi \rangle &=\int d^4q \int {d \hat \omega \over \pi}   f_\sigma(q + \hat \omega n) f_\sigma(q) {\sin \hat \omega T  \over \hat \omega } e^{i \hat \omega t_0}  \langle \langle J(q + \hat \omega n) \cE(\ho, n)  J(q) \rangle \rangle ,
 \ee
where it is natural to choose the wave function to be centered around a given momentum $P^\mu$ with some spread $1 / \sigma$,
\be
 f_\sigma(q) &\sim e^{- \sigma^2 \left[ (q^0-P^0)^2 + (\vec q - \vec P)^2 \right]} \,.
\ee 
In the coordinate space such a wave function is localized around the origin with a spread  $\sigma$ both in space and time.

\begin{figure}[t]
\centering
\includegraphics[scale=.55]{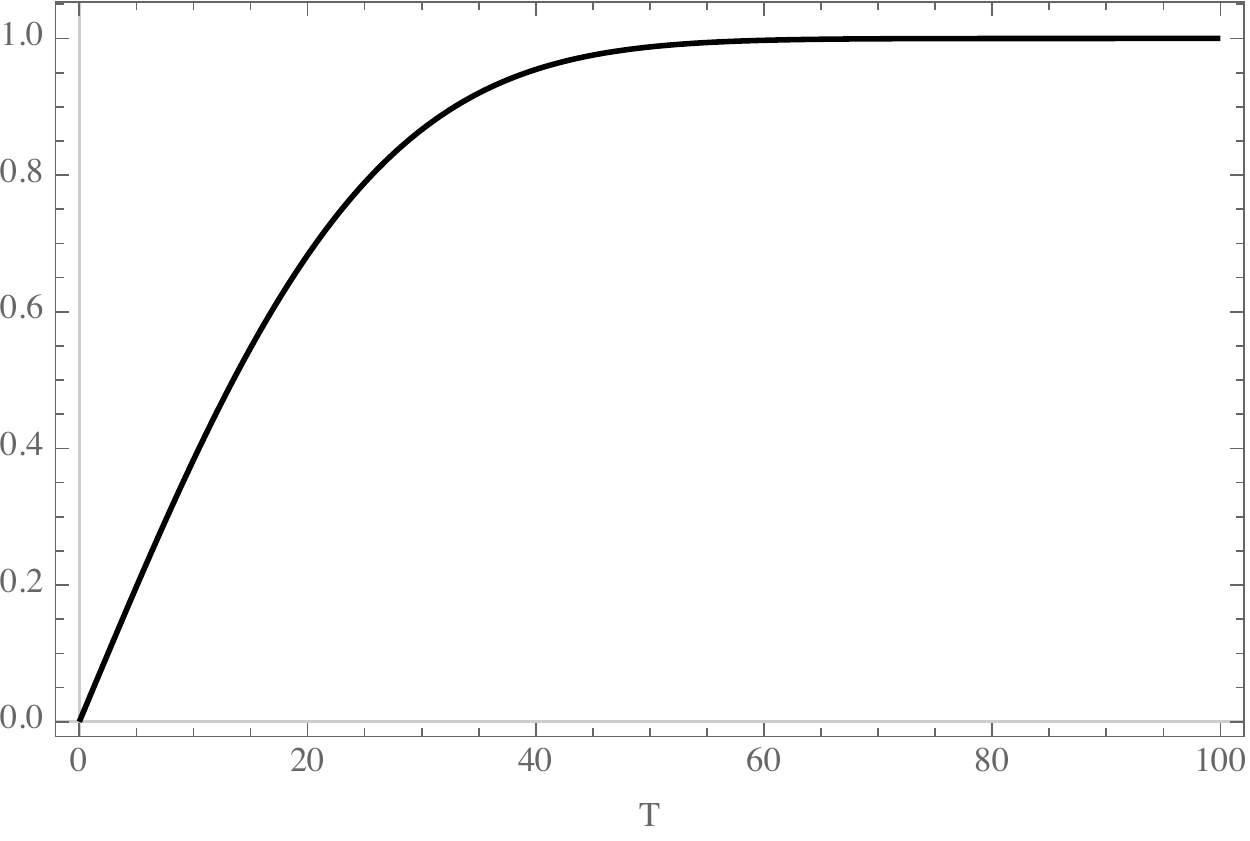} \qquad \includegraphics[scale=.567]{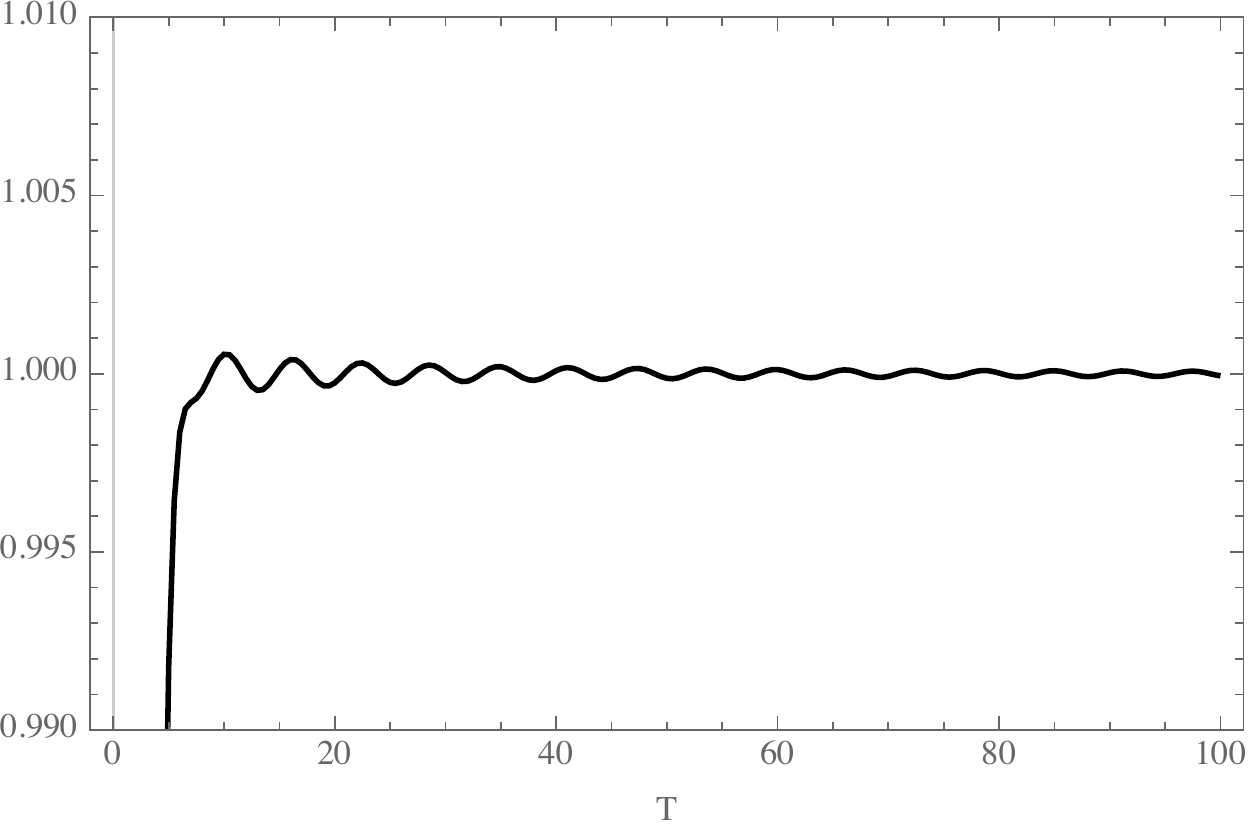} 
\caption{ Time dependence of the energy measurement for $P^\mu = (P^0, \vec 0)$ with $P^0=1$. We set $\sigma =10$ (the left panel) and
 for $\sigma =1$ (the right panel). In the latter case we observe small quantum oscillations around the asymptotic value. 
 }
\label{fig:timedep}
\end{figure}

Let us choose the source to have the scaling dimension $\Delta_\phi = 2$  for simplicity and use  \eqref{f+} and \eqref{f-}. We go to the center-of-mass frame and set $P^\mu = (P^0, \vec 0)$ with $P^0=1$.
We also consider a narrow wave packet so that $\sigma P^0 \gg 1$. In this limit, the integral over $q$ is localized at $q=P$. Finally, we set $t_0 =0$ since its effect is simply to shift the origin of the plots considered below. In this regime the result is captured by the following integral
\be
\label{eq:timedependent}
  \int_{-1}^\infty  {d \hat \omega \over \pi} e^{- 2 \sigma^2 \hat \omega^2} {\sin \hat \omega T  \over \hat \omega} \left( 1 + \hat \omega + {1 \over 6}  \hat \omega^2 \right) ,
\ee
where we used the fact that $ f_\sigma(P+\hat \omega n) =e^{- 2\sigma^2 \hat \omega^2} $. The lower limit on the integral comes from the condition that the time component of $P+\hat \omega n$ is non-negative, which is necessary for $  \langle \langle J(P + \hat \omega n) \cE(\ho, n)  J(P) \rangle \rangle$ to be non-zero.

We plot \eqref{eq:timedependent} in Figure~\ref{fig:timedep} for different values of $\sigma$. The response of the calorimeter starts at $0$ and then goes to its asymptotic value $1$ on time scales set by the size of the spread of the wave function in the coordinate space $\sigma$.
\noindent
As $\sigma$ decreases  two effects occur: first, the asymptotic value of the energy detector response is reached faster; second, quantum oscillations  develop around the asymptotic value $1$.\footnote{We say quantum because in a classical field theory, where the energy flux is locally positive, the one-point function is monotonically increasing as we increase $T$.}

\bibliographystyle{JHEP} 

\bibliography{papers}   

\providecommand{\href}[2]{#2}\begingroup\raggedright\begin{thebibliography}{10}

\bibitem{Hofman:2008ar}
D.~M. Hofman and J.~Maldacena, \emph{{Conformal collider physics: Energy and
  charge correlations}},
  \href{https://doi.org/10.1088/1126-6708/2008/05/012}{\emph{JHEP} {\bfseries
  05} (2008) 012} [\href{https://arxiv.org/abs/0803.1467}{{\ttfamily
  0803.1467}}].

\bibitem{Sveshnikov:1995vi}
N.~A. Sveshnikov and F.~V. Tkachov, \emph{{Jets and quantum field theory}},
  \href{https://doi.org/10.1016/0370-2693(96)00558-8}{\emph{Phys. Lett. B}
  {\bfseries 382} (1996) 403}
  [\href{https://arxiv.org/abs/hep-ph/9512370}{{\ttfamily hep-ph/9512370}}].

\bibitem{Korchemsky:1997sy}
G.~P. Korchemsky, G.~Oderda and G.~F. Sterman, \emph{{Power corrections and
  nonlocal operators}}, \href{https://doi.org/10.1063/1.53732}{\emph{AIP Conf.
  Proc.} {\bfseries 407} (1997) 988}
  [\href{https://arxiv.org/abs/hep-ph/9708346}{{\ttfamily hep-ph/9708346}}].

\bibitem{Korchemsky:1999kt}
G.~P. Korchemsky and G.~F. Sterman, \emph{{Power corrections to event shapes
  and factorization}},
  \href{https://doi.org/10.1016/S0550-3213(99)00308-9}{\emph{Nucl. Phys. B}
  {\bfseries 555} (1999) 335}
  [\href{https://arxiv.org/abs/hep-ph/9902341}{{\ttfamily hep-ph/9902341}}].

\bibitem{Hartman:2016lgu}
T.~Hartman, S.~Kundu and A.~Tajdini, \emph{{Averaged Null Energy Condition from
  Causality}}, \href{https://doi.org/10.1007/JHEP07(2017)066}{\emph{JHEP}
  {\bfseries 07} (2017) 066}
  [\href{https://arxiv.org/abs/1610.05308}{{\ttfamily 1610.05308}}].

\bibitem{Faulkner:2016mzt}
T.~Faulkner, R.~G. Leigh, O.~Parrikar and H.~Wang, \emph{{Modular Hamiltonians
  for Deformed Half-Spaces and the Averaged Null Energy Condition}},
  \href{https://doi.org/10.1007/JHEP09(2016)038}{\emph{JHEP} {\bfseries 09}
  (2016) 038} [\href{https://arxiv.org/abs/1605.08072}{{\ttfamily
  1605.08072}}].

\bibitem{Hatta:2012kn}
Y.~Hatta, E.~Iancu, A.~H. Mueller and D.~N. Triantafyllopoulos, \emph{{Jet
  evolution from weak to strong coupling}},
  \href{https://doi.org/10.1007/JHEP12(2012)114}{\emph{JHEP} {\bfseries 12}
  (2012) 114} [\href{https://arxiv.org/abs/1210.1534}{{\ttfamily 1210.1534}}].

\bibitem{Belin:2020lsr}
A.~Belin, D.~M. Hofman, G.~Mathys and M.~T. Walters, \emph{{On the stress
  tensor light-ray operator algebra}},
  \href{https://doi.org/10.1007/JHEP05(2021)033}{\emph{JHEP} {\bfseries 05}
  (2021) 033} [\href{https://arxiv.org/abs/2011.13862}{{\ttfamily
  2011.13862}}].

\bibitem{Martin:2020jlu}
A.~Martin and J.-M. Richard, \emph{{New result on phase shift analysis}},
  \href{https://doi.org/10.1103/PhysRevD.101.094014}{\emph{Phys. Rev. D}
  {\bfseries 101} (2020) 094014}
  [\href{https://arxiv.org/abs/2004.11156}{{\ttfamily 2004.11156}}].

\bibitem{Kravchuk:2018htv}
P.~Kravchuk and D.~Simmons-Duffin, \emph{{Light-ray operators in conformal
  field theory}}, \href{https://doi.org/10.1007/JHEP11(2018)102}{\emph{JHEP}
  {\bfseries 11} (2018) 102}
  [\href{https://arxiv.org/abs/1805.00098}{{\ttfamily 1805.00098}}].

\bibitem{Chen:2020vvp}
H.~Chen, I.~Moult, X.~Zhang and H.~X. Zhu, \emph{{Rethinking jets with energy
  correlators: Tracks, resummation, and analytic continuation}},
  \href{https://doi.org/10.1103/PhysRevD.102.054012}{\emph{Phys. Rev. D}
  {\bfseries 102} (2020) 054012}
  [\href{https://arxiv.org/abs/2004.11381}{{\ttfamily 2004.11381}}].

\bibitem{Dixon:2019uzg}
L.~J. Dixon, I.~Moult and H.~X. Zhu, \emph{{Collinear limit of the
  energy-energy correlator}},
  \href{https://doi.org/10.1103/PhysRevD.100.014009}{\emph{Phys. Rev. D}
  {\bfseries 100} (2019) 014009}
  [\href{https://arxiv.org/abs/1905.01310}{{\ttfamily 1905.01310}}].

\bibitem{Chen:2021gdk}
H.~Chen, I.~Moult and H.~X. Zhu, \emph{{Spinning Gluons from the QCD Light-Ray
  OPE}},  \href{https://arxiv.org/abs/2104.00009}{{\ttfamily 2104.00009}}.

\bibitem{Chen:2020adz}
H.~Chen, I.~Moult and H.~X. Zhu, \emph{{Quantum Interference in Jet
  Substructure from Spinning Gluons}},
  \href{https://doi.org/10.1103/PhysRevLett.126.112003}{\emph{Phys. Rev. Lett.}
  {\bfseries 126} (2021) 112003}
  [\href{https://arxiv.org/abs/2011.02492}{{\ttfamily 2011.02492}}].

\bibitem{Kologlu:2019bco}
M.~Kologlu, P.~Kravchuk, D.~Simmons-Duffin and A.~Zhiboedov, \emph{{Shocks,
  Superconvergence, and a Stringy Equivalence Principle}},
  \href{https://doi.org/10.1007/JHEP11(2020)096}{\emph{JHEP} {\bfseries 11}
  (2020) 096} [\href{https://arxiv.org/abs/1904.05905}{{\ttfamily
  1904.05905}}].

\bibitem{Belitsky:2013xxa}
A.~V. Belitsky, S.~Hohenegger, G.~P. Korchemsky, E.~Sokatchev and A.~Zhiboedov,
  \emph{{From correlation functions to event shapes}},
  \href{https://doi.org/10.1016/j.nuclphysb.2014.04.020}{\emph{Nucl. Phys. B}
  {\bfseries 884} (2014) 305}
  [\href{https://arxiv.org/abs/1309.0769}{{\ttfamily 1309.0769}}].

\bibitem{Belitsky:2013bja}
A.~V. Belitsky, S.~Hohenegger, G.~P. Korchemsky, E.~Sokatchev and A.~Zhiboedov,
  \emph{{Event shapes in $\mathcal{N} = 4$ super-Yang-Mills theory}},
  \href{https://doi.org/10.1016/j.nuclphysb.2014.04.019}{\emph{Nucl. Phys.}
  {\bfseries B884} (2014) 206}
  [\href{https://arxiv.org/abs/1309.1424}{{\ttfamily 1309.1424}}].

\bibitem{Belitsky:2013ofa}
A.~V. Belitsky, S.~Hohenegger, G.~P. Korchemsky, E.~Sokatchev and A.~Zhiboedov,
  \emph{{Energy-Energy Correlations in N=4 Supersymmetric Yang-Mills Theory}},
  \href{https://doi.org/10.1103/PhysRevLett.112.071601}{\emph{Phys. Rev. Lett.}
  {\bfseries 112} (2014) 071601}
  [\href{https://arxiv.org/abs/1311.6800}{{\ttfamily 1311.6800}}].

\bibitem{Henn:2019gkr}
J.~M. Henn, E.~Sokatchev, K.~Yan and A.~Zhiboedov, \emph{{Energy-energy
  correlation in $N$=4 super Yang-Mills theory at next-to-next-to-leading
  order}}, \href{https://doi.org/10.1103/PhysRevD.100.036010}{\emph{Phys. Rev.
  D} {\bfseries 100} (2019) 036010}
  [\href{https://arxiv.org/abs/1903.05314}{{\ttfamily 1903.05314}}].

\bibitem{Belitsky:2014zha}
A.~V. Belitsky, S.~Hohenegger, G.~P. Korchemsky and E.~Sokatchev, \emph{{N=4
  superconformal Ward identities for correlation functions}},
  \href{https://doi.org/10.1016/j.nuclphysb.2016.01.008}{\emph{Nucl. Phys. B}
  {\bfseries 904} (2016) 176}
  [\href{https://arxiv.org/abs/1409.2502}{{\ttfamily 1409.2502}}].

\bibitem{Korchemsky:2015ssa}
G.~P. Korchemsky and E.~Sokatchev, \emph{{Four-point correlation function of
  stress-energy tensors in $ \mathcal{N}=4 $ superconformal theories}},
  \href{https://doi.org/10.1007/JHEP12(2015)133}{\emph{JHEP} {\bfseries 12}
  (2015) 133} [\href{https://arxiv.org/abs/1504.07904}{{\ttfamily
  1504.07904}}].

\bibitem{contact}
G.~P. Korchemsky and A.~Zhiboedov, \emph{{On the light-ray algebra in conformal
  field theories}},  \href{https://arxiv.org/abs/2109.13269}{{\ttfamily
  2109.13269}}.

\bibitem{Alday:2017xua}
L.~F. Alday and A.~Bissi, \emph{{Loop Corrections to Supergravity on $AdS_5
  \times S^5$}},
  \href{https://doi.org/10.1103/PhysRevLett.119.171601}{\emph{Phys. Rev. Lett.}
  {\bfseries 119} (2017) 171601}
  [\href{https://arxiv.org/abs/1706.02388}{{\ttfamily 1706.02388}}].

\bibitem{Aprile:2017bgs}
F.~Aprile, J.~M. Drummond, P.~Heslop and H.~Paul, \emph{{Quantum Gravity from
  Conformal Field Theory}},
  \href{https://doi.org/10.1007/JHEP01(2018)035}{\emph{JHEP} {\bfseries 01}
  (2018) 035} [\href{https://arxiv.org/abs/1706.02822}{{\ttfamily
  1706.02822}}].

\bibitem{Alday:2017vkk}
L.~F. Alday and S.~Caron-Huot, \emph{{Gravitational S-matrix from CFT
  dispersion relations}},
  \href{https://doi.org/10.1007/JHEP12(2018)017}{\emph{JHEP} {\bfseries 12}
  (2018) 017} [\href{https://arxiv.org/abs/1711.02031}{{\ttfamily
  1711.02031}}].

\bibitem{Alday:2018kkw}
L.~F. Alday, \emph{{On genus-one string amplitudes on $AdS_5 \times S^5$}},
  \href{https://doi.org/10.1007/JHEP04(2021)005}{\emph{JHEP} {\bfseries 04}
  (2021) 005} [\href{https://arxiv.org/abs/1812.11783}{{\ttfamily
  1812.11783}}].

\bibitem{Penedones:2019tng}
J.~Penedones, J.~A. Silva and A.~Zhiboedov, \emph{{Nonperturbative Mellin
  Amplitudes: Existence, Properties, Applications}},
  \href{https://doi.org/10.1007/JHEP08(2020)031}{\emph{JHEP} {\bfseries 08}
  (2020) 031} [\href{https://arxiv.org/abs/1912.11100}{{\ttfamily
  1912.11100}}].

\bibitem{Qualls:2015qjb}
J.~D. Qualls, \emph{{Lectures on Conformal Field Theory}},
  \href{https://arxiv.org/abs/1511.04074}{{\ttfamily 1511.04074}}.

\bibitem{Witten:2018zxz}
E.~Witten, \emph{{APS Medal for Exceptional Achievement in Research: Invited
  article on entanglement properties of quantum field theory}},
  \href{https://doi.org/10.1103/RevModPhys.90.045003}{\emph{Rev. Mod. Phys.}
  {\bfseries 90} (2018) 045003}
  [\href{https://arxiv.org/abs/1803.04993}{{\ttfamily 1803.04993}}].

\bibitem{Korchemsky:2019nzm}
G.~P. Korchemsky, \emph{{Energy correlations in the end-point region}},
  \href{https://doi.org/10.1007/JHEP01(2020)008}{\emph{JHEP} {\bfseries 01}
  (2020) 008} [\href{https://arxiv.org/abs/1905.01444}{{\ttfamily
  1905.01444}}].

\bibitem{Kologlu:2019mfz}
M.~Kologlu, P.~Kravchuk, D.~Simmons-Duffin and A.~Zhiboedov, \emph{{The
  light-ray OPE and conformal colliders}},
  \href{https://doi.org/10.1007/JHEP01(2021)128}{\emph{JHEP} {\bfseries 01}
  (2021) 128} [\href{https://arxiv.org/abs/1905.01311}{{\ttfamily
  1905.01311}}].

\bibitem{Kravchuk:2020scc}
P.~Kravchuk, J.~Qiao and S.~Rychkov, \emph{{Distributions in CFT. Part I.
  Cross-ratio space}},
  \href{https://doi.org/10.1007/JHEP05(2020)137}{\emph{JHEP} {\bfseries 05}
  (2020) 137} [\href{https://arxiv.org/abs/2001.08778}{{\ttfamily
  2001.08778}}].

\bibitem{Braun:2003rp}
V.~M. Braun, G.~P. Korchemsky and D.~M\"uller, \emph{{The Uses of conformal
  symmetry in QCD}},
  \href{https://doi.org/10.1016/S0146-6410(03)90004-4}{\emph{Prog. Part. Nucl.
  Phys.} {\bfseries 51} (2003) 311}
  [\href{https://arxiv.org/abs/hep-ph/0306057}{{\ttfamily hep-ph/0306057}}].

\bibitem{Gary:2009ae}
M.~Gary, S.~B. Giddings and J.~Penedones, \emph{{Local bulk S-matrix elements
  and CFT singularities}},
  \href{https://doi.org/10.1103/PhysRevD.80.085005}{\emph{Phys. Rev. D}
  {\bfseries 80} (2009) 085005}
  [\href{https://arxiv.org/abs/0903.4437}{{\ttfamily 0903.4437}}].

\bibitem{Maldacena:2015iua}
J.~Maldacena, D.~Simmons-Duffin and A.~Zhiboedov, \emph{{Looking for a bulk
  point}}, \href{https://doi.org/10.1007/JHEP01(2017)013}{\emph{JHEP}
  {\bfseries 01} (2017) 013}
  [\href{https://arxiv.org/abs/1509.03612}{{\ttfamily 1509.03612}}].

\bibitem{Penedones:2010ue}
J.~Penedones, \emph{{Writing CFT correlation functions as AdS scattering
  amplitudes}}, \href{https://doi.org/10.1007/JHEP03(2011)025}{\emph{JHEP}
  {\bfseries 03} (2011) 025} [\href{https://arxiv.org/abs/1011.1485}{{\ttfamily
  1011.1485}}].

\bibitem{Aprile:2020luw}
F.~Aprile and P.~Vieira, \emph{{Large $p$ explorations. From SUGRA to big
  STRINGS in Mellin space}},
  \href{https://doi.org/10.1007/JHEP12(2020)206}{\emph{JHEP} {\bfseries 12}
  (2020) 206} [\href{https://arxiv.org/abs/2007.09176}{{\ttfamily
  2007.09176}}].

\bibitem{Chester:2019pvm}
S.~M. Chester, \emph{{Genus-2 holographic correlator on AdS$_{5}$ $times$
  S$^{5}$ from localization}},
  \href{https://doi.org/10.1007/JHEP04(2020)193}{\emph{JHEP} {\bfseries 04}
  (2020) 193} [\href{https://arxiv.org/abs/1908.05247}{{\ttfamily
  1908.05247}}].

\bibitem{Carmi:2019cub}
D.~Carmi and S.~Caron-Huot, \emph{{A Conformal Dispersion Relation:
  Correlations from Absorption}},
  \href{https://doi.org/10.1007/JHEP09(2020)009}{\emph{JHEP} {\bfseries 09}
  (2020) 009} [\href{https://arxiv.org/abs/1910.12123}{{\ttfamily
  1910.12123}}].

\bibitem{Mazac:2019shk}
D.~Maz\'a\v{c}, L.~Rastelli and X.~Zhou, \emph{{A Basis of Analytic Functionals
  for CFTs in General Dimension}},
  \href{https://arxiv.org/abs/1910.12855}{{\ttfamily 1910.12855}}.

\bibitem{Caron-Huot:2020adz}
S.~Caron-Huot, D.~Mazac, L.~Rastelli and D.~Simmons-Duffin, \emph{{Dispersive
  CFT Sum Rules}}, \href{https://doi.org/10.1007/JHEP05(2021)243}{\emph{JHEP}
  {\bfseries 05} (2021) 243}
  [\href{https://arxiv.org/abs/2008.04931}{{\ttfamily 2008.04931}}].

\bibitem{Mack:2009mi}
G.~Mack, \emph{{D-independent representation of Conformal Field Theories in D
  dimensions via transformation to auxiliary Dual Resonance Models. Scalar
  amplitudes}},  \href{https://arxiv.org/abs/0907.2407}{{\ttfamily 0907.2407}}.

\bibitem{Mack:2009gy}
G.~Mack, \emph{{D-dimensional Conformal Field Theories with anomalous
  dimensions as Dual Resonance Models}}, {\emph{Bulg. J. Phys.} {\bfseries 36}
  (2009) 214} [\href{https://arxiv.org/abs/0909.1024}{{\ttfamily 0909.1024}}].

\bibitem{Costa:2012cb}
M.~S. Costa, V.~Goncalves and J.~Penedones, \emph{{Conformal Regge theory}},
  \href{https://doi.org/10.1007/JHEP12(2012)091}{\emph{JHEP} {\bfseries 12}
  (2012) 091} [\href{https://arxiv.org/abs/1209.4355}{{\ttfamily 1209.4355}}].

\bibitem{Gopakumar:2021dvg}
R.~Gopakumar, A.~Sinha and A.~Zahed, \emph{{Crossing Symmetric Dispersion
  Relations for Mellin Amplitudes}},
  \href{https://doi.org/10.1103/PhysRevLett.126.211602}{\emph{Phys. Rev. Lett.}
  {\bfseries 126} (2021) 211602}
  [\href{https://arxiv.org/abs/2101.09017}{{\ttfamily 2101.09017}}].

\bibitem{Fitzpatrick:2015qma}
A.~L. Fitzpatrick, J.~Kaplan, M.~T. Walters and J.~Wang, \emph{{Eikonalization
  of Conformal Blocks}},
  \href{https://doi.org/10.1007/JHEP09(2015)019}{\emph{JHEP} {\bfseries 09}
  (2015) 019} [\href{https://arxiv.org/abs/1504.01737}{{\ttfamily
  1504.01737}}].

\bibitem{Bissi:2020woe}
A.~Bissi, G.~Fardelli and A.~Georgoudis, \emph{{All loop structures in
  Supergravity Amplitudes on $AdS_5 \times S^5$ from CFT}},
  \href{https://arxiv.org/abs/2010.12557}{{\ttfamily 2010.12557}}.

\bibitem{Amati:1990xe}
D.~Amati, M.~Ciafaloni and G.~Veneziano, \emph{{Higher Order Gravitational
  Deflection and Soft Bremsstrahlung in Planckian Energy Superstring
  Collisions}}, \href{https://doi.org/10.1016/0550-3213(90)90375-N}{\emph{Nucl.
  Phys. B} {\bfseries 347} (1990) 550}.

\bibitem{Cornalba:2007zb}
L.~Cornalba, M.~S. Costa and J.~Penedones, \emph{{Eikonal approximation in
  AdS/CFT: Resumming the gravitational loop expansion}},
  \href{https://doi.org/10.1088/1126-6708/2007/09/037}{\emph{JHEP} {\bfseries
  09} (2007) 037} [\href{https://arxiv.org/abs/0707.0120}{{\ttfamily
  0707.0120}}].

\bibitem{Goncalves:2014ffa}
V.~Gon\c{c}alves, \emph{{Four point function of $\mathcal{N}=4$ stress-tensor
  multiplet at strong coupling}},
  \href{https://doi.org/10.1007/JHEP04(2015)150}{\emph{JHEP} {\bfseries 04}
  (2015) 150} [\href{https://arxiv.org/abs/1411.1675}{{\ttfamily 1411.1675}}].

\bibitem{Binder:2019jwn}
D.~J. Binder, S.~M. Chester, S.~S. Pufu and Y.~Wang, \emph{{$ \mathcal{N} $ = 4
  Super-Yang-Mills correlators at strong coupling from string theory and
  localization}}, \href{https://doi.org/10.1007/JHEP12(2019)119}{\emph{JHEP}
  {\bfseries 12} (2019) 119}
  [\href{https://arxiv.org/abs/1902.06263}{{\ttfamily 1902.06263}}].

\bibitem{Gross:1987ar}
D.~J. Gross and P.~F. Mende, \emph{{String Theory Beyond the Planck Scale}},
  \href{https://doi.org/10.1016/0550-3213(88)90390-2}{\emph{Nucl. Phys. B}
  {\bfseries 303} (1988) 407}.

\bibitem{Cornalba:2006xk}
L.~Cornalba, M.~S. Costa, J.~Penedones and R.~Schiappa, \emph{{Eikonal
  Approximation in AdS/CFT: From Shock Waves to Four-Point Functions}},
  \href{https://doi.org/10.1088/1126-6708/2007/08/019}{\emph{JHEP} {\bfseries
  08} (2007) 019} [\href{https://arxiv.org/abs/hep-th/0611122}{{\ttfamily
  hep-th/0611122}}].

\bibitem{Cornalba:2006xm}
L.~Cornalba, M.~S. Costa, J.~Penedones and R.~Schiappa, \emph{{Eikonal
  Approximation in AdS/CFT: Conformal Partial Waves and Finite N Four-Point
  Functions}},
  \href{https://doi.org/10.1016/j.nuclphysb.2007.01.007}{\emph{Nucl. Phys. B}
  {\bfseries 767} (2007) 327}
  [\href{https://arxiv.org/abs/hep-th/0611123}{{\ttfamily hep-th/0611123}}].

\bibitem{DiVecchia:2019myk}
P.~Di~Vecchia, A.~Luna, S.~G. Naculich, R.~Russo, G.~Veneziano and C.~D. White,
  \emph{{A tale of two exponentiations in ${\cal N}=8$ supergravity}},
  \href{https://doi.org/10.1016/j.physletb.2019.134927}{\emph{Phys. Lett. B}
  {\bfseries 798} (2019) 134927}
  [\href{https://arxiv.org/abs/1908.05603}{{\ttfamily 1908.05603}}].

\bibitem{DiVecchia:2019kta}
P.~Di~Vecchia, S.~G. Naculich, R.~Russo, G.~Veneziano and C.~D. White, \emph{{A
  tale of two exponentiations in $ \mathcal{N} $ = 8 supergravity at subleading
  level}}, \href{https://doi.org/10.1007/JHEP03(2020)173}{\emph{JHEP}
  {\bfseries 03} (2020) 173}
  [\href{https://arxiv.org/abs/1911.11716}{{\ttfamily 1911.11716}}].

\bibitem{Arkani-Hamed:2015bza}
N.~Arkani-Hamed and J.~Maldacena, \emph{{Cosmological Collider Physics}},
  \href{https://arxiv.org/abs/1503.08043}{{\ttfamily 1503.08043}}.

\bibitem{Chang:2020qpj}
C.-H. Chang, M.~Kologlu, P.~Kravchuk, D.~Simmons-Duffin and A.~Zhiboedov,
  \emph{{Transverse spin in the light-ray OPE}},
  \href{https://arxiv.org/abs/2010.04726}{{\ttfamily 2010.04726}}.

\bibitem{Besken:2020snx}
M.~Be\c{s}ken, J.~De~Boer and G.~Mathys, \emph{{On Local and Integrated
  Stress-Tensor Commutators}},
  \href{https://arxiv.org/abs/2012.15724}{{\ttfamily 2012.15724}}.

\bibitem{Bousso:2015mna}
R.~Bousso, Z.~Fisher, S.~Leichenauer and A.~C. Wall, \emph{{Quantum focusing
  conjecture}}, \href{https://doi.org/10.1103/PhysRevD.93.064044}{\emph{Phys.
  Rev. D} {\bfseries 93} (2016) 064044}
  [\href{https://arxiv.org/abs/1506.02669}{{\ttfamily 1506.02669}}].

\bibitem{Bousso:2015wca}
R.~Bousso, Z.~Fisher, J.~Koeller, S.~Leichenauer and A.~C. Wall, \emph{{Proof
  of the Quantum Null Energy Condition}},
  \href{https://doi.org/10.1103/PhysRevD.93.024017}{\emph{Phys. Rev. D}
  {\bfseries 93} (2016) 024017}
  [\href{https://arxiv.org/abs/1509.02542}{{\ttfamily 1509.02542}}].

\bibitem{Balakrishnan:2017bjg}
S.~Balakrishnan, T.~Faulkner, Z.~U. Khandker and H.~Wang, \emph{{A General
  Proof of the Quantum Null Energy Condition}},
  \href{https://doi.org/10.1007/JHEP09(2019)020}{\emph{JHEP} {\bfseries 09}
  (2019) 020} [\href{https://arxiv.org/abs/1706.09432}{{\ttfamily
  1706.09432}}].

\bibitem{Chen:2019bpb}
H.~Chen, M.-X. Luo, I.~Moult, T.-Z. Yang, X.~Zhang and H.~X. Zhu, \emph{{Three
  point energy correlators in the collinear limit: symmetries, dualities and
  analytic results}},
  \href{https://doi.org/10.1007/JHEP08(2020)028}{\emph{JHEP} {\bfseries 08}
  (2020) 028} [\href{https://arxiv.org/abs/1912.11050}{{\ttfamily
  1912.11050}}].

\bibitem{Bork:2014eqa}
L.~V. Bork, \emph{{On form factors in $ \mathcal{N}=4 $ SYM theory and
  polytopes}}, \href{https://doi.org/10.1007/JHEP12(2014)111}{\emph{JHEP}
  {\bfseries 12} (2014) 111} [\href{https://arxiv.org/abs/1407.5568}{{\ttfamily
  1407.5568}}].

\bibitem{Bianchi:2018peu}
L.~Bianchi, A.~Brandhuber, R.~Panerai and G.~Travaglini, \emph{{Form factor
  recursion relations at loop level}},
  \href{https://doi.org/10.1007/JHEP02(2019)182}{\emph{JHEP} {\bfseries 02}
  (2019) 182} [\href{https://arxiv.org/abs/1812.09001}{{\ttfamily
  1812.09001}}].

\bibitem{Chicherin:2015bza}
D.~Chicherin, R.~Doobary, B.~Eden, P.~Heslop, G.~P. Korchemsky and
  E.~Sokatchev, \emph{{Bootstrapping correlation functions in N=4 SYM}},
  \href{https://doi.org/10.1007/JHEP03(2016)031}{\emph{JHEP} {\bfseries 03}
  (2016) 031} [\href{https://arxiv.org/abs/1506.04983}{{\ttfamily
  1506.04983}}].

\bibitem{Chicherin:2014uca}
D.~Chicherin, R.~Doobary, B.~Eden, P.~Heslop, G.~P. Korchemsky, L.~Mason
  et~al., \emph{{Correlation functions of the chiral stress-tensor multiplet in
  $ \mathcal{N}=4 $ SYM}},
  \href{https://doi.org/10.1007/JHEP06(2015)198}{\emph{JHEP} {\bfseries 06}
  (2015) 198} [\href{https://arxiv.org/abs/1412.8718}{{\ttfamily 1412.8718}}].

\bibitem{Fleury:2019ydf}
T.~Fleury and R.~Pereira, \emph{{Non-planar data of $ \mathcal{N} $ = 4 SYM}},
  \href{https://doi.org/10.1007/JHEP03(2020)003}{\emph{JHEP} {\bfseries 03}
  (2020) 003} [\href{https://arxiv.org/abs/1910.09428}{{\ttfamily
  1910.09428}}].

\bibitem{Korchemsky:2018hnb}
G.~P. Korchemsky, \emph{{Exact scattering amplitudes in conformal fishnet
  theory}}, \href{https://doi.org/10.1007/JHEP08(2019)028}{\emph{JHEP}
  {\bfseries 08} (2019) 028}
  [\href{https://arxiv.org/abs/1812.06997}{{\ttfamily 1812.06997}}].

\bibitem{Gillioz:2020mdd}
M.~Gillioz, M.~Meineri and J.~Penedones, \emph{{A scattering amplitude in
  Conformal Field Theory}},
  \href{https://doi.org/10.1007/JHEP11(2020)139}{\emph{JHEP} {\bfseries 11}
  (2020) 139} [\href{https://arxiv.org/abs/2003.07361}{{\ttfamily
  2003.07361}}].

\bibitem{Gonzo:2020xza}
R.~Gonzo and A.~Pokraka, \emph{{Light-ray operators, detectors and
  gravitational event shapes}},
  \href{https://doi.org/10.1007/JHEP05(2021)015}{\emph{JHEP} {\bfseries 05}
  (2021) 015} [\href{https://arxiv.org/abs/2012.01406}{{\ttfamily
  2012.01406}}].

\bibitem{Costa:2011mg}
M.~S. Costa, J.~Penedones, D.~Poland and S.~Rychkov, \emph{{Spinning Conformal
  Correlators}}, \href{https://doi.org/10.1007/JHEP11(2011)071}{\emph{JHEP}
  {\bfseries 11} (2011) 071} [\href{https://arxiv.org/abs/1107.3554}{{\ttfamily
  1107.3554}}].

\bibitem{Dobrev:1977qv}
V.~K. Dobrev, G.~Mack, V.~B. Petkova, S.~G. Petrova and I.~T. Todorov,
  \emph{{Harmonic Analysis on the n-Dimensional Lorentz Group and Its
  Application to Conformal Quantum Field Theory}}, vol.~63. 1977,
  \href{https://doi.org/10.1007/BFb0009678}{10.1007/BFb0009678}.

\end{thebibliography}\endgroup
     
\end{document}